\documentclass[journal]{IEEEtran}

\usepackage{docmute}
\usepackage{ifpdf}
\usepackage{cite}
\usepackage{amsthm}
\usepackage{amssymb}
\usepackage{booktabs}
\usepackage{url}
\usepackage{color}
\usepackage{multirow}
\usepackage{graphics}
\usepackage{siunitx}
\usepackage{amsmath,graphicx}
\usepackage{amsfonts}
\usepackage{algorithm, algpseudocode}
\usepackage[geometry]{ifsym}
\usepackage{array}
\usepackage{tabularx}
\usepackage[normalem]{ulem}
\usepackage{xcolor}
\usepackage{afterpage}

\def\x{{\mathbf x}}
\def\y{{\mathbf y}}
\def\z{{\mathbf z}}

\def\argmin{\mathop{\rm argmin}\limits}

\def\tilHr{\widetilde{\mathbf{h}}_{\mathrm{r}}}
\def\tilHt{\widetilde{\mathbf{h}}_{\mathrm{t}}}
\def\tilH{\widetilde{\mathbf{h}}}

\def\hatHr{\widehat{\mathbf{h}}_{\mathrm{r}}}
\def\hatHt{\widehat{\mathbf{h}}_{\mathrm{t}}}
\def\Hr{\mathbf{h}_{\mathrm{r}}}

\def\Lr{\mathbf{l}_{\mathrm{r}}}
\def\Lt{\mathbf{l}_{\mathrm{t}}}

\def\nh {\mathbf{n}_{\mathrm{h}}}
\def\nl {\mathbf{n}_{\mathrm{l}}}
\def\sh {\mathbf{s}_{\mathrm{h}}}
\def\sl {\mathbf{s}_{\mathrm{l}}}
\def\th {\mathbf{t}_{\mathrm{h}}}
\def\tl {\mathbf{t}_{\mathrm{l}}}

\def\tilshr{\widetilde{\mathbf{s}}_{\mathrm{hr}}}
\def\tilslr{\widetilde{\mathbf{s}}_{\mathrm{lr}}}
\def\tilslt{\widetilde{\mathbf{s}}_{\mathrm{lt}}}
\def\tilthr{\widetilde{\mathbf{t}}_{\mathrm{hr}}}
\def\tiltlr{\widetilde{\mathbf{t}}_{\mathrm{lr}}}
\def\tiltlt{\widetilde{\mathbf{t}}_{\mathrm{lt}}}

\def\Nh{N_{\mathrm{h}}}
\def\Nl{N_{\mathrm{l}}}
\def\Wh{W_{\mathrm{h}}}

\def\Hh{H_{\mathrm{h}}}

\def\epsh{\varepsilon_{\mathrm{h}}}
\def\epsl{\varepsilon_{\mathrm{l}}}
\def\etah{\eta_{\mathrm{h}}}
\def\etal{\eta_{\mathrm{l}}}
\def\zetah{\zeta_{\mathrm{h}}}
\def\zetal{\zeta_{\mathrm{l}}}
\def\sigmah{\sigma_{\mathrm{h}}}
\def\sigmal{\sigma_{\mathrm{l}}}
\def\rh{r_{\mathrm{h}}}
\def\rl{r_{\mathrm{l}}}
\def\sh{ s_{\mathrm{h}}}
\def\sl{ s_{\mathrm{l}}}
\def\eh{e_{\mathrm{h}}}
\def\el{e_{\mathrm{l}}}

\def\onenorm#1{\| #1 \|_1}
\def\twonorm#1{\| #1 \|_2}
\def\qnorm#1{\| #1 \|_q}

\def\onetwonorm#1{\| #1 \|_{1,2}}

\def \R{\mathbb{R}}
\def \D{\mathbf{D}}
\def \W{\mathbf{W}}
\def \S{\mathbf{S}}
\def \B{\mathbf{B}}
\def \I{\mathbf{I}}

\def \M{\mathbf{M}}

\def \Dt{\mathbf{D}^{\top}}
\def \Wt{\mathbf{W}^{\top}}

\def \x{\mathbf{x}}
\def \y{\mathbf{y}}
\def \z{\mathbf{z}}
\def \z{\mathbf{z}}
\def \u{\mathbf{u}}

\def \prox{\mathrm{prox}}

\begin{document}
\bstctlcite{IEEEexample:BSTcontrol}
\title{Temporally-Similar Structure-Aware \\ Spatiotemporal Fusion of Satellite Images}

\author{Ryosuke~Isono~\IEEEmembership{Student~Member,~IEEE,}~Shunsuke~Ono,~\IEEEmembership{Member,~IEEE,}
\thanks{R. Isono is with the Department of Computer Science, Institute of Science Tokyo, Yokohama, 226-8503, Japan (e-mail: isono.r.1f44@m.isct.ac.jp).}
\thanks{S. Ono is with the Department of Computer Science, Institute of Science Tokyo, Yokohama, 226-8503, Japan (e-mail: ono@comp.isct.ac.jp).}
\thanks{This work was supported in part by JST FOREST under Grant JPMJFR232M, JST AdCORP under Grant JPMJKB2307, JST ACT-X under Grant JPMJAX24C1, and JST BOOST under Grant JPMJBS2430, and in part by JSPS KAKENHI under Grant 22H03610, 22H00512, 23H01415, 23K17461, 24K03119, 24K22291, 25H01296, and 25K03136.}}

\markboth{IEEE TRANSACTIONS ON GEOSCIENCE AND REMOTE SENSING}%
{Shell \MakeLowercase{\textit{et al.}}: Bare Demo of IEEEtran.cls for Journals}

\maketitle

\begin{abstract}
This paper proposes a spatiotemporal (ST) fusion framework robust against diverse noise for satellite images, named Temporally-Similar Structure-Aware ST fusion (TSSTF). ST fusion is a promising approach to address the trade-off between the spatial and temporal resolution of satellite images. In real-world scenarios, observed satellite images are severely degraded by noise due to measurement equipment and environmental conditions. Consequently, some recent studies have focused on enhancing the robustness of ST fusion methods against noise. However, existing noise-robust ST fusion approaches often fail to capture fine spatial structure, leading to oversmoothing and artifacts. To address this issue, TSSTF introduces two key mechanisms: Temporally-Guided Total Variation (TGTV) and Temporally-Guided Edge Constraint (TGEC). TGTV is a weighted total variation-based regularization that promotes spatial piecewise smoothness while preserving structural details, guided by a reference high spatial resolution image acquired on a nearby date. TGEC enforces consistency in edge locations between two temporally adjacent images, while allowing for spectral variations. We formulate the ST fusion task as a constrained optimization problem incorporating TGTV and TGEC, and develop an efficient algorithm based on a preconditioned primal-dual splitting method. Experimental results demonstrate that TSSTF performs comparably to state-of-the-art methods under noise-free conditions and outperforms them under noisy conditions. 
\end{abstract}


\begin{IEEEkeywords}
Spatiotemporal fusion, remote sensing, total variation, constrained optimization, primal-dual splitting method
\end{IEEEkeywords}

\IEEEpeerreviewmaketitle

\vspace{-2mm}
\section{Introduction}
\label{sec:intro}
\IEEEPARstart{T}{he} analysis of temporal image series is necessary and important in many remote sensing applications, such as vegetation/crop monitoring and estimation \cite{vegetation}, evapotranspiration estimation \cite{evapotranspiration}, atmosphere monitoring \cite{atmosphere}, land-cover/land-use change detection \cite{landuse}, surface dynamic mapping \cite{mapping}, ecosystem monitoring \cite{ecosystem}, soil water content analysis \cite{soil}, and detailed analysis of human-nature interactions \cite{human}. These applications require time series of high spatial resolution images to properly model the ground surface. In addition, time series of high temporal resolution images are also needed to capture the changes in the ground surface that occur over short periods of time. 

However, there is a trade-off between the temporal and spatial resolution of satellite sensors, and no single sensor can satisfy both requirements. For example, the Landsat sensors can acquire images with a high spatial resolution of 30-m,  but they have a revisit period of up to 16 dates. On the other hand, the Moderate resolution Imaging Spectroradiometer (MODIS) sensors can acquire images for the same scene at least once per date, but the images are at a low spatial resolution of 500-m~\cite{STARFM}. Therefore, the simultaneous acquisition of image series of high spatial and high temporal resolution is a major challenge in the remote sensing community~\cite{STfusion_survey}. The direct solution to this challenge is super-resolution~\cite{LIIF, MetaSR}, which estimates an unobserved high spatial resolution (HR) image from the single corresponding low spatial resolution (LR) image. However, its effectiveness is limited in many practical cases, as the spatial resolution gap between two satellite images is often quite large. 

\textit{Spatiotemporal fusion (ST fusion)} addresses this challenge by leveraging HR and LR image pairs acquired on reference dates temporally close to the target date. Specifically, the unobserved HR image on the target date is estimated by integrating the spatial structure extracted from reference HR images with the spectral changes inferred from a temporal LR image series.

In ideal situations where reliable reference images are available, accurate ST fusion would be easily acheived because the correct spatial structure and spectral changes are readily available. However, in real-world applications, such situations are very rare. This is because satellite images are often contaminated with various types of noise including random noise, outliers, missing values, and stripe noise~\cite{noise, sparsenoise,stripenoise1}, due to the measurement equipment and/or environmental conditions. ST fusion methods that do not account for such noise would produce noisy target HR images, which will have a significant negative impact on subsequent processing. Therefore, from a practical point of view, it is crucial to develop a noise-robust ST fusion method.

\begin{figure*}[t]
	\begin{center}
        \begin{minipage}{1\hsize}
			\centerline{\includegraphics[width=\hsize]{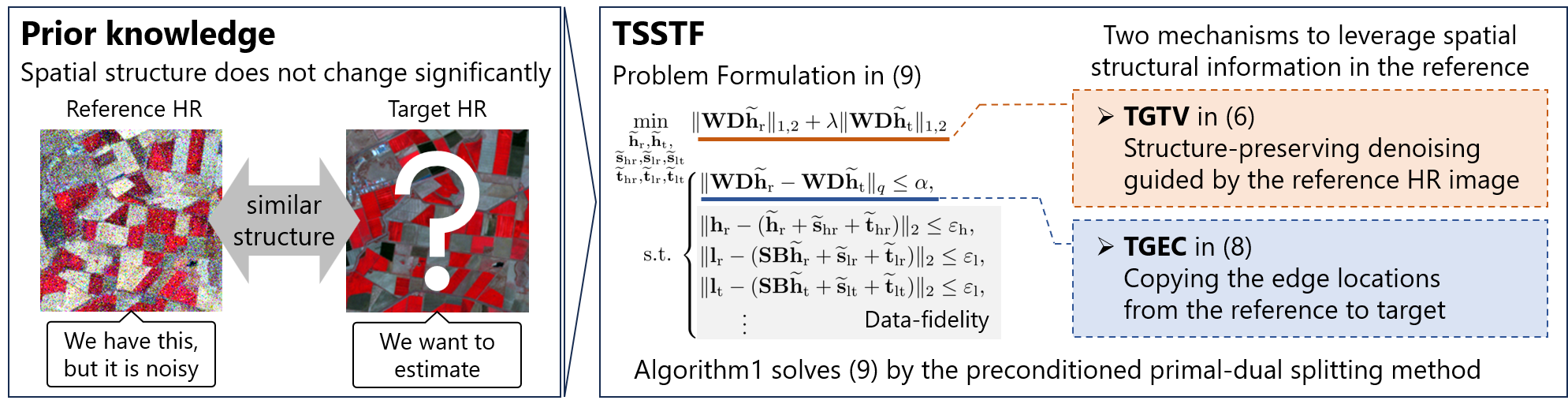}} 
		\end{minipage}
		\end{center}
    \vspace{-3mm}
	\caption{Illustration of our TSSTF.}
 \label{fig: TSSTF illustration}
\end{figure*}

\vspace{-2mm}
\subsection{Prior Research}
\label{ssec: prior research}

ST fusion has been extensively studied over the past decades. From a methodological perspective, existing ST fusion techniques can be broadly classified into three groups: rule-based, learning-based, and optimization-based approaches.
Rule-based methods estimate the target HR image using explicitly defined rules, such as weighted averaging
or spectral unmixing
~\cite{STARFM, ESTARFM, FSDAF, VIPSTF, MMT, UBDF}. They are the most traditional and have laid the foundation for subsequent developments in this field. 
Recently, learning-based methods have been proposed by leveraging machine learning models to learn a direct mapping from observed HR-LR pairs to the target HR image~\cite{random_forest, CNN, aritifical_NN, extreme, RSFN, GAN-STFM, DISTF, SWIN, ECPW}. By utilizing a large amount of extra training data, these methods are effective at capturing complex land cover structure and nonlinear temporal variations withouht explicitly modeling them.
In contrast, optimization-based methods formulate the ST fusion task as an optimization problem and solve it either analytically or using optimization algorithms~\cite{dict, RobOt, ROSTF, Bayesian1, Bayesian2}. A key advantage of these methods is that they do not require any external training data; instead, they rely solely on the observed HR and LR images to perform the fusion in a self-contained and model-driven manner.

Although a wide variety of ST fusion methods have been proposed, most are designed without explicitly accounting for noise. Consequently, they may fail in real-world scenarios where measurement noise is inevitable. To the best of our knowledge, only two methods have been developed with a specific focus on robustness to noise: a learning-based method called robust spatiotemporal fusion network (RSFN)~\cite{RSFN}, and an optimization-based method called robust optimization-based spatiotemporal fusion (ROSTF)~\cite{ROSTF}.

RSFN mitigates the impact of noise by employing convolutional neural networks (CNNs), generative adversarial networks (GANs), and an attention mechanism. It takes as input two reference HR images acquired before and after the target date, along with a LR image on the target date. The attention mechanism is designed to suppress the influence of noisy pixels. Specifically, it compares the two reference HR images and places greater reliance on the less noisy one in each spatial region. However, this strategy can be effective only when the noisy pixels do not appear at the same locations across the two references. In practice, this assumption rarely holds because noise often affects entire images. Moreover, RSFN assumes that the noise follows a Gaussian distribution, limiting its applicability to other noise types commonly observed in satellite data. Additionally, the requirement for future observations does not align with the fundamental goal of ST fusion, which aims to estimate the HR image at the target date using only currently available or earlier observations.

To achieve more practical ST fusion, ROSTF was developed to handle mixed Gaussian and sparse noise affecting the entire image, requiring only a single past HR-LR image pair as well as a LR image on the target date. It introduces an observation model that explicitly characterizes how such noise contaminates satellite images. By incorporating this observation model along with prior knowledge of a temporal satellite image series, ROSTF formulates ST fusion and noise removal as a unified optimization problem. This formulation allows the two tasks to be performed simultaneously, enabling noise-robust ST fusion. However, a significant limitation still remains: the loss of original spatial structure. The noise removal process in ROSTF relies on total variation (TV) regularization~\cite{TV1,TV2,HTV}, a widely used technique in image denoising. While TV effectively promotes piecewise-smoothness in the estimated image, this regularization operates independently of the intrinsic spatial structure in the underlying noise-free image. Consequently, it cannot distinguish between noise and meaningful fine structure, such as edges, that should be preserved. As a result, ROSTF tends to oversmooth the image and introduce unnatural artifacts, degrading spatial fidelity.

\vspace{-2mm}
\subsection{Contributions and Paper Organization}
\label{ssec: contributions}
Now, a natural question arises: \textit{Is it possible to achieve noise-robust ST fusion while preserving the original spatial structure?} To answer this, we propose the Temporally-Similar Structure-Aware ST fusion (TSSTF) framework. When the reference and target dates are temporally close, the corresponding HR images are expected to have similar spatial structure. Exploiting this prior knowledge, TSSTF accurately extracts the underlying spatial structure from the noisy reference image and effectively reflects it to the estimated target HR image. The TSSTF framework is built upon the following two key mechanisms as illiustrated in Fig.~\ref{fig: TSSTF illustration}.
\begin{itemize}
    \item \textit{Temporally-Guided Total Variation:} We employ a weighted total variation-based regularization specialized for ST fusion, termed temporally-guided total variation (TGTV). TGTV evaluates neighborhood differences with adaptive weights derived from the reference HR image, unlike standard total variation, which treats them all uniformly. Minimizing TGTV promotes spatial piecewise-smoothness while preserving the original spatial structure.
    
    \item \textit{Temporally-Guided Edge Constraint: } When the reference and target HR images share similar spatial structure, the locations of their edges are likely to coincide. On the other hand, the edge intensities may vary due to temporal changes in the spectral brightness of the surrounding regions. Then, we develop a constraint, termed the temporally-guided edge constraint (TGEC), that enforces the consistency of edge locations while allowing for variations in edge intensities.
\end{itemize}

By incorporating these mechanisms, we formulate ST fusion as a constrained optimization problem. Specifically, the objective function consists of TGTV, while the constraints include TGEC along with data-fidelity conditions. To solve this problem, we construct an efficient algorithm based on a preconditioned primal-dual splitting method (P-PDS)~\cite{P-PDS} with an operator-norm-based design method of variable-wise diagonal preconditioning technique~(OVDP)~\cite{P-PDS_OVDP}. 


The main contributions of this paper are given as follows.
\begin{itemize}
    \item[1)] \textit{Improved Robustness to Noise:}
    We achieve improved robustness in ST fusion while faithfully preserving the original spatial structure by integrating the two temporally-similar structure-aware mechanisms, TGTV and TGEC, into a unified optimization framework. The effectiveness of the proposed TSSTF is extensively validated through experiments considering various noise types, including Gaussian, sparse, stripe, and Poisson noise.

    \item[2)] \textit{Ease of Parameter Selection:} Our framework reduces the effort required for manual parameter tuning. We employ a constrained optimization approach, where the parameters associated with each constraint are independent and physically interpretable~\cite{constrained1,constrained3}. This allows for systematic adjustment based on prior information. In addition, the adopted optimization algorithm, P-PDS with OVDP, automatically determines appropriate stepsizes according to the problem structure, eliminating the tedious manual selection typically required by standard optimization algorithms.
    
    \item[3)] \textit{Adaptive Constraint Level Tuning Strategy:} The appropriate constraint level for TGEC is highly dependent on data characteristics, such as land-cover changes and spatial complexity, and cannot be accurately determined in advance under noisy conditions. To address this, we introduce an adaptive adjustment strategy that identifies the appropriate constraint level during the optimization process. This strategy not only reduces the effort for parameter tuning but also leads to stabilized algorithmic behavior and accelerated convergence by effectively constraining the solution space.
    
    
\end{itemize}

In the following sections, we first cover the mathematical preliminaries for our method in Sec.~\ref{sec:preli} and then move on to the establishment of our method in Sec.~\ref{sec:proposed}. In Sec.~\ref{sec:exp}, we demonstrate the performance of TSSTF through comparative experiments. Experimental results show that TSSTF performs comparably to several state-of-the-art ST fusion methods for noiseless images and better than them for noisy images. Additionally, we provide a recommended set of parameters for TSSTF, which can be determined based on prior information. Sec.~\ref{sec:discussion} discusses the computational cost, performance under significant land-cover changes, and performance under cloud contamination. Finally, we conclude this paper in Sec.~\ref{sec:conclusion}.

The preliminary version of this work, without considering sparse noise, mathematical details, comprehensive experimental comparison, deeper discussion, has appeared in conference proceedings~\cite{ICASSP}.

\section{Preliminaries}
\label{sec:preli}
\vspace{-1mm}

\subsection{Notations}
\label{ssec:Notation}
\vspace{-0.5mm}
Let $\mathbb{R}$ be the set of all real numbers. Vectors and matrices are denoted by bold lower and upper case letters, respectively. 
We represent a multispectral image with spatial resolution $W \times H$ and $B$ spectral bands as a vector $\mathbf{x}\in \mathbb{R}^{WHB}$, where $x_{i,j,b}$ denotes the pixel value at the location $(i,j)$ in the $b$-th spectral band.
Let $\Gamma_0(\mathbb{R}^{WHB})$ be the set of all proper lower-semicontinuous convex functions defined on $\mathbb{R}^{WHB}$.
The $\ell_1$-norm, the $\ell_2$-norm, and the mixed $\ell_{1,2}$-norm of $\mathbf{x}$ are defined as $\onenorm{\x}:=\sum_{i,j,b}|x_{i,j,b}|$, $\twonorm{\x}:=\sqrt{\sum_{i,j,b}|x_{i,j,b}|^{2}}$, and $\onetwonorm{\x}:=\sum_{i,j}\sqrt{\sum_{b}|x_{i,j,b}|^{2}}$, respectively.
For an image $\x \in \R^{WHB}$, let $\D_1$, $\D_2$, $\D_3$, and $\D_4\in \mathbb{R}^{WHB \times WHB}$ denote the matrices for computing the differences between each pixel value and its four neighborhood pixel values as follows:
\begin{align}
\label{eq: difference matrix}
    [\D_1 \x]_{i,j,b} = x_{i+1,j,b} - x_{i,j,b}, 
    \;& [\D_2 \x]_{i,j,b} = x_{i+1,j-1,b} - x_{i,j,b}, \nonumber\\
    [\D_3 \x]_{i,j,b} = x_{i,j-1,b} - x_{i,j,b}, 
    \;& [\D_4 \x]_{i,j,b} = x_{i-1,j-1,b} - x_{i,j,b}. \nonumber
    \vspace{-0.5mm}
\end{align}
Then, let $\D := ({\D_1}^{\top} ~ {\D_2}^{\top}~ {\D_3}^{\top} ~ {\D_4}^{\top})^{\top} \in \mathbb{R}^{4WHB \times WHB}$.
The hyperslab with the center $\omega$ and radius $\alpha$ is denoted as
\begin{equation}
	\label{eq: hyperslab}
	S_{\alpha}^{\omega} := \{ \mathbf{z} | \,| \omega - \mathbf{1}^{\top}\mathbf{z} | \leq \alpha  \} \nonumber.
\end{equation}
The norm ball with the center $\mathbf{c}$ and radius $\varepsilon$  is denoted as 
\begin{equation}
	\label{eq: lqnorm_ball}
	B_{q}^{\mathbf{c}, \varepsilon} := \{ \mathbf{z} | \| \mathbf{z} - \mathbf{c} \|_q \leq \varepsilon \} \nonumber,
\end{equation}
for the $\ell_1$-norm, $\ell_2$-norm, and mixed $\ell_{1,2}$-norm.
The indicator function $\iota_C : \mathbb{R}^N \rightarrow (-\infty,\infty]$ of a nonempty closed convex set $C$ is defined as
\begin{eqnarray}
	\label{eq: indicator func}
	\iota_C := 
	\begin{cases}
		0, & \mathrm{if} \, \mathbf{x} \in C, \\
		\infty, & \mathrm{otherwise}.
	\end{cases}
\end{eqnarray}

\subsection{Proximal Tools}
\label{ssec: prox}
\vspace{-0.5mm}
The optimization problem of TSSTF that will be formulated in Sec.~\ref{ssec: formulation} consists of nonsmooth convex functions. To solve such a problem, we introduce the notion of the {\it proximity operator} of $f \in \Gamma_0(\mathbb{R}^{WHB})$ with a parameter $\gamma > 0$ as follows: 
\begin{equation}
	\label{eq: prox}
	\mathrm{prox}_{\gamma f}: \mathbb{R}^{WHB} \rightarrow \mathbb{R}^{WHB} : 
	\mathbf{x} \mapsto \argmin_{\mathbf{y} \in \mathbb{R}^{WHB}} f(\mathbf{y}) + \frac{1}{2\gamma}\|\mathbf{x} - \mathbf{y}\|_2^2. \nonumber
\end{equation}


Below, we show the specific proximity operators of the functions that we use in this paper.
The proximity operator of the mixed $\ell_{1,2}$-norm is given by
\begin{equation}
	\label{eq: prox_l1,2norm}
	[\mathrm{prox}_{\gamma \|\cdot\|_{1,2}}(\mathbf{x})]_{i,j,b}  = \max \left\{ 1 - \frac{\gamma}{\sqrt{\sum_{b'=1}^{B} |x_{i,j,b'}|^2}},0 \right\} x_{i,j,b}. \nonumber
\end{equation}
The proximity operator of the indicator function $\iota_C$ is equivalent to the projection onto the set $C$, given by
\begin{equation}
	\label{eq: prox_indicator}
	\mathrm{prox}_{\gamma \iota_C}(\mathbf{x}) = P_C(\mathbf{x}) := \argmin_{\mathbf{y} \in C} \|\mathbf{x} - \mathbf{y}\|_2. \nonumber
\end{equation}
The projection onto the hyperslab is expressed as follows: 
\begin{eqnarray}
	\label{eq: prox_hyperslab}
	&P_{S_{\alpha}^{\omega}}(\mathbf{x})& =
	\begin{cases}
		\mathbf{x} + \frac{\theta_1 - \mathbf{1}^{\top}\mathbf{x}}{WHB}\mathbf{1}, \; \; & \mathrm{if} \, \mathbf{1}^{\top}\mathbf{x} < \theta_1, \\
		\mathbf{x} + \frac{\theta_2 - \mathbf{1}^{\top}\mathbf{x}}{WHB}\mathbf{1}, \; \; & \mathrm{if} \, \mathbf{1}^{\top}\mathbf{x} > \theta_2, \\
		\mathbf{x}, & \mathrm{otherwise},
	\end{cases} \nonumber \\ \nonumber
\end{eqnarray}
where $\theta_1 = \omega - \alpha$ and $\theta_2 = \omega + \alpha$.
The projections onto the $\ell_2$-norm ball and the $\ell_1$-norm ball are calculated by 
\begin{eqnarray}
	\label{eq: prox_l2ball}
	&P_{B_{2}^{\mathbf{c}, \varepsilon}}(\mathbf{x})& =
	\begin{cases}
		\mathbf{x}, & \mathrm{if} \, \mathbf{x} \in \iota_{B_{2}^{\mathbf{c}, \varepsilon}}, \\
		\mathbf{c} + \frac{\varepsilon (\mathbf{x} - \mathbf{c})}{\| \mathbf{x} - \mathbf{c} \|_2}, & \mathrm{otherwise},
	\end{cases} \nonumber
\end{eqnarray}
and a fast $\ell_1$-ball projection algorithm~\cite{ell1ball_projection}, respectively.
The projection onto the mixed $\ell_{1,2}$-norm ball with the center $\mathbf{0}\in\mathbb{R}^{HWB}$ is calculated by
\begin{eqnarray}
	\label{eq: prox_l1,2ball}
	&[P_{B_{1,2}^{\mathbf{0}, \varepsilon}}(\mathbf{x})]_{i,j}& =
	\begin{cases}
		\mathbf{0}\in\mathbb{R}^{B}, & \mathrm{if} \, \|[\mathbf{x}]_{i,j}\|_2=0, \\
		\xi_{i,j} \frac{[\mathbf{x}]_{i,j}}{\|[\mathbf{x}]_{i,j}\|_2}, & \mathrm{otherwise},
	\end{cases} \nonumber
\end{eqnarray}
where $[\mathbf{x}]_{i,j}=(x_{i,j,1},\cdots,x_{i,j,B})\in\mathbb{R}^{B}$ and $\mathbf{\xi} = P_{B_{1}^{\mathbf{0}, \varepsilon}}(\|[\mathbf{x}]_{1,1}\|_2,\cdots,\|[\mathbf{x}]_{W,H}\|_2) \in \mathbb{R}^{WH}$.

\section{Proposed Method}
\label{sec:proposed}
In this section, we first organize the problem setting and introduce the observation models for high-resolution (HR) and low-resolution (LR) images. Then, we establish two main mechanisms, temporally-guided total variation (TGTV) and temporally-guided edge constraint (TGEC). After that, we formulate ST fusion as an optimization problem based on these mechanisms and develop an algorithm to solve it.

\subsection{Problem Setting}
\label{ssec: problem setting}
In this paper, we consider a practical scenario where a single HR-LR image pair acquired on a past reference date is available, together with an LR image acquired on the target date. This setting is realistic because the availability of multiple HR-LR pairs is often limited in real-world applications due to various factors such as cloud contamination, temporal inconsistency in image acquisition, or limited data accessibility. Therefore, ST fusion methods that require only a single reference HR-LR pair are applicable to a wider range of real-world cases than those requiring multiple reference pairs, although such a situation is obviously challenging~\cite{one-pair}.

\begin{figure}[t]
	\begin{center}
		\scalebox{1.0}{
        \begin{minipage}{1\hsize}
			\centerline{\includegraphics[width=\hsize]{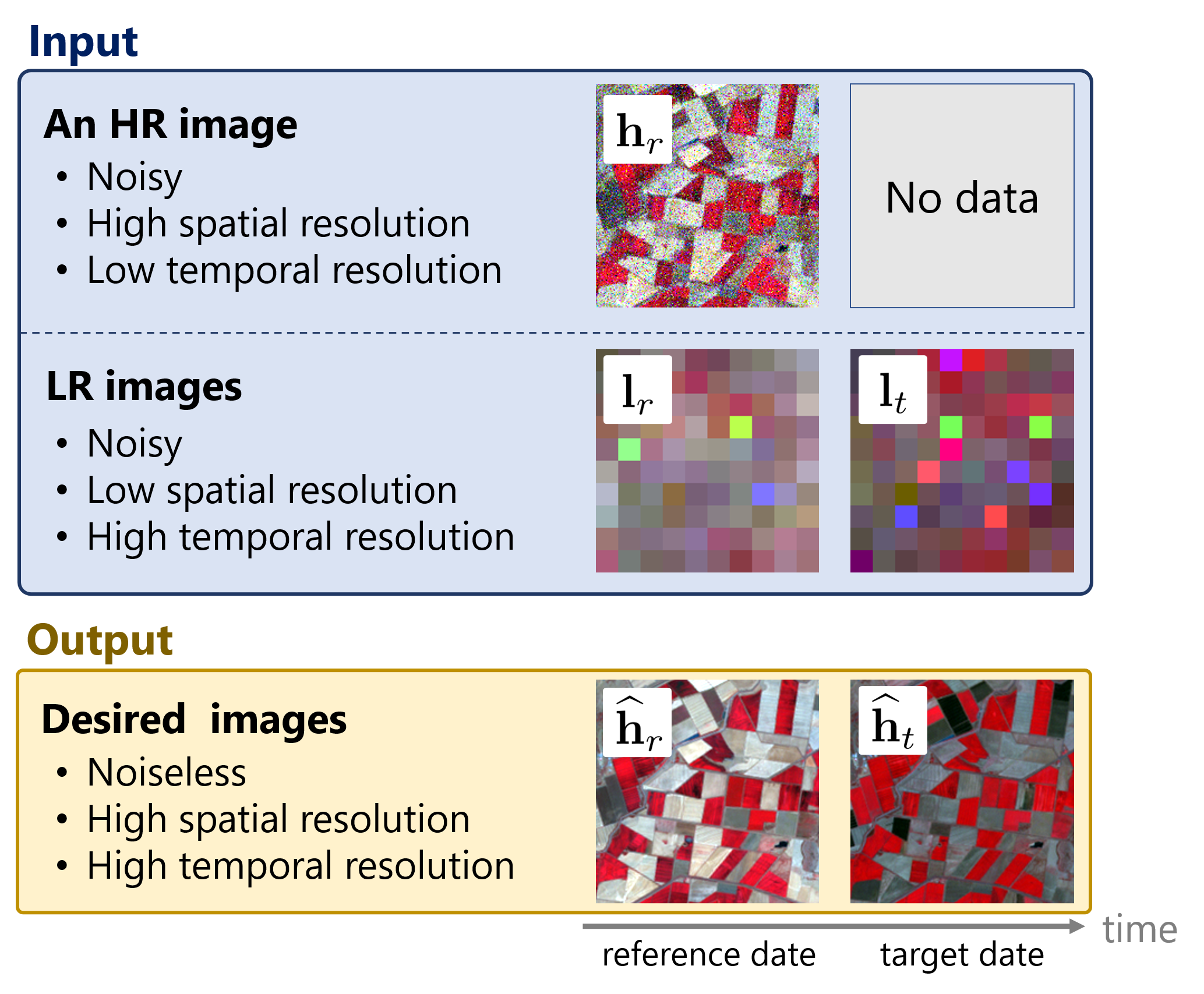}} 
		\end{minipage}
		}
		\end{center}
		\vspace{-5mm}
	\caption{Our problem setting}
 \label{fig: problemsetting}
\end{figure}

Let the HR image on the reference date, the LR image on the reference date, and the LR image on the target date be $\Hr\in \mathbb{R}^{W_h H_h B}$, $\Lr\in \mathbb{R}^{W_l H_l B}$, and $\Lt\in \mathbb{R}^{W_l H_l B}$, respectively. As shown in Fig.~\ref{fig: problemsetting}, our goal is to estimate $\hatHt\in \mathbb{R}^{W_h H_h B}$ (the noiseless HR image on the target date) based on these three observed images, while simultaneously denoising $\Hr$, i.e., estimating $\hatHr$ (the noiseless HR image on the reference date). For simplicity of description, we use $\Nh:=W_h H_h$ and $\Nl:=W_l H_l$ as needed.

\subsection{Observation Models}
\label{ssec: observation models}
Let $\widehat{\mathbf{h}} \in \mathbb{R}^{\Nh B}$ and $\widehat{\mathbf{l}} \in \mathbb{R}^{\Nl B}$ be a noiseless HR image and a noiseless LR image, respectively, acquired on the same date. We define the observation models for the observed HR image $\mathbf{h} \in \mathbb{R}^{\Nh B}$ and LR image $\mathbf{l} \in \mathbb{R}^{\Nl B}$ as follows: 
\begin{align}
\label{eq: noise model}
  \mathbf{h} &= \widehat{\mathbf{h}} + \nh + \sh + \th, \nonumber\\
  \mathbf{l} &=  \widehat{\mathbf{l}} + \nl + \sl + \tl,
\end{align}
where $\nh$ and $\nl$ represent random noise components (e.g., Gaussian or Poisson noise), $\sh$ and $\sl$ denote sparse noise components modeling outliers and missing values, and $\th$ and $\tl$ correspond to stripe noise components. 

In addition, $\widehat{\mathbf{l}}$ can be approximated by the image obtained by blurring and down-sampling $\widehat{\mathbf{h}}$, known as a typical super-resolution model \cite{relationshipmodel}, as follows:
\begin{eqnarray}
\label{eq: relationship model}
  \widehat{\mathbf{l}} = \mathbf{S}\mathbf{B}\widehat{\mathbf{h}} + \mathbf{m},
\end{eqnarray}
where $\mathbf{B} \in \mathbb{R}^{\Nh B \times \Nh B}$ is the spatial spread transform matrix introduced in \cite{blurring}, $\mathbf{S}\in \mathbb{R}^{\Nl B \times \Nh B}$ is the down-sampling matrix, and $\mathbf{m} \in \mathbb{R}^{\Nl B}$ is the modeling error. This model has been widely used in the ST fusion literature \cite{integrated}.

\begin{figure*}[t]
	\begin{center}
		\scalebox{1.0}{
        \begin{minipage}{1\hsize}
			\centerline{\includegraphics[width=\hsize]{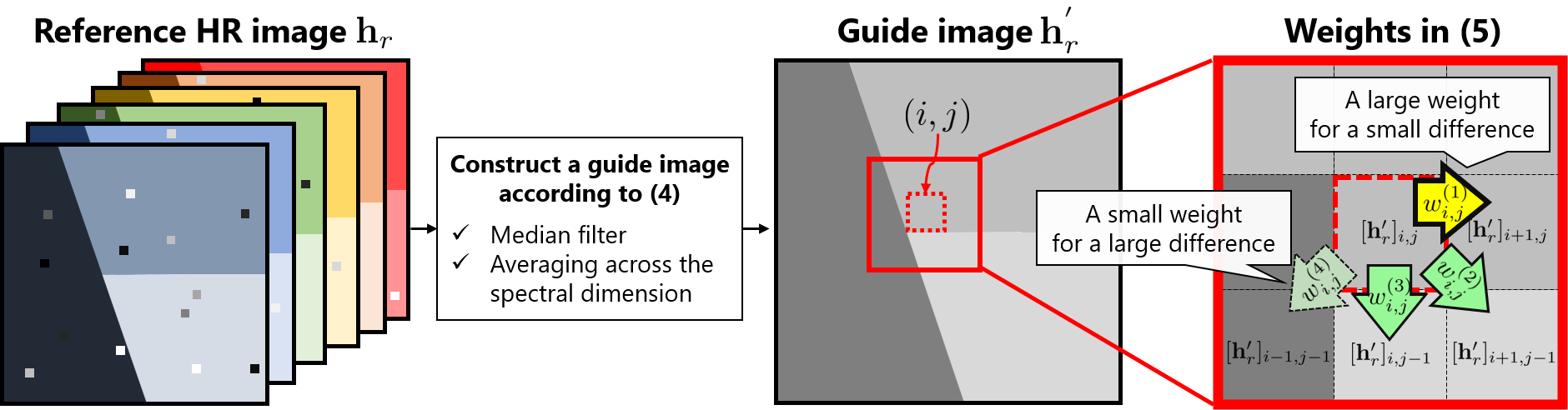}} 
		\end{minipage}
		}
		\end{center}
	\caption{Illustration of the weights in (\ref{eq: weight}). A noise-attenuated guide image $\Hr^{\prime}$ in (\ref{eq: guide image}) is first constructed from the reference HR image. The weights are then computed from this guide image according to (\ref{eq: weight}). In the right panel, each arrow corresponds to a weight; the darker the arrow, the greater the weight.}
 \label{fig: weights illustration}
\end{figure*}
\begin{figure*}[t]
	\begin{center}
        \begin{minipage}{1\hsize}
			\centerline{\includegraphics[width=\hsize]{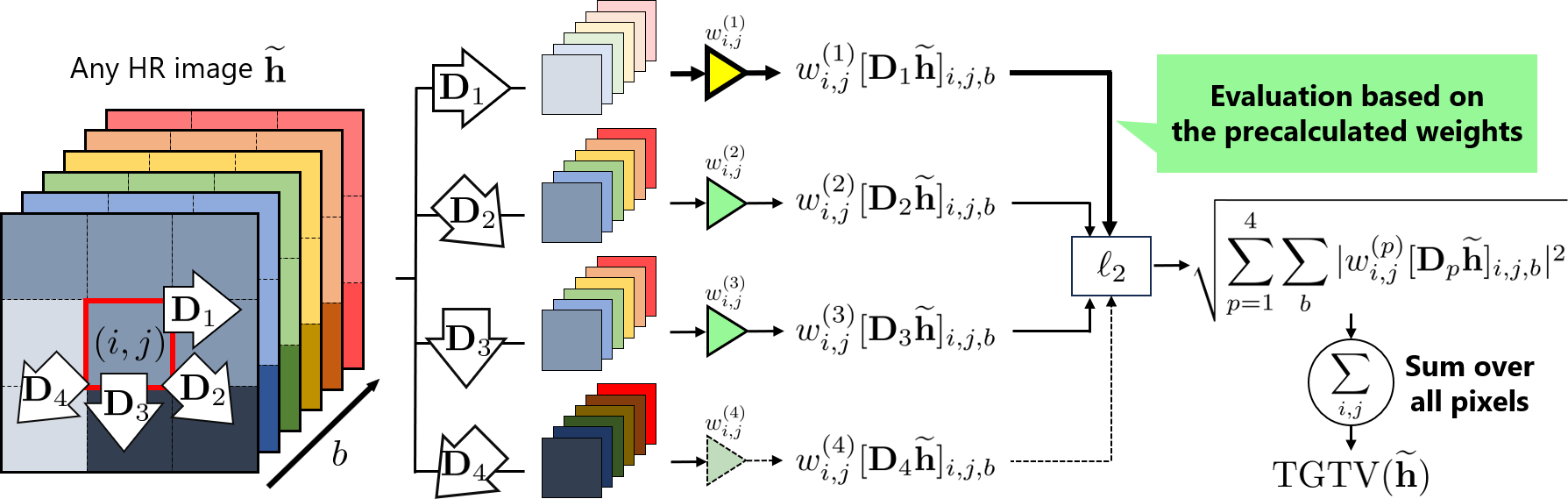}} 
		\end{minipage}
		\end{center}
	\caption{Illustration of TGTV. TGTV evaluates the four neighborhood differences with adaptive weights derived from the guided HR image $\Hr'$.}
 \label{fig: TGTV_illustration}
\end{figure*}

\subsection{Temporally-Guided Total Variation}
\label{ssec: TGTV}
To acheive simultaneous denoising and ST fusion, we introcuce a regularization function imposed on the HR images. Total variation (TV) regularization is widely used for denoising, due to its ability to promote piecewise smoothness~\cite{TV1, TV2}. However, standard TV regularization does not explicitly account for the underlying spatial structure in the image, which can result in over-smoothing. To overcome this limitation, we propose a regularization function that effectively captures the spatial structure of the target region by leveraging a reference HR image $\Hr$.

In practical scenarios, $\Hr$ is often contaminated by noise, which hinders accurate extraction of spatial structure. Therefore, we construct a guide image $\Hr'\in \R^{\Nh}$ as follows:
\begin{equation} 
  \label{eq: guide image} \vspace{-1.5mm} 
  \Hr' := \frac{1}{B} \sum_{b=1}^{B} \mathrm{Med}([\Hr]_b), \vspace{-0.5mm} 
\end{equation}
where $[\Hr]_b\in\R^{\Nh}$ denotes the $b$-th spectral band of $\Hr$.
Specifically, we first apply a median filter $\mathrm{Med}(\cdot):\R^{\Nh}\rightarrow\R^{\Nh}$ to each spectral band of $\Hr$. The filtered bands are then averaged across the spectral dimension to yeild a grayscale guide image. This guide image preserves the dominant spatial structure of $\Hr$ while effectively reducing noise effects.

Based on the guide image $\Hr'$, we define a weight $w_{i,j}^{(p)}$ for each pixel location $(i,j)$ and direction $p\in\{1,2,3,4\}$ as follows:
\begin{equation}
\label{eq: weight}
    w_{i,j}^{(p)} = \exp\left(-\frac{1}{\delta^2}|[\D_{p}\Hr']_{i,j}|^{2}\right),
\end{equation}
where $\delta$ is a parameter controlling the sensitivity of the weights to spatial differences. As shown in Fig.~\ref{fig: weights illustration}, this formulation assigns larger weights to directions where the local difference in $\Hr'$ is small, and smaller weights where the difference is large, thereby reflecting the underlying spatial structure. To further enhance structural selectivity, we introduce an additional parameter $k \in \{1,2,3,4\}$, and set to zero the $k$ smallest values among ${w_{i,j}^{(1)}, w_{i,j}^{(2)}, w_{i,j}^{(3)}, w_{i,j}^{(4)}}$ for each $(i,j)$. This operation effectively suppresses the influence of directions with minimal structural similarity.

For each direction $p$, we arrange the weights $\{w_{i,j}^{(p)}\}$ diagonally to make $\W^{(p)} := \mathrm{diag}(w_{1,1}^{(p)}, \dots ,w_{W_h,H_h}^{(p)})\in \R^{\Nh \times \Nh}$, and then replicate $\W^{(p)}$ across the spectral bands to form $\overline{\W}^{(p)}:=\mathrm{diag}(\W^{(p)}, \dots, \W^{(p)})\in \R^{\Nh B \times \Nh B}$. Finally, the overall weight matrix $\W$ is defined by stacking the four matrices $\overline{\W}^{(p)}$ as follows:
\begin{equation}
    \W := \mathrm{diag}(\overline{\W}^{(1)},\overline{\W}^{(2)}, \overline{\W}^{(3)}, \overline{\W}^{(4)}) \in \R^{4\Nh B \times 4\Nh B}. \nonumber
\end{equation}

Using this weight matrix, we define our proposed regularization function, temporally-guided total variation (TGTV), for any HR image $\tilH \in \mathbb{R}^{\Nh B}$ as follows:
\begin{equation}
\label{eq: TGTV}
    \mathrm{TGTV}(\tilH) := \| \W\D\tilH\|_{1,2} 
    =\sum_{i,j}\sqrt{\sum_{b}\sum_{p=1}^{4}|w_{i,j}^{(p)}[\D_{p}\tilH]_{i,j,b}|^2}.
\end{equation}
As shown in Fig.~\ref{fig: TGTV_illustration}, TGTV evaluates the four neighborhood differences with adaptive weights derived from the guided image $\Hr'$. Therefore, minimizing TGTV encourages the estimated image to exhibit spatial smoothness in regions where the guided image $\Hr'$ is homogeneous, while preserving edges and structural discontinuities where $\Hr'$ exhibits strong spatial differences. As a result, TGTV promotes piecewise smoothness while preserving the intrinsic spatial structure.

\begin{figure*}[t]
	\begin{center}
		\scalebox{0.9}{
        \begin{minipage}{1\hsize}
			\centerline{\includegraphics[width=\hsize]{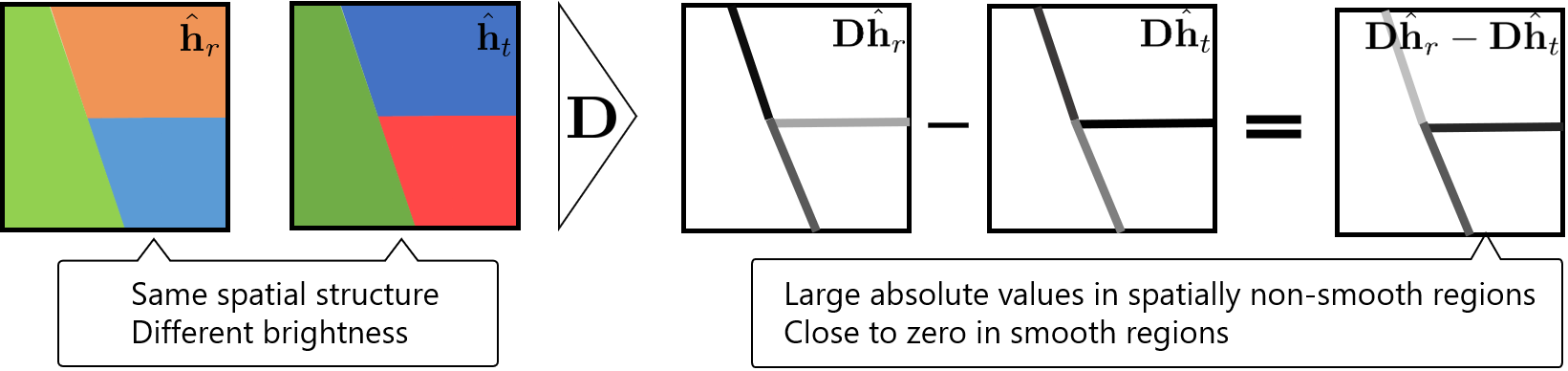}} 
		\end{minipage}
		}
		\end{center}
	\caption{Illustration of edge similarity. When the reference and target dates are temporally close, the locations of edges are likely to coincide. On the other hand, edge intensities may differ due to changes in the spectral brightness of surrounding regions over time.}
 \label{fig: edgesimilarity}
\end{figure*}

\subsection{Temporally-Guided Edge Constraint}
\label{ssec: TGEC}
When the reference and target dates are temporally close, the corresponding HR images are expected to share similar spatial structure, which means that the locations of edges are likely to coincide. 
Based on this prior knowledge, we propose a constraint that enforces the edges in the reference and target HR images to appear at the same spatial locations. 

A simple way to incorporate this prior knowledge is to directly compare the edge intensities of the two images and constrain their difference to be small. This can be expressed as the following constraint:
\begin{equation}
  \label{eq: normal edge constraint}
  \| \D\tilHr - \D\tilHt \|_{q}\leq\alpha,
\end{equation}
where $\tilHr$ and $\tilHt$ denote the estimated reference and target HR images, respectively, $q$ indicates the norm used to measure the difference, and $\alpha$ is a small positive constant that controls the allowable discrepancy.

While this constraint has been shown to copy edges from the reference HR image to the target HR image~\cite{ROSTF}, it suffers from a critical limitation: it does not account for temporal variations in edge intensities between the two images. Such intensity variations are caused by changes in the spectral brightness of surrounding regions over time. Consequently, the components of $\D\hatHr - \D\hatHt$ tend to have large absolute values in spatially non-smooth regions, while they are close to zero in spatially smooth regions, as illustrated in Fig.~\ref{fig: edgesimilarity}. Nevertheless, the constraint in \eqref{eq: normal edge constraint} evaluates all components of $\D\tilHr - \D\tilHt$ equally, enforcing them to approach zero uniformly, which can introduce undesired artifacts.

To address this issue, we incorporate the same weighting scheme used in TGTV as follows: 
\begin{equation}
  \label{eq: TGEC}
  \| \W(\D\tilHr - \D\tilHt) \|_{q}\leq\alpha.
\end{equation}
The matrix $\W$ assigns larger weights to smooth regions and smaller values to non-smooth regions based on the guide image $\Hr'$ as defined in \eqref{eq: weight}. Thus, this constraint encourages the components of $\D\tilHr - \D\tilHt$ to approach zero preferentially in smooth regions, while allowing flexibility in non-smooth regions. As a result, the constraint can tolerate temporal variations in edge intensities while still enforcing consistency in the edge locations. We refer to the constraint as the temporally-guided edge constraint (TGEC).

Regarding the choice of the norm \( q \) in TGEC, we evaluated several candidates, including the \( \ell_1 \) norm, the \( \ell_2 \) norm, and the mixed \( \ell_{1,2} \) norm. Among them, the mixed \( \ell_{1,2} \) norm was found to yield the best performance in our experiments. For detailed comparisons, please refer to Sec.~\ref{ssec:norm q of tgec}.

\subsection{Problem Formulation}
\label{ssec: formulation}

Based on the TGTV and TGEC mechanisms, we formulate the ST fusion task as the following constrained optimization problem:
\begin{align}
\label{eq: our ST fusion}
	\min_{\substack{\tilHr,\tilHt, \\ \tilshr,\tilslr,\tilslt \\ \tilthr,\tiltlr,\tiltlt}} \:
	& \|\W\D\tilHr\|_{1,2} + 
	\lambda\|\W\D\tilHt\|_{1,2} \\ 
	\mathrm{s.t.} \:
	& \begin{cases}
        \qnorm{\W\D\tilHr - \W\D\tilHt} \leq \alpha, \\
        \left| \mathbf{1}^{\top}[\mathbf{l}_{r}]_{b}/{\Nl} - \mathbf{1}^{\top}[\widetilde{\mathbf{h}}_{r}]_{b}/{\Nh}\right| \leq \beta_{b} \, (b = 1, \ldots, B), \\
        \left| \mathbf{1}^{\top}[\mathbf{l}_{t}]_{b}/{\Nl} - \mathbf{1}^{\top}[\widetilde{\mathbf{h}}_{t}]_{b}/{\Nh}\right| \leq \beta_{b} \, (b = 1, \ldots, B), \\
        \twonorm{\Hr - (\tilHr + \tilshr + \tilthr)}\leq\epsh, \\
        \twonorm{\Lr - (\S\B\tilHr + \tilslr + \tiltlr)}\leq\epsl,\\
        \twonorm{\Lt - (\S\B\tilHt + \tilslt + \tiltlt)}\leq\epsl,\\
        \onenorm{\tilshr} \leq \etah, 
        \onenorm{\tilslr} \leq \etal, 
        \onenorm{\tilslt} \leq \etal,\\
        \onenorm{\tilthr} \leq \zetah, 
        \onenorm{\tiltlr} \leq \zetal, 
        \onenorm{\tiltlt} \leq \zetal,\\
        \D_3\tilthr = \mathbf{0},
        \D_3\tiltlr = \mathbf{0},
        \D_3\tiltlt = \mathbf{0}, \nonumber
	\end{cases}
\end{align}
where $\lambda$ is a balancing parameter. The variables $\tilHr$ and $\tilHt$ represent the estimates of the latent noise-free HR images $\hatHr$ and $\hatHt$, respectively. The variables $\tilshr$, $\tilslr$ and $\tilslt$ correspond to the estimates of sparse noise components, whereas $\tilthr$, $\tiltlr$ and $\tiltlt$ denote the estimates of stripe noise components superimposed on $\Hr$, $\Lr$ and $\Lt$, respectively. The objective function and the first constraint are TGTV and TGEC, respectively. The other constraints play roles in data-fidelity and noise characterization, as described below.
\begin{itemize}
    
    \item The second and third constraints ensure that the HR and LR images on the same date have similar average spectral brightness per band~\cite{ROSTF}. This holds when the HR and LR sensors have the same spectral resolution, which is a common assumption in the context of ST fusion.
    
    \item The forth to sixth constraints serve as data-fidelity based on the observation models in \eqref{eq: noise model} and \eqref{eq: relationship model}.
    
    \item The seventh to ninth constraints characterize the sparse noise using the $\ell_1$-norms. 

    \item The last six constraints model the stripe noise components as being sparse and constant along the stripe direction~\cite{stripenoise1}, which is assumed to be vertical in this study.
\end{itemize}

Using constraints instead of adding terms to the objective function in this way simplifies the parameter setting\cite{constrained1,constrained3}: we can determine the appropriate parameters for each constraint independently because they are decoupled. The detailed setting of these parameters is discussed in Sec.~\ref{ssec:experimental setup}.

\subsection{Optimization}
\label{ssec: optimization}
The algorithm for solving \eqref{eq: our ST fusion} is developed based on the preconditioned primal-dual splitting method (P-PDS)~\cite{P-PDS} with the operator-norm-based design method of the variable-wise diagonal preconditioning technique~(OVDP)~\cite{P-PDS_OVDP}. The overall algorithm is summarized in Algorithm~\ref{algo: PDS_for_OptForm}. The explanation of P-PDS with OVDP and the derivation of Algorithm~\ref{algo: PDS_for_OptForm} are provided in Appendices~\ref{append:p-pds} and~\ref{append:algorithm_derivation}, respectively.

We should note that the algorithm does not strictly solve the original optimization problem in \eqref{eq: our ST fusion}. Although the first constraint in \eqref{eq: our ST fusion}, i.e., TGEC, includes a parameter $\alpha$ treated as a fixed constant, the algorithm instead updates $\alpha$ at each iteration (line 20) as follows:
\begin{equation}
  \label{eq: alpha setting}
  \alpha^{(n)} \leftarrow c_{\alpha}\| \W\D\tilHr^{(n+1)} \|_{q} \cdot \|\Lr-\Lt\|_{1}/\Nl,
\end{equation}
where $c_{\alpha}$ is a hyperparameter. 

TGEC is designed to promote similarity in edge locations across time. 
However, the appropriate value of the threshold $\alpha$ in TGEC depends not only on edge alignment but also on the following two factors:
\begin{itemize}
  \item the degree of temporal changes in spectral brightness,
  \item the edge strength and density of the target area.
\end{itemize}
While the first factor can be estimated from the input LR images via $\|\Lr-\Lt\|_{1}/\Nl$, the second factor cannot be reliably measured from the observed HR image $\Hr$ due to strong noise contamination.
To address this, we exploit the progressively denoised intermediate estimate $\tilHr^{(n+1)}$ to compute $\|\W\D\tilHr^{(n+1)} \|_{q}$, which is considered to reflect the edge strength and density of the target area as the noise is reduced.
As a result, $\alpha^{(n)}$ is refined adaptively during the optimization, allowing the constraint strictness to be adjusted appropriately over iterations.

From a theoretical perspective, if $\alpha$ is fixed at any intermediate iteration, the algorithm can be regarded as solving a convex optimization problem with a fixed constraint, and it is guaranteed to converge to its optimal solution. Experimentally, we observe that the variation in $\alpha^{(n)}$ becomes negligible as the iterations proceed, and the algorithm exhibits convergent behavior (see detail in Sec.~\ref{ssec: algorithm convergence}). Therefore, in practical applications, the algorithm can be used reliably without the need to explicitly fix $\alpha$ during the optimization process.
The selection of the hyperparameter $c_{\alpha}$ will be discussed in detail in Sec.~\ref{ssec:coefficient c_alpha of tgec}.

\begin{algorithm}[t]
  \caption{P-PDS-based solver for (\ref{eq: our ST fusion})}
  \label{algo: PDS_for_OptForm}
  \begin{algorithmic}[1]
      \Require{$\lambda, c_{\alpha}, q, \beta_b, \epsh, \epsl, \etah, \etal, \zetah, \zetal$}
      \Ensure{$\tilHr^{(n)}, \tilHt^{(n)}, \tilshr^{(n)}, \tilslr^{(n)}, \tilslt^{(n)}, \tilthr^{(n)}, \tiltlr^{(n)}, \tiltlt^{(n)}$}
      \State Initialize $\tilHr^{(0)}, \tilHt^{(0)}, \tilshr^{(0)}, \tilslr^{(0)}, \tilslt^{(0)}, \tilthr^{(0)}, \tiltlr^{(0)}, \tiltlt^{(0)}, \mathbf{z}_{j}^{(0)}(j=1,\dots,9)$;
      \State Set $\gamma_{1,i} (i=1,\cdots, 8), \gamma_{2,j} (j=1,\cdots, 9),$ as in \eqref{eq: stepsizes setting};
      \While {until a stopping criterion is not satisfied}
      \State $\u_{r} \leftarrow \Dt\Wt\z_{1}^{(n)} + \Dt\Wt\z_{3}^{(n)} + \z_{4}^{(n)} +  \B^{\top}\S^{\top}\z_{5}^{(n)}$
      \State $\u_{t} \leftarrow \Dt\Wt\z_{2}^{(n)} - \Dt\Wt\z_{3}^{(n)} +  \B^{\top}\S^{\top}\z_{6}^{(n)}$
      \State $\tilHr^{(n+1)} \leftarrow \tilHr^{(n)} - \gamma_{1,1}\u_{r}$
      \State $\tilHt^{(n+1)}\leftarrow \tilHt^{(n)}-\gamma_{1,2}\u_{t}$
      \State $[\tilHr^{(n+1)}]_{b}\leftarrow P_{S_{N_h\beta_b}^{N_h\phi_{b}}}([\tilHr^{(n+1)}]_{b}), \quad (b=1,\cdots,B)$
      \State $[\tilHt^{(n+1)}]_{b}\leftarrow P_{S_{N_h\beta_b}^{N_h\psi_{b}}}([\tilHt^{(n+1)}]_{b}), \quad (b=1,\cdots,B)$
      \State $\tilshr^{(n+1)} \leftarrow P_{B_{1}^{0,\etah}}(\tilshr^{(n)} - \gamma_{1,3} \z_{4})$
      \State $\tilslr^{(n+1)} \leftarrow P_{B_{1}^{0,\etal}}(\tilslr^{(n)} - \gamma_{1,4} \z_{5})$
      \State $\tilslt^{(n+1)} \leftarrow P_{B_{1}^{0,\etal}}(\tilslt^{(n)} - \gamma_{1,5}\z_{6})$
      \State $\tilthr^{(n+1)} \leftarrow P_{B_{1}^{0,\zetah}}(\tilthr^{(n)} - \gamma_{1,6} (\z_{4} + \D_3^{\top}\z_7))$
      \State $\tiltlr^{(n+1)} \leftarrow P_{B_{1}^{0,\zetal}}(\tiltlr^{(n)} - \gamma_{1,7} (\z_{5} + \D_3^{\top}\z_7))$
      \State $\tiltlt^{(n+1)} \leftarrow P_{B_{1}^{0,\zetal}}(\tiltlt^{(n)} - \gamma_{1,8} (\z_{6} + \D_3^{\top}\z_7))$
      \State $(\tilHr^{\prime}, \tilHt^{\prime}) \leftarrow 2(\tilHr^{(n+1)}, \tilHt^{(n+1)}) - (\tilHr^{(n)}, \tilHt^{(n)})$
      \State $(\tilshr^{\prime},\tilthr^{\prime}) \leftarrow 2(\tilshr^{(n+1)}, \tilthr^{(n+1)}) - (\tilshr^{(n)}, \tilthr^{(n)})$
      \State $(\tilslr^{\prime},\tiltlr^{\prime}) \leftarrow 2(\tilslr^{(n+1)}, \tiltlr^{(n+1)}) - (\tilslr^{(n)}, \tiltlr^{(n)})$
      \State $(\tilslt^{\prime},\tiltlt^{\prime}) \leftarrow 2(\tilslt^{(n+1)}, \tiltlt^{(n+1)}) - (\tilslt^{(n)}, \tiltlt^{(n)})$
      \State $\alpha^{(n)} \leftarrow c_{\alpha}\| \W\D\tilHr^{(n+1)} \|_{q} \cdot \|\Lr-\Lt\|_{1}/N_{l} $
      \State $\z_{1}^{(n+1)} \leftarrow \z_{1}^{(n)} + \gamma_{2,1}\W\D\tilHr^{\prime}$
      \State $\z_{2}^{(n+1)} \leftarrow \z_{2}^{(n)} + \gamma_{2,2}\W\D\tilHt^{\prime}$
      \State $\z_{3}^{(n+1)} \leftarrow \z_{3}^{(n)} + \gamma_{2,3}(\W\D\tilHr^{\prime} - \W\D\tilHt^{\prime})$
      \State $\z_{4}^{(n+1)} \leftarrow \z_{4}^{(n)} + \gamma_{2,4}(\tilHr^{\prime} + \tilshr^{\prime} + \tilshr^{\prime})$
      \State $\z_{5}^{(n+1)} \leftarrow \z_{5}^{(n)} + \gamma_{2,5}(\S\B\tilHr^{\prime} + \tilslr^{\prime} + \tilslr^{\prime})$
      \State $\z_{6}^{(n+1)} \leftarrow \z_{6}^{(n)} + \gamma_{2,6}(\S\B\tilHt^{\prime} + \tiltlt^{\prime} + \tiltlt^{\prime})$
      \State $\z_{7}^{(n+1)} \leftarrow \z_{7}^{(n)} + \gamma_{2,7}\D_3\tilthr^{\prime}$
      \State $\z_{8}^{(n+1)} \leftarrow \z_{8}^{(n)} + \gamma_{2,8}\D_3\tilthr^{\prime}$
      \State $\z_{9}^{(n+1)} \leftarrow \z_{9}^{(n)} + \gamma_{2,9}\D_3\tilthr^{\prime}$
      \State $\z_{1}^{(n+1)} \leftarrow \z_{1}^{(n+1)} - \gamma_{2,1}\prox_{\frac{1}{\gamma_{2,1}}\|\cdot\|_{1,2}}(\frac{1}{\gamma_{2,1}}\z_{1}^{(n+1)})$
      \State $\z_{2}^{(n+1)} \leftarrow \z_{2}^{(n+1)} - \gamma_{2,2}\prox_{\frac{\lambda}{\gamma_{2,2}}\|\cdot\|_{1,2}}(\frac{1}{\gamma_{2,2}}\z_{2}^{(n+1)})$
      \State $\z_{3}^{(n+1)} \leftarrow \z_{3}^{(n+1)} - \gamma_{2,3}P_{B_{q}^{\mathbf{0},\alpha^{(n)}}}(\frac{1}{\gamma_{2,3}}\mathbf{z}_{3}^{(n+1)})$
      \State $\z_{4}^{(n+1)} \leftarrow \z_{4}^{(n+1)} - \gamma_{2,4}P_{B_{2}^{\Hr,\epsh}}(\frac{1}{\gamma_{2,4}}\mathbf{z}_{4}^{(n+1)})$
      \State $\z_{5}^{(n+1)} \leftarrow \z_{5}^{(n+1)} - \gamma_{2,5}P_{B_{2}^{\Lr,\epsl}}(\frac{1}{\gamma_{2,5}}\mathbf{z}_{5}^{(n+1)})$
      \State $\z_{6}^{(n+1)} \leftarrow \z_{6}^{(n+1)} - \gamma_{2,6}P_{B_{2}^{\Lt,\epsl}}(\frac{1}{\gamma_{2,6}}\mathbf{z}_{6}^{(n+1)})$
      \State $n\leftarrow n+1$
      \EndWhile
  \end{algorithmic}
\end{algorithm}

\section{Experiments}
\label{sec:exp}
\begin{figure*}[t]
\begin{center}
  \begin{minipage}{0.01\hsize}
    \centerline{}
  \end{minipage}
  \begin{minipage}{0.155\hsize}
    \centerline{Site1}
  \end{minipage}
  \begin{minipage}{0.155\hsize}
    \centerline{Site2}
  \end{minipage}
  \begin{minipage}{0.155\hsize}
    \centerline{Site3}
  \end{minipage}
  \begin{minipage}{0.155\hsize}
    \centerline{Site4}
  \end{minipage}
  \begin{minipage}{0.155\hsize}
    \centerline{Site5}
  \end{minipage}
  \begin{minipage}{0.155\hsize}
    \centerline{Site6}
  \end{minipage} \par

  \begin{minipage}{0.01\hsize}
    \centerline{}
  \end{minipage}
  \begin{minipage}{0.075\hsize}
    \centerline{\raisebox{0.6ex}{reference}}
  \end{minipage}
  \begin{minipage}{0.075\hsize}
    \centerline{target}
  \end{minipage}
  \begin{minipage}{0.075\hsize}
    \centerline{\raisebox{0.6ex}{reference}}
  \end{minipage}
  \begin{minipage}{0.075\hsize}
    \centerline{target}
  \end{minipage}
  \begin{minipage}{0.075\hsize}
    \centerline{\raisebox{0.6ex}{reference}}
  \end{minipage}
  \begin{minipage}{0.075\hsize}
    \centerline{target}
  \end{minipage}
  \begin{minipage}{0.075\hsize}
    \centerline{\raisebox{0.6ex}{reference}}
  \end{minipage}
  \begin{minipage}{0.075\hsize}
    \centerline{target}
  \end{minipage}
  \begin{minipage}{0.075\hsize}
    \centerline{\raisebox{0.6ex}{reference}}
  \end{minipage}
  \begin{minipage}{0.075\hsize}
    \centerline{target}
  \end{minipage}
  \begin{minipage}{0.075\hsize}
    \centerline{\raisebox{0.6ex}{reference}}
  \end{minipage}
  \begin{minipage}{0.075\hsize}
    \centerline{target}
  \end{minipage} \par

  \begin{minipage}{0.01\hsize}
    \centerline{\rotatebox{90}{HR}}
  \end{minipage}
  \begin{minipage}{0.075\hsize}
    \centerline{\includegraphics[width=\hsize]{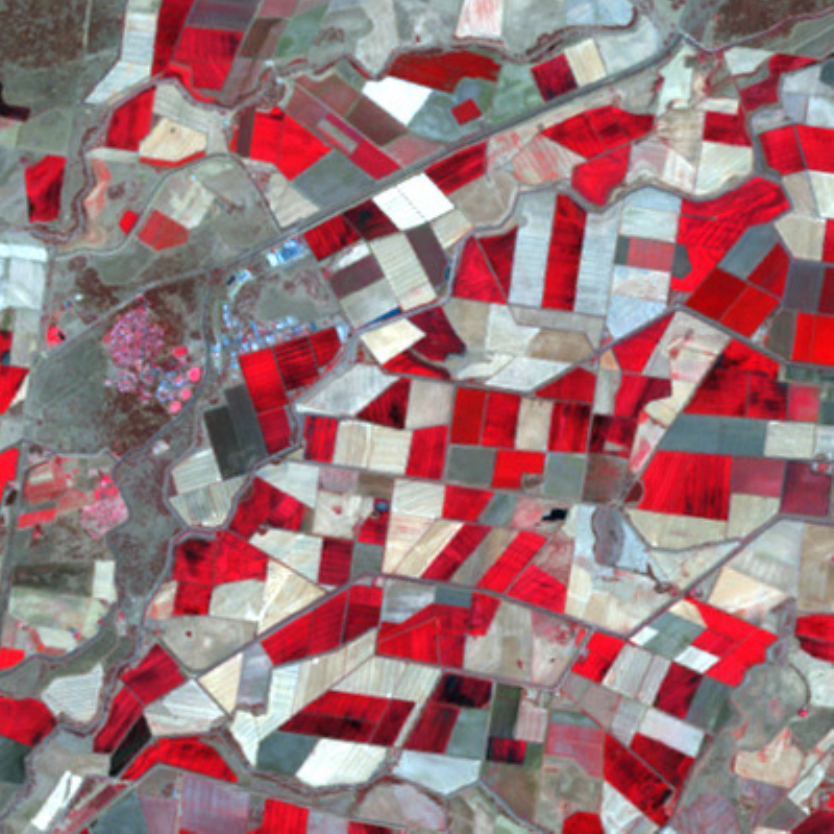}}
  \end{minipage}
  \begin{minipage}{0.075\hsize}
    \centerline{\includegraphics[width=\hsize]{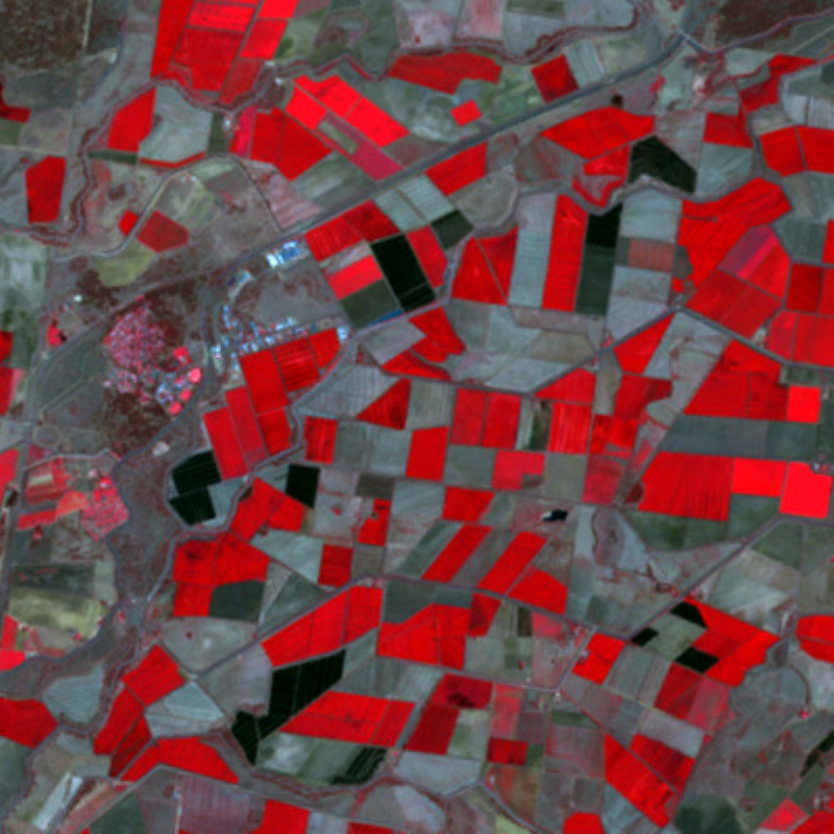}}
  \end{minipage}
  \begin{minipage}{0.075\hsize}
    \centerline{\includegraphics[width=\hsize]{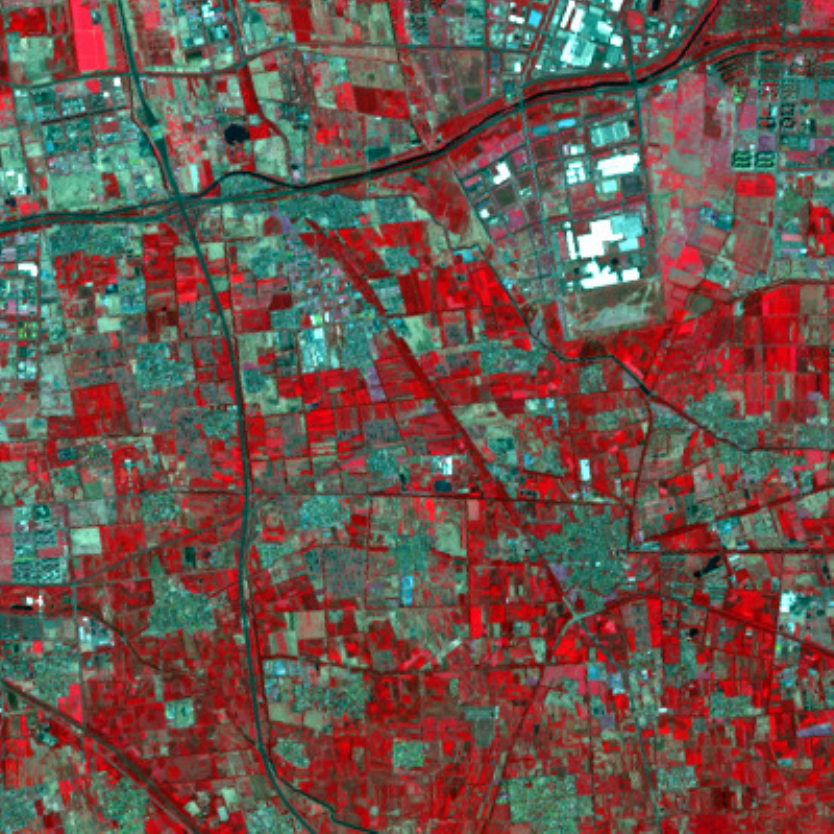}}
  \end{minipage}
  \begin{minipage}{0.075\hsize}
    \centerline{\includegraphics[width=\hsize]{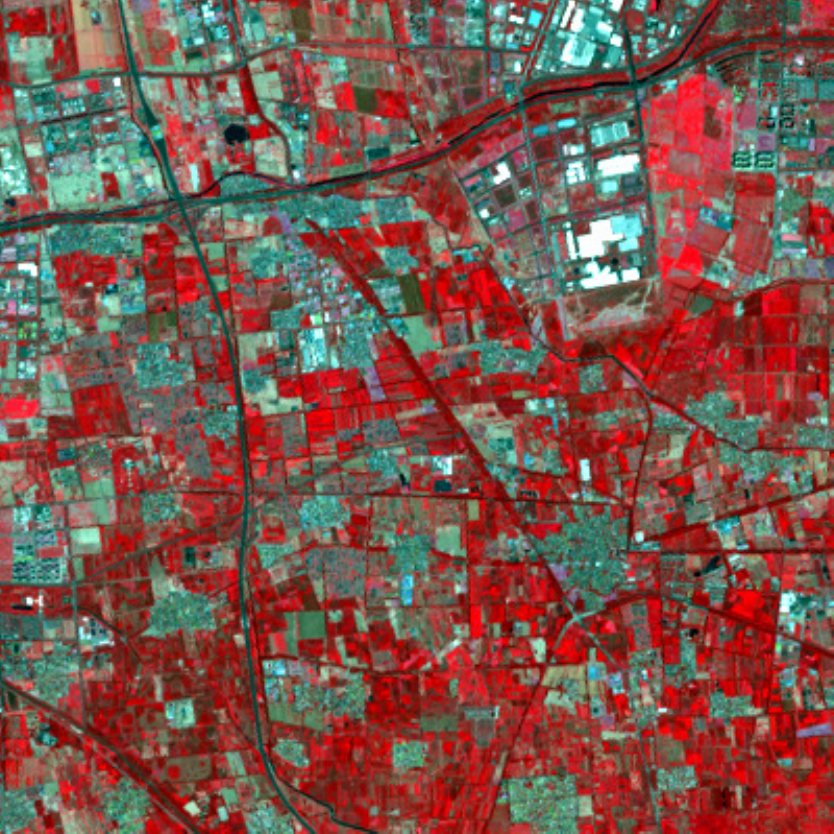}}
  \end{minipage}
  \begin{minipage}{0.075\hsize}
    \centerline{\includegraphics[width=\hsize]{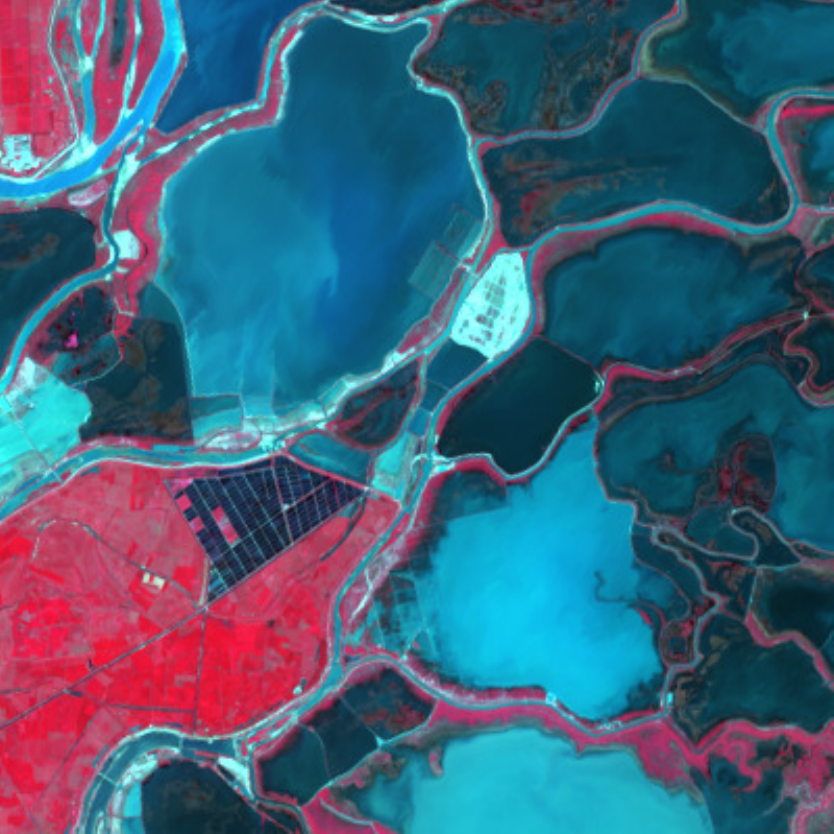}}
  \end{minipage}
  \begin{minipage}{0.075\hsize}
    \centerline{\includegraphics[width=\hsize]{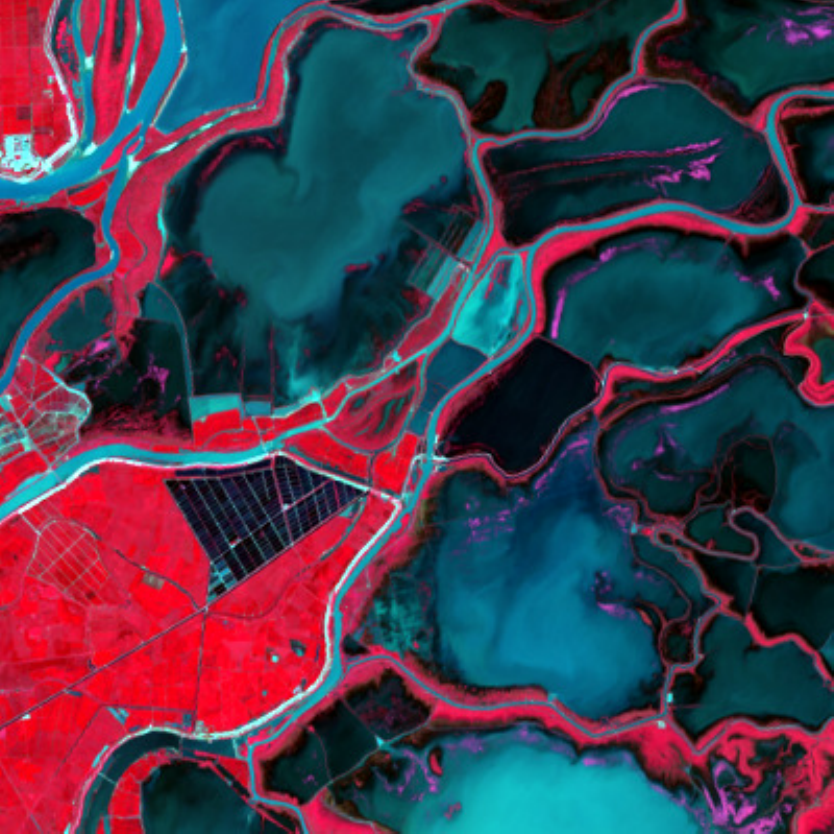}}
  \end{minipage}
  \begin{minipage}{0.075\hsize}
    \centerline{\includegraphics[width=\hsize]{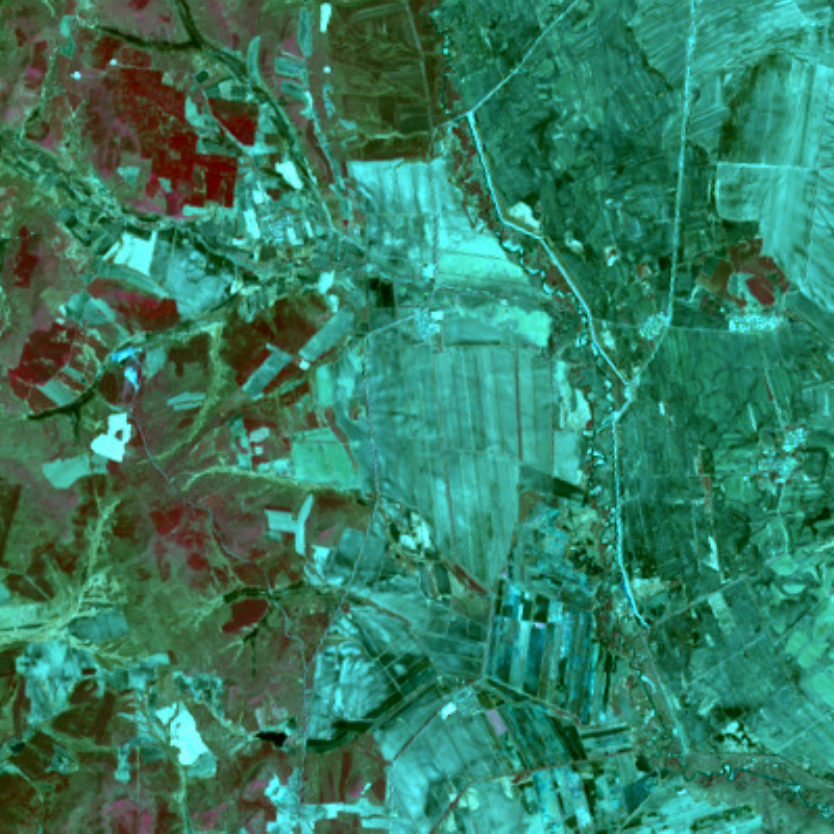}}
  \end{minipage}
  \begin{minipage}{0.075\hsize}
    \centerline{\includegraphics[width=\hsize]{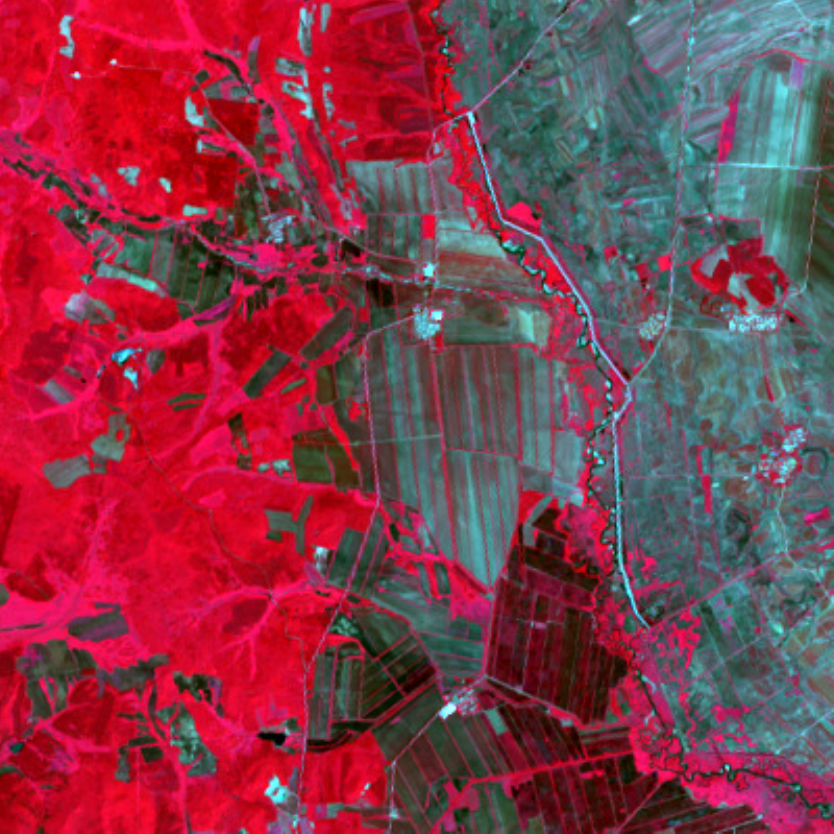}}
  \end{minipage}
  \begin{minipage}{0.075\hsize}
    \centerline{\includegraphics[width=\hsize]{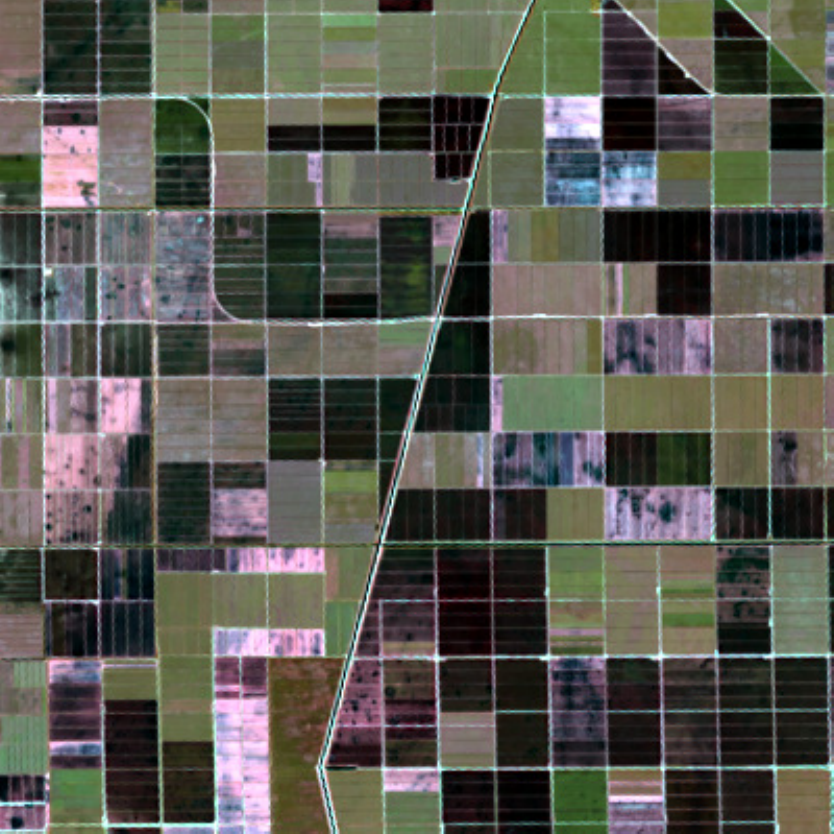}}
  \end{minipage}
  \begin{minipage}{0.075\hsize}
    \centerline{\includegraphics[width=\hsize]{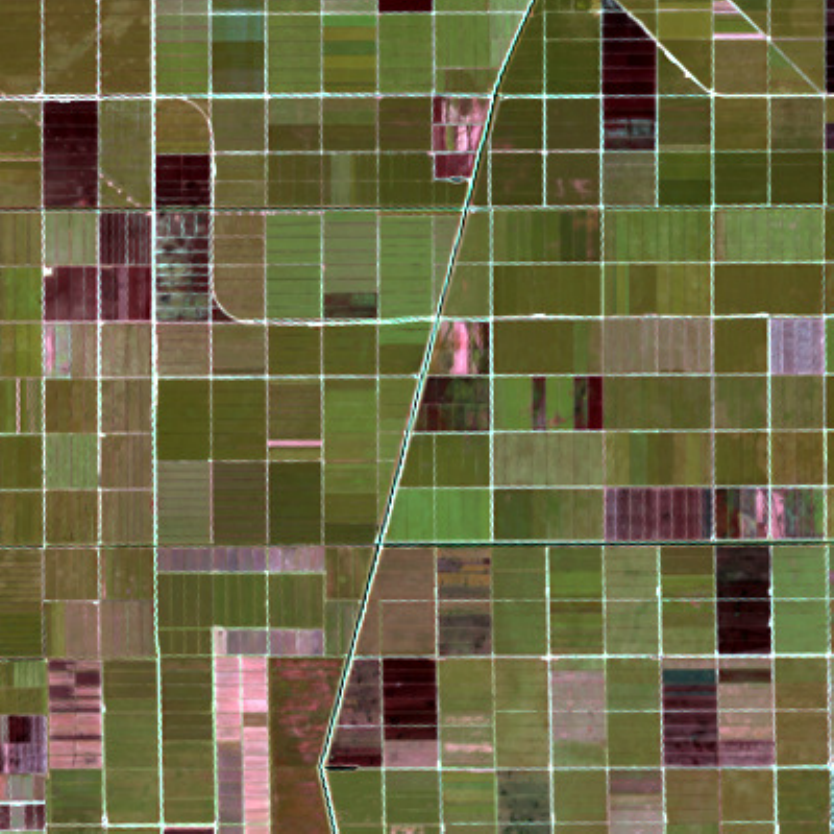}}
  \end{minipage} 
  \begin{minipage}{0.075\hsize}
    \centerline{\includegraphics[width=\hsize]{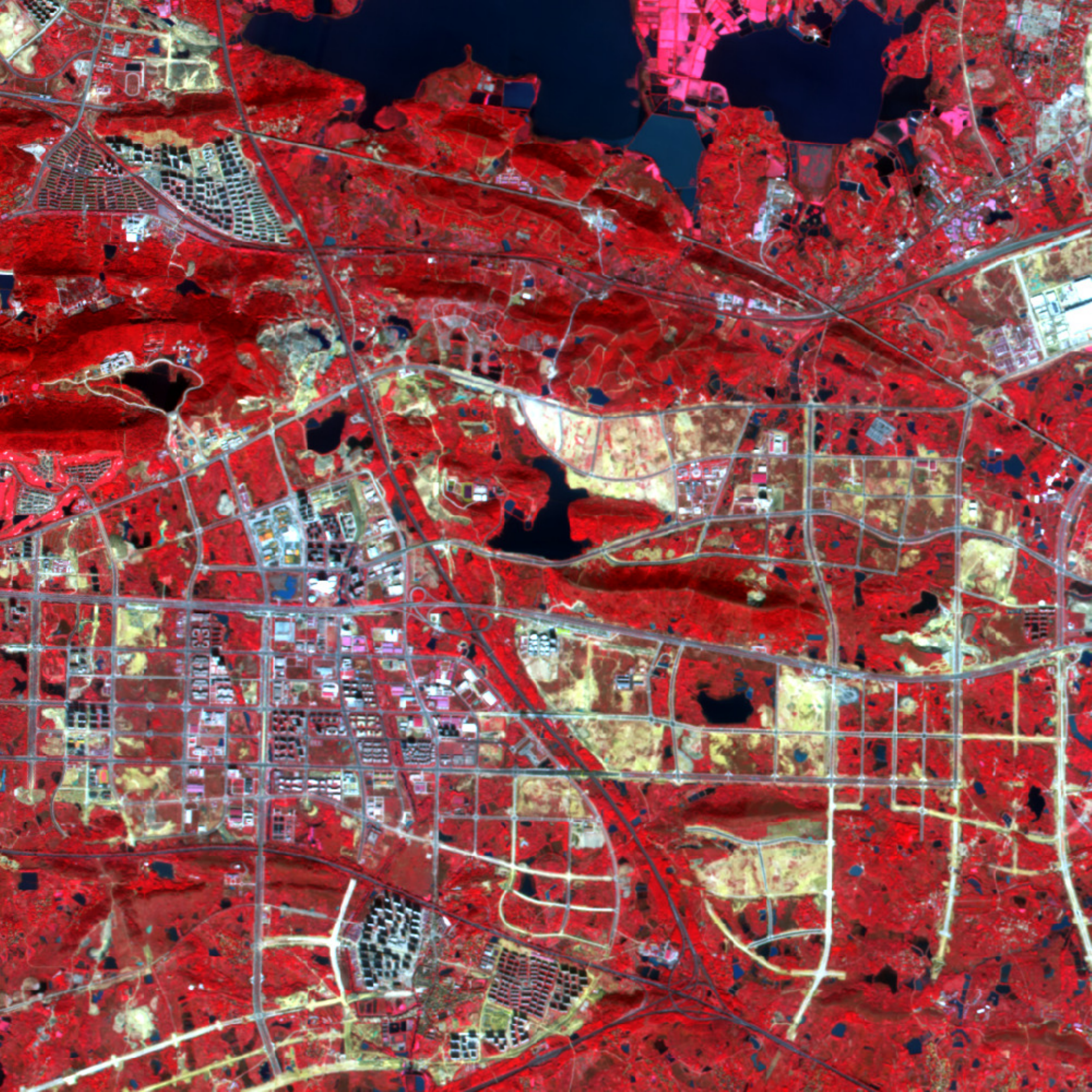}}
  \end{minipage}
  \begin{minipage}{0.075\hsize}
    \centerline{\includegraphics[width=\hsize]{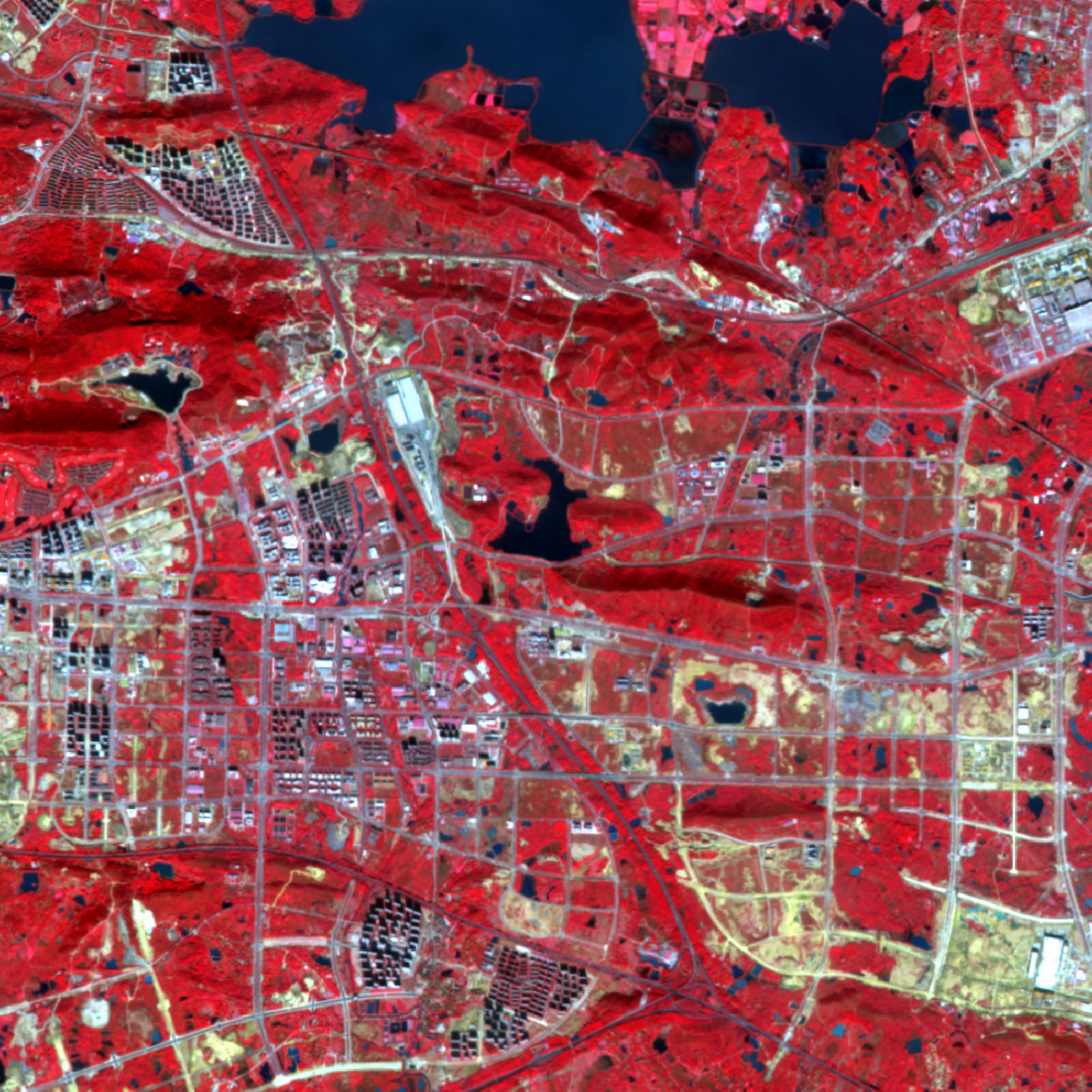}}
  \end{minipage}\par

  \vspace{1mm}

  \begin{minipage}{0.01\hsize}
    \centerline{\rotatebox{90}{LR}}
  \end{minipage}
  \begin{minipage}{0.075\hsize}
    \centerline{\includegraphics[width=\hsize]{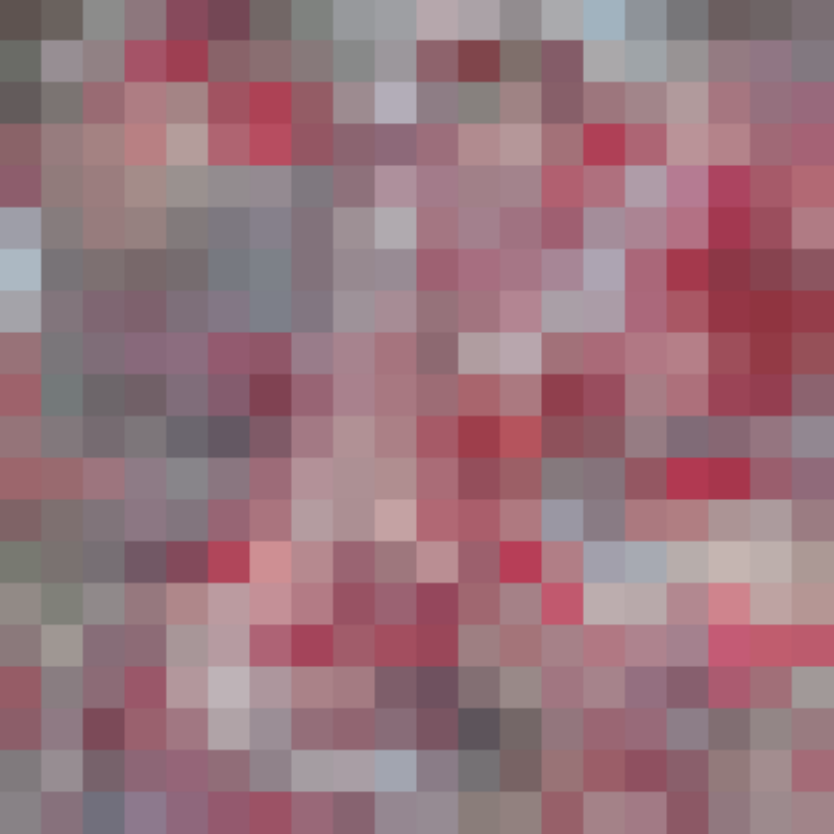}}
  \end{minipage}
  \begin{minipage}{0.075\hsize}
    \centerline{\includegraphics[width=\hsize]{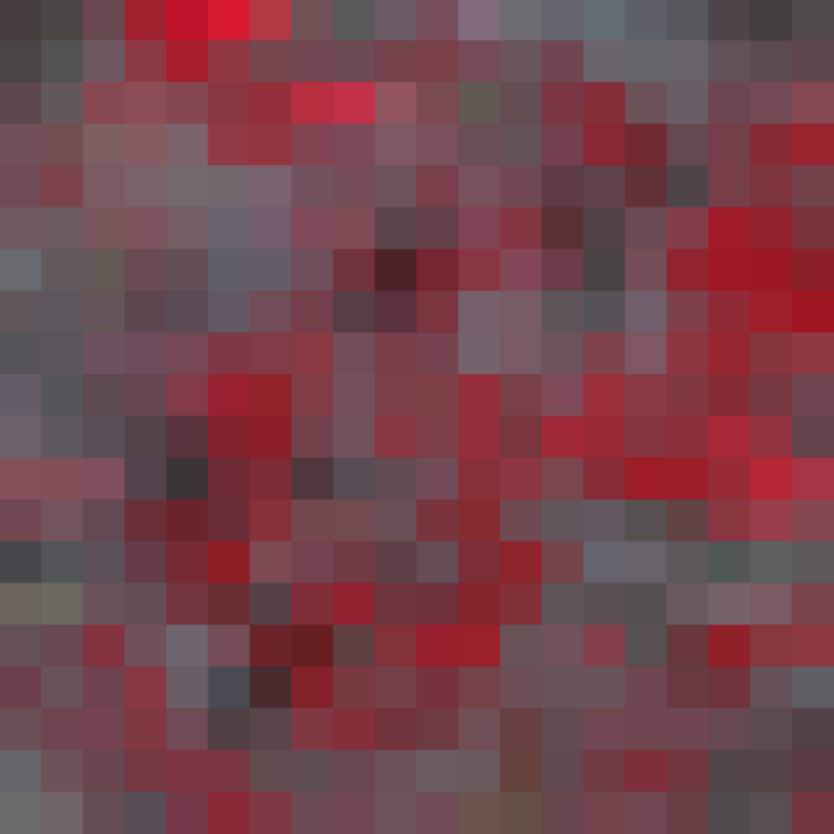}}
  \end{minipage}
  \begin{minipage}{0.075\hsize}
    \centerline{\includegraphics[width=\hsize]{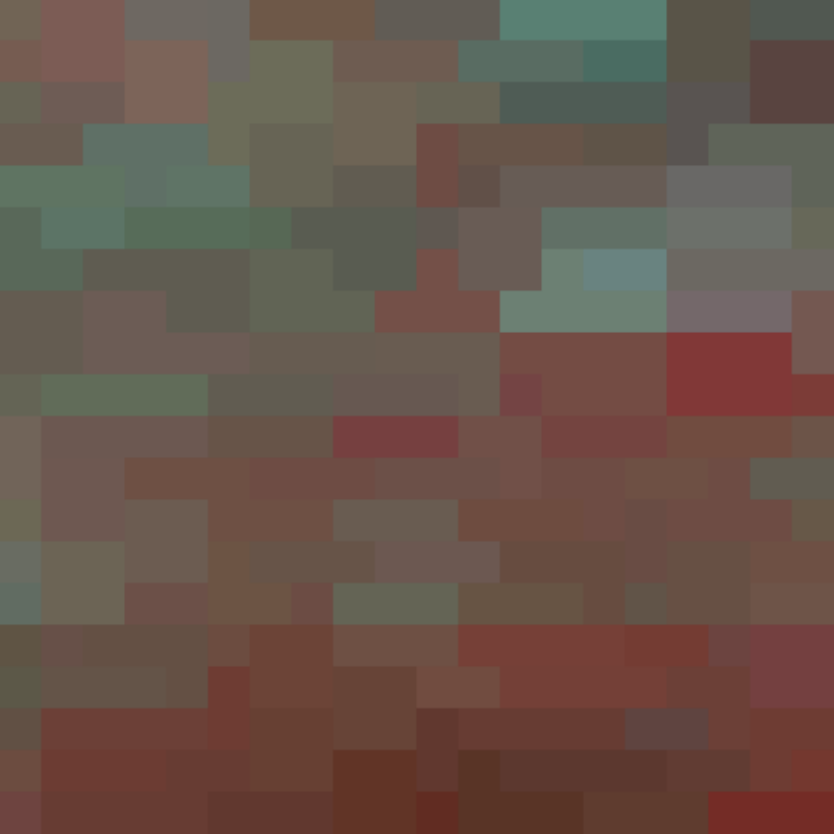}}
  \end{minipage}
  \begin{minipage}{0.075\hsize}
    \centerline{\includegraphics[width=\hsize]{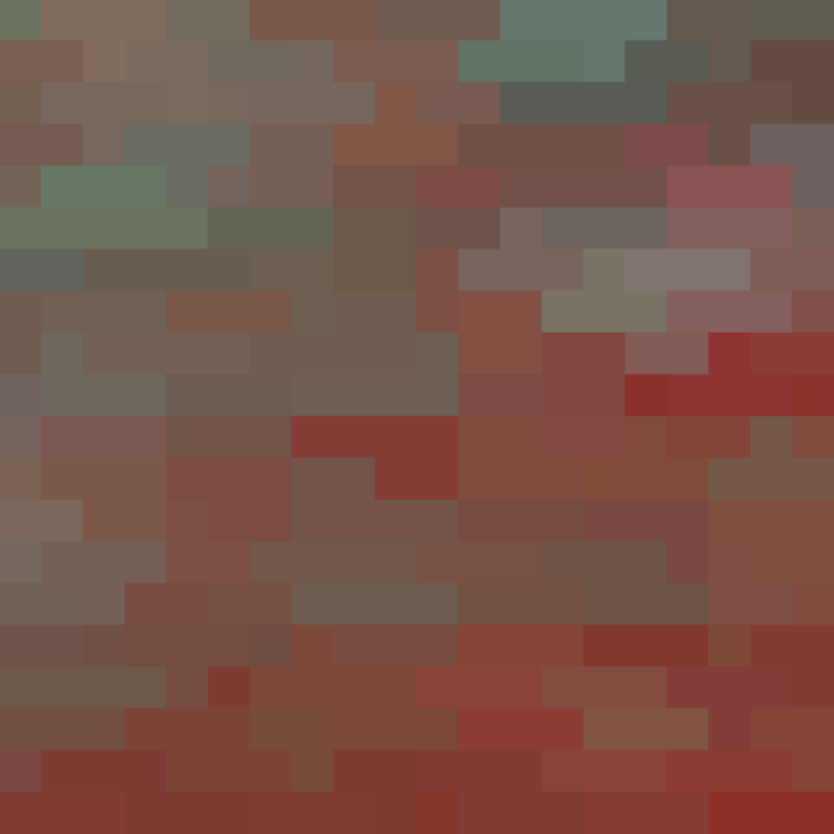}}
  \end{minipage}
  \begin{minipage}{0.075\hsize}
    \centerline{\includegraphics[width=\hsize]{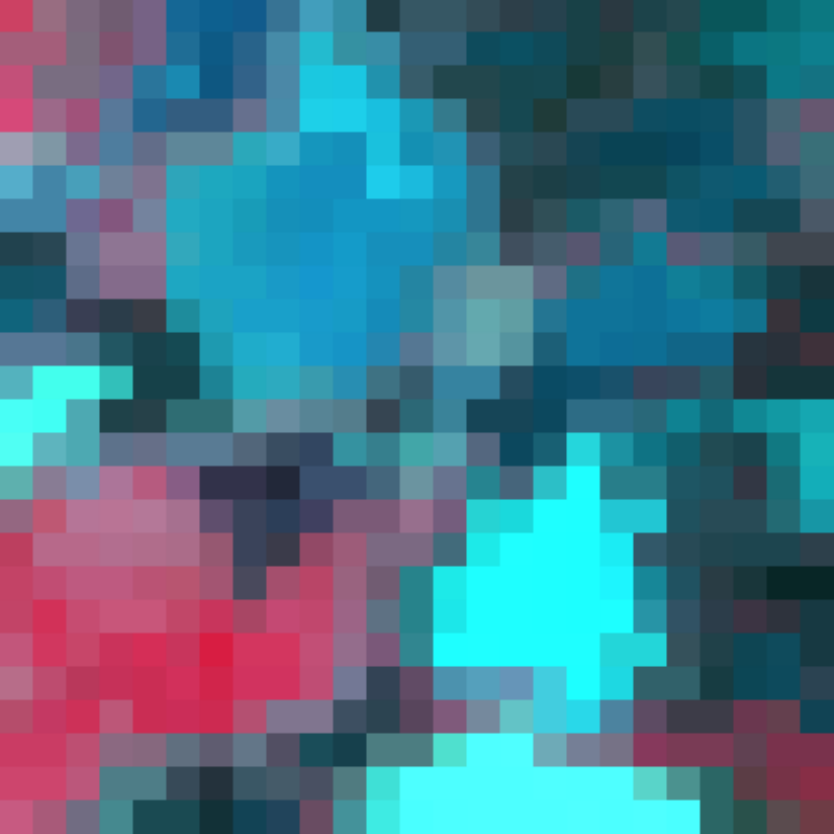}}
  \end{minipage}
  \begin{minipage}{0.075\hsize}
    \centerline{\includegraphics[width=\hsize]{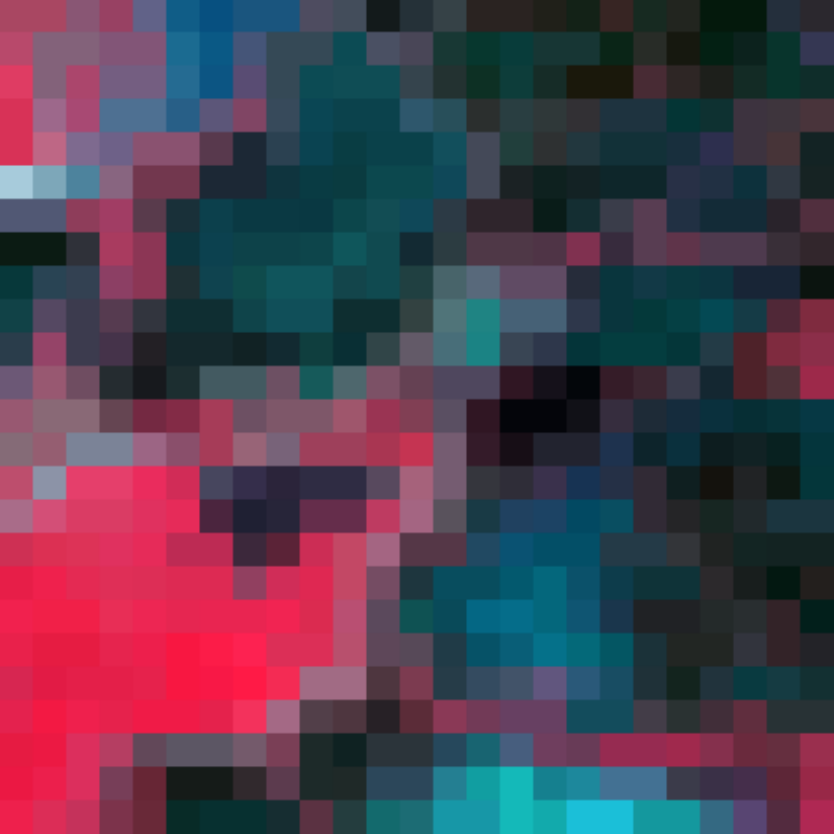}}
  \end{minipage}
  \begin{minipage}{0.075\hsize}
    \centerline{\includegraphics[width=\hsize]{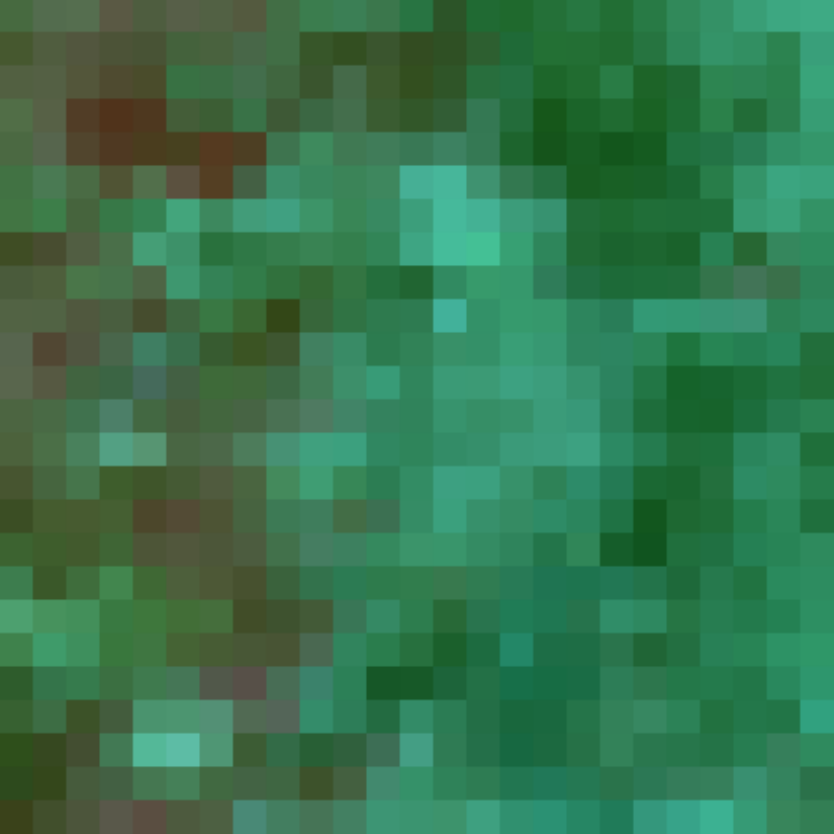}}
  \end{minipage}
  \begin{minipage}{0.075\hsize}
    \centerline{\includegraphics[width=\hsize]{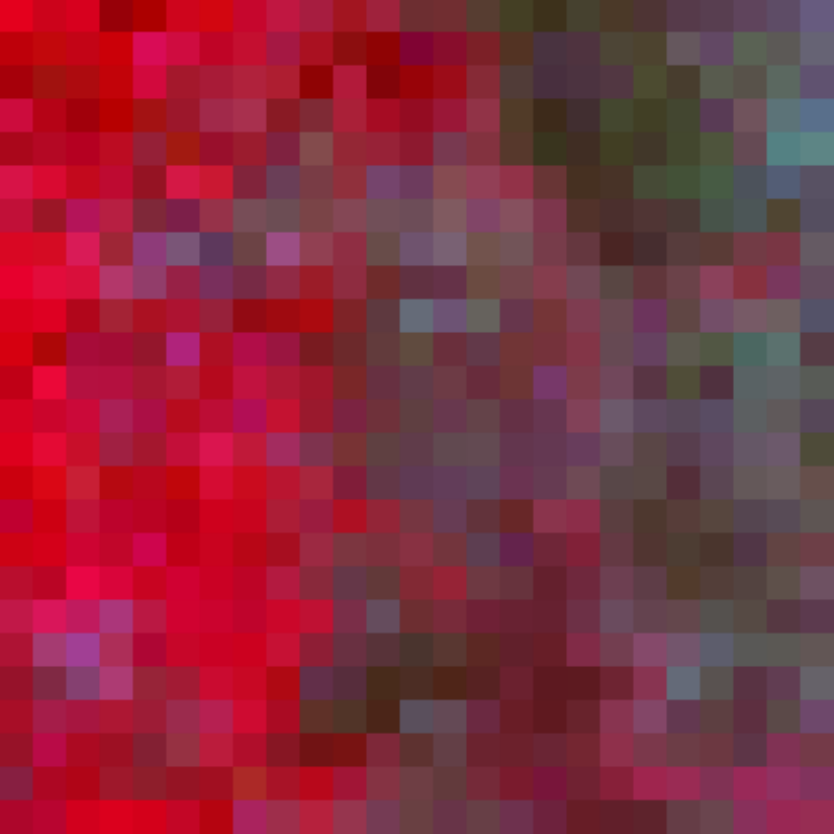}}
  \end{minipage}
  \begin{minipage}{0.075\hsize}
    \centerline{\includegraphics[width=\hsize]{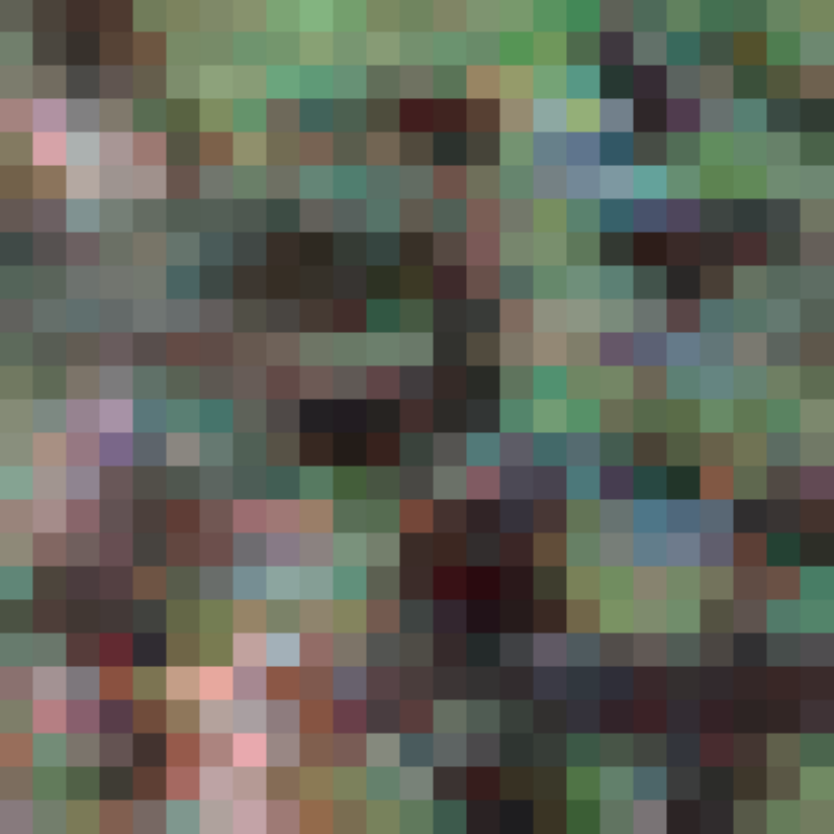}}
  \end{minipage}
  \begin{minipage}{0.075\hsize}
    \centerline{\includegraphics[width=\hsize]{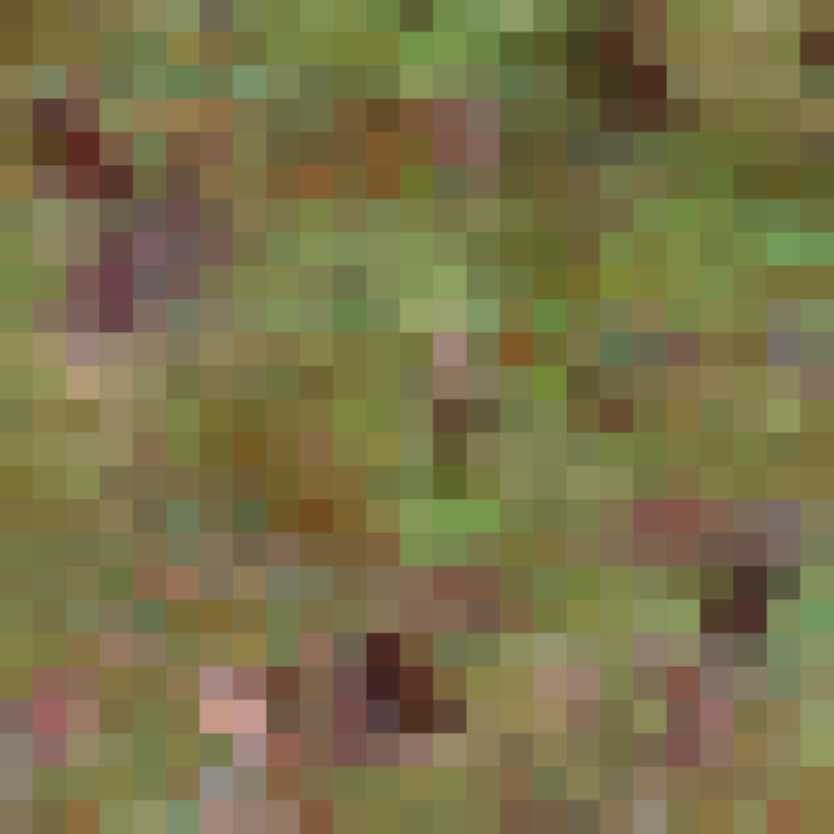}}
  \end{minipage}
  \begin{minipage}{0.075\hsize}
    \centerline{\includegraphics[width=\hsize]{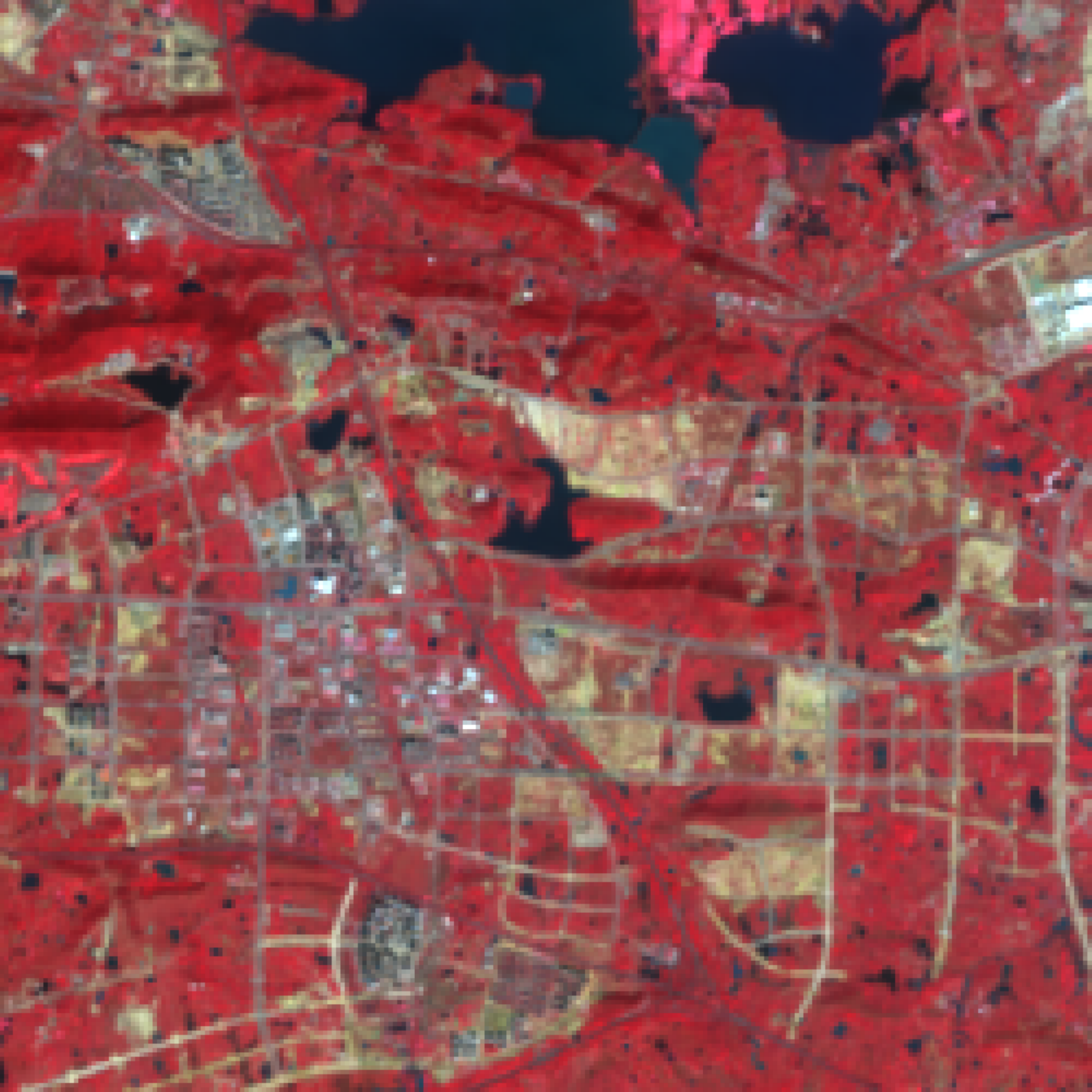}}
  \end{minipage}
  \begin{minipage}{0.075\hsize}
    \centerline{\includegraphics[width=\hsize]{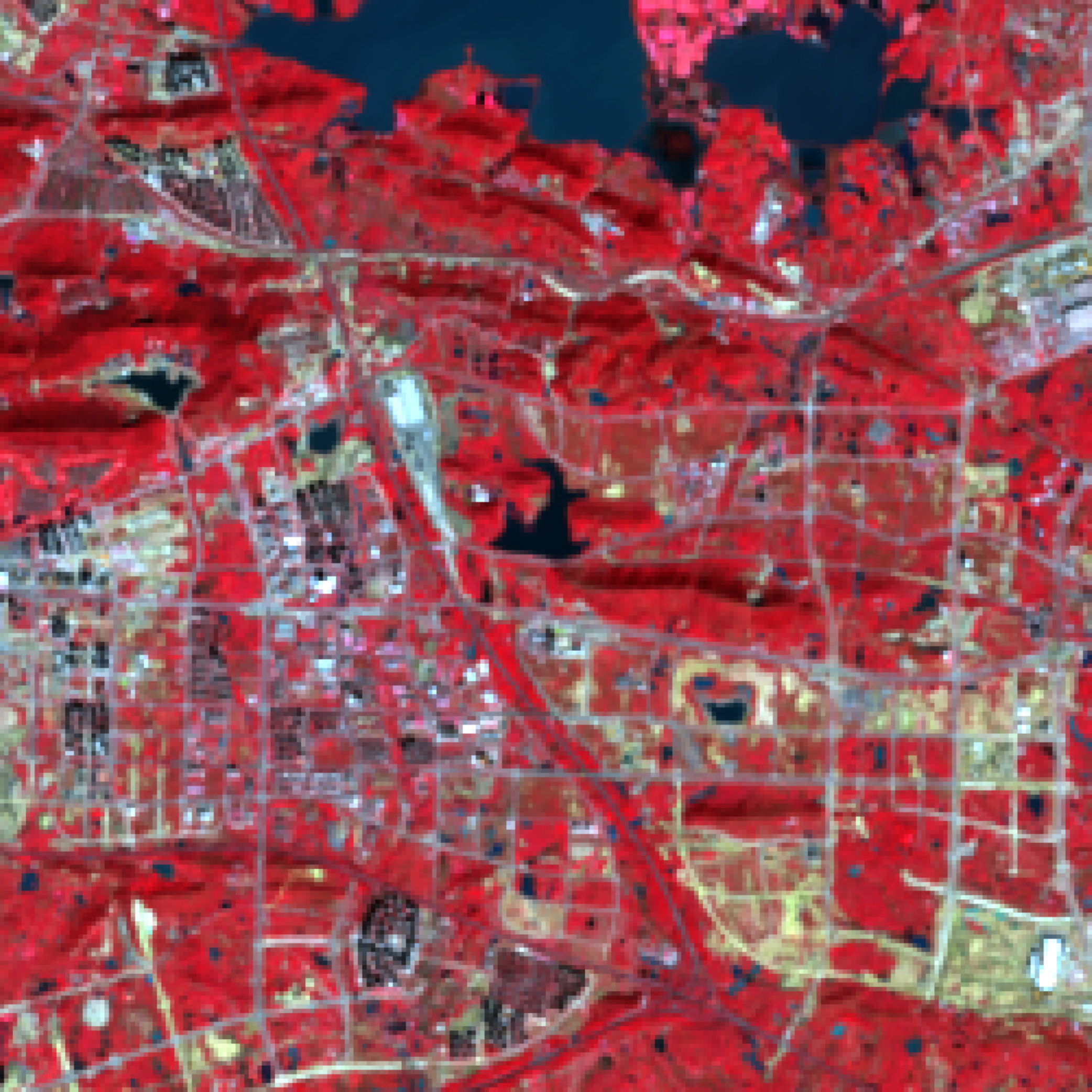}}
  \end{minipage}
\end{center}
  \caption{Visualization of the real HR and LR satellite images used for six experimental sites (Site1–Site6). For each site, reference and target images are shown side-by-side.}
  \label{fig: dataset}
\end{figure*}

We demonstrate the effectiveness of our ST fusion method, TSSTF, through comprehensive experiments using simulated and real data for six sites.
Our experiments aim to verify the following three items.
\begin{itemize}
    \item TSSTF is as effective as state-of-the-art ST fusion methods in noiseless cases and outperforms them in noisy cases. We conducted comparative experiments on nine cases of noise contamination. The experimental results for simulated data and real data are presented in Sec.~\ref{ssec: semisim results} and Sec.~\ref{ssec: real results}, respectively.
    \item The parameters of TSSTF can be set independently of target regions or noise levels. As detailed in Sec.~\ref{ssec:facilitation of parameter selection}, we extensively varied each parameter and found a recommended value that yields consistently high performance.
    \item The algorithm converges stably. We analyze the behavior of the TSSTF algorithm in Sec.~\ref{ssec: algorithm convergence}, demonstrating that the adaptive update of $\alpha^{(n)}$ stabilizes during optimization and that the algorithm converges without explicitly fixing $\alpha$ at an intermediate iteration.
\end{itemize}

\begin{table}[t]
\centering
\caption{Noise Cases}
\label{table: noise cases}
\vspace{-2mm}
\scalebox{0.9}{
    \begin{tabular}{lccccc}
    \toprule
    & & Gaussian & Salt-and-pepper & Stripe & Poisson\\
    \midrule
    \multirow{2}{*}{Case1}
    & HR & -- & -- & -- & -- \\
    & LR & -- & -- & -- & -- \\
    \midrule
    \multirow{2}{*}{Case2}
    & HR & $\sigmah = 0.05$ & -- & -- & -- \\
    & LR & -- & -- & -- & -- \\
    \midrule
    \multirow{2}{*}{Case3}
    & HR & $\sigmah = 0.05$ & -- & -- & -- \\
    & LR & $\sigmal = 0.01$ & -- & -- & -- \\
    \midrule
    \multirow{2}{*}{Case4}
    & HR & $\sigmah = 0.05$ & $\rh = 0.05$ & -- & -- \\
    & LR & -- & -- & -- & -- \\
    \midrule
    \multirow{2}{*}{Case5}
    & HR & $\sigmah = 0.05$ & $\rh = 0.05$ & -- & -- \\
    & LR & $\sigmal = 0.01$ & $\rl = 0.01$ & -- & -- \\
    \midrule
    \multirow{2}{*}{Case6}
    & HR & $\sigmah = 0.05$ & -- & $\sh = 0.05$ & -- \\
    & LR & -- & -- & -- & -- \\
    \midrule
    \multirow{2}{*}{Case7}
    & HR & $\sigmah = 0.05$ & -- & $\sh = 0.05$ & -- \\
    & LR & $\sigmal = 0.01$ & -- & $\sl = 0.01$ & -- \\
    \midrule
    \multirow{2}{*}{Case8}
    & HR & $\sigmah = 0.05$ & -- & -- & $\eh = 200$ \\
    & LR & -- & -- & -- & -- \\
    \midrule
    \multirow{2}{*}{Case9}
    & HR & $\sigmah = 0.05$ & -- & -- & $\eh = 200$ \\
    & LR & $\sigmal = 0.01$ & -- & -- & $\el = 800$ \\
    \bottomrule
    \end{tabular}
}
\end{table}

\subsection{Experimental Setup}
\label{ssec:experimental setup}

\subsubsection{Data Description}
\label{ssec:data description}
We evaluated our method using real satellite images from six different geographical sites, which we denote as Site1 to Site6. For Sites1–5, the high-resolution (HR) and low-resolution (LR) images are acquired by Landsat and MODIS, respectively, whereas for Site6 they are acquired by Gaofen and Landsat. The paired images at each site are shown in Fig.~\ref{fig: dataset}. The data for Site1, Site2, Sites3–5, and Site6 are provided in \cite{Site1data}, \cite{Site2data}, \cite{Site345data}, and \cite{Site6data}, respectively.

We also conducted experiments on simulated data. In the case of real satellite observations, radiometric and geometric inconsistencies exist between different sensors. This means that the pure fusion capability of each method cannot be accurately evaluated in experiments using real data because these inconsistencies affect performance, as also addressed in \cite{FSDAF}. Therefore, we additionally conducted experiments on simulated data generated based on our observation model. Specifically, the simulated LR images were generated from the corresponding real HR images using the model in \eqref{eq: relationship model} with no modelling error ($\mathbf{m}=\mathbf{0}$), while the real HR images were used as HR images.

\begin{table}[t]
	\begin{center}
		\caption{The Parameter Setting}
		\label{table: parameters}
            \vspace{-2mm}
		\begin{tabular}{cc}
			\toprule
            Parameter & Value / Setting \\
            \midrule
			\vspace{1mm}
            $\delta$ 
            &  0.1
            \vspace{1mm} \\
            $k$ 
            &  2
            \vspace{1mm} \\
			$c_\alpha$ 
            & 5 
            \vspace{1mm} \\
            $q$ 
            & the mixed $\ell_{1,2}$-norm
            \vspace{1mm} \\
            $ \lambda $ 
            & 1
            \vspace{1mm} \\
            $\beta_b$ 
            &$\left | \frac{1}{\Nl}\mathbf{1}^{\top}[\mathbf{l}_{r}]_{b}- \frac{1}{\Nh}\mathbf{1}^{\top}[\mathbf{h}_{r}]_{b} \right |$
            \vspace{1mm} \\
            $\epsh$ (w/o Poisson noise) 
            & $0.98 \sqrt{\sigma_h^2 \Nh B (1 - r_h)}$
            \vspace{1mm} \\
            $\epsh$ (w/ Poisson noise) 
            & $0.98 \sqrt{(\frac{1}{e_h}\mathbf{1}^{\top} \Hr + \sigma_h^2 \Nh B) (1 - r_h)}$
            \vspace{1mm} \\
            $\epsl$ 
            & $\| \Lr - \mathbf{S}\mathbf{B}\Hr\|_2$
            \vspace{1mm} \\
            $\etah$ 
            & $0.49 \Nh B r_h$
            \vspace{1mm} \\
            $\etal$ 
            & $0.49 \Nl B r_l$
            \vspace{1mm} \\
            $\zetah$
            & $0.49 \times 0.2 \Nh B s_h$
            \vspace{1mm} \\
            $\zetal$
            & $0.49 \times 0.2 \Nl B s_l$
            \vspace{1mm} \\
            \bottomrule
		\end{tabular}
	\end{center}
\end{table}
\begin{table*}[t]
    \begin{center}
    \caption{The PSNR, MSSIM, CC, and SAM Results in the Experiments With Simulated Data}
    \label{table: simulated metrics}
    \vspace{-2mm}
    \centering
        \begin{tabular}{p{0.7cm} p{0.7cm} >{\centering\arraybackslash}p{1.3cm} >{\centering\arraybackslash}p{1.3cm} >{\centering\arraybackslash}p{1.3cm} >{\centering\arraybackslash}p{1.3cm} >{\centering\arraybackslash}p{1.3cm} >{\centering\arraybackslash}p{1.3cm} >{\centering\arraybackslash}p{1.3cm} >{\centering\arraybackslash}p{1.3cm} >{\centering\arraybackslash}p{1.3cm}}
            \toprule
            Case & Metric & STARFM & VIPSTF & RSFN-1 & RSFN-2 & RobOt & SwinSTFM & ROSTF & ECPW & TSSTF \\
            \midrule
            \multirow{4}{*}{Case1} & PSNR & \uline{28.57} & 28.47 & 23.83 & 25.34 & 28.47 & 25.67 & 28.16 & 24.88 & \textbf{29.83} \\
            & MSSIM & 0.7608 & \uline{0.7683} & 0.6986 & 0.7270 & 0.7664 & 0.6987 & 0.7599 & 0.7035 & \textbf{0.7931} \\
            & CC & 0.9422 & \uline{0.9463} & 0.8365 & 0.8819 & 0.9352 & 0.8960 & 0.9326 & 0.8416 & \textbf{0.9510} \\
            & SAM & \textbf{0.1146} & 0.1202 & 0.2351 & 0.2156 & \uline{0.1173} & 0.1668 & 0.1261 & 0.2234 & 0.1223 \\
            \midrule
            \multirow{4}{*}{Case2} & PSNR & 23.50 & 25.25 & 23.71 & 25.30 & 23.48 & 25.53 & \uline{28.30} & 24.77 & \textbf{29.21} \\
            & MSSIM & 0.3679 & 0.5031 & 0.6763 & 0.7198 & 0.3806 & 0.6686 & \uline{0.7396} & 0.6821 & \textbf{0.7507} \\
            & CC & 0.8918 & 0.9104 & 0.8343 & 0.8815 & 0.8884 & 0.8928 & \uline{0.9367} & 0.8405 & \textbf{0.9480} \\
            & SAM & 0.2421 & 0.2029 & 0.2387 & 0.2173 & 0.2405 & 0.1725 & \textbf{0.1284} & 0.2276 & \uline{0.1296} \\
            \midrule
            \multirow{4}{*}{Case3} & PSNR & 23.47 & 25.43 & 23.71 & 25.29 & 23.41 & 25.49 & \uline{28.19} & 24.78 & \textbf{28.99} \\
            & MSSIM & 0.3669 & 0.5173 & 0.6761 & 0.7197 & 0.3799 & 0.6676 & \uline{0.7382} & 0.6826 & \textbf{0.7467} \\
            & CC & 0.8913 & 0.9123 & 0.8342 & 0.8813 & 0.8876 & 0.8926 & \uline{0.9397} & 0.8404 & \textbf{0.9462} \\
            & SAM & 0.2432 & 0.1991 & 0.2390 & 0.2173 & 0.2430 & 0.1732 & \textbf{0.1312} & 0.2278 & \uline{0.1335} \\
            \midrule
            \multirow{4}{*}{Case4} & PSNR & 16.36 & 19.73 & 22.71 & 24.89 & 16.79 & 24.44 & \uline{25.69} & 24.26 & \textbf{27.85} \\
            & MSSIM & 0.1614 & 0.2811 & 0.5660 & 0.6782 & 0.1774 & 0.5435 & \uline{0.6867} & 0.5983 & \textbf{0.7142} \\
            & CC & 0.6656 & 0.7503 & 0.8169 & 0.8755 & 0.6825 & 0.8712 & \uline{0.8933} & 0.8291 & \textbf{0.9327} \\
            & SAM & 0.3385 & 0.2745 & 0.2608 & 0.2278 & 0.3291 & 0.2081 & \uline{0.1799} & 0.2429 & \textbf{0.1454} \\
            \midrule
            \multirow{4}{*}{Case5} & PSNR & 16.12 & 22.74 & 22.48 & 24.61 & 16.37 & 23.76 & \uline{25.64} & 24.24 & \textbf{25.90} \\
            & MSSIM & 0.1541 & 0.4424 & 0.5637 & 0.6750 & 0.1746 & 0.5294 & \uline{0.6865} & 0.5974 & \textbf{0.6877} \\
            & CC & 0.6552 & 0.8479 & 0.8100 & 0.8709 & 0.6609 & 0.8577 & \uline{0.8936} & 0.8283 & \textbf{0.9016} \\
            & SAM & 0.3632 & 0.2254 & 0.2620 & 0.2292 & 0.3534 & 0.2189 & \textbf{0.1794} & 0.2442 & \uline{0.1877} \\
            \midrule
            \multirow{4}{*}{Case6} & PSNR & 22.94 & 24.87 & 23.64 & 25.29 & 22.93 & 25.41 & \uline{28.06} & 24.71 & \textbf{29.04} \\
            & MSSIM & 0.3464 & 0.4837 & 0.6669 & \uline{0.7170} & 0.3604 & 0.6579 & 0.7161 & 0.6725 & \textbf{0.7451} \\
            & CC & 0.8813 & 0.9029 & 0.8335 & 0.8814 & 0.8780 & 0.8909 & \uline{0.9325} & 0.8390 & \textbf{0.9464} \\
            & SAM & 0.2564 & 0.2129 & 0.2408 & 0.2179 & 0.2537 & 0.1759 & \uline{0.1330} & 0.2305 & \textbf{0.1313} \\
            \midrule
            \multirow{4}{*}{Case7} & PSNR & 22.81 & 25.05 & 23.61 & 25.25 & 22.71 & 25.24 & \uline{27.42} & 24.69 & \textbf{28.39} \\
            & MSSIM & 0.3446 & 0.5054 & 0.6666 & \uline{0.7163} & 0.3594 & 0.6546 & 0.7049 & 0.6730 & \textbf{0.7364} \\
            & CC & 0.8784 & 0.9090 & 0.8325 & 0.8807 & 0.8726 & 0.8893 & \uline{0.9255} & 0.8382 & \textbf{0.9398} \\
            & SAM & 0.2597 & 0.2056 & 0.2413 & 0.2184 & 0.2614 & 0.1777 & \uline{0.1426} & 0.2305 & \textbf{0.1418} \\
            \midrule
            \multirow{4}{*}{Case8} & PSNR & 22.73 & 24.69 & 23.70 & 25.29 & 22.75 & 25.47 & \uline{28.27} & 24.76 & \textbf{29.09} \\
            & MSSIM & 0.3202 & 0.4613 & 0.6683 & 0.7173 & 0.3336 & 0.6561 & \uline{0.7337} & 0.6758 & \textbf{0.7431} \\
            & CC & 0.8747 & 0.8980 & 0.8338 & 0.8814 & 0.8722 & 0.8914 & \uline{0.9364} & 0.8398 & \textbf{0.9466} \\
            & SAM & 0.2652 & 0.2188 & 0.2396 & 0.2177 & 0.2622 & 0.1744 & \textbf{0.1294} & 0.2285 & \uline{0.1311} \\
            \midrule
            \multirow{4}{*}{Case9} & PSNR & 22.64 & 25.18 & 23.68 & 25.26 & 22.54 & 25.38 & \uline{27.99} & 24.75 & \textbf{28.61} \\
            & MSSIM & 0.3179 & 0.4983 & 0.6683 & 0.7170 & 0.3319 & 0.6537 & \uline{0.7301} & 0.6761 & \textbf{0.7344} \\
            & CC & 0.8729 & 0.9048 & 0.8333 & 0.8807 & 0.8676 & 0.8902 & \uline{0.9378} & 0.8396 & \textbf{0.9422} \\
            & SAM & 0.2677 & 0.2066 & 0.2397 & 0.2177 & 0.2688 & 0.1756 & \textbf{0.1339} & 0.2283 & \uline{0.1390} \\
            \bottomrule
        \end{tabular}
    \end{center}
\end{table*}

To verify the noise robustness, we tested nine cases with different noise settings, as summarized in Table~\ref{table: noise cases}. Case1 represents the noise-free baseline. In the remaining eight cases, Gaussian noise was consistently applied to the high-resolution (HR) images. We categorized these scenarios into four groups based on the noise mixture: pure Gaussian (Cases~2–3), mixed salt-and-pepper (Cases~4–5), mixed vertical stripe with intensity uniformly random in the range $[-0.2, 0.2]$ (Cases~6–7), and mixed Poisson noise (Cases~8–9). Within each group, the first scenario considers noise only in the HR data, while the second scenario introduces additional noise to the LR data. We configured our noise parameters so that the contamination levels in the HR images are higher than those in the LR images. This follows the physical principle that HR images typically exhibit more severe noise than LR images because the light intensity received per pixel decreases with increased spatial resolution~\cite{noise_pixel_relationship}. In the table, $\sigma$, $r$, $s$, and $e$ denote the Gaussian standard deviation, salt-and-pepper noise ratio, stripe noise ratio, and Poisson scale coefficient, respectively, where the subscripts $\mathrm{h}$ and $\mathrm{l}$ correspond to the HR and LR images.

\subsubsection{Pareameter Settings}
\label{ssec:parameter settings}

The parameters in our method were set as shown in Table~\ref{table: parameters}. Among them, $\delta$, k, $c_{\alpha}$, and $q$ were heuristically determined by empirical tuning to achieve favorable fusion performance consistently across various sites and noise conditions. The details of these parameter searches are provided in Sec.~\ref{ssec:facilitation of parameter selection}.
The maximum number of iterations for Algorithm~\ref{algo: PDS_for_OptForm} was fixed at 10000. The stopping criterion of Algorithm~\ref{algo: PDS_for_OptForm} was defined as $\|\tilHr^{(n)} - \tilHr^{(n-1)}\|_2/\| \tilHr^{(n-1)} \|_2<10^{-5}$ ,$\|\tilHt^{(n)} - \tilHt^{(n-1)}\|_2/\| \tilHt^{(n-1)} \|_2<10^{-5}$, $\| \Lr - \mathbf{S}\mathbf{B}\tilHr^{(n)}\|_2 \leq \epsl$, and $\| \Lt - \mathbf{S}\mathbf{B}\tilHt^{(n)}\|_2 \leq \epsl$. 

The spatial spread transform matrix $\B$ in \eqref{eq: relationship model} was implemented as a blurring operator with an averaging filter, where the kernel size corresponds to the spatial resolution ratio between the HR and LR images. Accordingly, the downsampling matrix $\S$ in \eqref{eq: relationship model} was designed to extract the center pixel value within the window defined by this ratio.

\begin{figure*}[ht]
    \begin{center}
        \scriptsize 
        \setlength{\tabcolsep}{0pt} 

        \begin{minipage}{0.1\hsize}
            \centerline{\includegraphics[width=\hsize]{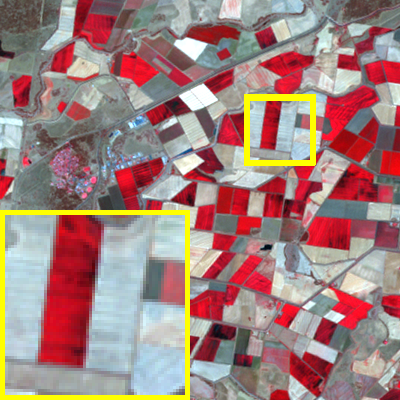}}
        \end{minipage}
        \begin{minipage}{0.1\hsize}
            \centerline{\includegraphics[width=\hsize]{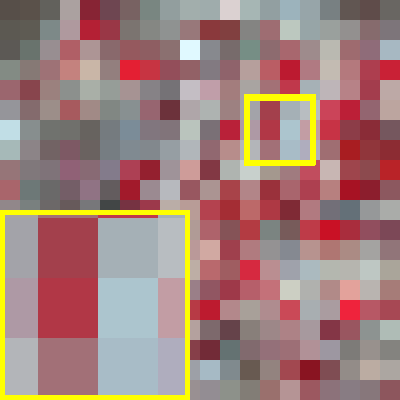}}
        \end{minipage}
        \begin{minipage}{0.1\hsize}
            \centerline{\includegraphics[width=\hsize]{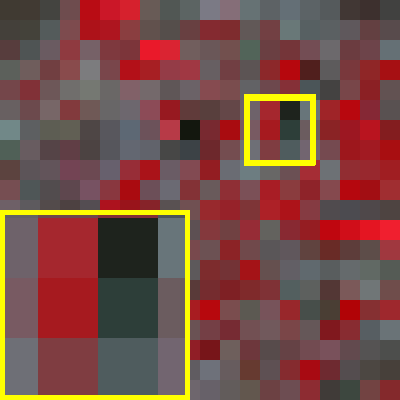}}
        \end{minipage}
        \begin{minipage}{0.1\hsize}
            \centerline{\includegraphics[width=\hsize]{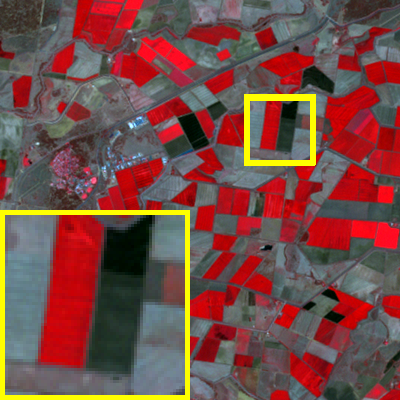}}
        \end{minipage} \\
        \vspace{1mm}
        \begin{minipage}{0.1\hsize} 
            \centerline{$\Hr$}
        \end{minipage}
        \begin{minipage}{0.1\hsize} 
            \centerline{$\Lr$}
        \end{minipage}
        \begin{minipage}{0.1\hsize} 
            \centerline{$\Lt$}
        \end{minipage}
        \begin{minipage}{0.1\hsize} 
            \centerline{Ground-truth}
        \end{minipage} \\
        \vspace{1mm}

        \begin{minipage}{0.01\hsize}
            \centerline{\rotatebox{90}{Case1}}
        \end{minipage}
        \begin{minipage}{0.1\hsize}
            \centerline{\includegraphics[width=\hsize]{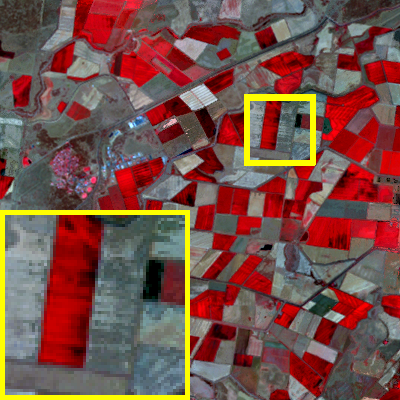}}
        \end{minipage}
        \begin{minipage}{0.1\hsize}
            \centerline{\includegraphics[width=\hsize]{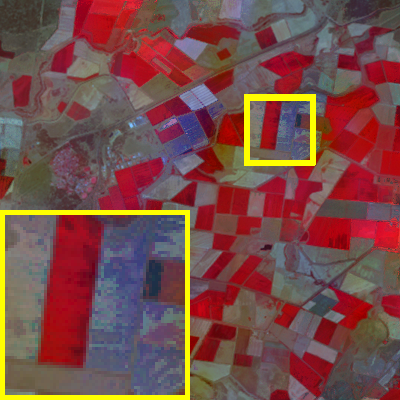}}
        \end{minipage}
        \begin{minipage}{0.1\hsize}
            \centerline{\includegraphics[width=\hsize]{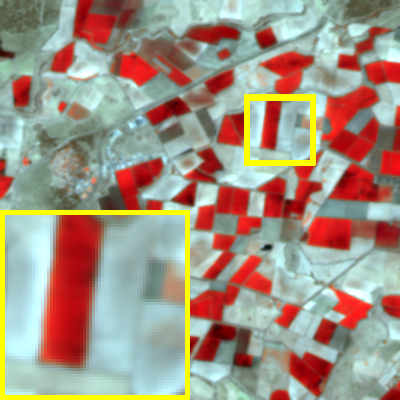}}
        \end{minipage}
        \begin{minipage}{0.1\hsize}
            \centerline{\includegraphics[width=\hsize]{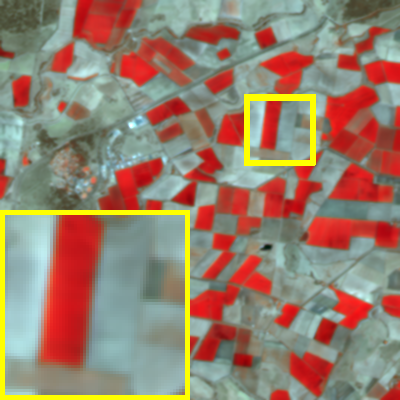}}
        \end{minipage}
        \begin{minipage}{0.1\hsize}
            \centerline{\includegraphics[width=\hsize]{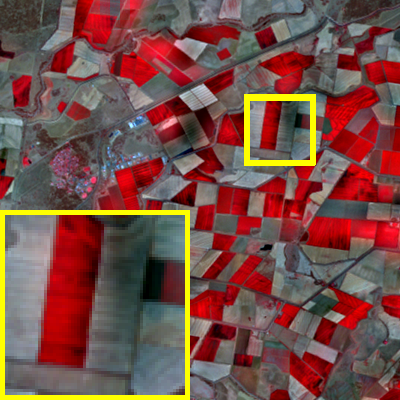}}
        \end{minipage}
        \begin{minipage}{0.1\hsize}
            \centerline{\includegraphics[width=\hsize]{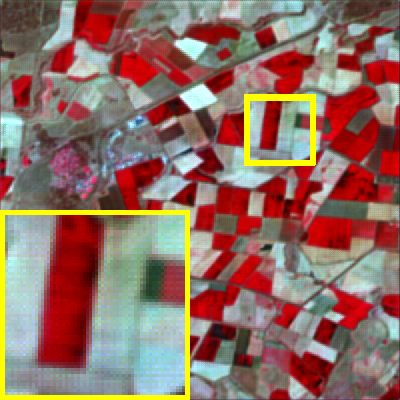}}
        \end{minipage}
        \begin{minipage}{0.1\hsize}
            \centerline{\includegraphics[width=\hsize]{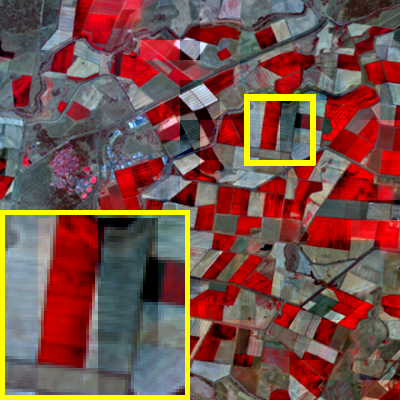}}
        \end{minipage}
        \begin{minipage}{0.1\hsize} 
            \centerline{\includegraphics[width=\hsize]{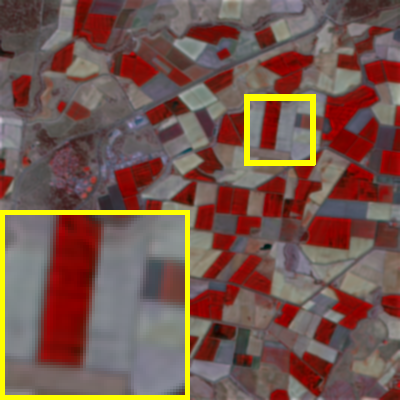}}
        \end{minipage}
        \begin{minipage}{0.1\hsize}
            \centerline{\includegraphics[width=\hsize]{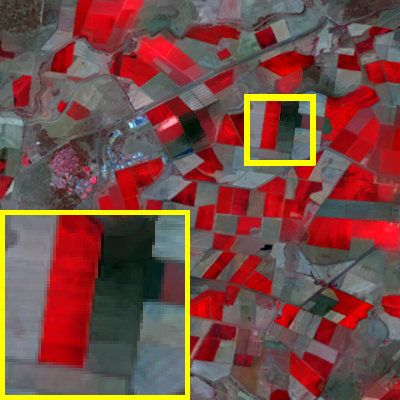}}
        \end{minipage} \\
        \vspace{1mm}

        \begin{minipage}{0.01\hsize}
            \centerline{\rotatebox{90}{Case3}}
        \end{minipage}
        \begin{minipage}{0.1\hsize}
            \centerline{\includegraphics[width=\hsize]{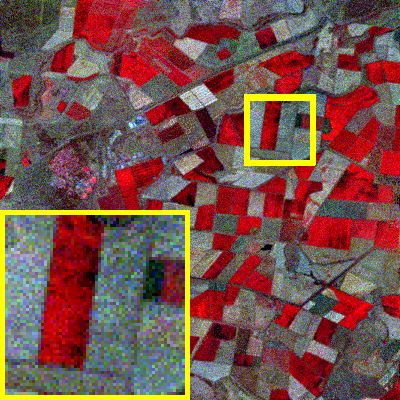}}
        \end{minipage}
        \begin{minipage}{0.1\hsize}
            \centerline{\includegraphics[width=\hsize]{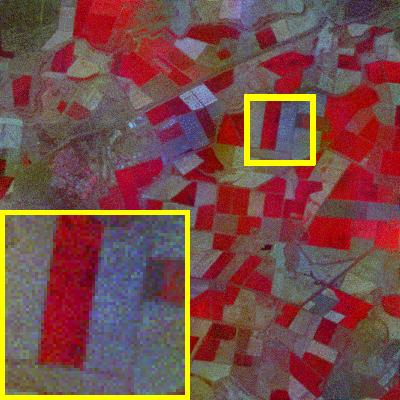}}
        \end{minipage}
        \begin{minipage}{0.1\hsize}
            \centerline{\includegraphics[width=\hsize]{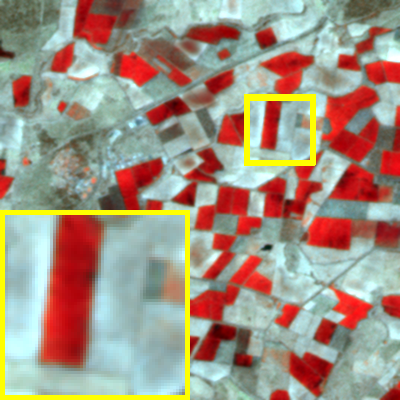}}
        \end{minipage}
        \begin{minipage}{0.1\hsize}
            \centerline{\includegraphics[width=\hsize]{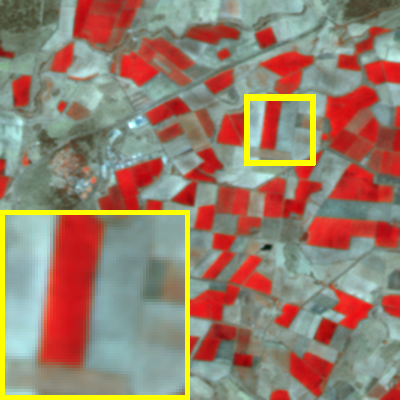}}
        \end{minipage}
        \begin{minipage}{0.1\hsize}
            \centerline{\includegraphics[width=\hsize]{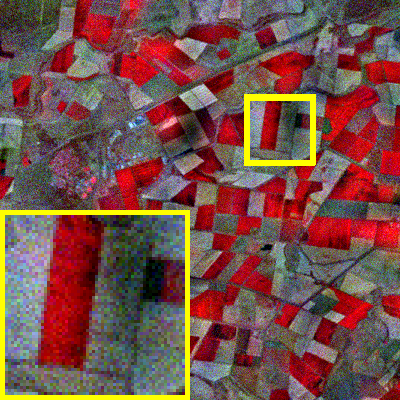}}
        \end{minipage}
        \begin{minipage}{0.1\hsize}
            \centerline{\includegraphics[width=\hsize]{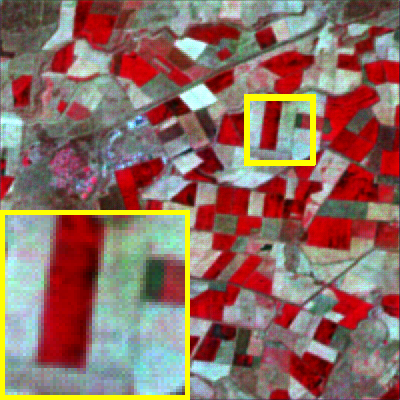}}
        \end{minipage}
        \begin{minipage}{0.1\hsize}
            \centerline{\includegraphics[width=\hsize]{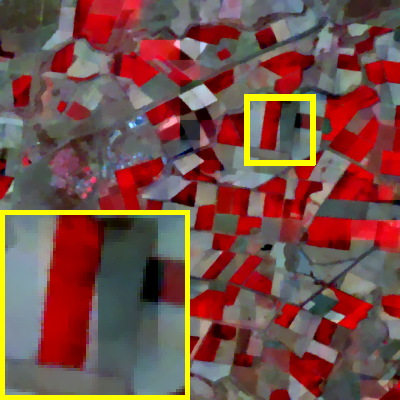}}
        \end{minipage}
        \begin{minipage}{0.1\hsize} 
            \centerline{\includegraphics[width=\hsize]{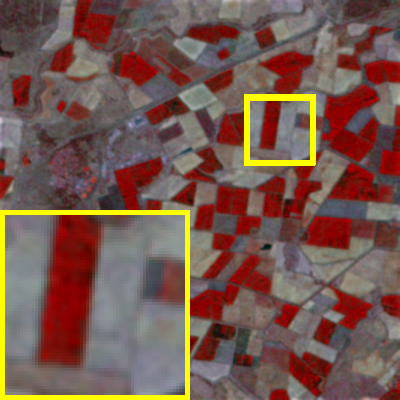}}
        \end{minipage}
        \begin{minipage}{0.1\hsize}
            \centerline{\includegraphics[width=\hsize]{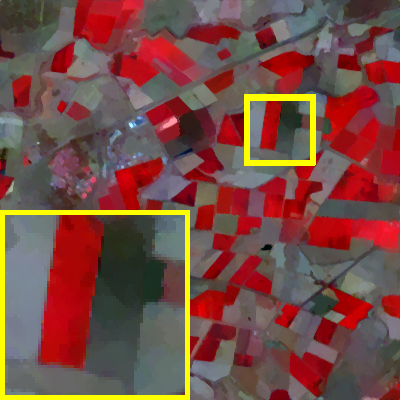}}
        \end{minipage} \\
        \vspace{1mm}

        \begin{minipage}{0.01\hsize}
            \centerline{\rotatebox{90}{Case5}}
        \end{minipage}
        \begin{minipage}{0.1\hsize}
            \centerline{\includegraphics[width=\hsize]{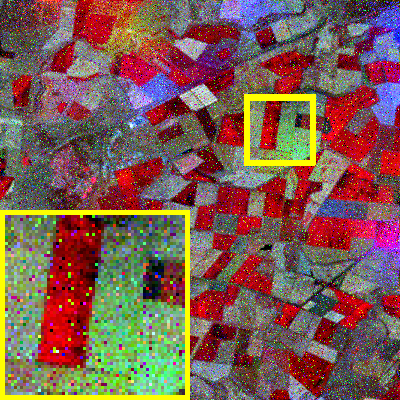}}
        \end{minipage}
        \begin{minipage}{0.1\hsize}
            \centerline{\includegraphics[width=\hsize]{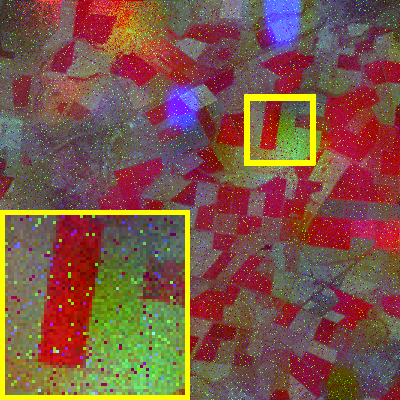}}
        \end{minipage}
        \begin{minipage}{0.1\hsize}
            \centerline{\includegraphics[width=\hsize]{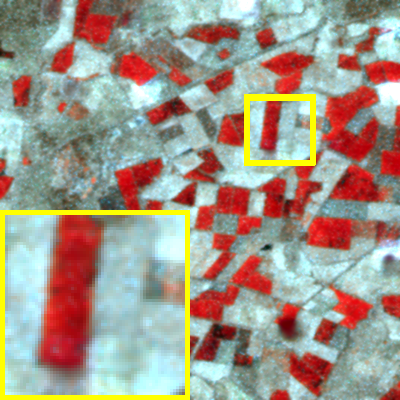}}
        \end{minipage}
        \begin{minipage}{0.1\hsize}
            \centerline{\includegraphics[width=\hsize]{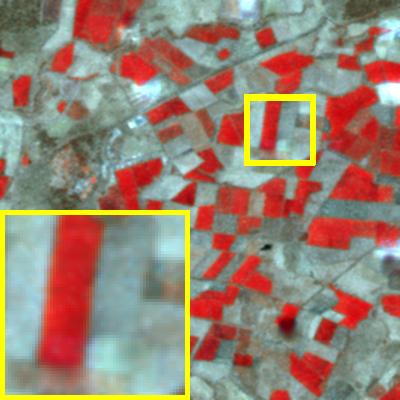}}
        \end{minipage}
        \begin{minipage}{0.1\hsize}
            \centerline{\includegraphics[width=\hsize]{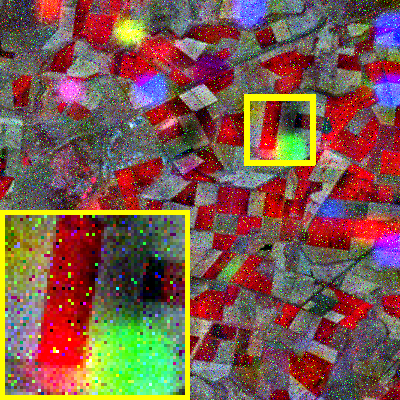}}
        \end{minipage}
        \begin{minipage}{0.1\hsize}
            \centerline{\includegraphics[width=\hsize]{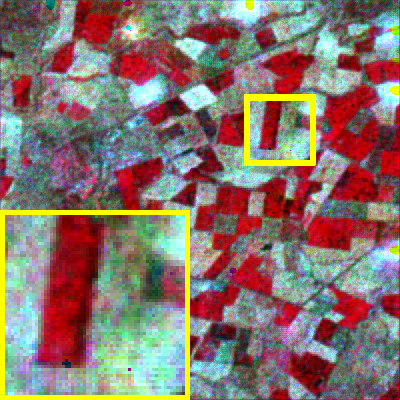}}
        \end{minipage}
        \begin{minipage}{0.1\hsize}
            \centerline{\includegraphics[width=\hsize]{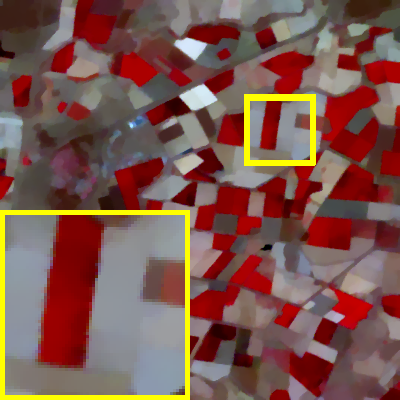}}
        \end{minipage}
        \begin{minipage}{0.1\hsize} 
            \centerline{\includegraphics[width=\hsize]{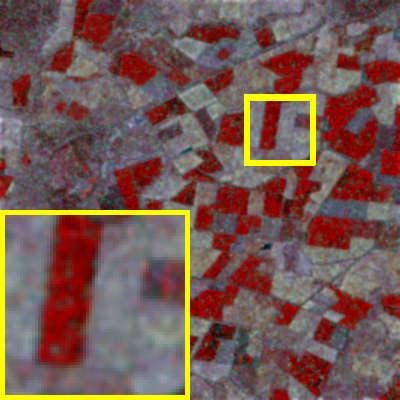}}
        \end{minipage}
        \begin{minipage}{0.1\hsize}
            \centerline{\includegraphics[width=\hsize]{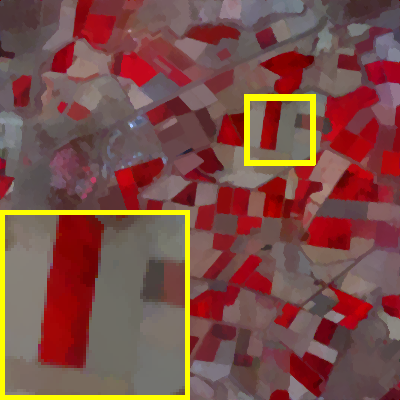}}
        \end{minipage} \\
        \vspace{1mm}

        \begin{minipage}{0.01\hsize}
            \centerline{\rotatebox{90}{Case7}}
        \end{minipage}
        \begin{minipage}{0.1\hsize}
            \centerline{\includegraphics[width=\hsize]{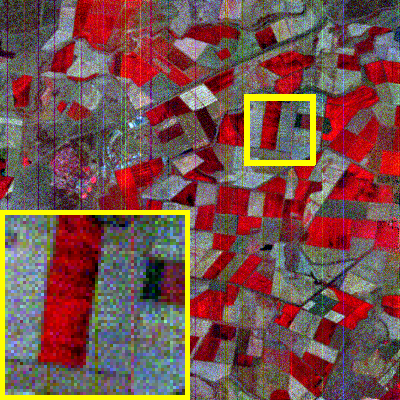}}
        \end{minipage}
        \begin{minipage}{0.1\hsize}
            \centerline{\includegraphics[width=\hsize]{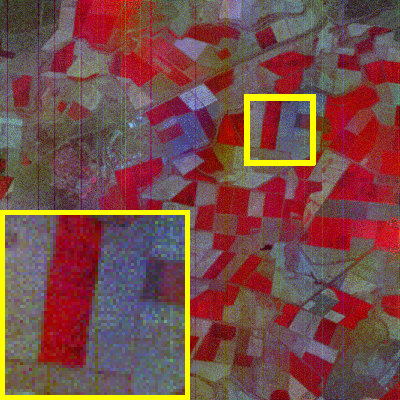}}
        \end{minipage}
        \begin{minipage}{0.1\hsize}
            \centerline{\includegraphics[width=\hsize]{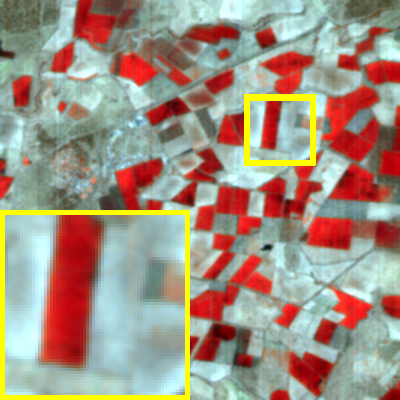}}
        \end{minipage}
        \begin{minipage}{0.1\hsize}
            \centerline{\includegraphics[width=\hsize]{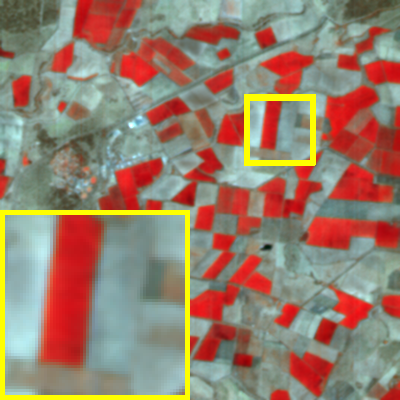}}
        \end{minipage}
        \begin{minipage}{0.1\hsize}
            \centerline{\includegraphics[width=\hsize]{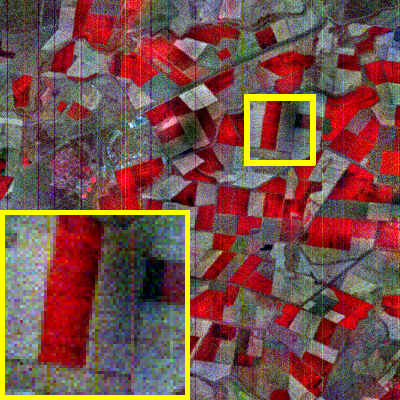}}
        \end{minipage}
        \begin{minipage}{0.1\hsize}
            \centerline{\includegraphics[width=\hsize]{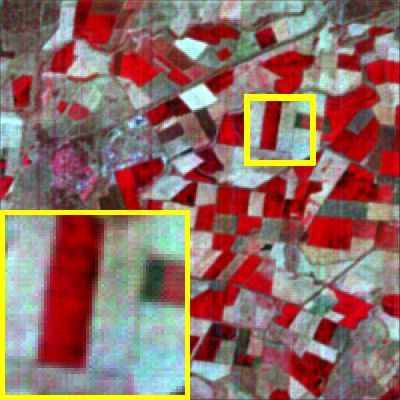}}
        \end{minipage}
        \begin{minipage}{0.1\hsize}
            \centerline{\includegraphics[width=\hsize]{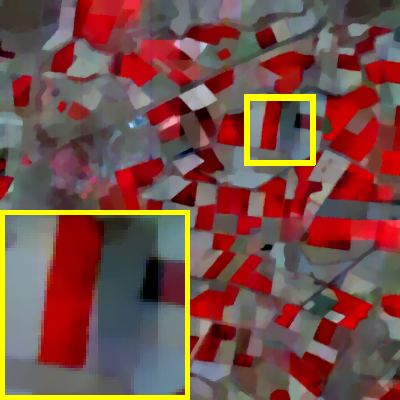}}
        \end{minipage}
        \begin{minipage}{0.1\hsize} 
            \centerline{\includegraphics[width=\hsize]{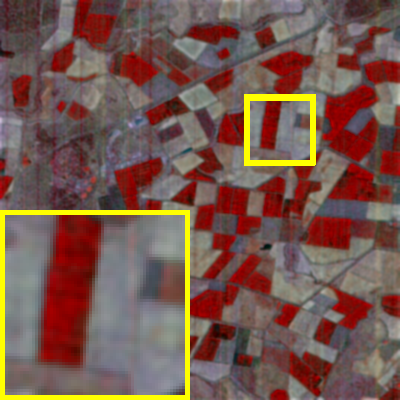}}
        \end{minipage}
        \begin{minipage}{0.1\hsize}
            \centerline{\includegraphics[width=\hsize]{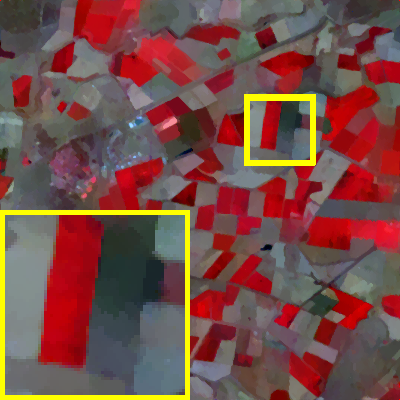}}
        \end{minipage} \\
        \vspace{1mm}

        \begin{minipage}{0.01\hsize}
            \centerline{\rotatebox{90}{Case9}}
        \end{minipage}
        \begin{minipage}{0.1\hsize}
            \centerline{\includegraphics[width=\hsize]{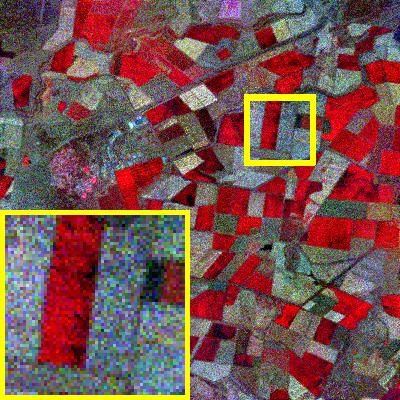}}
        \end{minipage}
        \begin{minipage}{0.1\hsize}
            \centerline{\includegraphics[width=\hsize]{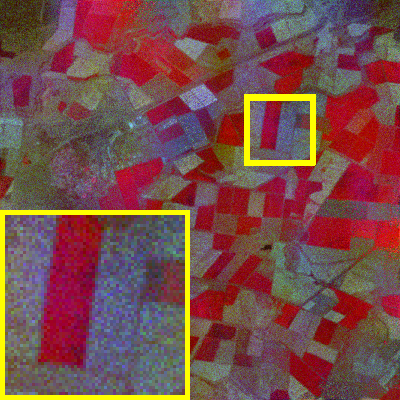}}
        \end{minipage}
        \begin{minipage}{0.1\hsize}
            \centerline{\includegraphics[width=\hsize]{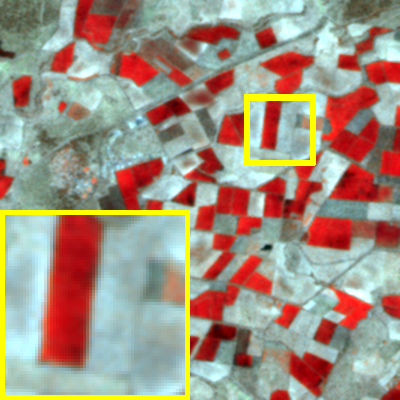}}
        \end{minipage}
        \begin{minipage}{0.1\hsize}
            \centerline{\includegraphics[width=\hsize]{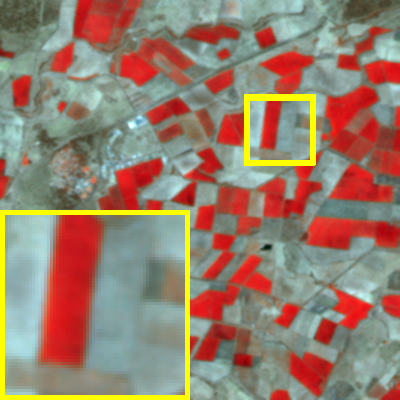}}
        \end{minipage}
        \begin{minipage}{0.1\hsize}
            \centerline{\includegraphics[width=\hsize]{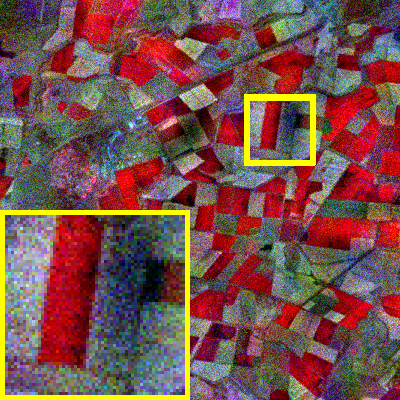}}
        \end{minipage}
        \begin{minipage}{0.1\hsize}
            \centerline{\includegraphics[width=\hsize]{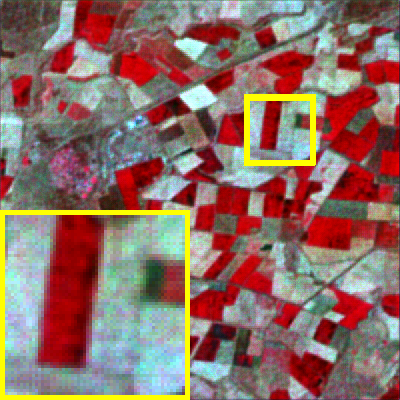}}
        \end{minipage}
        \begin{minipage}{0.1\hsize}
            \centerline{\includegraphics[width=\hsize]{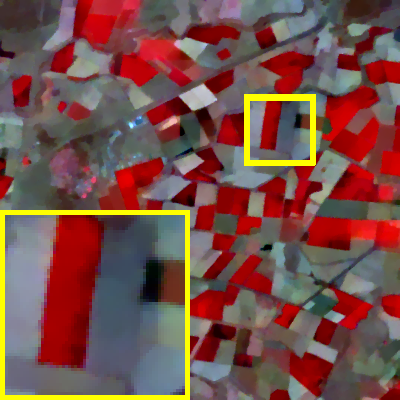}}
        \end{minipage}
        \begin{minipage}{0.1\hsize} 
            \centerline{\includegraphics[width=\hsize]{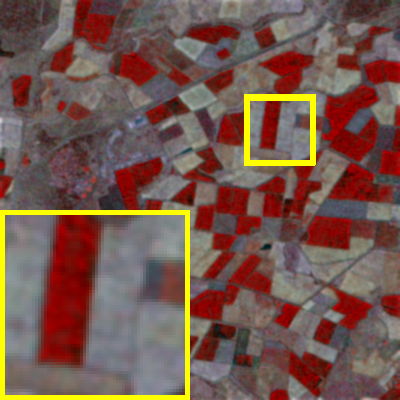}}
        \end{minipage}
        \begin{minipage}{0.1\hsize}
            \centerline{\includegraphics[width=\hsize]{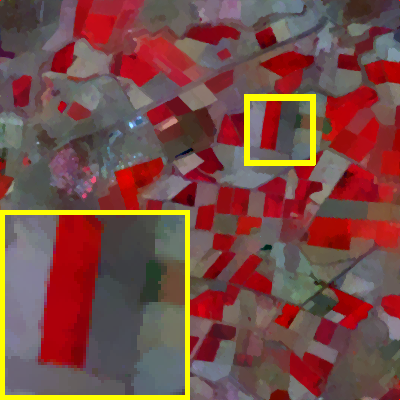}}
        \end{minipage} \\
        
        \vspace{1mm}
        \begin{minipage}{0.01\hsize}
            \centerline{}
        \end{minipage}
        \begin{minipage}{0.1\hsize} 
            \centerline{STARFM}
        \end{minipage}
        \begin{minipage}{0.1\hsize} 
            \centerline{VIPSTF}
        \end{minipage}
        \begin{minipage}{0.1\hsize} 
            \centerline{RSFN-1}
        \end{minipage}
        \begin{minipage}{0.1\hsize} 
            \centerline{RSFN-2}
        \end{minipage}
        \begin{minipage}{0.1\hsize} 
            \centerline{RobOt}
        \end{minipage}
        \begin{minipage}{0.1\hsize} 
            \centerline{SwinSTFM}
        \end{minipage}
        \begin{minipage}{0.1\hsize} 
            \centerline{ROSTF}
        \end{minipage}
        \begin{minipage}{0.1\hsize} 
            \centerline{ECPW}
        \end{minipage}
        \begin{minipage}{0.1\hsize} 
            \centerline{\textbf{TSSTF}}
        \end{minipage}\\
    \end{center}
    \vspace{-3mm}
    \caption{ST fusion results for the Site1 simulated data. The top row shows the noiseless input images ($\Hr$, $\Lr$, and $\Lt$), and ground truth~$\widehat{\mathbf{h}}_{t}$. The remaining rows correspond to Case1, Case3, Case5, Case7, and Case9, respectively.}
    \label{fig: Site1 SemiSim results}
\end{figure*}

\subsubsection{Evaluation Metrics}
\label{ssec:evaluation metrics}
For the quantitative evaluation, we used the following four metrics: the peak-to-noise ratio (PSNR):
\begin{equation}
    \mathrm{PSNR} = 10\log_{10}\left(\frac{1}{\mathrm{\sqrt{\frac{1}{\Nh B} \|\tilHt - \hatHt\|_2^2}}}\right), \nonumber
\end{equation}
where $\tilHt$ and $\hatHt$ denote an estimated HR image and a ground-truth HR image, respectively, on the target date; the mean structural  similarity overall bands (MSSIM)~\cite{SSIM}:
\begin{equation}
    \mathrm{MSSIM} = \frac{1}{B}\sum_{b = 1}^{B}\mathrm{SSIM}\left([\tilHt]_{b}, [\hatHt]_{b} \right); \nonumber
\end{equation}
the correlation coefficient (CC):
\begin{equation}
    \mathrm{CC} = \frac{\mathrm{cov}(\tilHt,\hatHt)}{\mathrm{std}(\tilHt)\cdot\mathrm{std}(\hatHt)},\nonumber
\end{equation}
where $\mathrm{cov}(\tilHt,\hatHt)$ denotes the covariance of $\tilHt$ and $\hatHt$, and $\mathrm{std}(\tilHt)$ and $\mathrm{std}(\hatHt)$ denote the standard deviations of $\tilHt$ and $\hatHt$, respectively; and the spectral angle mapper (SAM)~\cite{SAM}:
\begin{equation}
    \mathrm{SAM} = \frac{1}{\Nh}\sum_{\substack{i=1 \\ j=1}}^{\Wh, \Hh}\mathrm{arccos}\left(\frac{\sum_{b=1}^{B}[\tilHt]_{i,j,b}[\hatHt]_{i,j,b}}{\sqrt{\sum_{b=1}^{B}[\tilHt]_{i,j,b}^2\sum_{b=1}^{B}[\hatHt]_{i,j,b}^2}}\right).\nonumber
\end{equation}
PSNR was calculated to measure the difference between the estimated image and the ground-truth at the pixel level. MSSIM was used to evaluate the similarity of the overall structure. CC shows the strength of the linear relationship between the estimated image and the ground-truth. SAM was used to measure spectral fidelity. Higher PSNR, MSSIM, and CC indicate better estimation performance, while lower SAM indicates better performance.

\begin{table*}[t]
    \begin{center}
    \caption{The PSNR, MSSIM, CC, and SAM Results in the Experiments With Real Data}
    \label{table: real metrics}
    \vspace{-2mm}
    \centering
        \begin{tabular}{p{0.7cm} p{0.7cm} >{\centering\arraybackslash}p{1.3cm} >{\centering\arraybackslash}p{1.3cm} >{\centering\arraybackslash}p{1.3cm} >{\centering\arraybackslash}p{1.3cm} >{\centering\arraybackslash}p{1.3cm} >{\centering\arraybackslash}p{1.3cm} >{\centering\arraybackslash}p{1.3cm} >{\centering\arraybackslash}p{1.3cm} >{\centering\arraybackslash}p{1.3cm}}
            \toprule
            Case & Metric & STARFM & VIPSTF & RSFN-1 & RSFN-2 & RobOt & SwinSTFM & ROSTF & ECPW & TSSTF \\
            \midrule
            \multirow{4}{*}{Case1} & PSNR & 26.97 & \uline{27.09} & 23.66 & 25.15 & 26.83 & 25.32 & 26.99 & 24.75 & \textbf{27.72} \\
            & MSSIM & 0.7315 & 0.7370 & 0.6971 & 0.7257 & 0.7418 & 0.6961 & \uline{0.7430} & 0.7010 & \textbf{0.7677} \\
            & CC & 0.9249 & \textbf{0.9310} & 0.8348 & 0.8804 & 0.9159 & 0.8923 & 0.9212 & 0.8388 & \uline{0.9279} \\
            & SAM & \uline{0.1443} & 0.1480 & 0.2356 & 0.2162 & 0.1480 & 0.1679 & \textbf{0.1428} & 0.2247 & 0.1558 \\
            \midrule
            \multirow{4}{*}{Case2} & PSNR & 22.91 & 24.96 & 23.62 & 25.14 & 22.89 & 25.18 & \uline{26.65} & 24.66 & \textbf{27.12} \\
            & MSSIM & 0.3591 & 0.5130 & 0.6752 & \uline{0.7187} & 0.3778 & 0.6659 & 0.7115 & 0.6800 & \textbf{0.7200} \\
            & CC & 0.8789 & 0.8988 & 0.8338 & 0.8802 & 0.8726 & 0.8888 & \uline{0.9199} & 0.8376 & \textbf{0.9230} \\
            & SAM & 0.2641 & 0.2176 & 0.2393 & 0.2180 & 0.2568 & 0.1741 & \textbf{0.1526} & 0.2284 & \uline{0.1656} \\
            \midrule
            \multirow{4}{*}{Case3} & PSNR & 22.89 & 25.20 & 23.61 & 25.14 & 22.81 & 25.15 & \uline{26.62} & 24.66 & \textbf{27.10} \\
            & MSSIM & 0.3580 & 0.5368 & 0.6752 & \uline{0.7187} & 0.3765 & 0.6649 & 0.7112 & 0.6800 & \textbf{0.7187} \\
            & CC & 0.8786 & 0.9016 & 0.8336 & 0.8801 & 0.8700 & 0.8885 & \uline{0.9196} & 0.8379 & \textbf{0.9227} \\
            & SAM & 0.2654 & 0.2119 & 0.2389 & 0.2178 & 0.2609 & 0.1745 & \textbf{0.1532} & 0.2285 & \uline{0.1664} \\
            \midrule
            \multirow{4}{*}{Case4} & PSNR & 16.22 & 20.16 & 22.70 & 24.81 & 16.67 & 24.16 & \uline{25.43} & 24.19 & \textbf{26.35} \\
            & MSSIM & 0.1592 & 0.3126 & 0.5653 & 0.6773 & 0.1771 & 0.5407 & \uline{0.6796} & 0.5966 & \textbf{0.6959} \\
            & CC & 0.6646 & 0.7471 & 0.8174 & 0.8747 & 0.6792 & 0.8665 & \uline{0.8927} & 0.8267 & \textbf{0.9095} \\
            & SAM & 0.3582 & 0.2824 & 0.2610 & 0.2280 & 0.3410 & 0.2110 & \uline{0.1888} & 0.2449 & \textbf{0.1786} \\
            \midrule
            \multirow{4}{*}{Case5} & PSNR & 16.02 & 22.66 & 22.45 & 24.56 & 16.30 & 23.47 & \uline{25.41} & 24.16 & \textbf{25.42} \\
            & MSSIM & 0.1528 & 0.4838 & 0.5636 & 0.6741 & 0.1743 & 0.5260 & \uline{0.6785} & 0.5968 & \textbf{0.6818} \\
            & CC & 0.6562 & 0.8386 & 0.8084 & 0.8689 & 0.6590 & 0.8527 & \uline{0.8916} & 0.8259 & \textbf{0.8923} \\
            & SAM & 0.3808 & 0.2360 & 0.2621 & 0.2290 & 0.3632 & 0.2205 & \textbf{0.1921} & 0.2453 & \uline{0.1994} \\
            \midrule
            \multirow{4}{*}{Case6} & PSNR & 22.44 & 24.66 & 23.53 & 25.10 & 22.44 & 25.06 & \uline{26.38} & 24.58 & \textbf{27.01} \\
            & MSSIM & 0.3389 & 0.4962 & 0.6654 & \uline{0.7153} & 0.3584 & 0.6544 & 0.6859 & 0.6707 & \textbf{0.7156} \\
            & CC & 0.8693 & 0.8922 & 0.8322 & 0.8796 & 0.8641 & 0.8863 & \uline{0.9143} & 0.8362 & \textbf{0.9212} \\
            & SAM & 0.2777 & 0.2260 & 0.2410 & 0.2185 & 0.2681 & 0.1781 & \textbf{0.1591} & 0.2313 & \uline{0.1668} \\
            \midrule
            \multirow{4}{*}{Case7} & PSNR & 22.24 & 24.58 & 23.50 & 25.06 & 22.22 & 24.88 & \uline{26.11} & 24.58 & \textbf{26.90} \\
            & MSSIM & 0.3331 & 0.5200 & 0.6666 & \textbf{0.7158} & 0.3567 & 0.6528 & 0.6803 & 0.6704 & \uline{0.7120} \\
            & CC & 0.8663 & 0.8915 & 0.8298 & 0.8791 & 0.8588 & 0.8852 & \uline{0.9081} & 0.8357 & \textbf{0.9188} \\
            & SAM & 0.2860 & 0.2240 & 0.2409 & 0.2182 & 0.2765 & 0.1795 & \textbf{0.1686} & 0.2311 & \uline{0.1723} \\
            \midrule
            \multirow{4}{*}{Case8} & PSNR & 22.24 & 24.53 & 23.60 & 25.13 & 22.27 & 25.14 & \uline{26.60} & 24.64 & \textbf{27.02} \\
            & MSSIM & 0.3133 & 0.4771 & 0.6669 & \textbf{0.7159} & 0.3326 & 0.6533 & 0.7051 & 0.6738 & \uline{0.7125} \\
            & CC & 0.8630 & 0.8877 & 0.8328 & 0.8798 & 0.8582 & 0.8872 & \uline{0.9189} & 0.8371 & \textbf{0.9216} \\
            & SAM & 0.2864 & 0.2310 & 0.2400 & 0.2182 & 0.2762 & 0.1763 & \textbf{0.1541} & 0.2291 & \uline{0.1668} \\
            \midrule
            \multirow{4}{*}{Case9} & PSNR & 22.17 & 24.94 & 23.59 & 25.12 & 22.09 & 25.05 & \uline{26.55} & 24.63 & \textbf{26.98} \\
            & MSSIM & 0.3110 & 0.5215 & 0.6668 & \textbf{0.7157} & 0.3301 & 0.6511 & 0.7049 & 0.6735 & \uline{0.7108} \\
            & CC & 0.8613 & 0.8945 & 0.8324 & 0.8796 & 0.8532 & 0.8856 & \uline{0.9181} & 0.8372 & \textbf{0.9208} \\
            & SAM & 0.2884 & 0.2182 & 0.2399 & 0.2182 & 0.2831 & 0.1780 & \textbf{0.1553} & 0.2294 & \uline{0.1691} \\
            \bottomrule
        \end{tabular}
    \end{center}
\end{table*}
\begin{figure*}[ht]
    \begin{center}
        \scriptsize 
        \setlength{\tabcolsep}{0pt} 

        \begin{minipage}{0.1\hsize}
            \centerline{\includegraphics[width=\hsize]{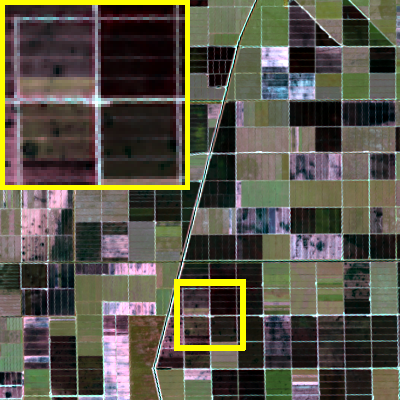}}
        \end{minipage}
        \begin{minipage}{0.1\hsize}
            \centerline{\includegraphics[width=\hsize]{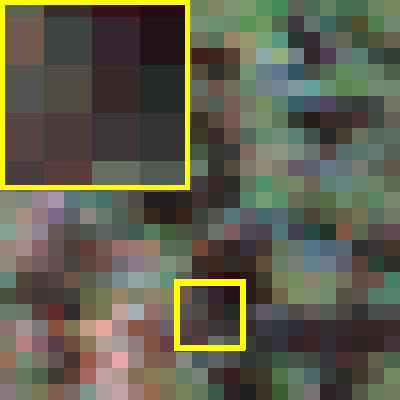}}
        \end{minipage}
        \begin{minipage}{0.1\hsize}
            \centerline{\includegraphics[width=\hsize]{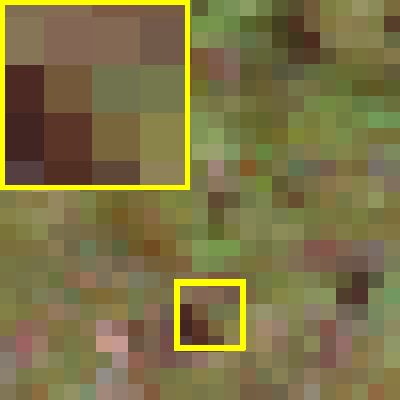}}
        \end{minipage}
        \begin{minipage}{0.1\hsize}
            \centerline{\includegraphics[width=\hsize]{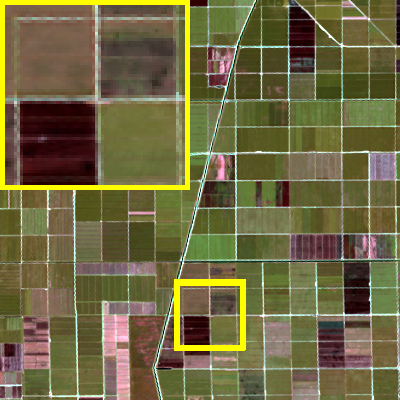}}
        \end{minipage} \\
        \vspace{1mm}
        \begin{minipage}{0.1\hsize} 
            \centerline{$\Hr$}
        \end{minipage}
        \begin{minipage}{0.1\hsize} 
            \centerline{$\Lr$}
        \end{minipage}
        \begin{minipage}{0.1\hsize} 
            \centerline{$\Lt$}
        \end{minipage}
        \begin{minipage}{0.1\hsize} 
            \centerline{Ground-truth}
        \end{minipage} \\
        \vspace{1mm}

        \begin{minipage}{0.01\hsize}
            \centerline{\rotatebox{90}{Case1}}
        \end{minipage}
        \begin{minipage}{0.1\hsize}
            \centerline{\includegraphics[width=\hsize]{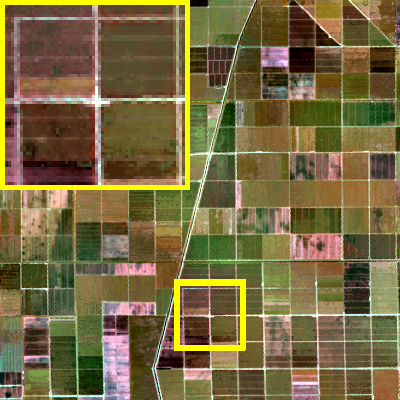}}
        \end{minipage}
        \begin{minipage}{0.1\hsize}
            \centerline{\includegraphics[width=\hsize]{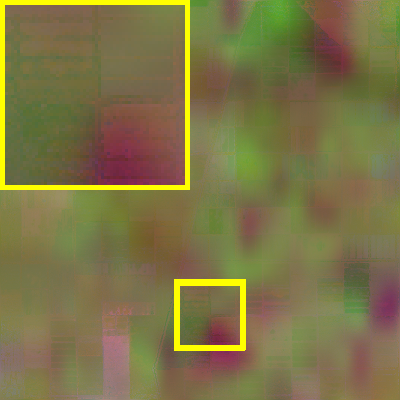}}
        \end{minipage}
        \begin{minipage}{0.1\hsize}
            \centerline{\includegraphics[width=\hsize]{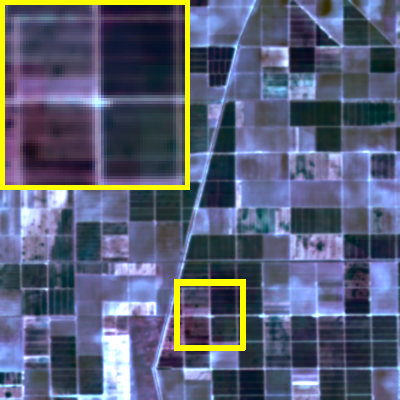}}
        \end{minipage}
        \begin{minipage}{0.1\hsize}
            \centerline{\includegraphics[width=\hsize]{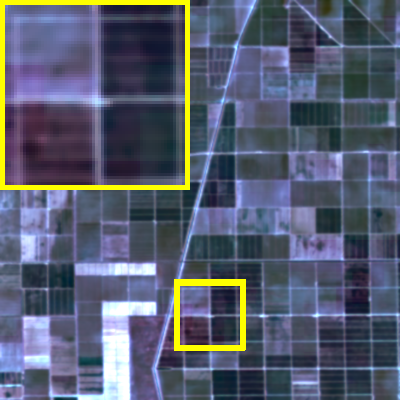}}
        \end{minipage}
        \begin{minipage}{0.1\hsize}
            \centerline{\includegraphics[width=\hsize]{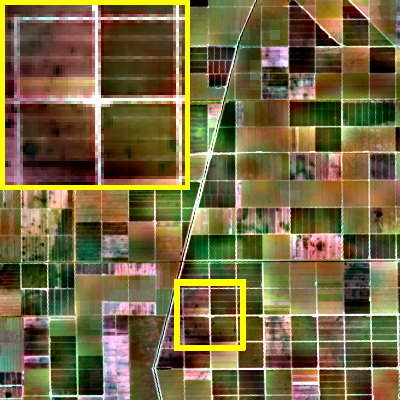}}
        \end{minipage}
        \begin{minipage}{0.1\hsize}
            \centerline{\includegraphics[width=\hsize]{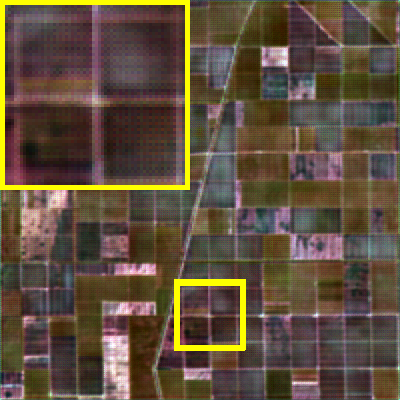}}
        \end{minipage}
        \begin{minipage}{0.1\hsize}
            \centerline{\includegraphics[width=\hsize]{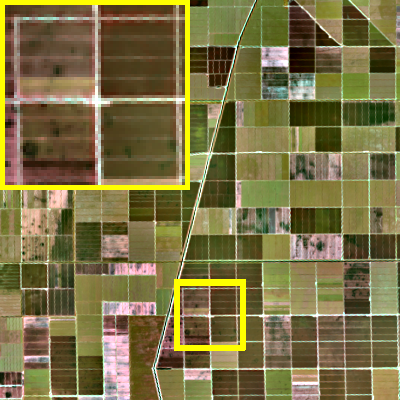}}
        \end{minipage}
        \begin{minipage}{0.1\hsize} 
            \centerline{\includegraphics[width=\hsize]{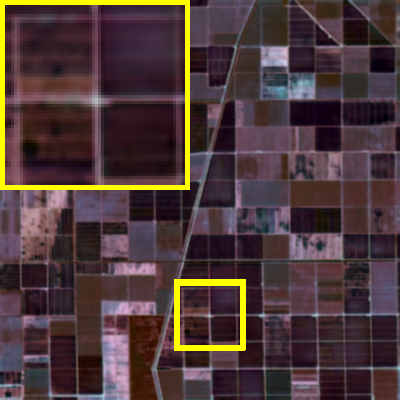}}
        \end{minipage}
        \begin{minipage}{0.1\hsize}
            \centerline{\includegraphics[width=\hsize]{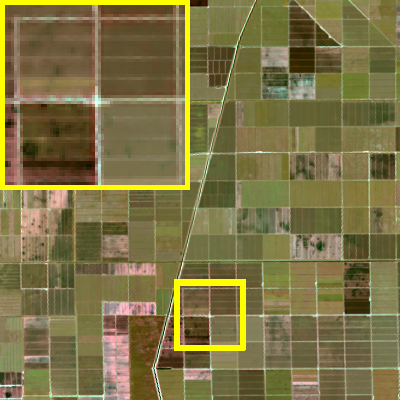}}
        \end{minipage} \\
        \vspace{1mm}

        \begin{minipage}{0.01\hsize}
            \centerline{\rotatebox{90}{Case3}}
        \end{minipage}
        \begin{minipage}{0.1\hsize}
            \centerline{\includegraphics[width=\hsize]{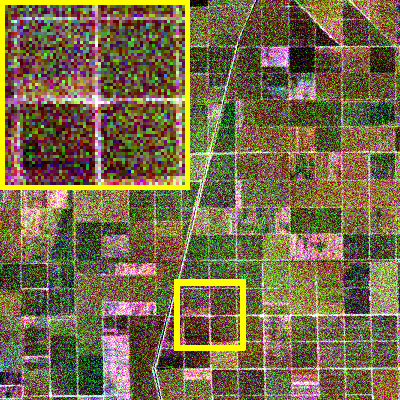}}
        \end{minipage}
        \begin{minipage}{0.1\hsize}
            \centerline{\includegraphics[width=\hsize]{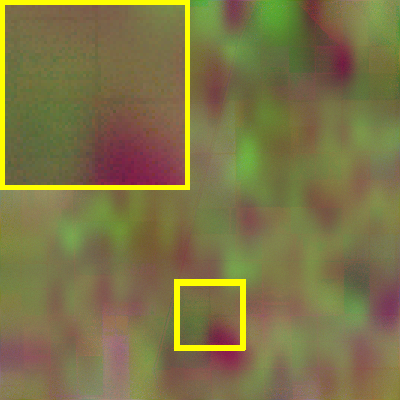}}
        \end{minipage}
        \begin{minipage}{0.1\hsize}
            \centerline{\includegraphics[width=\hsize]{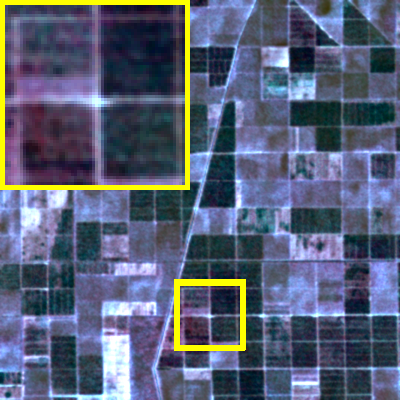}}
        \end{minipage}
        \begin{minipage}{0.1\hsize}
            \centerline{\includegraphics[width=\hsize]{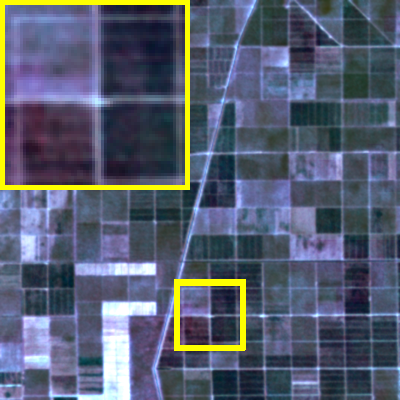}}
        \end{minipage}
        \begin{minipage}{0.1\hsize}
            \centerline{\includegraphics[width=\hsize]{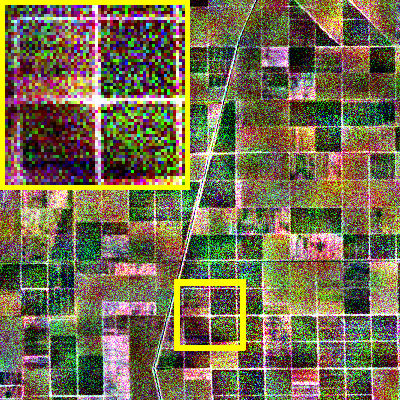}}
        \end{minipage}
        \begin{minipage}{0.1\hsize}
            \centerline{\includegraphics[width=\hsize]{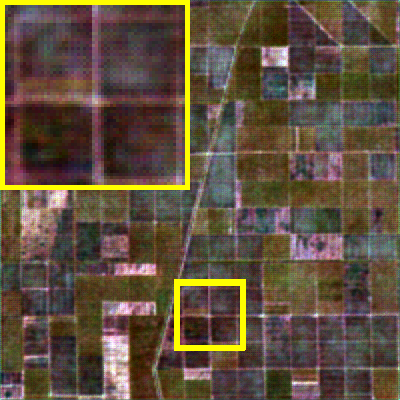}}
        \end{minipage}
        \begin{minipage}{0.1\hsize}
            \centerline{\includegraphics[width=\hsize]{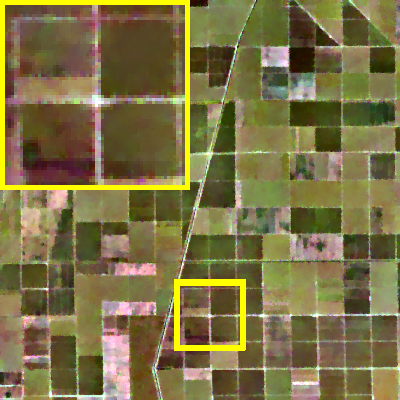}}
        \end{minipage}
        \begin{minipage}{0.1\hsize} 
            \centerline{\includegraphics[width=\hsize]{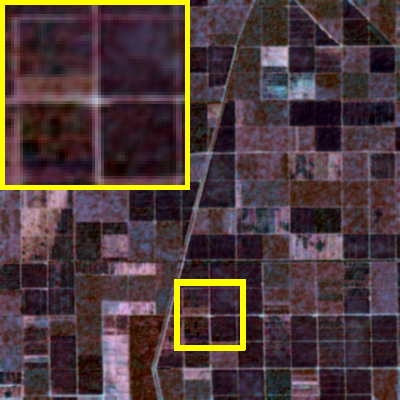}}
        \end{minipage}
        \begin{minipage}{0.1\hsize}
            \centerline{\includegraphics[width=\hsize]{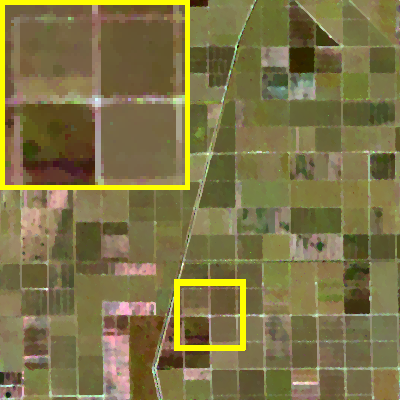}}
        \end{minipage} \\
        \vspace{1mm}

        \begin{minipage}{0.01\hsize}
            \centerline{\rotatebox{90}{Case5}}
        \end{minipage}
        \begin{minipage}{0.1\hsize}
            \centerline{\includegraphics[width=\hsize]{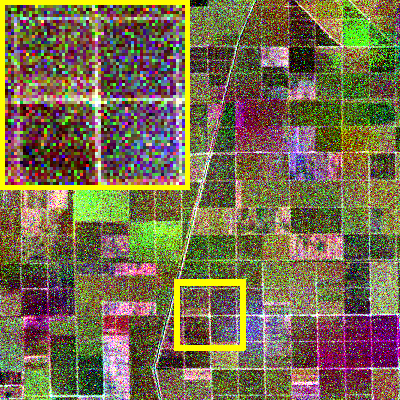}}
        \end{minipage}
        \begin{minipage}{0.1\hsize}
            \centerline{\includegraphics[width=\hsize]{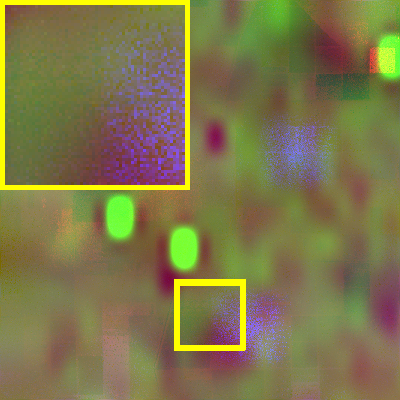}}
        \end{minipage}
        \begin{minipage}{0.1\hsize}
            \centerline{\includegraphics[width=\hsize]{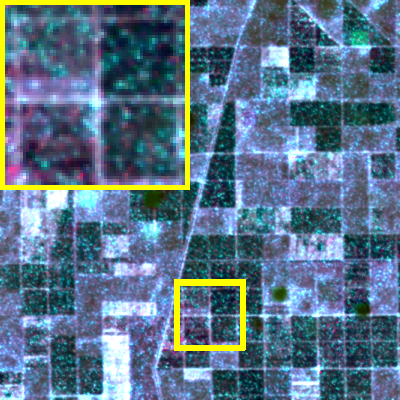}}
        \end{minipage}
        \begin{minipage}{0.1\hsize}
            \centerline{\includegraphics[width=\hsize]{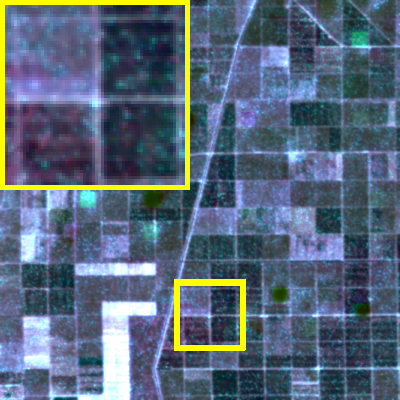}}
        \end{minipage}
        \begin{minipage}{0.1\hsize}
            \centerline{\includegraphics[width=\hsize]{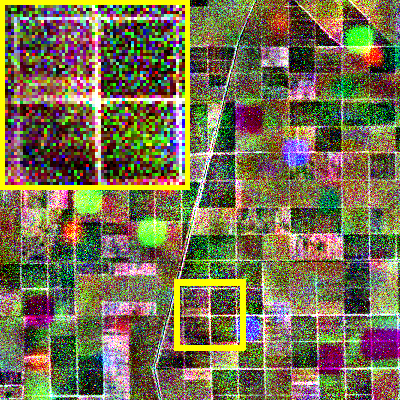}}
        \end{minipage}
        \begin{minipage}{0.1\hsize}
            \centerline{\includegraphics[width=\hsize]{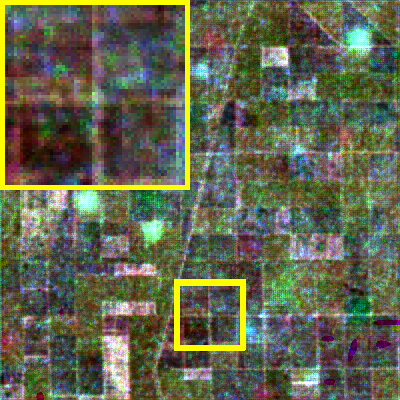}}
        \end{minipage}
        \begin{minipage}{0.1\hsize}
            \centerline{\includegraphics[width=\hsize]{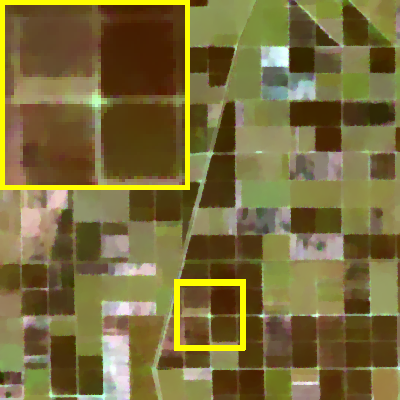}}
        \end{minipage}
        \begin{minipage}{0.1\hsize} 
            \centerline{\includegraphics[width=\hsize]{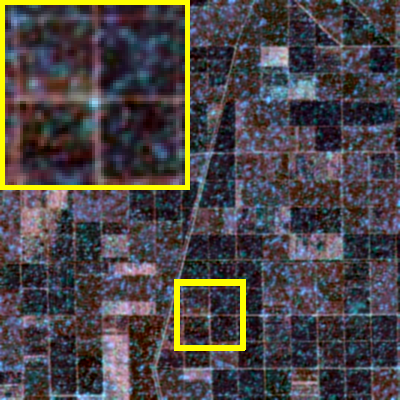}}
        \end{minipage}
        \begin{minipage}{0.1\hsize}
            \centerline{\includegraphics[width=\hsize]{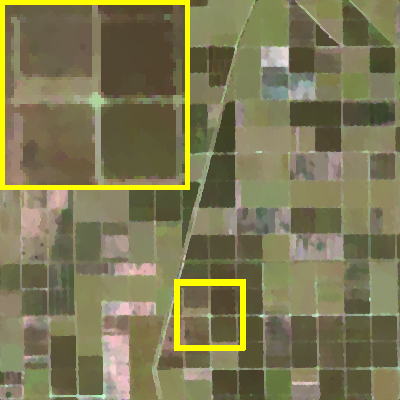}}
        \end{minipage} \\
        \vspace{1mm}

        \begin{minipage}{0.01\hsize}
            \centerline{\rotatebox{90}{Case7}}
        \end{minipage}
        \begin{minipage}{0.1\hsize}
            \centerline{\includegraphics[width=\hsize]{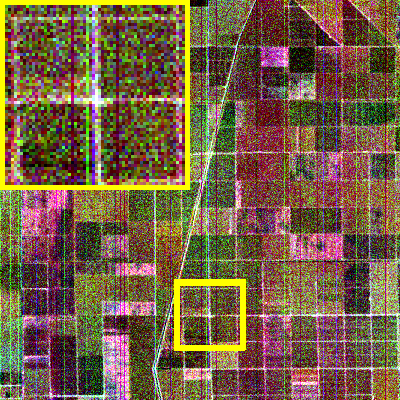}}
        \end{minipage}
        \begin{minipage}{0.1\hsize}
            \centerline{\includegraphics[width=\hsize]{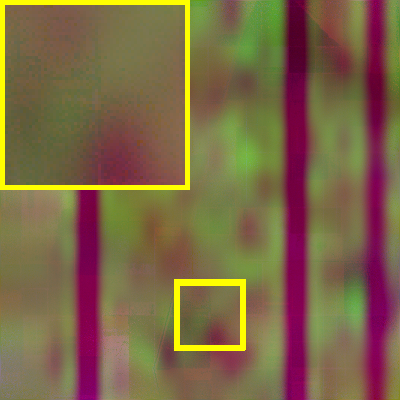}}
        \end{minipage}
        \begin{minipage}{0.1\hsize}
            \centerline{\includegraphics[width=\hsize]{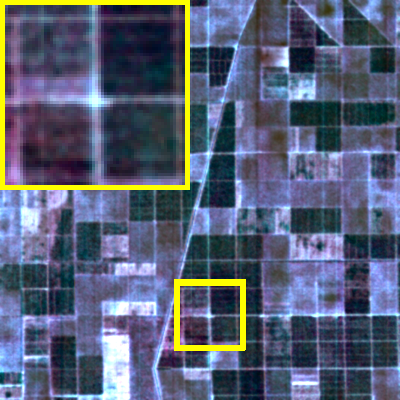}}
        \end{minipage}
        \begin{minipage}{0.1\hsize}
            \centerline{\includegraphics[width=\hsize]{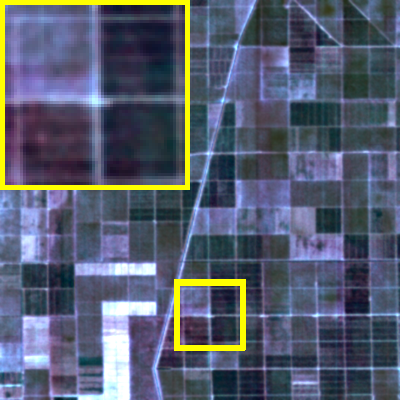}}
        \end{minipage}
        \begin{minipage}{0.1\hsize}
            \centerline{\includegraphics[width=\hsize]{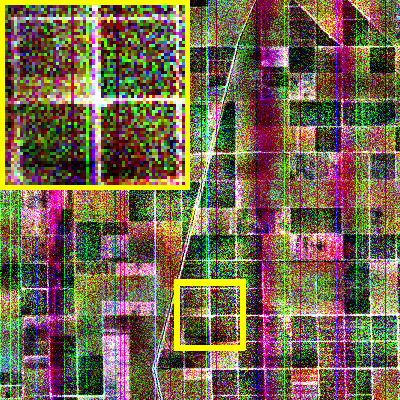}}
        \end{minipage}
        \begin{minipage}{0.1\hsize}
            \centerline{\includegraphics[width=\hsize]{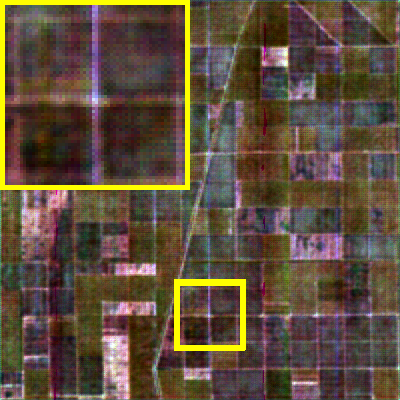}}
        \end{minipage}
        \begin{minipage}{0.1\hsize}
            \centerline{\includegraphics[width=\hsize]{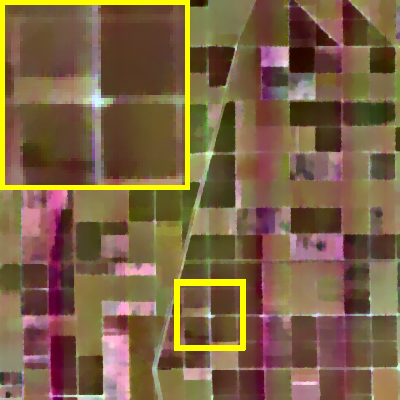}}
        \end{minipage}
        \begin{minipage}{0.1\hsize} 
            \centerline{\includegraphics[width=\hsize]{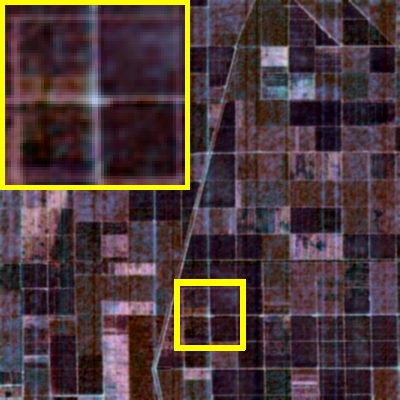}}
        \end{minipage}
        \begin{minipage}{0.1\hsize}
            \centerline{\includegraphics[width=\hsize]{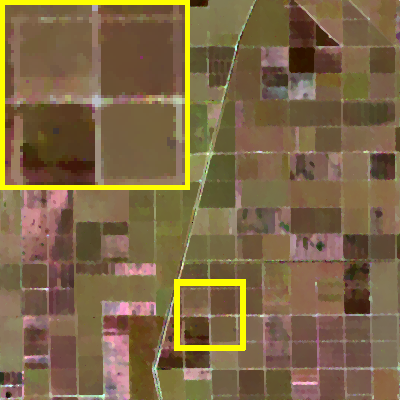}}
        \end{minipage} \\
        \vspace{1mm}

        \begin{minipage}{0.01\hsize}
            \centerline{\rotatebox{90}{Case9}}
        \end{minipage}
        \begin{minipage}{0.1\hsize}
            \centerline{\includegraphics[width=\hsize]{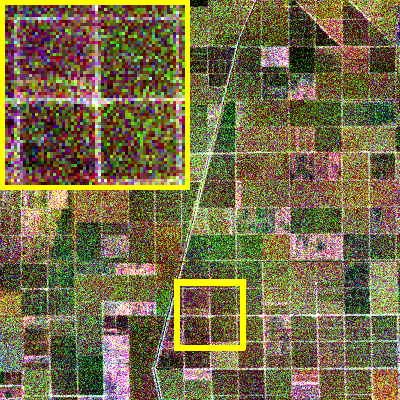}}
        \end{minipage}
        \begin{minipage}{0.1\hsize}
            \centerline{\includegraphics[width=\hsize]{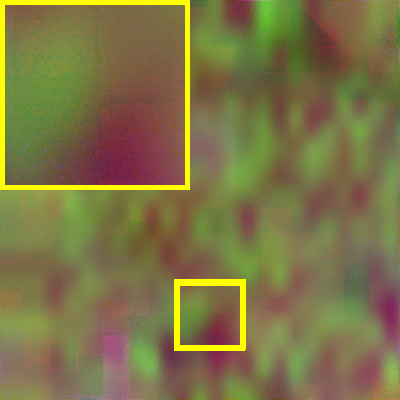}}
        \end{minipage}
        \begin{minipage}{0.1\hsize}
            \centerline{\includegraphics[width=\hsize]{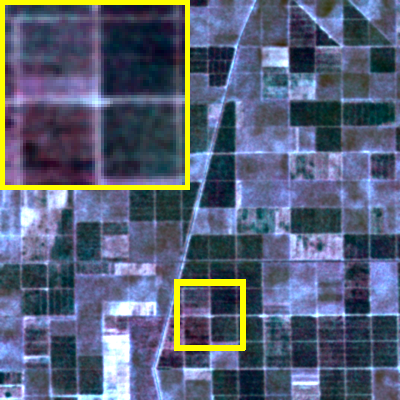}}
        \end{minipage}
        \begin{minipage}{0.1\hsize}
            \centerline{\includegraphics[width=\hsize]{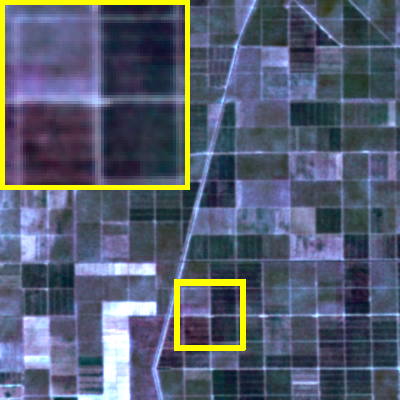}}
        \end{minipage}
        \begin{minipage}{0.1\hsize}
            \centerline{\includegraphics[width=\hsize]{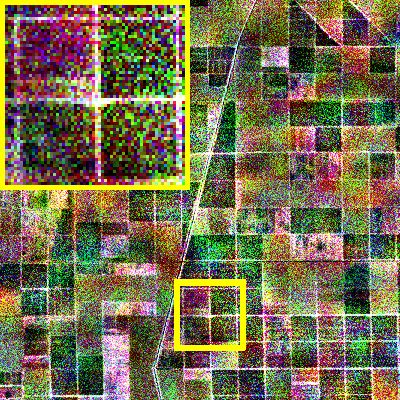}}
        \end{minipage}
        \begin{minipage}{0.1\hsize}
            \centerline{\includegraphics[width=\hsize]{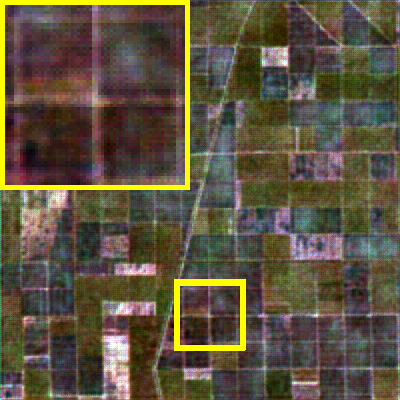}}
        \end{minipage}
        \begin{minipage}{0.1\hsize}
            \centerline{\includegraphics[width=\hsize]{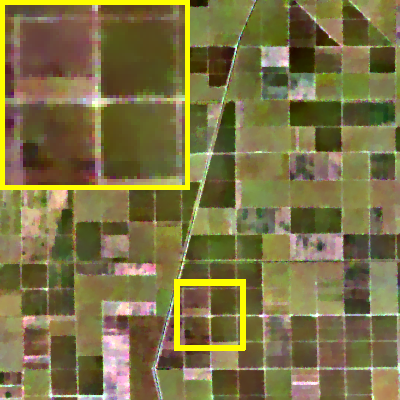}}
        \end{minipage}
        \begin{minipage}{0.1\hsize} 
            \centerline{\includegraphics[width=\hsize]{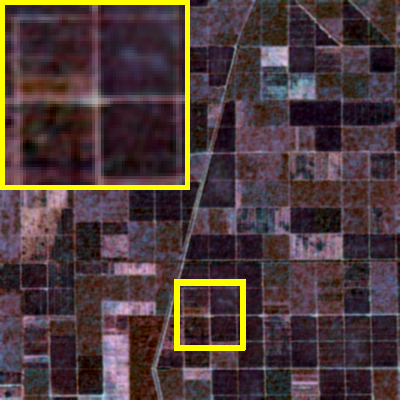}}
        \end{minipage}
        \begin{minipage}{0.1\hsize}
            \centerline{\includegraphics[width=\hsize]{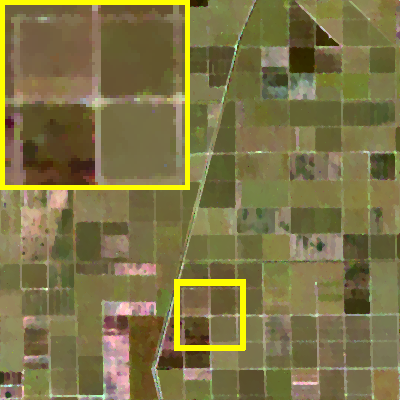}}
        \end{minipage} \\
        
        \vspace{1mm}
        \begin{minipage}{0.01\hsize}
            \centerline{}
        \end{minipage}
        \begin{minipage}{0.1\hsize} 
            \centerline{STARFM}
        \end{minipage}
        \begin{minipage}{0.1\hsize} 
            \centerline{VIPSTF}
        \end{minipage}
        \begin{minipage}{0.1\hsize} 
            \centerline{RSFN-1}
        \end{minipage}
        \begin{minipage}{0.1\hsize} 
            \centerline{RSFN-2}
        \end{minipage}
        \begin{minipage}{0.1\hsize} 
            \centerline{RobOt}
        \end{minipage}
        \begin{minipage}{0.1\hsize} 
            \centerline{SwinSTFM}
        \end{minipage}
        \begin{minipage}{0.1\hsize} 
            \centerline{ROSTF}
        \end{minipage}
        \begin{minipage}{0.1\hsize} 
            \centerline{ECPW}
        \end{minipage}
        \begin{minipage}{0.1\hsize} 
            \centerline{\textbf{TSSTF}}
        \end{minipage}\\
    \end{center}
    \vspace{-3mm}
    \caption{ST fusion results for the Site5 real data. The top row shows the input images, i.e., $\Hr$, $\Lr$, and $\Lt$, and ground truth~$\widehat{\mathbf{h}}_{t}$. The remaining rows correspond to Case1, Case3, Case5, Case7, and Case9, respectively.}
    \label{fig: Site5 Real results}
\end{figure*}

\subsubsection{Comparison Methods}
\label{ssec:comparison methods}
Our method was compared with STARFM~\cite{STARFM}, VIPSTF~\cite{VIPSTF}, RSFN~\cite{RSFN}, RobOt~\cite{RobOt}, SwinSTFM~\cite{SWIN}, ROSTF~\cite{ROSTF}, and ECPW~\cite{ECPW}. STARFM and VIPSTF are rule-based methods, while SwinSTFM, RSFN, and ECPW are learning-based methods. RobOt, ROSTF, and the proposed TSSTF belong to optimization-based methods. For each existing method, we used the parameter values recommended in the corresponding references. 

We trained all learning-based methods using the Landsat-MODIS image pairs provided in \cite{Site1data} acquired at the same geographic locations as Site~1. For the single-reference methods (SwinSTFM and ECPW), we allocated 12 samples for training and four samples for validation, excluding the sample used for testing. In contrast, for RSFN, we allocated 11 samples for training and three samples for validation, as the number of feasible samples is inherently reduced by its dual-reference requirement (i.e., data from both preceding and subsequent reference dates). 
Furthermore, to ensure a fair comparison with single-reference approaches, including TSSTF, we evaluated the trained RSFN model under two testing configurations. The first, denoted as RSFN-1, uses the same data as both the preceding and following references to align with the single-reference constraint. The second, denoted as RSFN-2, follows the original design by utilizing reference data from two distinct dates, where the secondary reference is kept noise-free.


Among the existing methods, only RSFN and ROSTF are explicitly designed to handle noise. However, their robustness has been validated only under specific conditions in their original studies: RSFN focused on localized Gaussian noise in HR images, whereas ROSTF addressed mixed Gaussian and sparse noise in HR images. In contrast, we evaluate the performance of our TSSTF under more complex and diverse noise scenarios, including cases where both HR and LR images are noisy, to demonstrate its broader practical utility.

\subsection{Experimental Results with Simulated Data}
\label{ssec: semisim results}
Table~\ref{table: simulated metrics} shows the average PSNR, MSSIM, CC, and SAM results across all sites for the experiments with simulated data. The best and second-best results are highlighted in bold and underlined, respectively. 
Under the noise-free condition (Case1), our TSSTF achieves performance comparable to or better than existing methods across all sites. This demonstrates that TSSTF is effective as a fundamental ST fusion framework, even in the absence of noise. 

Under noise contamination (Cases2–9), the performance of rule-based and optimization-based methods that do not account for noise, such as STARFM, VIPSTF, and RobOt, drops significantly. This is because these methods trust noisy pixel values directly during the fusion process. 
In contrast, learning-based methods, including RSFN, SwinSTFM, and ECPW, are inherently robust, maintaining relatively stable performance regardless of whether they incorporate explicit noise-handling mechanisms. This can be attributed to their ability to extract regional features and high-level structural representations rather than relying on individual noisy pixels, which makes them less susceptible to noise. 
Nevertheless, the unified frameworks that simultaneously perform denoising and ST fusion, such as ROSTF and our TSSTF, consistently outperform these methods, achieving the best or second-best scores across all metrics in all noise cases. Notably, TSSTF outperforms ROSTF in most conditions and metrics. This consistent superiority indicates that the proposed TGTV and TGEC mechanisms effectively capture and preserve the intrinsic spatial structure from noisy references, leading to improved spatial and spectral fidelity.

Fig.~\ref{fig: Site1 SemiSim results} shows the estimated results for the Site1 simulated data, where there is a significant spectral change between the reference HR image~$\Hr$ and the ground-truth. We first focus on the results in the noiseless case (Case1). RSFN-1, RSFN-2, and SwinSTFM fail to capture the spectral change, resulting in brightness that is biased toward $\Hr$ rather than the ground-truth. VIPSTF suffers from severe spectral distortion. While STARFM, RobOt, ECPW and ROSTF offer relatively better estimates, they still exhibit noticeable artifacts in fine details or inaccurate spectral estimation. TSSTF successfully preserves both edge sharpness and spectral fidelity, yielding the most visually accurate fusion result among all methods.

We next examine the noisy cases, including Cases3, 5, 7, and 9, where both of the HR and LR images are contaminated with noise. STARFM, VIPSTF, and RobOt produce unstable results with noise, as they rely on noisy pixel values directly. While RSFN-1, RSFN-2, SwinSTFM, and ECPW are less affected by noise, they fail to capture the spectral changes sufficiently from the reference date to the target date especially in the zoomed-in region. While ROSTF provides robust estimates, it generates unnatural block artifacts that do not exist in the ground-truth. In contrast, TSSTF consistently yields the most accurate visual results by effectively suppressing the influence of various types of noise while faithfully capturing the intrinsic edge structure.

\subsection{Experimental Results with Real Data}
\label{ssec: real results}
Table~\ref{table: real metrics} shows the average PSNR, MSSIM, CC, and SAM results across all sites for the experiments with real data. Due to radiometric and geometric inconsistencies between the HR and LR sensors, the performance of all methods degrades compared to the results on the simulated data. In the noise-free case (Case1), TSSTF achieves the best PSNR and MSSIM scores and the second-best CC score. In the noisy cases, TSSTF consistently yields the best or the second-best scores across all metrics. These results confirm that TSSTF is both effective and robust under real-world conditions.

Fig.~\ref{fig: Site5 Real results} shows the estimated results for the Site5 real data. In Case1, VIPSTF produces overly blurred results. RSFN-1, RSFN-2, SwinSTFM, and ECPW fail to predict spectral brightness. RobOt and ROSTF generate relatively reasonable estimates; however, they struggle to accurately capture the spectral changes from the reference image $\Hr$ to the ground-truth, especially in the region enclosed by the yellow box. STARFM and the proposed TSSTF achieve faithful reconstruction in terms of spatial structure and spectral consistency, closely matching the ground-truth image.

Under the noisy conditions, STARFM, VIPSTF, and RobOt reflect the noise contained in the input images, resulting in noisy outputs. RSFN-1, RSFN-2, SwinSTFM, and ECPW fail to accurately predict the spectral brightness, similar to their performance in Case1. ROSTF suppresses noise to some extent but it suffers from spectral distortion around edges, as its spatially uniform total variation regularization fails to distinguish between noise and intrinsic spatial structure. Furthermore, in Case7, ROSTF is significantly affected by the stripe noise present in the LR images, which remains visible in their resulting images. The proposed TSSTF consistently produces the most accurate results across all cases, effectively mitigating the influence of diverse noise while faithfully reconstructing both spatial structure and spectral characteristics.

\subsection{Facilitation of Parameter Selection}
\label{ssec:facilitation of parameter selection}
We have four parameters ($\delta, k, q, c_{\alpha}$), which were heuristically determined by empirical tuning. In this section, we present the details of comprehensive parameter searches conducted using all the data introduced in Sec.~\ref{ssec:data description}. We provide a single recommended value for each parameter, which is robust against diverse data characteristics, such as land cover types, temporal changes in spectral brightness, and noise levels. During the search for each parameter, the other parameters were fixed to the values listed in Table~\ref{table: parameters}.

\subsubsection{Parameter $\delta$ in weights}
\label{ssec:parameter delta of weights}
\begin{figure}[t]
    \begin{center}
        \begin{minipage}{0.49\hsize}
            \centerline{\includegraphics[width=\hsize]{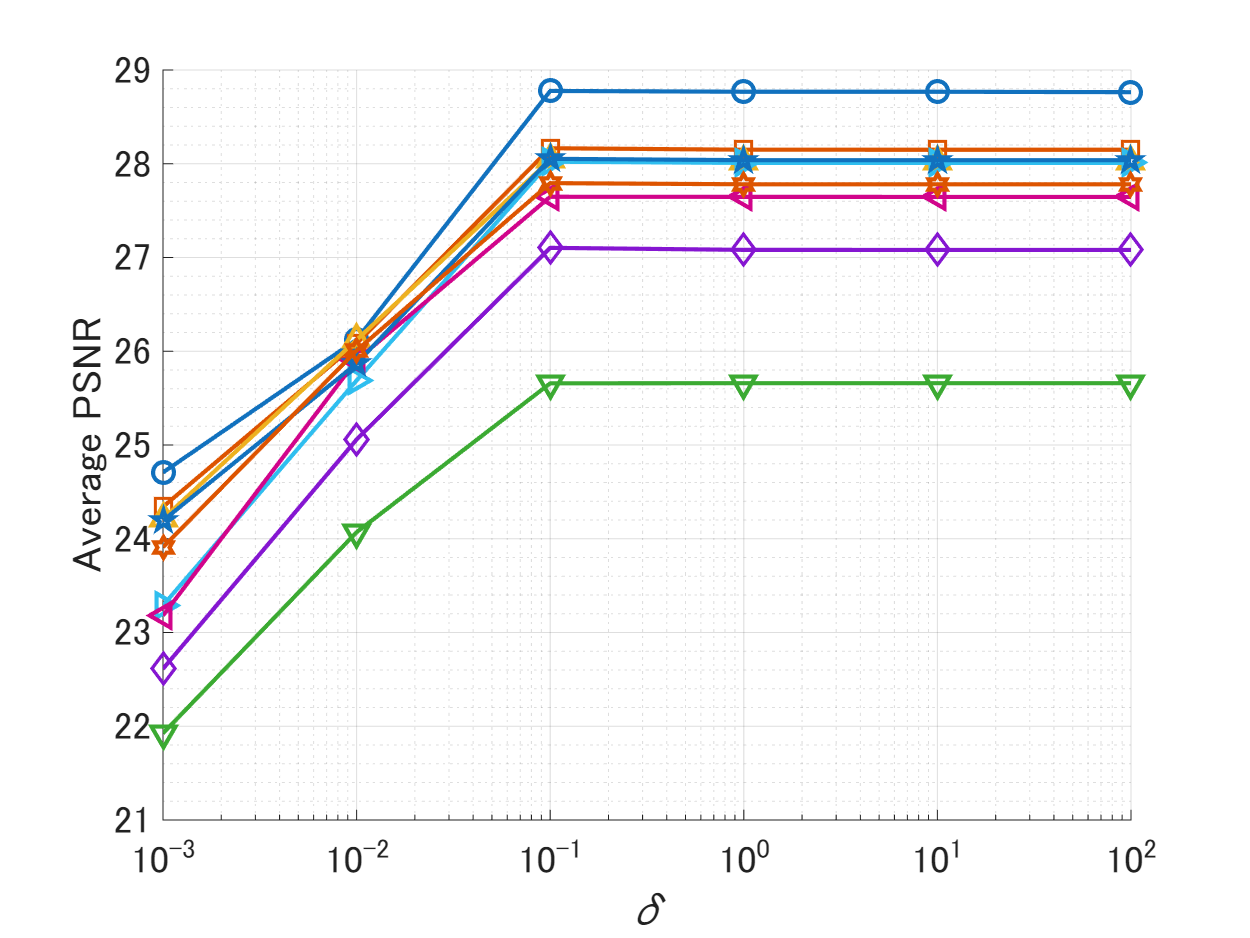}}
        \end{minipage}
        \hfill
        \begin{minipage}{0.49\hsize}
            \centerline{\includegraphics[width=\hsize]{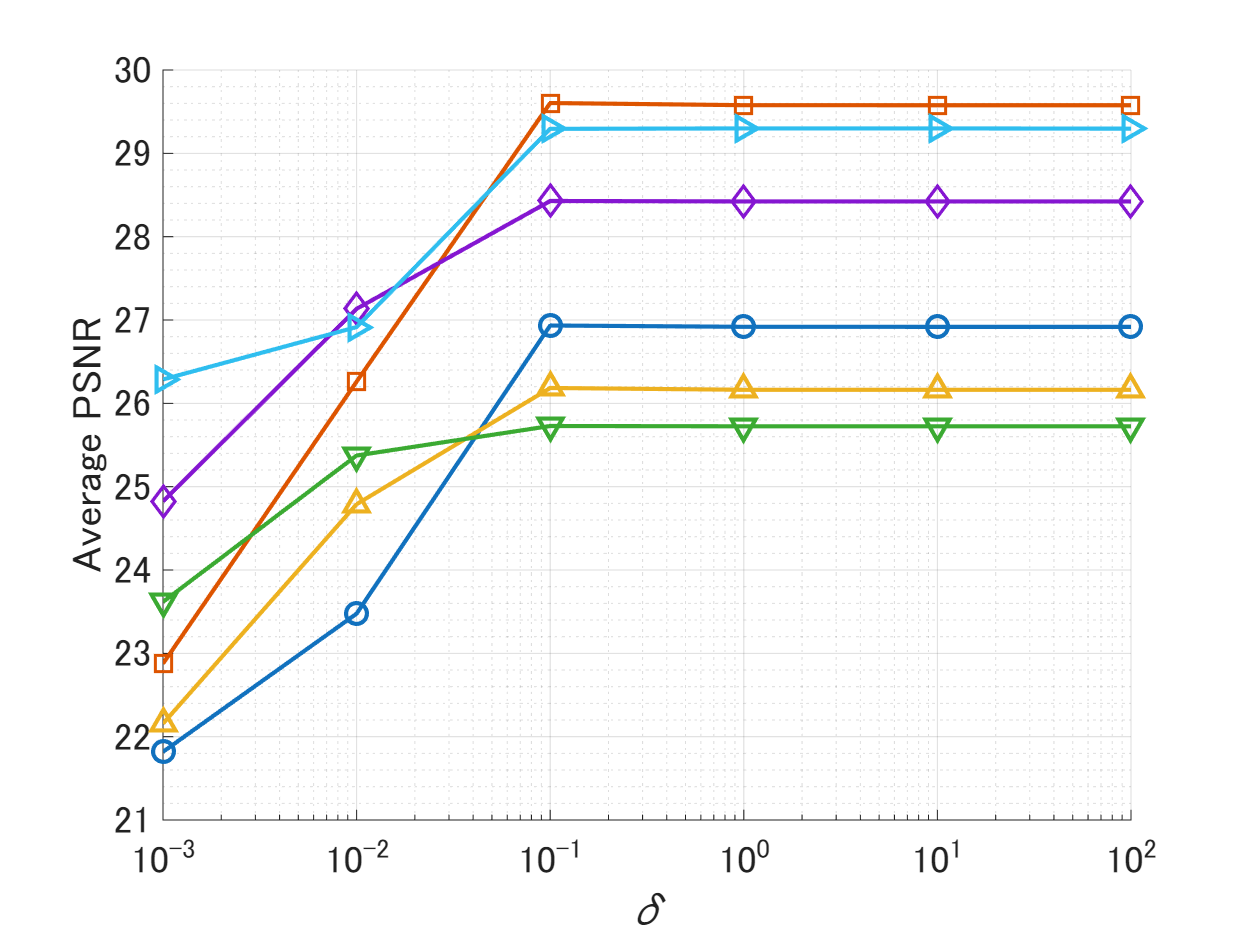}}
        \end{minipage} \\
        \begin{minipage}{0.49\hsize}
            \centerline{\includegraphics[width=\hsize]{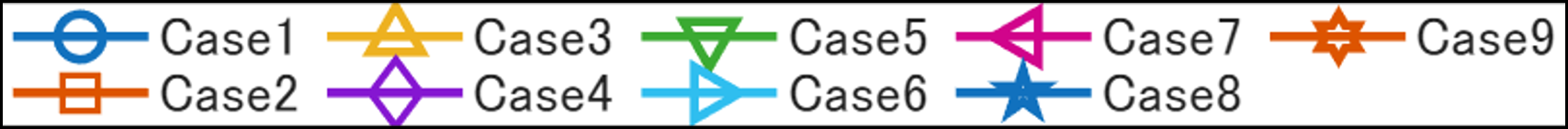}}
        \end{minipage}
        \hfill
        \begin{minipage}{0.49\hsize}
            \centerline{\includegraphics[width=\hsize]{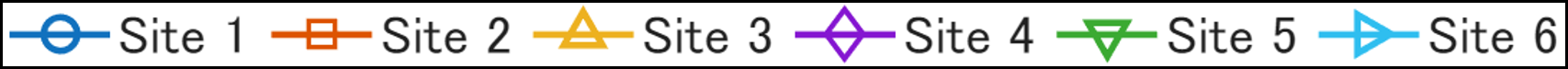}}
        \end{minipage} \\
        \vspace{1mm}
        \begin{minipage}{0.49\hsize}
            \centerline{(a)}
        \end{minipage}
        \hfill
        \begin{minipage}{0.49\hsize}
            \centerline{(b)}
        \end{minipage} \\
    \end{center}
    \vspace{-3mm}
    \caption{Comparison of the average PSNR according to $\delta$ under different conditions. The performance of TSSTF for each noise case (a) and for each site (b).}
    \label{fig: psnr_vs_delta_comparison}
\end{figure}

The parameter $\delta$, defined in \eqref{eq: weight}, controls the distribution of the weights. A larger $\delta$ leads to less variation among the weights, whereas a smaller $\delta$ results in greater variation. To identify an appropriate value for $\delta$, we conducted experiments using discrete values spanning from $10^{-3}$ to $10^{2}$ (, i.e., $10^{-3}, 10^{-2}, 10^{-1}, 10^{0}, 10^{1}$, and $10^{2}$). The impact of $\delta$ on performance was evaluated by averaging PSNR across all sites for each noise case (Fig.~\ref{fig: psnr_vs_delta_comparison}~(a)) and across all noise cases for each site (Fig.~\ref{fig: psnr_vs_delta_comparison}~(b)). As observed in both plots, the performance consistently peaks at $\delta = 10^{-1}$ and then either remains stable or slightly decreases. Importantly, this trend holds across different sites and noise levels. Therefore, we recommend to set $\delta = 10^{-1}$ as the default value.

\subsubsection{Parameter $k$ in weights}
\label{ssec:parameter k of weights}
\begin{figure}[t]
    \begin{center}
        \begin{minipage}{0.49\hsize}
            \centerline{\includegraphics[width=\hsize]{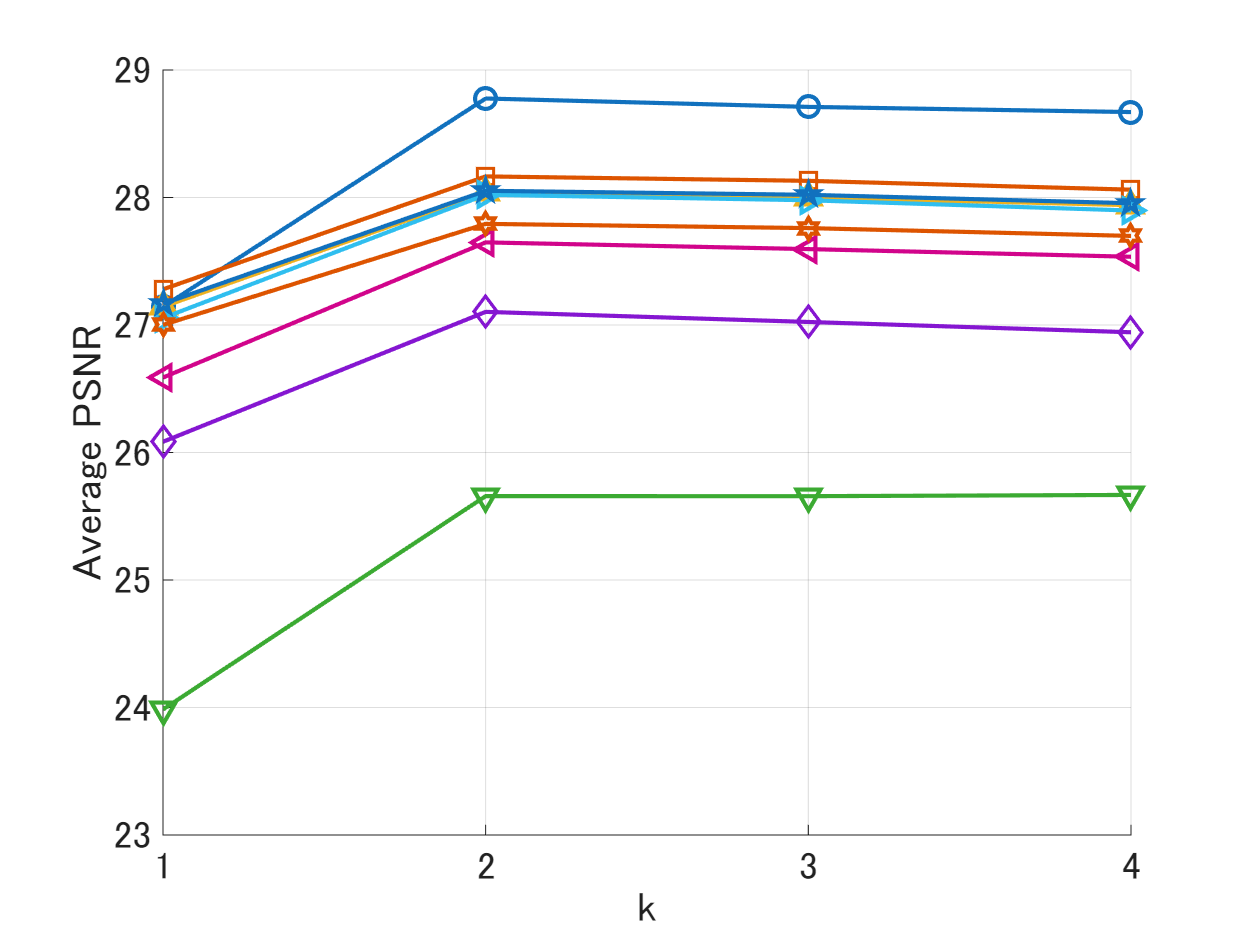}}
        \end{minipage}
        \hfill
        \begin{minipage}{0.49\hsize}
            \centerline{\includegraphics[width=\hsize]{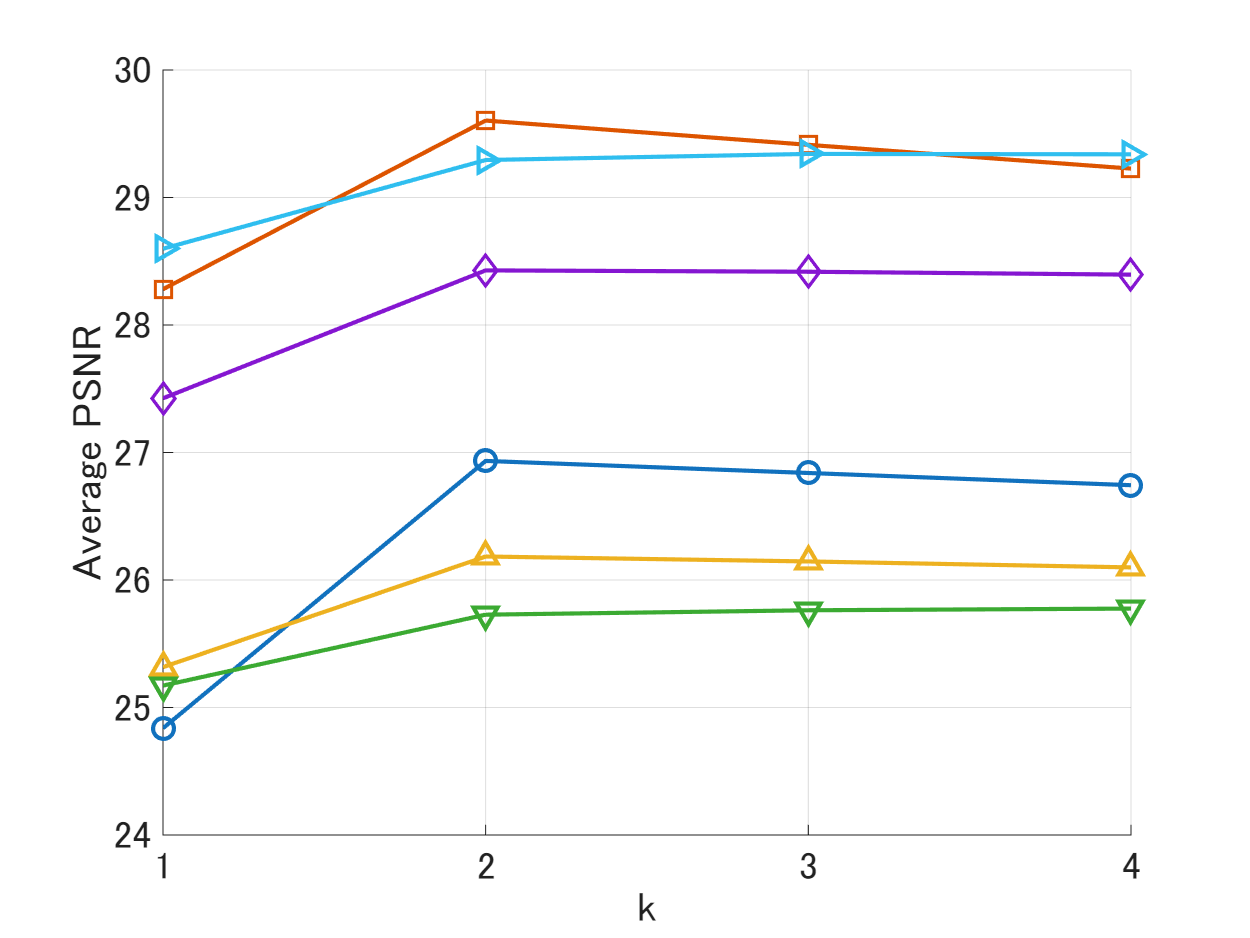}}
        \end{minipage} \\
        \begin{minipage}{0.49\hsize}
            \centerline{\includegraphics[width=\hsize]{img/legend_noisecase.png}}
        \end{minipage}
        \hfill
        \begin{minipage}{0.49\hsize}
            \centerline{\includegraphics[width=\hsize]{img/legend_Site.png}}
        \end{minipage} \\
        \vspace{1mm}
        \begin{minipage}{0.49\hsize}
            \centerline{(a)}
        \end{minipage}
        \hfill
        \begin{minipage}{0.49\hsize}
            \centerline{(b)}
        \end{minipage} \\
    \end{center}
    \vspace{-3mm}
    \caption{Comparison of the average PSNR according to $k$ under different conditions. The performance of TSSTF for each noise case (a) and for each site (b).}
    \label{fig: psnr_vs_k_comparison}
\end{figure}

The parameter $k$ specifies how many of the four directional weights $\{w_{i,j}^{(p)}\}_{p=1}^{4}$ are forced to zero at each pixel~$(i,j)$. We evaluated $k \in \{1,2,3,4\}$ and examined the resulting PSNR averaged over all sites for each noise case (Fig.~\ref{fig: psnr_vs_k_comparison}~(a)) and over all noise cases for each site (Fig.~\ref{fig: psnr_vs_k_comparison}~(b)). Across diverse noise conditions and sites, the PSNR generally peaks or maintains high values at $k=2$, indicating that this setting offers robust performance regardless of data characteristics.

\subsubsection{Norm $q$ in TGEC}
\label{ssec:norm q of tgec}
\begin{table*}[t]
\centering
\caption{Average PSNR Results With the $\ell_1$, $\ell_2$, and Mixed $\ell_{1,2}$-Norm in TGEC Across Different Noise Cases and Sites}
\label{table: psnr_vs_q}
\vspace{-2mm}
\begin{tabular}{l|ccccccccc|cccccc}
\toprule
    & Case1 & Case2 & Case3 & Case4 & Case5 & Case6 & Case7 & Case8 & Case9 & Site1 & Site2 & Site3 & Site4 & Site5 & Site6 \\
    \midrule
    $\ell_1$ & 28.17 & 27.79 & 27.67 & 26.73 & 25.26 & 27.65 & 27.25 & 27.69 & 27.42 & 26.33 & \textbf{29.60} & 25.56 & 28.13 & 25.05 & 29.08 \\
    $\ell_2$ & 28.63 & 27.98 & 27.85 & 26.88 & 25.46 & 27.82 & 27.42 & 27.86 & 27.60 & 26.46 & 29.53 & 25.94 & 28.38 & 25.51 & 29.19 \\
    $\ell_{1,2}$ & \textbf{28.78} & \textbf{28.17} & \textbf{28.04} & \textbf{27.10} & \textbf{25.66} & \textbf{28.02} & \textbf{27.65} & \textbf{28.05} & \textbf{27.79} & \textbf{26.93} & \textbf{29.60} & \textbf{26.18} & \textbf{28.43} & \textbf{25.73} & \textbf{29.29} \\
\bottomrule
\end{tabular}
\end{table*}
We considered three options for the TGEC norm \( q \) in \eqref{eq: TGEC}, namely the \( \ell_1 \)-norm, the \( \ell_2 \)-norm, and the mixed \( \ell_{1,2} \)-norm. 
Table~\ref{table: psnr_vs_q} shows the average PSNR results with the three types of TGEC norm across different noise conditions and sites. Since TSSTF consistently yields the best PSNR with the mixed \( \ell_{1,2} \)-norm, we recommend this setting for practical applications.

\subsubsection{Coefficient $c_{\alpha}$ in TGEC}
\label{ssec:coefficient c_alpha of tgec}
\begin{figure}[t]
    \begin{center}
        \begin{minipage}{0.49\hsize}
            \centerline{\includegraphics[width=\hsize]{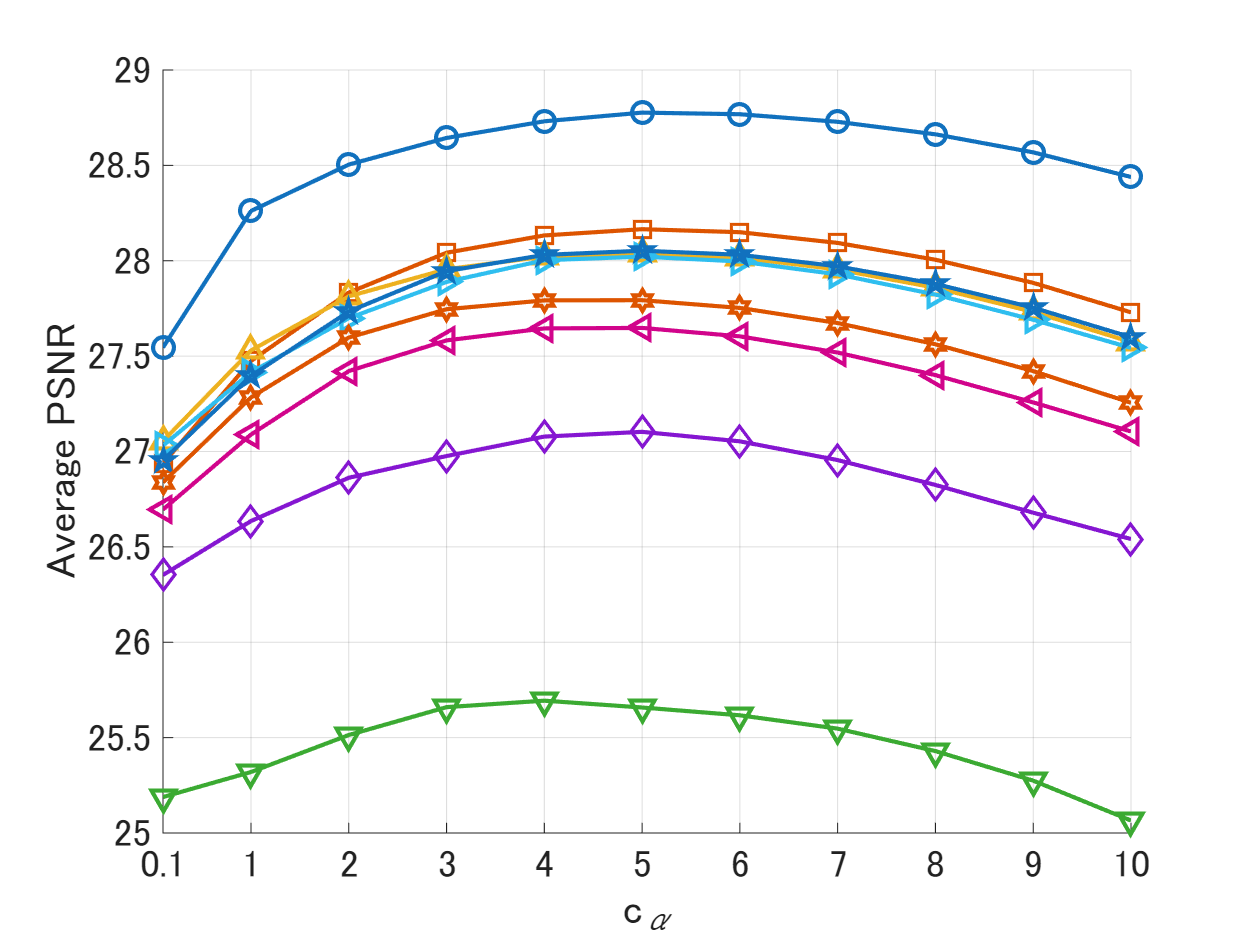}}
        \end{minipage}
        \hfill
        \begin{minipage}{0.49\hsize}
            \centerline{\includegraphics[width=\hsize]{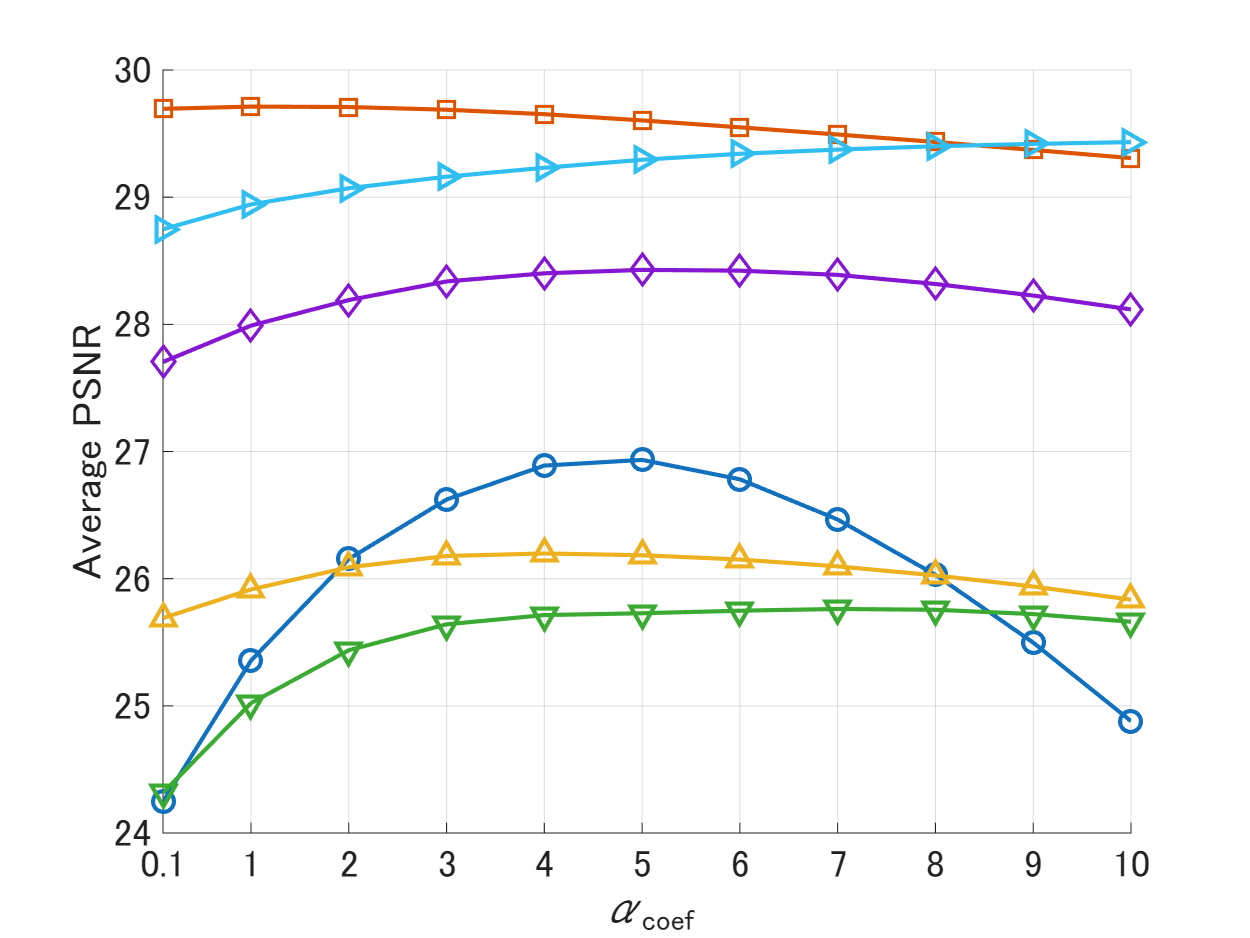}}
        \end{minipage} \\
        \begin{minipage}{0.49\hsize}
            \centerline{\includegraphics[width=\hsize]{img/legend_noisecase.png}}
        \end{minipage}
        \hfill
        \begin{minipage}{0.49\hsize}
            \centerline{\includegraphics[width=\hsize]{img/legend_Site.png}}
        \end{minipage} \\
        \vspace{1mm}
        \begin{minipage}{0.49\hsize}
            \centerline{(a)}
        \end{minipage}
        \hfill
        \begin{minipage}{0.49\hsize}
            \centerline{(b)}
        \end{minipage} \\
    \end{center}
    \vspace{-3mm}
    \caption{Comparison of the average PSNR according to $c_{\alpha}$ under different conditions. The performance of TSSTF for each noise case (a) and for each site (b).}
    \label{fig: psnr_vs_alpha_comparison}
\end{figure}

We have a parameter $c_{\alpha}$ in \eqref{eq: alpha setting} to control the threshhold $\alpha$ for TGEC, which regulates the degree of edge location similarity. We evaluated the performance of TSSTF with values of $c_{\alpha}$ ranging from 0.1 to 10 (specifically, 0.1, 1, 2, 3, 4, 5, 6, 7, 8, 9, and 10). Fig.~\ref{fig: psnr_vs_alpha_comparison}~(a) presents the average PSNR results across all sites for each noise case, while Fig.~\ref{fig: psnr_vs_alpha_comparison}~(b) shows the average PSNR results across all noise cases for each site. These results demonstrate that TSSTF consistently achieves high performance when $c_{\alpha}$ is set around $5$, regardless of input data. 

\subsection{Algoritghm Convergence}
\label{ssec: algorithm convergence}
\begin{figure*}[t]
    \begin{center}
        \begin{minipage}{0.01\hsize}
			\centerline{\rotatebox{90}{$\alpha^{(n)}$}}
		\end{minipage}
        \begin{minipage}{0.19\hsize}
            \centerline{\includegraphics[width=\hsize]{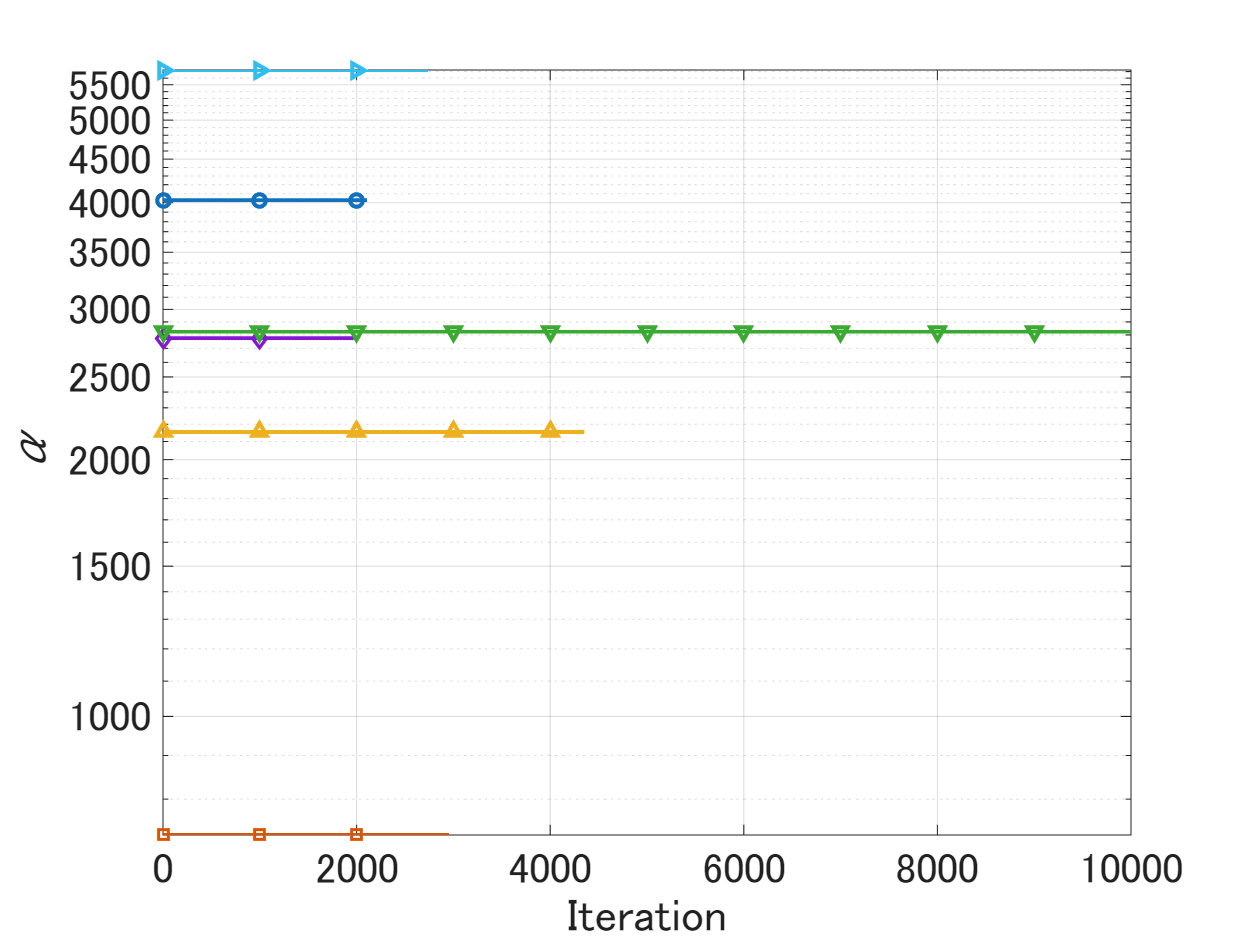}}
        \end{minipage}
        \hfill
        \begin{minipage}{0.19\hsize}
            \centerline{\includegraphics[width=\hsize]{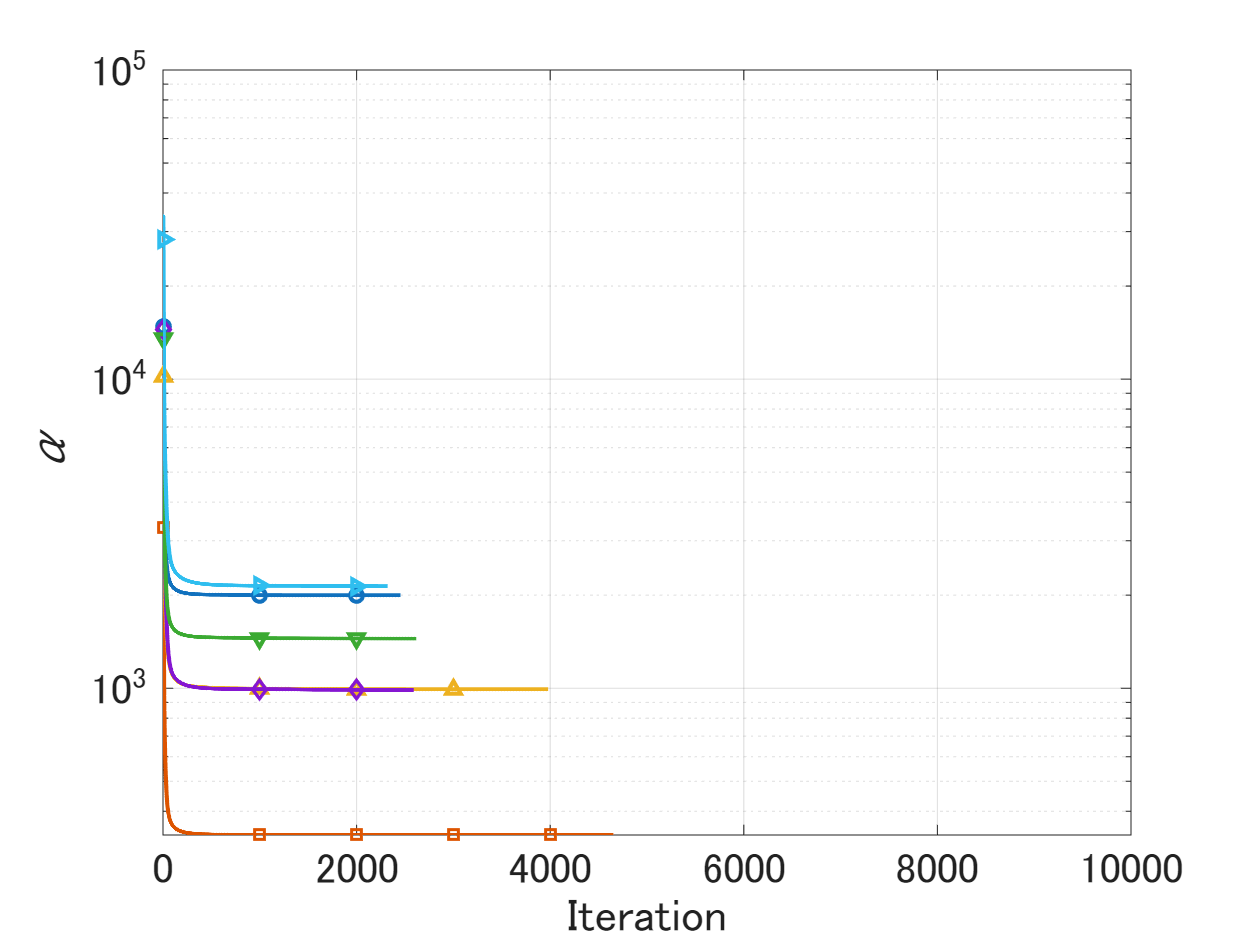}}
        \end{minipage}
        \hfill
        \begin{minipage}{0.19\hsize}
            \centerline{\includegraphics[width=\hsize]{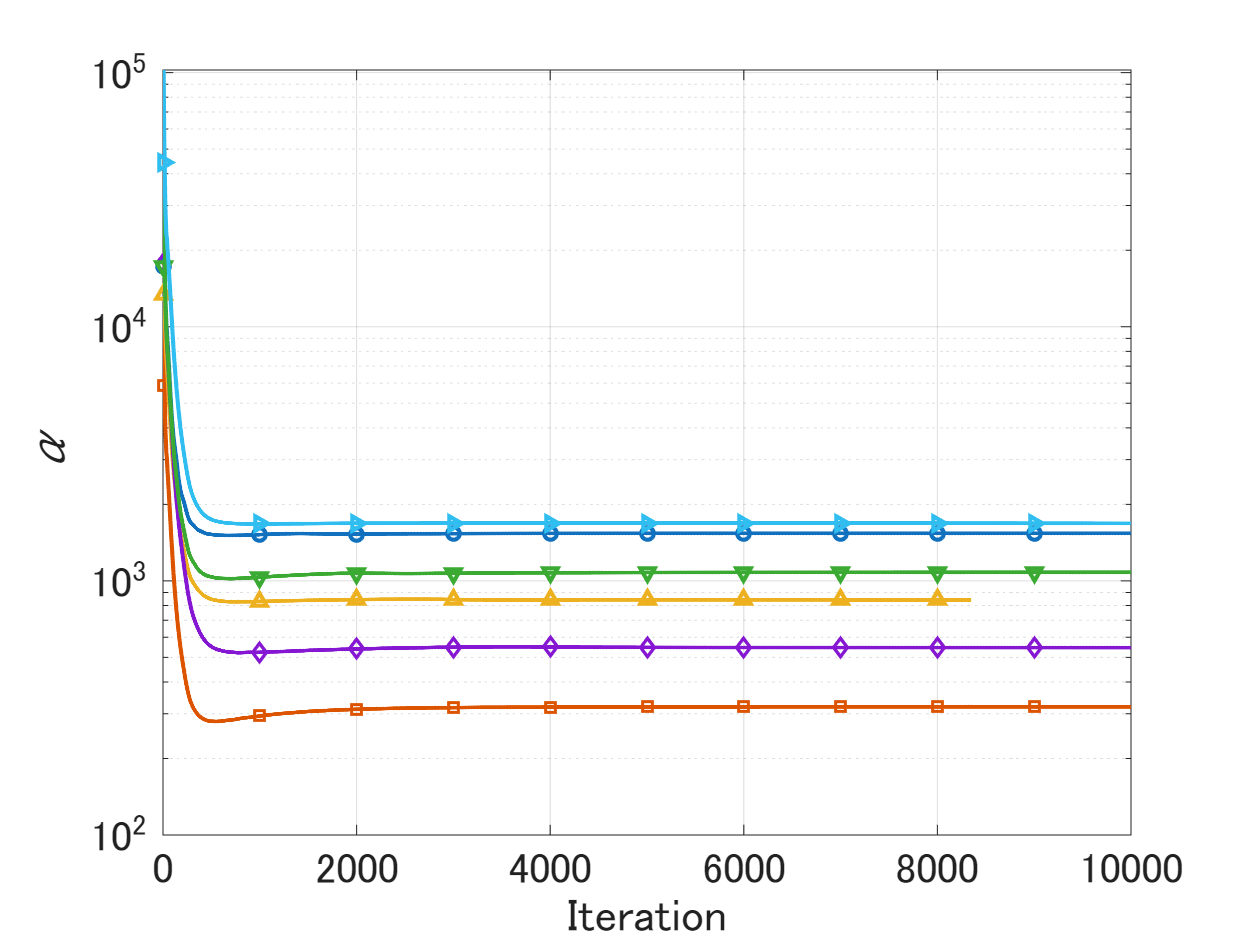}}
        \end{minipage}
        \hfill
        \begin{minipage}{0.19\hsize}
            \centerline{\includegraphics[width=\hsize]{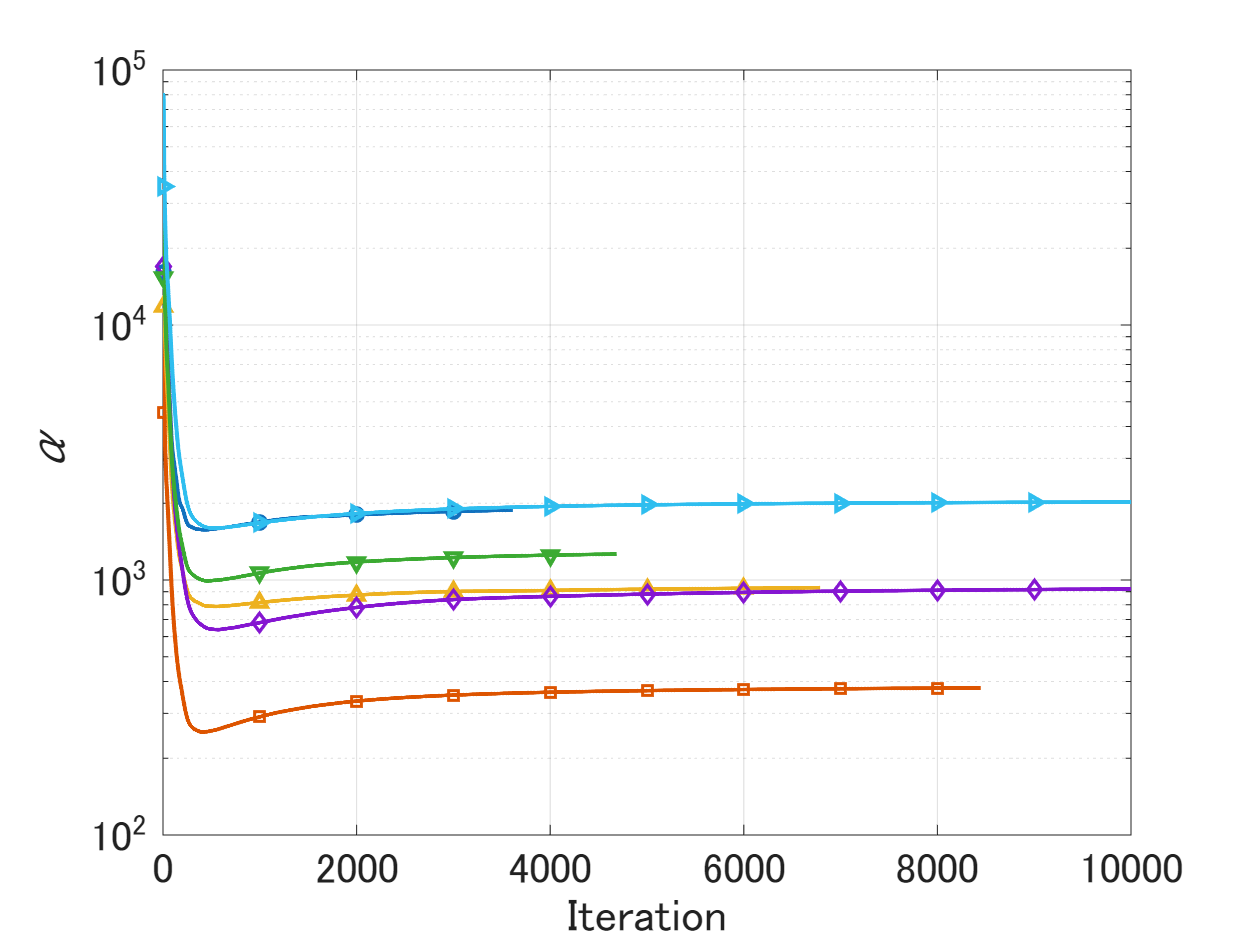}}
        \end{minipage}
        \hfill
        \begin{minipage}{0.19\hsize}
            \centerline{\includegraphics[width=\hsize]{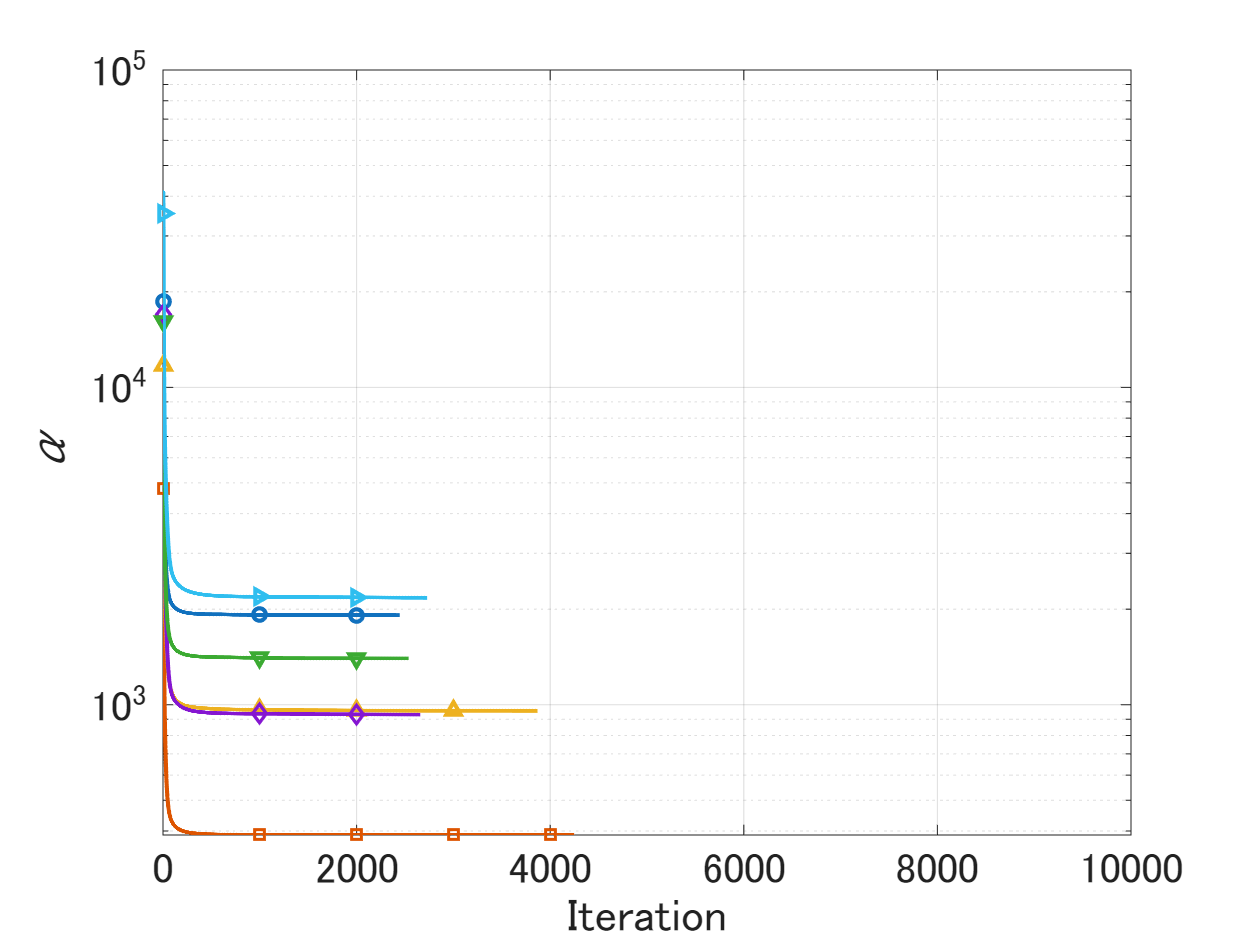}}
        \end{minipage} \\
        \begin{minipage}{0.01\hsize}
			\centerline{\rotatebox{90}{Update Error for $\tilHt$}}
		\end{minipage}
        \begin{minipage}{0.19\hsize}
            \centerline{\includegraphics[width=\hsize]{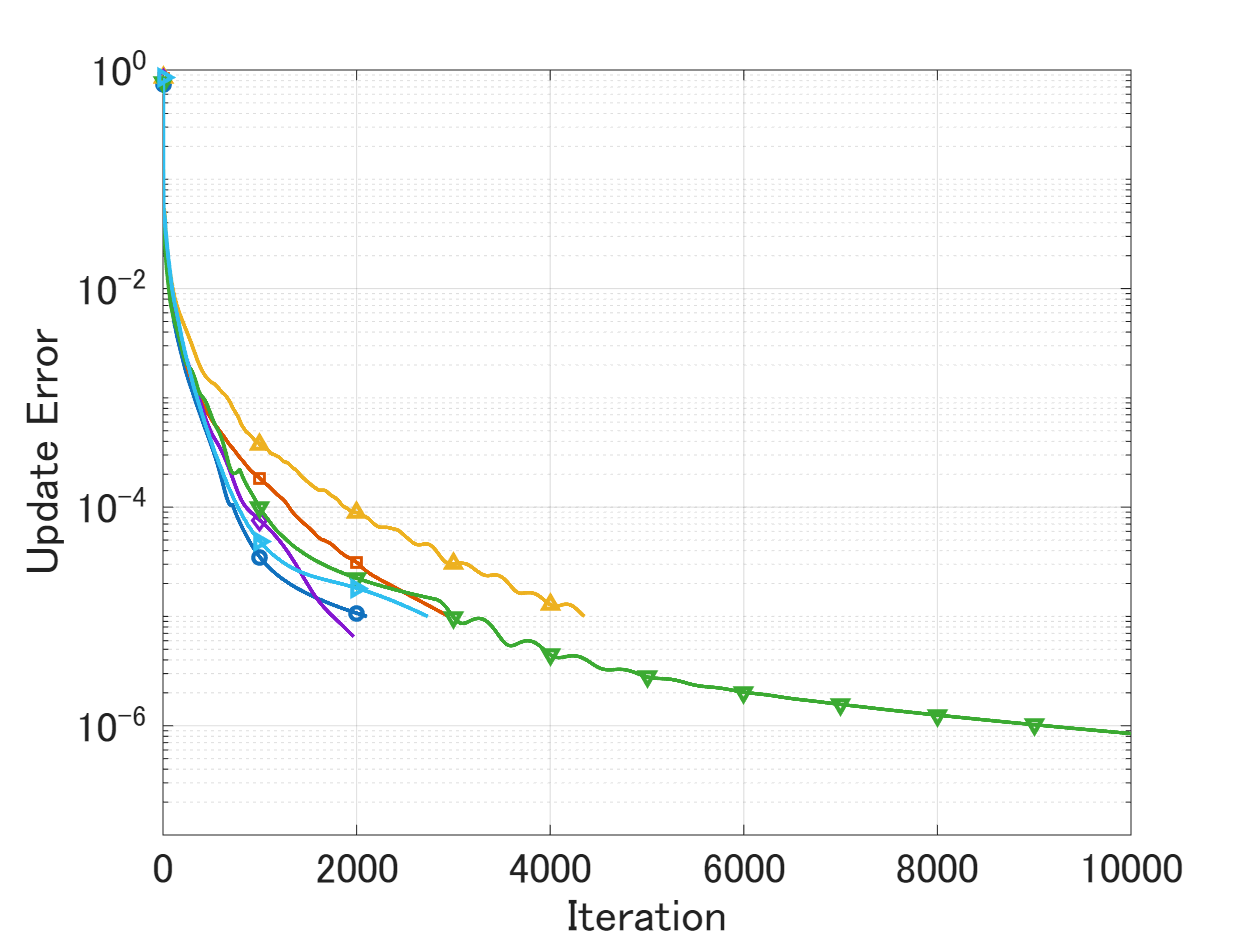}}
        \end{minipage}
        \hfill
        \begin{minipage}{0.19\hsize}
            \centerline{\includegraphics[width=\hsize]{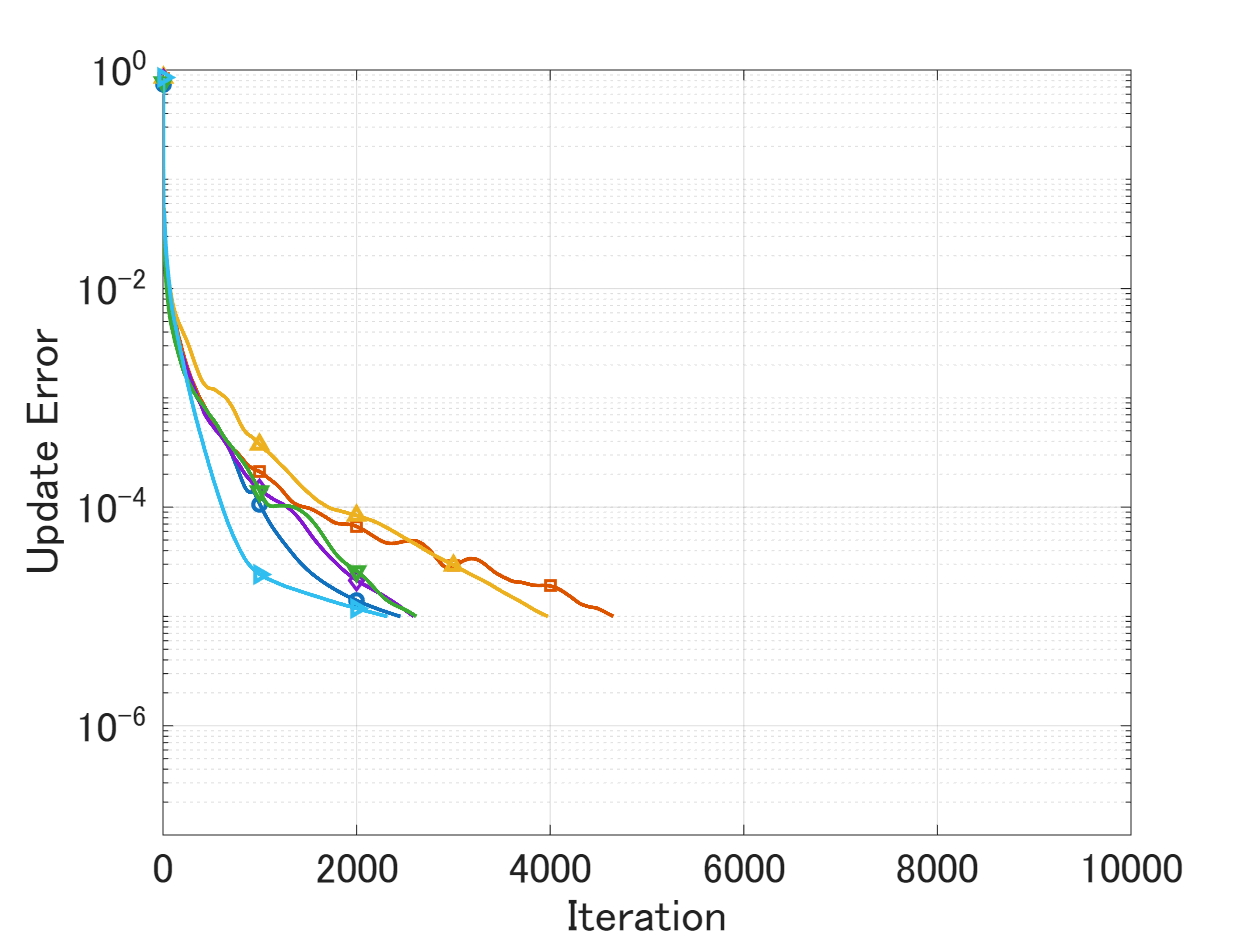}}
        \end{minipage}
        \hfill
        \begin{minipage}{0.19\hsize}
            \centerline{\includegraphics[width=\hsize]{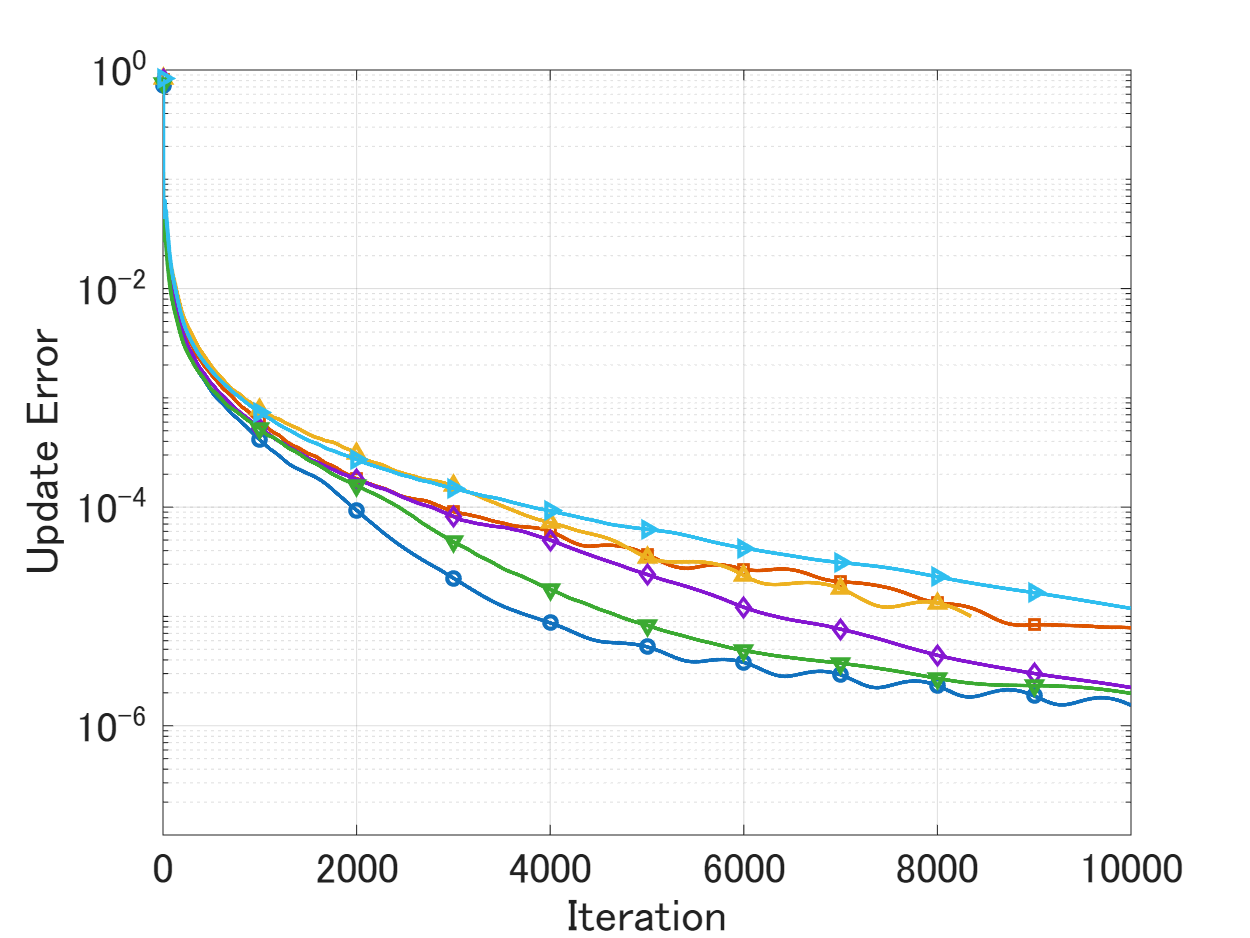}}
        \end{minipage}
        \hfill
        \begin{minipage}{0.19\hsize}
            \centerline{\includegraphics[width=\hsize]{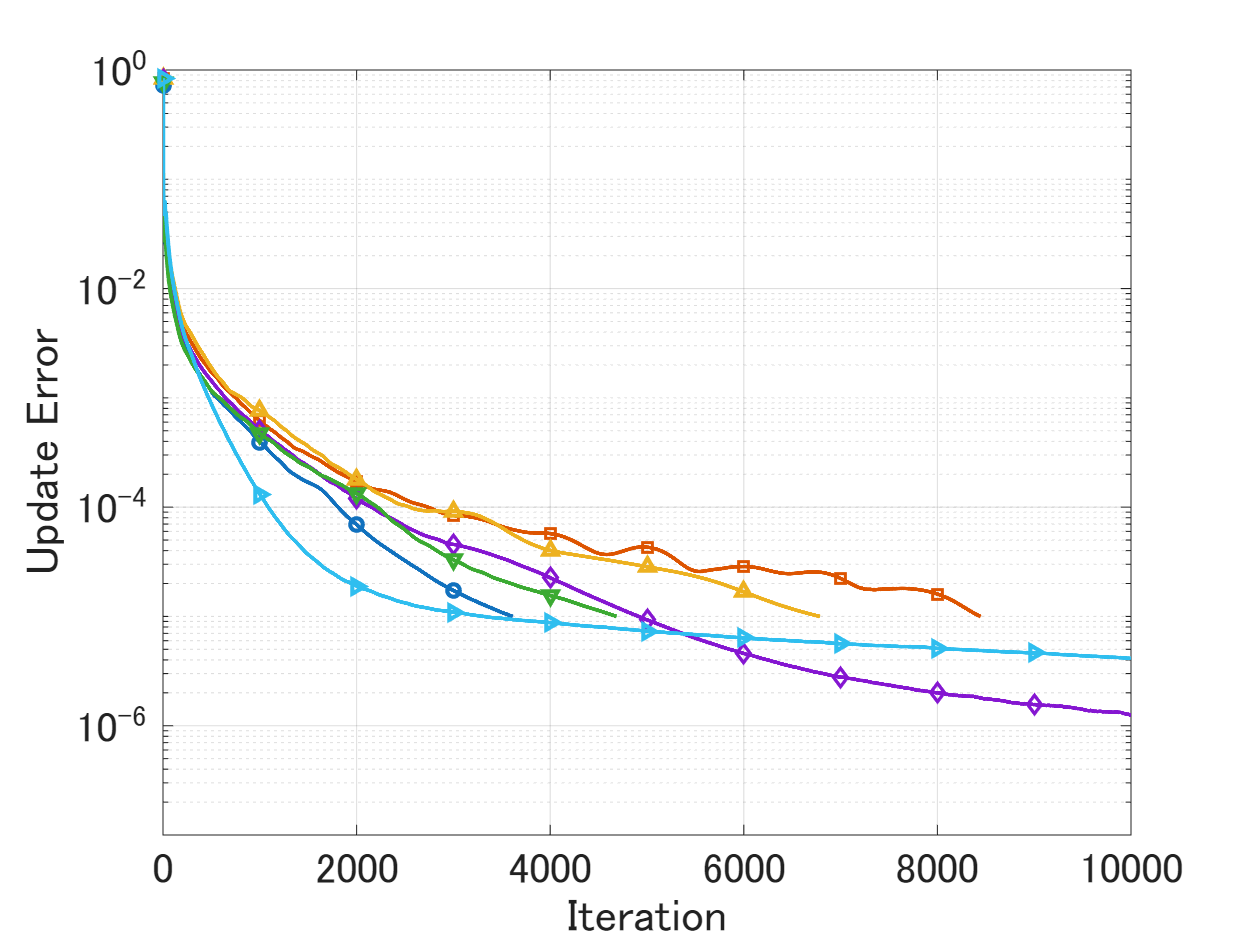}}
        \end{minipage}
        \hfill
        \begin{minipage}{0.19\hsize}
            \centerline{\includegraphics[width=\hsize]{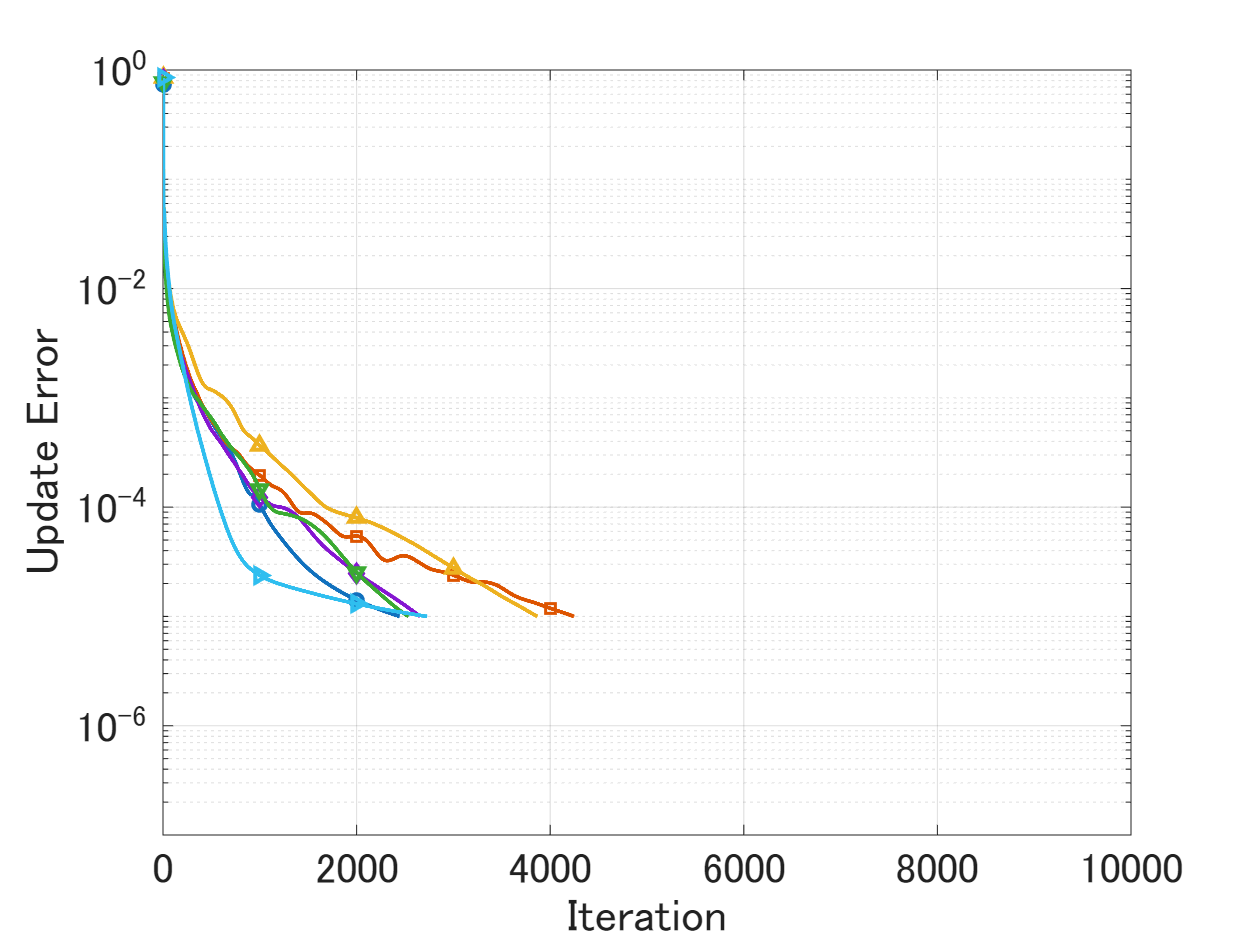}}
        \end{minipage} \\
        \begin{minipage}{0.01\hsize}
			\centerline{\rotatebox{90}{PSNR}}
		\end{minipage}
        \begin{minipage}{0.19\hsize}
            \centerline{\includegraphics[width=\hsize]{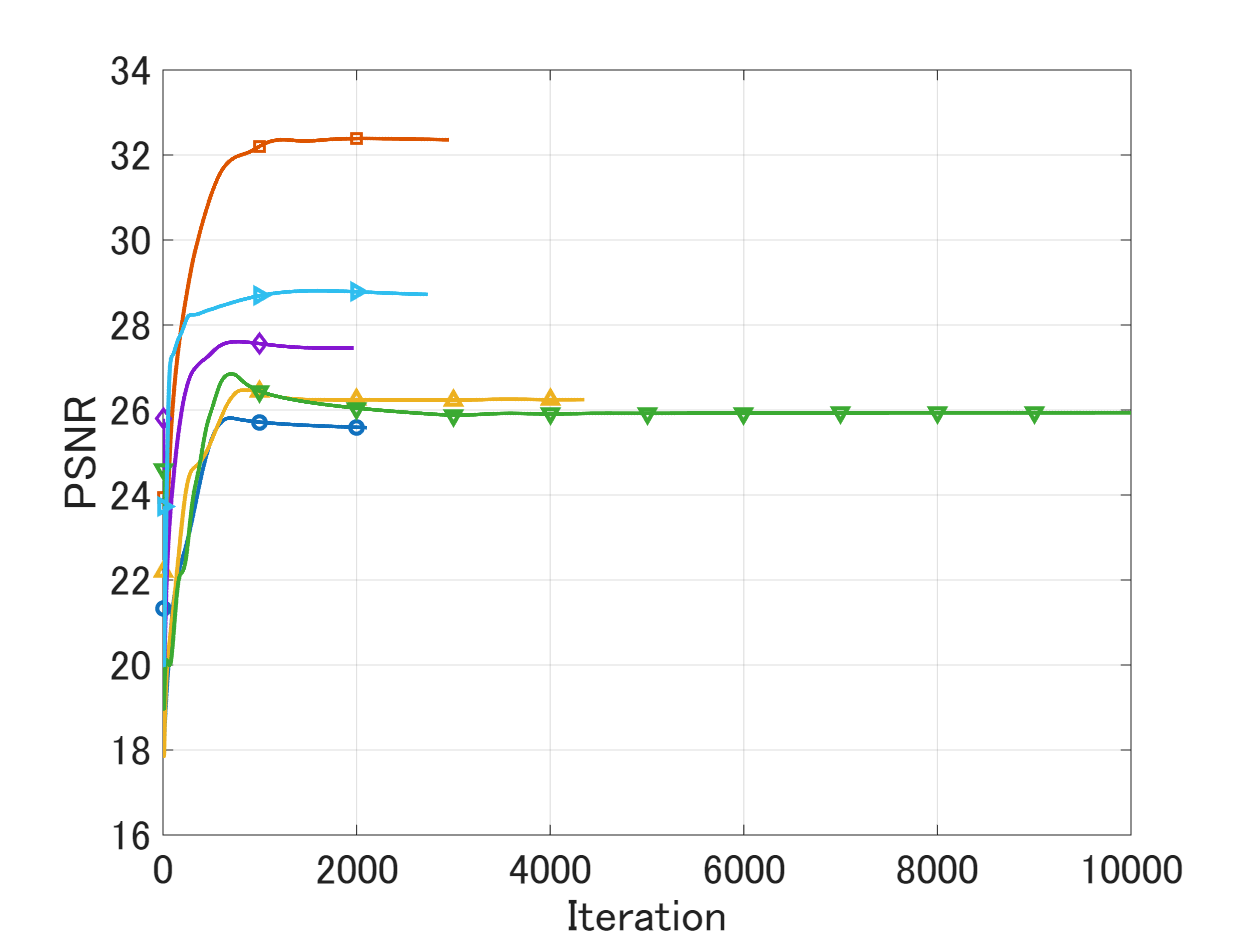}}
        \end{minipage}
        \hfill
        \begin{minipage}{0.19\hsize}
            \centerline{\includegraphics[width=\hsize]{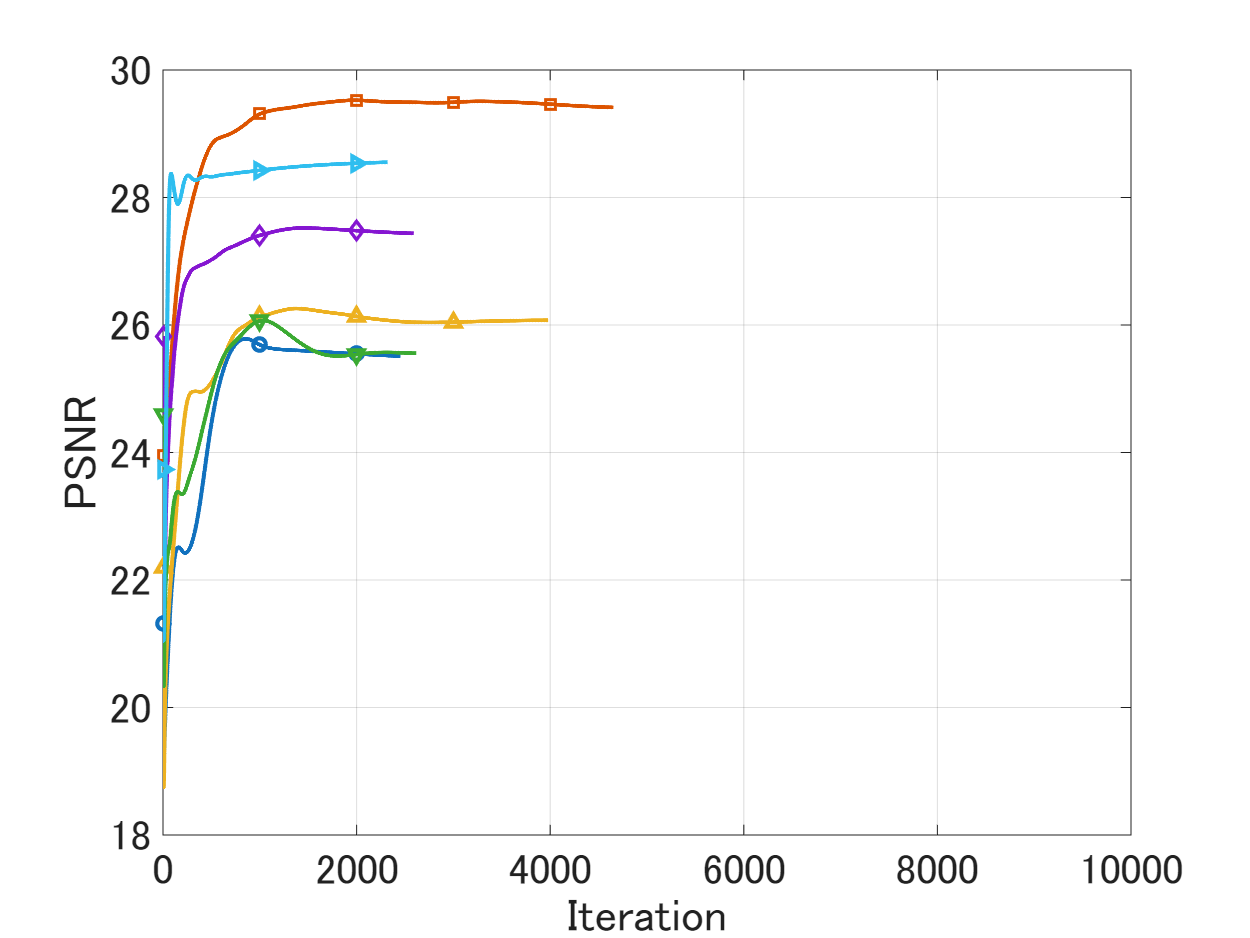}}
        \end{minipage}
        \hfill
        \begin{minipage}{0.19\hsize}
            \centerline{\includegraphics[width=\hsize]{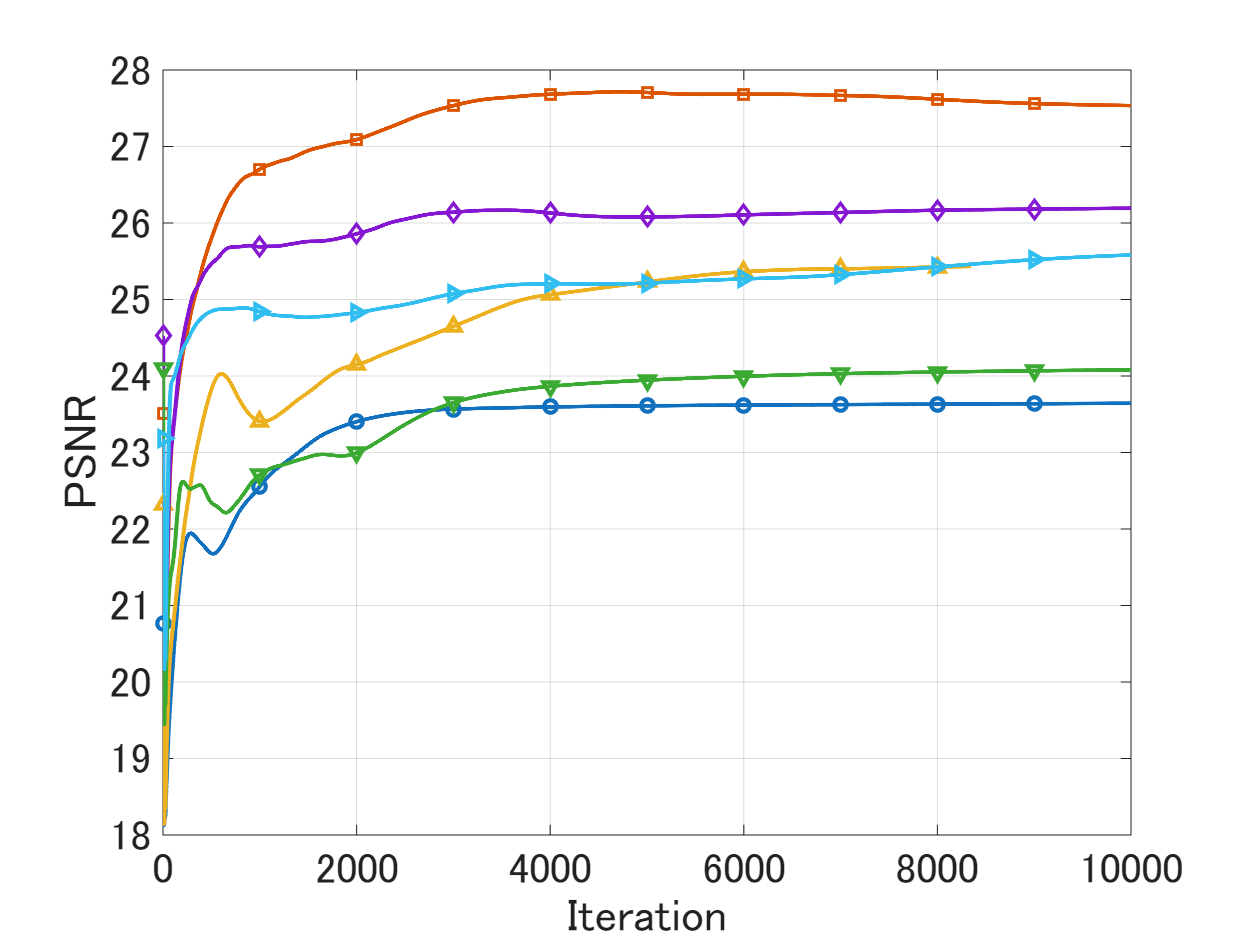}}
        \end{minipage}
        \hfill
        \begin{minipage}{0.19\hsize}
            \centerline{\includegraphics[width=\hsize]{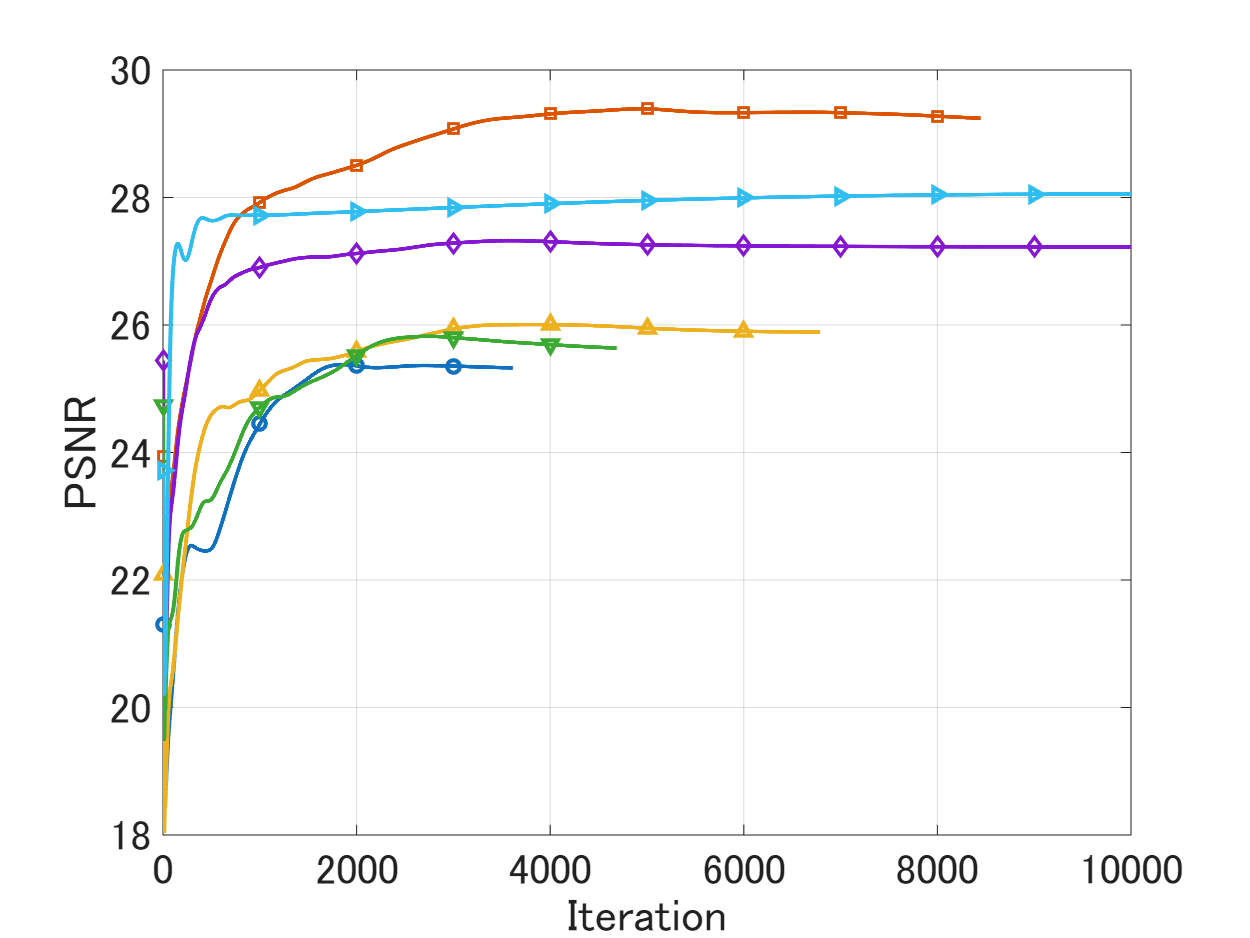}}
        \end{minipage}
        \hfill
        \begin{minipage}{0.19\hsize}
            \centerline{\includegraphics[width=\hsize]{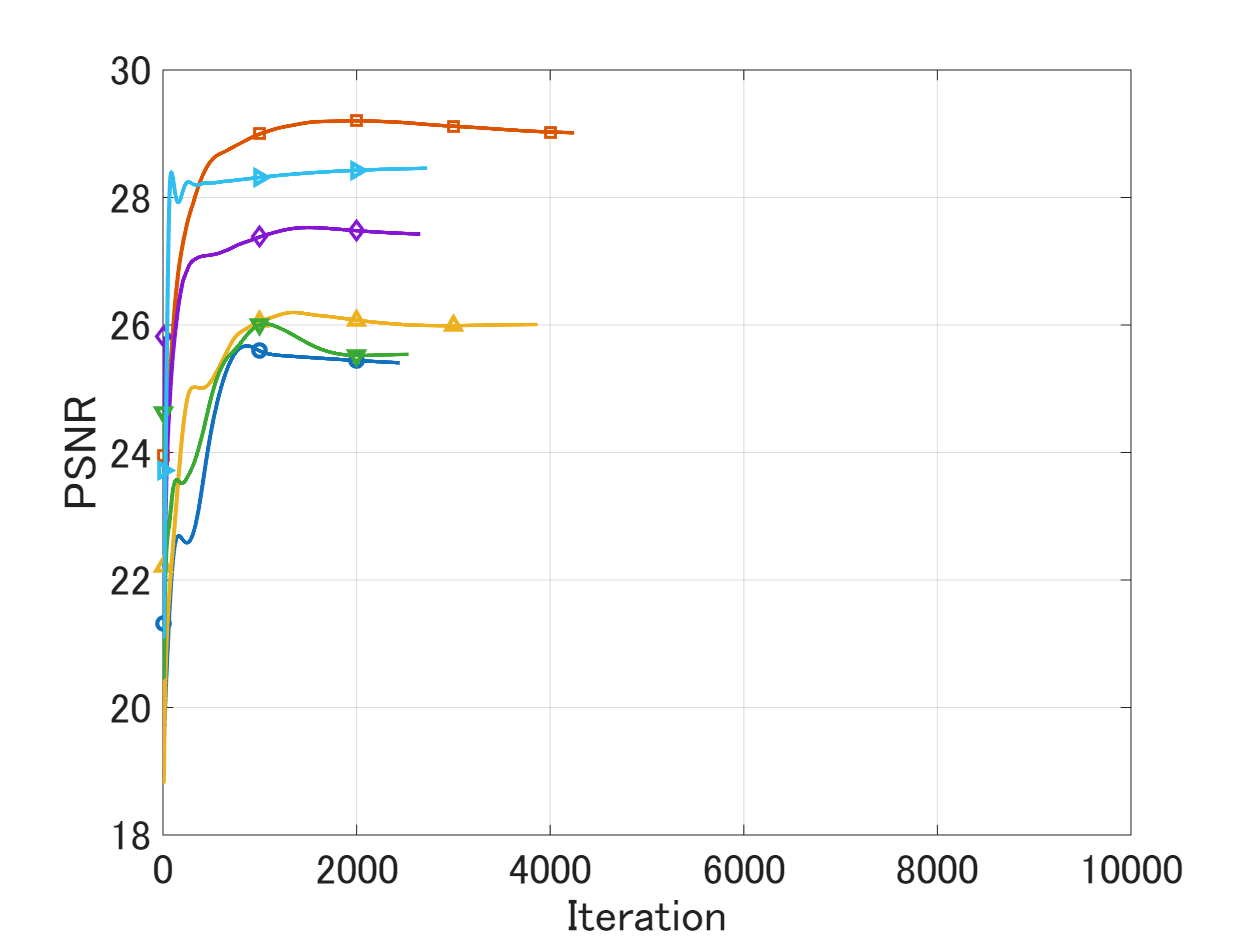}}
        \end{minipage} \\
        \vspace{1mm}
        \begin{minipage}{0.01\hsize}
			\centerline{}
		\end{minipage}
        \begin{minipage}{0.19\hsize}
            \centerline{Case1}
        \end{minipage}
        \hfill
        \begin{minipage}{0.19\hsize}
            \centerline{Case3}
        \end{minipage}
        \hfill
        \begin{minipage}{0.19\hsize}
            \centerline{Case5}
        \end{minipage}
        \hfill
        \begin{minipage}{0.19\hsize}
            \centerline{Case7}
        \end{minipage}
        \hfill
        \begin{minipage}{0.19\hsize}
            \centerline{Case9}
        \end{minipage} \\
        \vspace{2mm}
        \begin{minipage}{0.4\hsize}
            \centerline{\includegraphics[width=\hsize]{img/legend_Site.png}}
        \end{minipage}\\
    \end{center}
    \vspace{-3mm}
    \caption{Behavior of the TSSTF algorithm for real data across different noise cases. The top, middle, and bottom rows show the transitions of $\alpha^{(n)}$, the update error for $\tilHt$, and the PSNR values, respectively.}
    \label{fig: convergence_plot}
\end{figure*}

We investigate the dynamic behavior of the TSSTF algorithm during the optimization process. Specifically, we track three quantities at each iteration $n$:
\begin{itemize}
    \item the adaptive threshold $\alpha^{(n)}$ for TGEC defined in \eqref{eq: alpha setting},
    \item the update error for $\tilHt$, defined as $\|\tilHt^{(n)} - \tilHt^{(n-1)}\|_2/\|\tilHt^{(n-1)}\|_2$,
    \item the PSNR values, calculated between the intermediate estimate $\tilHt^{(n)}$ and the ground-truth $\hatHt$.
\end{itemize}
Fig.~\ref{fig: convergence_plot} shows their transitions for real data under five noise cases (Case1, Case3, Case5, Case7, Case9). Each curve corresponds to one site.

In the noise-free setting (Case1), $\alpha^{(n)}$ remains constant because the observed reference HR image $\Hr$ is fully trusted and the estimate $\tilHr$ therefore is not updated. 
Under noise contamination (Case3, Case5, Case7, Case9), $\alpha^{(n)}$ is adaptively updated by the rule in \eqref{eq: alpha setting}. After at least 2,000 iterations, however, these adjustments cease almost entirely and $\alpha^{(n)}$ stabilizes. From that point on, the algorithm can be regarded as solving a constrained optimization problem with a fixed $\alpha$, and it is expected to converge to its optimal solution.

This stable evolution of $\alpha^{(n)}$ arises because the intermediate estimate $\tilHr^{(n+1)}$ is gradually denoised through iterations. After a sufficient number of iterations, the quantity $\| \W \D \tilHr^{(n+1)} \|_{1,2}$ settles to a value that reflects the intrinsic edge strength and density of the target scene. Although the appropriate strictness of TGEC cannot be determined in advance due to noise, this adaptive setting avoids the need to pre-select a fixed $\alpha$, allowing the algorithm to self-tune across diverse noise conditions.

As shown in the middle and bottom rows of Fig.~\ref{fig: convergence_plot}, the update error decreases almost monotonically in all cases, and the PSNR curve flattens out after approximately 3,000 iterations. These results confirm that the proposed algorithm reliably converges without explicitly fixing $\alpha$ at an intermediate iteration.

\section{Discussion}
\label{sec:discussion}
\subsection{Computational Cost}
\label{ssec: computational cost}

\begin{table}[t]
\centering
\caption{Computational Time (in seconds)}
\label{table: running time}
\vspace{-2mm}
\begin{tabular}{ccccc}
\toprule
    STARFM & VIPSTF & RobOt & ROSTF & TSSTF \\
    \midrule
    88.33 & 86.74 & 0.1150 & 712.7 & 272.4 \\
\bottomrule
\end{tabular}
\end{table}

We measured the actual running times using MATLAB (R2025b) on a Windows 11 computer equipped with an Intel Core Ultra 9 285K processor, 32 GB of RAM, and an NVIDIA GeForce RTX 5090. We used the Landsat-MODIS data pairs (Sites1-5), each with spatial dimensions of $400\times400$ and six spectral bands.
Table~\ref{table: running time} shows the average running times of the non-deep learning-based comparison methods (STARFM, VIPSTF, RobOt, and ROSTF) and the proposed TSSTF.

\begin{figure}[t]
    \begin{center}
        \begin{minipage}{0.01\hsize}
			\centerline{\rotatebox{90}{Update Error for $\tilHt$}}
		\end{minipage}
        \begin{minipage}{0.48\hsize}
            \centerline{\includegraphics[width=\hsize]{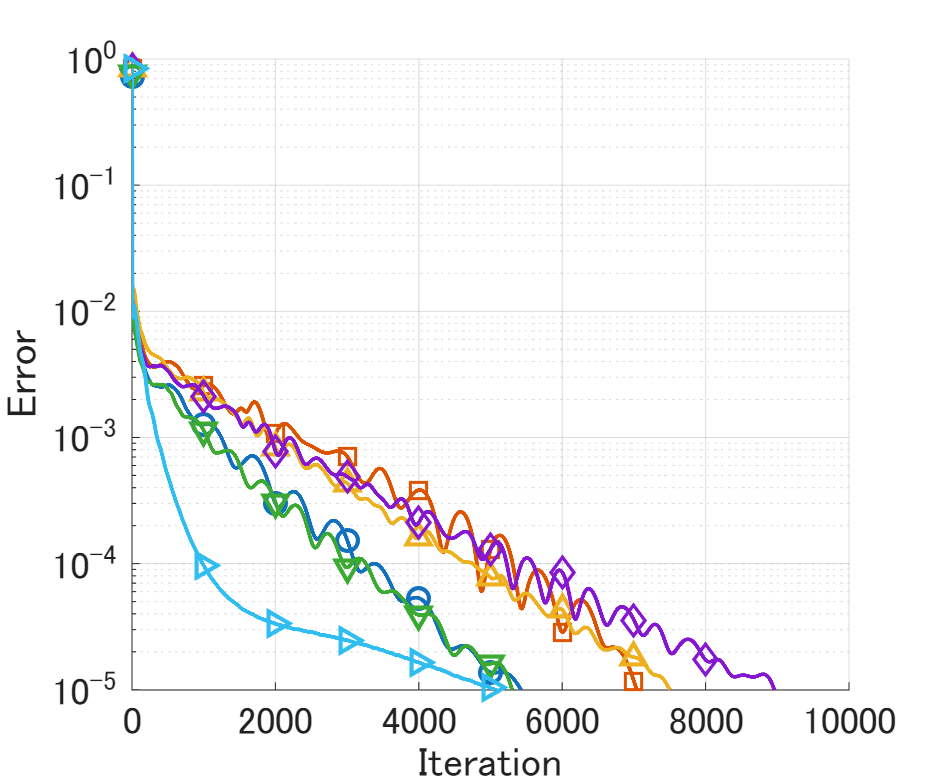}}
        \end{minipage}
        \hfill
        \begin{minipage}{0.48\hsize}
            \centerline{\includegraphics[width=\hsize]{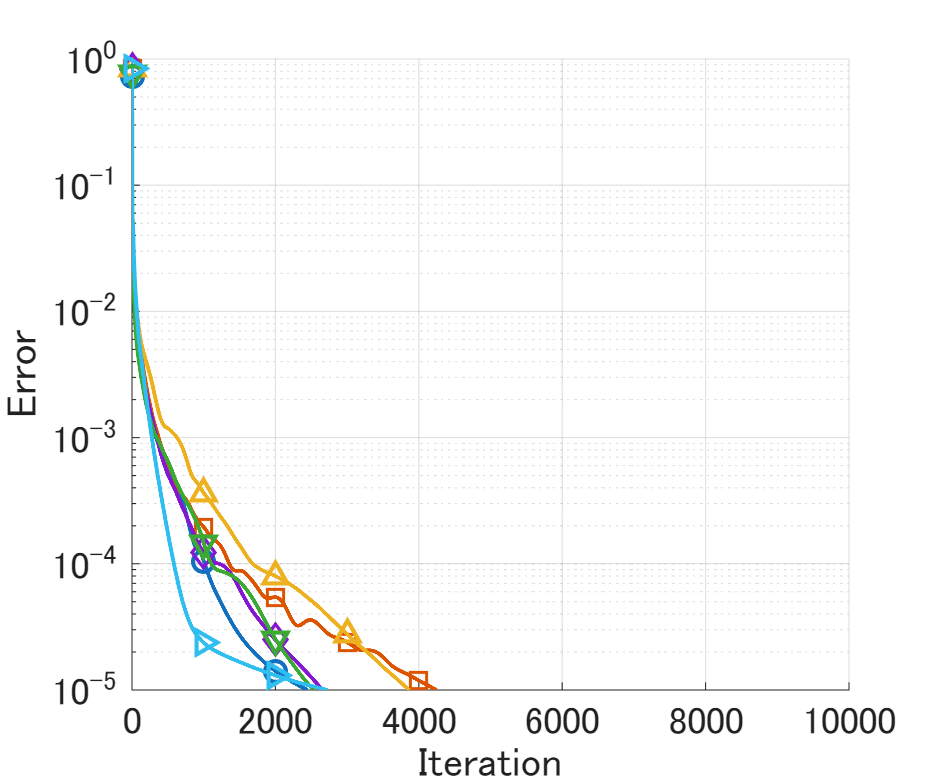}}
        \end{minipage} \\
        \begin{minipage}{0.01\hsize}
			\centerline{\rotatebox{90}{PSNR}}
		\end{minipage}
        \begin{minipage}{0.48\hsize}
            \centerline{\includegraphics[width=\hsize]{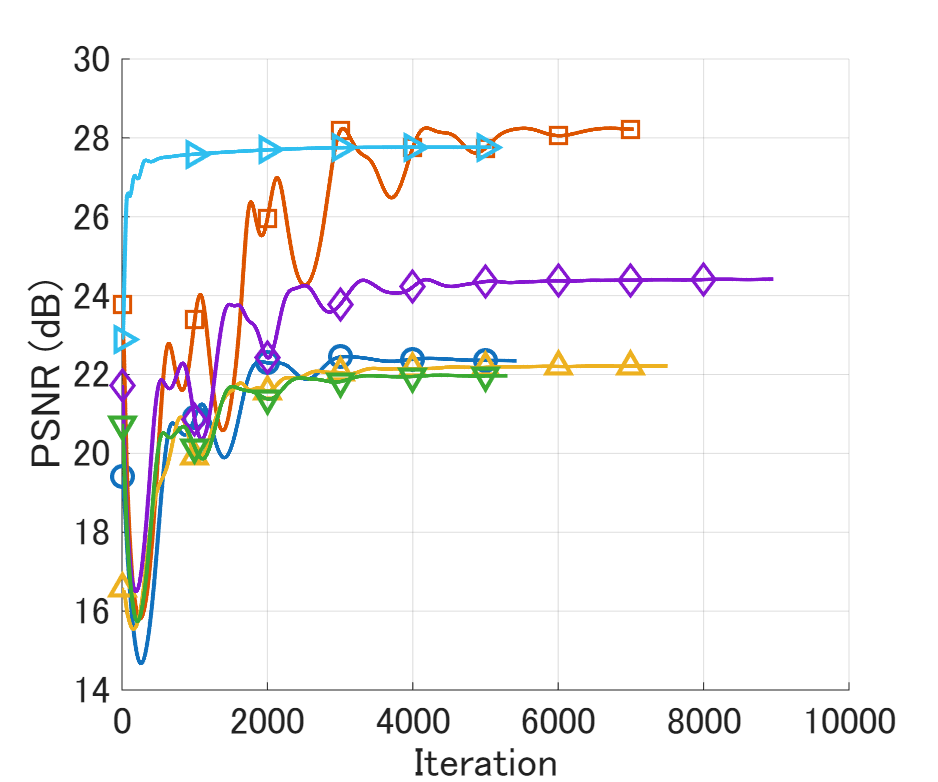}}
        \end{minipage}
        \hfill
        \begin{minipage}{0.48\hsize}
            \centerline{\includegraphics[width=\hsize]{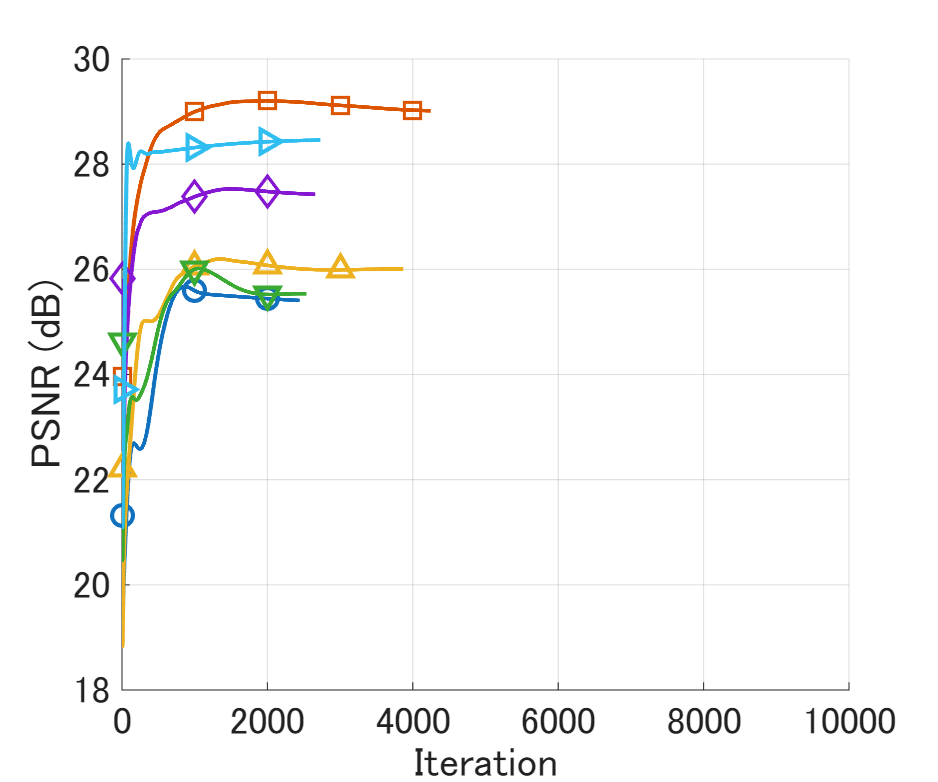}}
        \end{minipage} \\
        \vspace{1mm}\begin{minipage}{0.01\hsize}
			\centerline{}
		\end{minipage}
        \begin{minipage}{0.48\hsize}
            \centerline{(a)}
        \end{minipage}
        \hfill
        \begin{minipage}{0.48\hsize}
            \centerline{(b)}
        \end{minipage} \\
        \begin{minipage}{1\hsize}
            \centerline{\includegraphics[width=\hsize]{img/legend_Site.png}}
        \end{minipage} \\
    \end{center}
    \vspace{-3mm}
    \caption{The algorithm behavior of (a) TSSTF and (b) ROSTF under Case9 for real data. The top and bottom rows show the transitions of the update error for $\tilHt$ and the PSNR values, respectively. Each curve represents each site data.}
    \label{fig: TSSTF_ROSTF_convergence_comparison}
\end{figure}

Among all compared methods, RobOt exhibited the shortest running time. This is likely because the Least Absolute Shrinkage and Selection Operator problem in RobOt has a closed-form solution in our single-reference-date experimental setting. The rule-based methods, STARFM and VIPSTF, also ran faster than the iterative optimization-requiring methods ROSTF and TSSTF. However, it is notable that TSSTF ran more than twice as fast as ROSTF despite sharing similar optimization frameworks. This is because TSSTF demonstrates more stable algorithm behavior and requires fewer iterations to converge as shown in Fig.~\ref{fig: TSSTF_ROSTF_convergence_comparison}. Such a faster convergence is mainly due to our adaptive parameter adjustment strategy for TGEC, which effectively constrains the solution space and accelerates the optimization process.

\subsection{Performance Under Significant Land-Cover Changes}
\label{ssec: performance under significant land-cover changes}
The proposed TSSTF relies on the assumption that the spatial structure of the HR image remains relatively stable between the reference and target dates. In practice, however, significant land-cover changes, such as rapid urbanization, infrastructure development, or natural disasters, can occur, potentially violating this assumption. To rigorously evaluate the robustness of our method against violations of this assumption, we conducted both quantitative and qualitative analyses.

\subsubsection{Quantitative Analysis}
We quantitatively investigated how land-cover changes affect the fusion performance. Specifically, we partitioned the images into $50 \times 50$ non-overlapping patches. For each patch, we calculated the MSSIM value between the reference HR patch and the ground-truth target HR patch, denoted as $\mathrm{MSSIM}_{\text{change}}$. This metric quantifies the degree of land-cover change, where a lower $\mathrm{MSSIM}_{\text{change}}$ indicates more significant land-cover changes. Subsequently, we also measured the MSSIM between the estimated target HR patch and the ground-truth target HR patch, denoted as $\mathrm{MSSIM}_{\text{est}}$, to evaluate the spatial reconstruction accuracy for each patch.

\begin{figure*}[t]
    \begin{center}
        \scriptsize 
        \setlength{\tabcolsep}{0pt} 

        \begin{minipage}{0.1\hsize}
            \centerline{\includegraphics[width=\hsize]{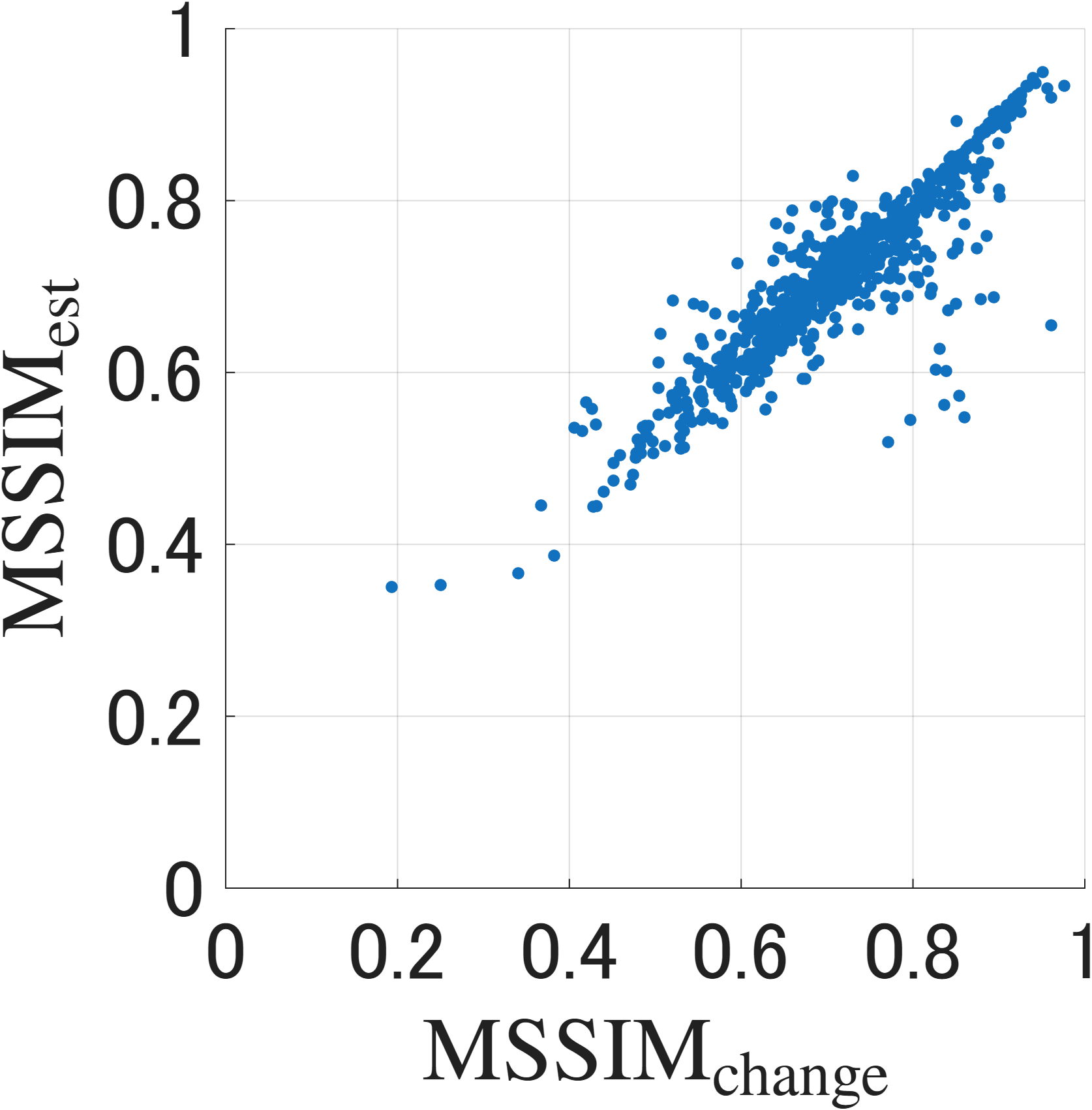}}
        \end{minipage}
        \begin{minipage}{0.1\hsize}
            \centerline{\includegraphics[width=\hsize]{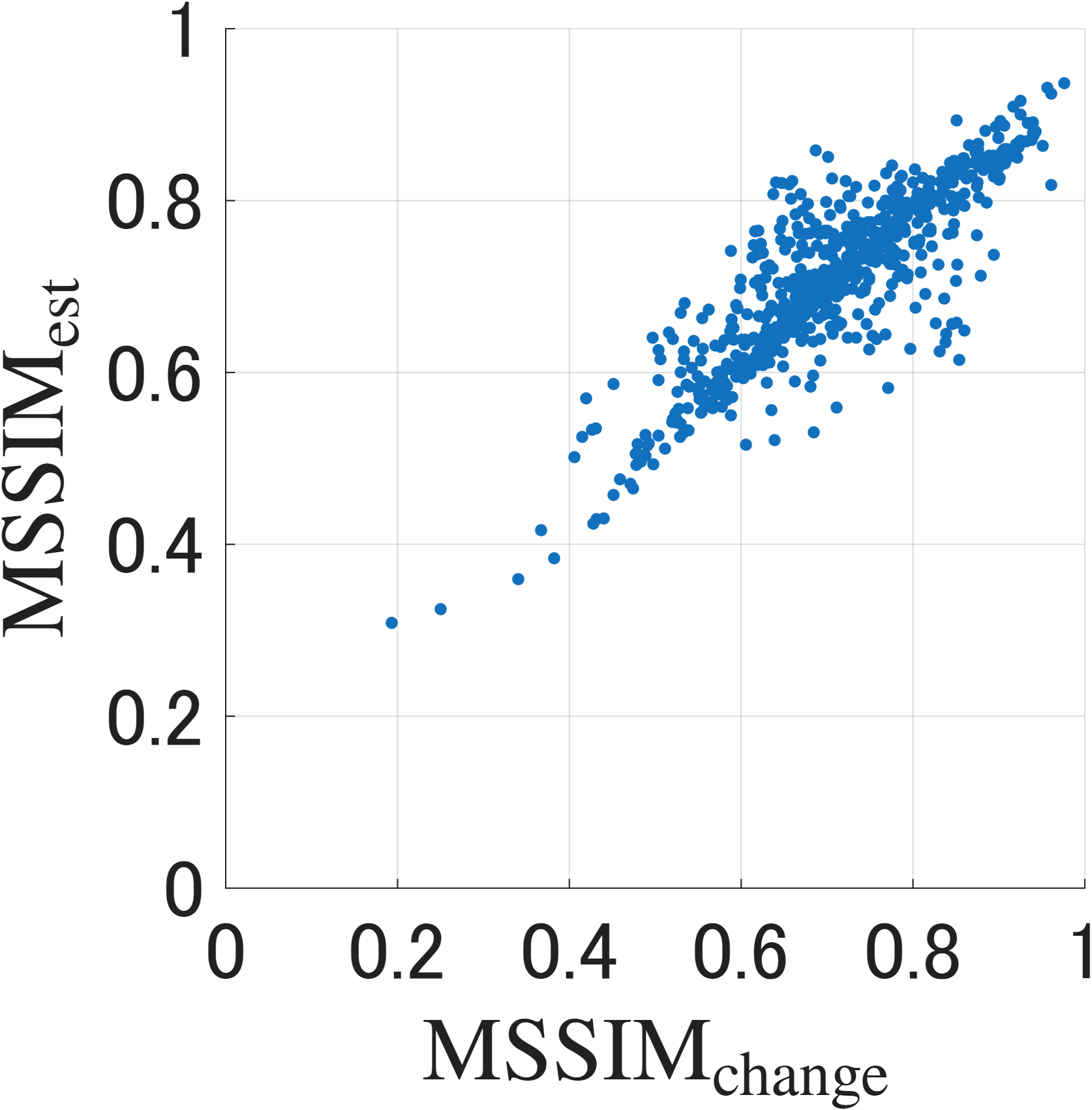}}
        \end{minipage}
        \begin{minipage}{0.1\hsize}
            \centerline{\includegraphics[width=\hsize]{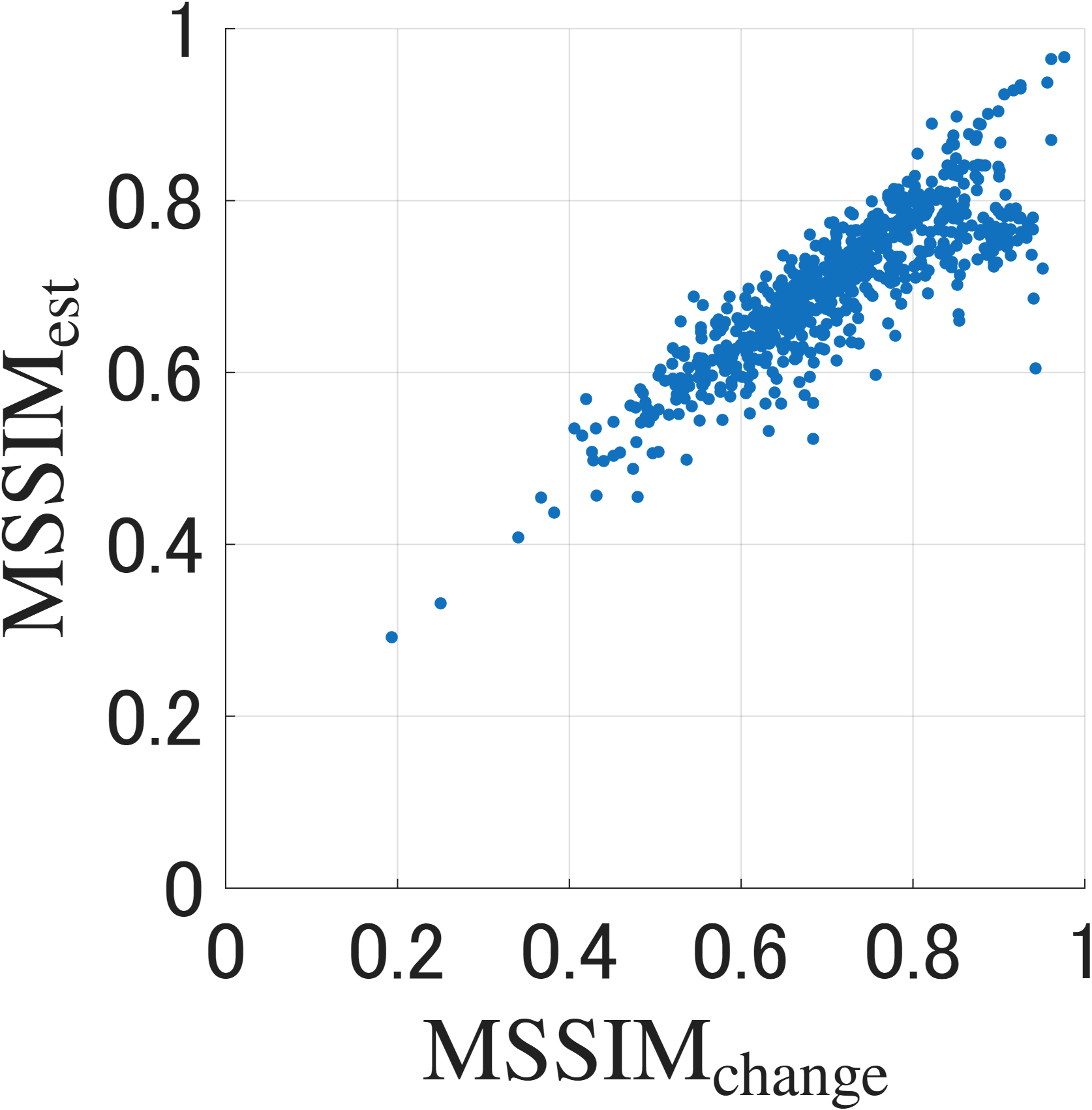}}
        \end{minipage}
        \begin{minipage}{0.1\hsize}
            \centerline{\includegraphics[width=\hsize]{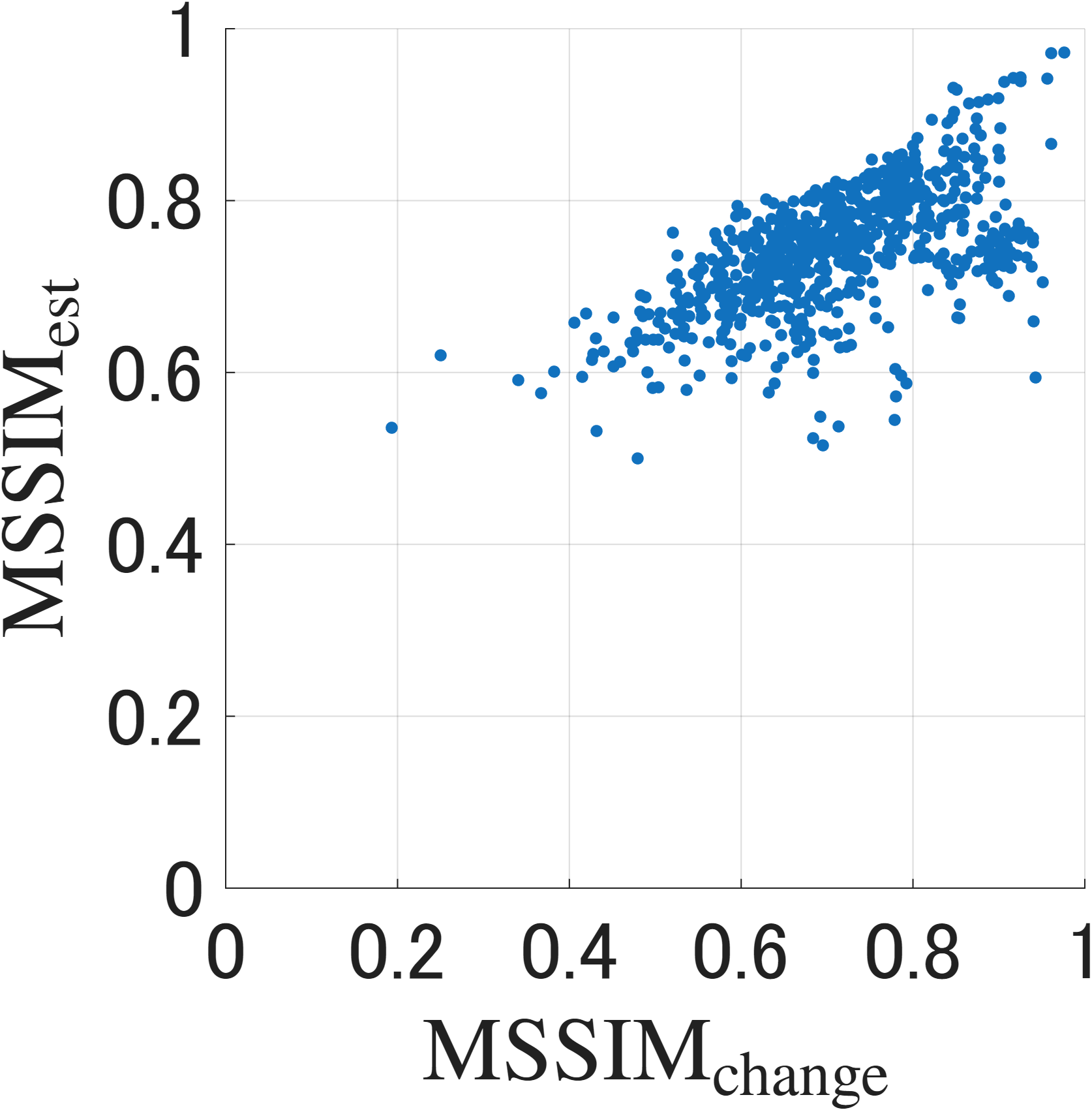}}
        \end{minipage}
        \begin{minipage}{0.1\hsize}
            \centerline{\includegraphics[width=\hsize]{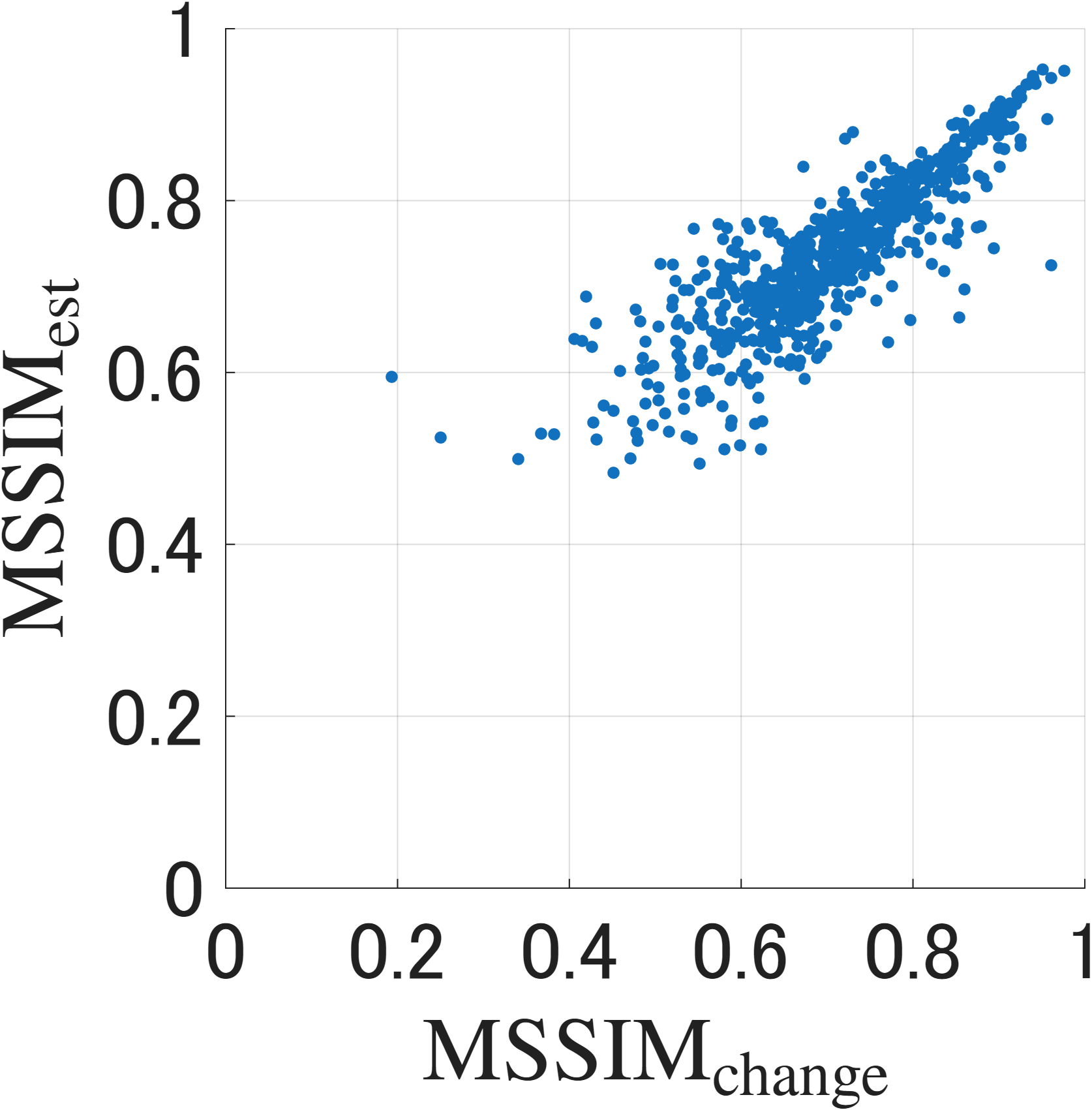}}
        \end{minipage}
        \begin{minipage}{0.1\hsize}
            \centerline{\includegraphics[width=\hsize]{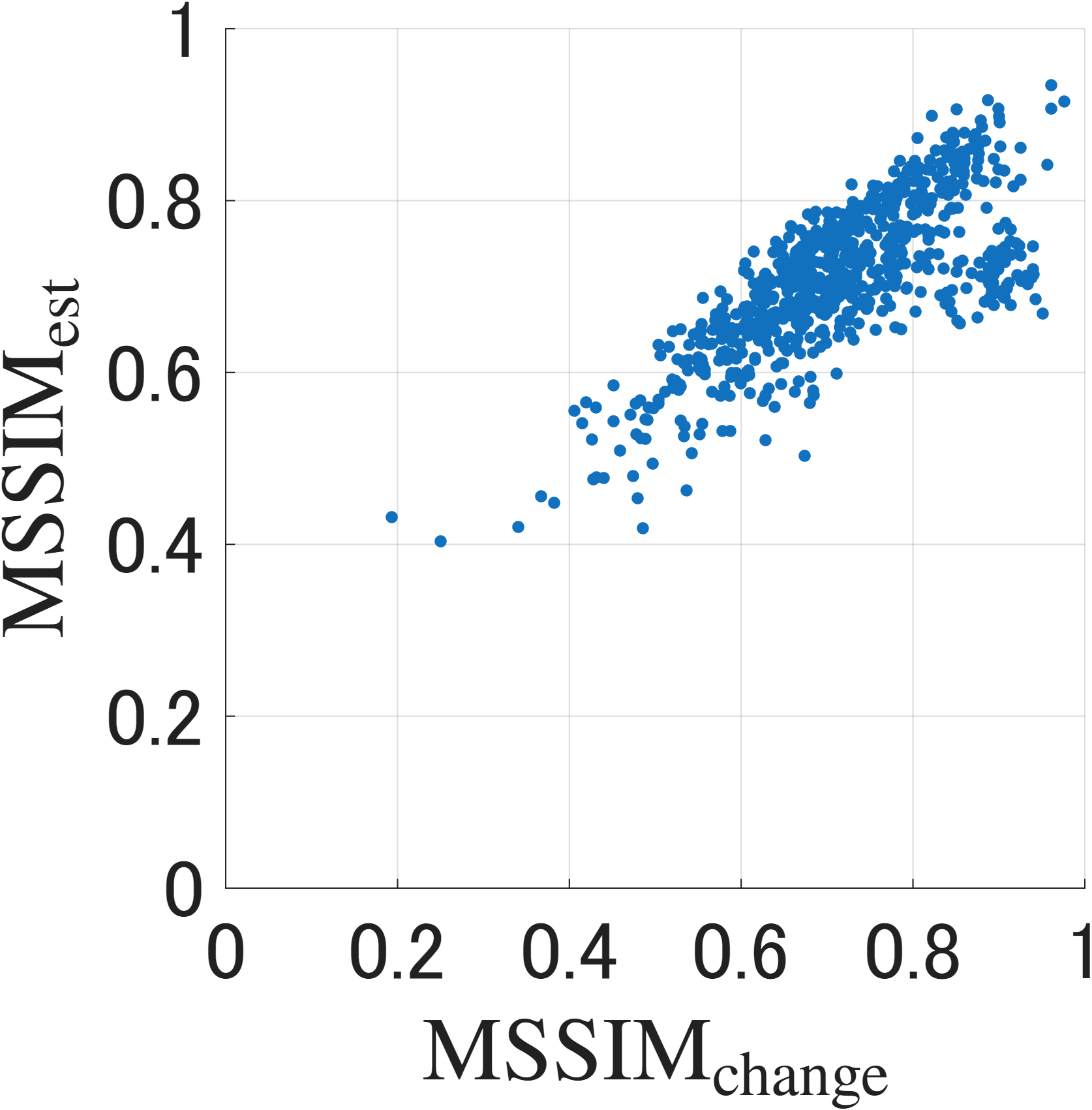}}
        \end{minipage}
        \begin{minipage}{0.1\hsize}
            \centerline{\includegraphics[width=\hsize]{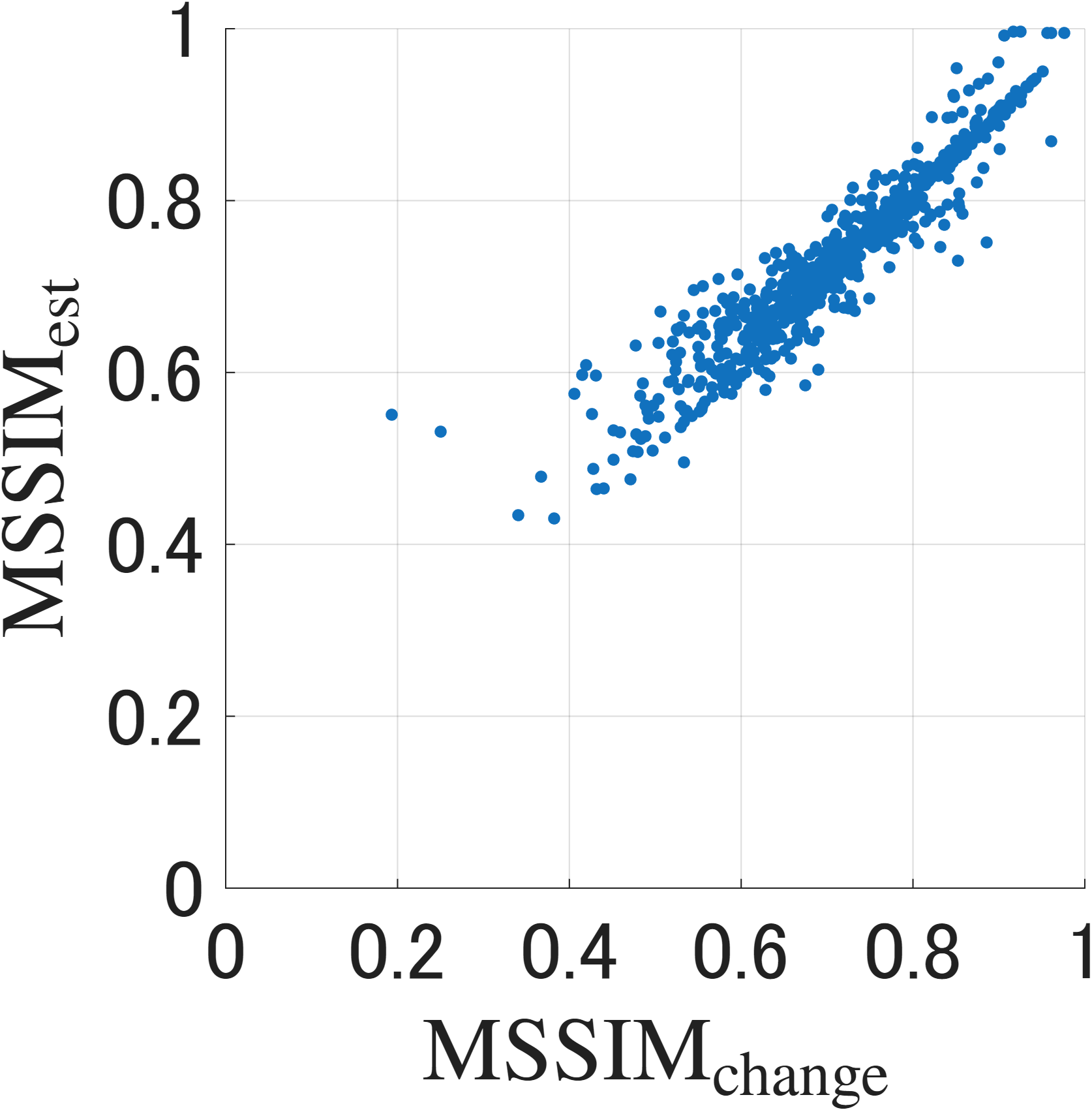}}
        \end{minipage}
        \begin{minipage}{0.1\hsize}
            \centerline{\includegraphics[width=\hsize]{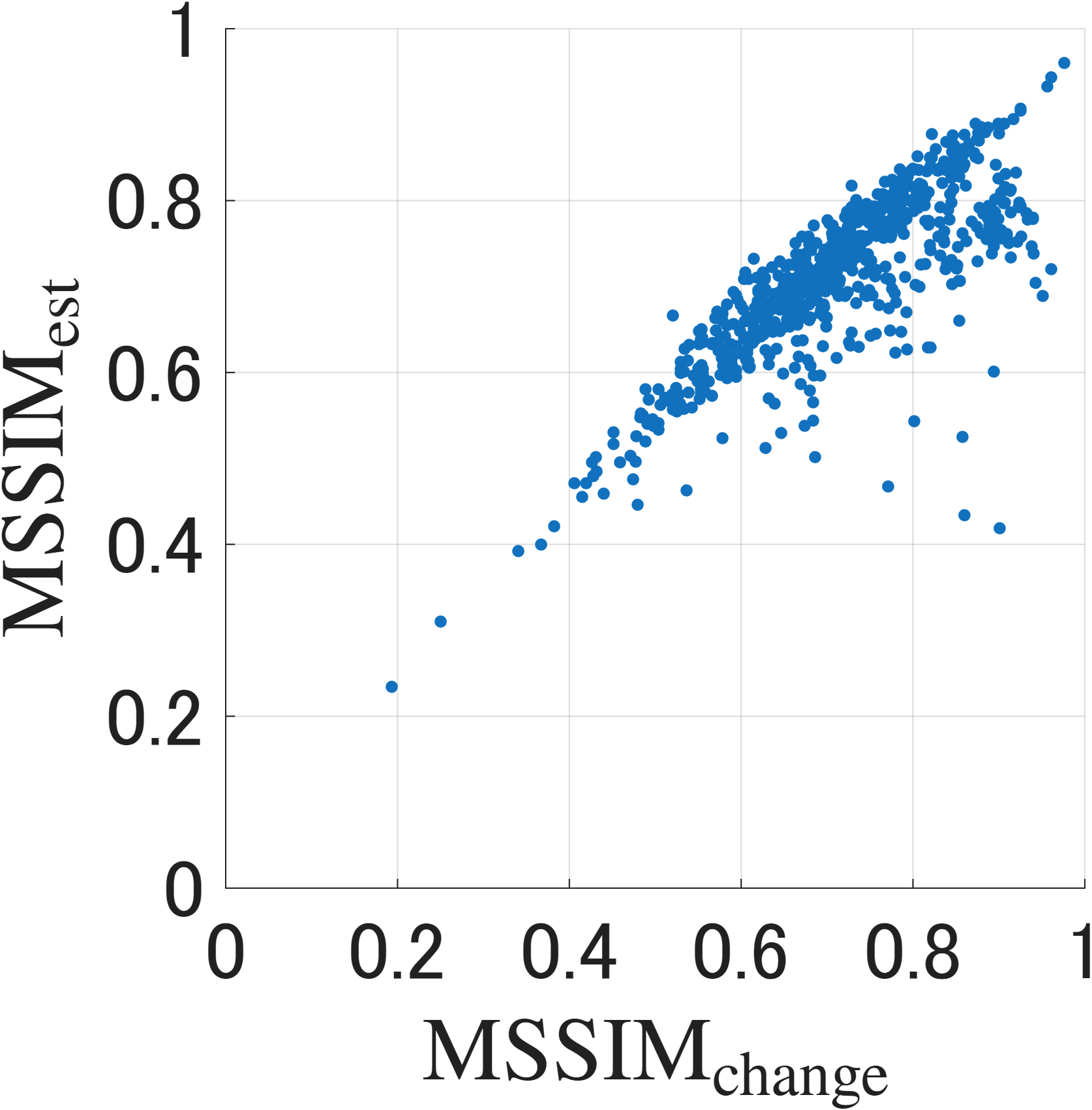}}
        \end{minipage}
        \begin{minipage}{0.1\hsize}
            \centerline{\includegraphics[width=\hsize]{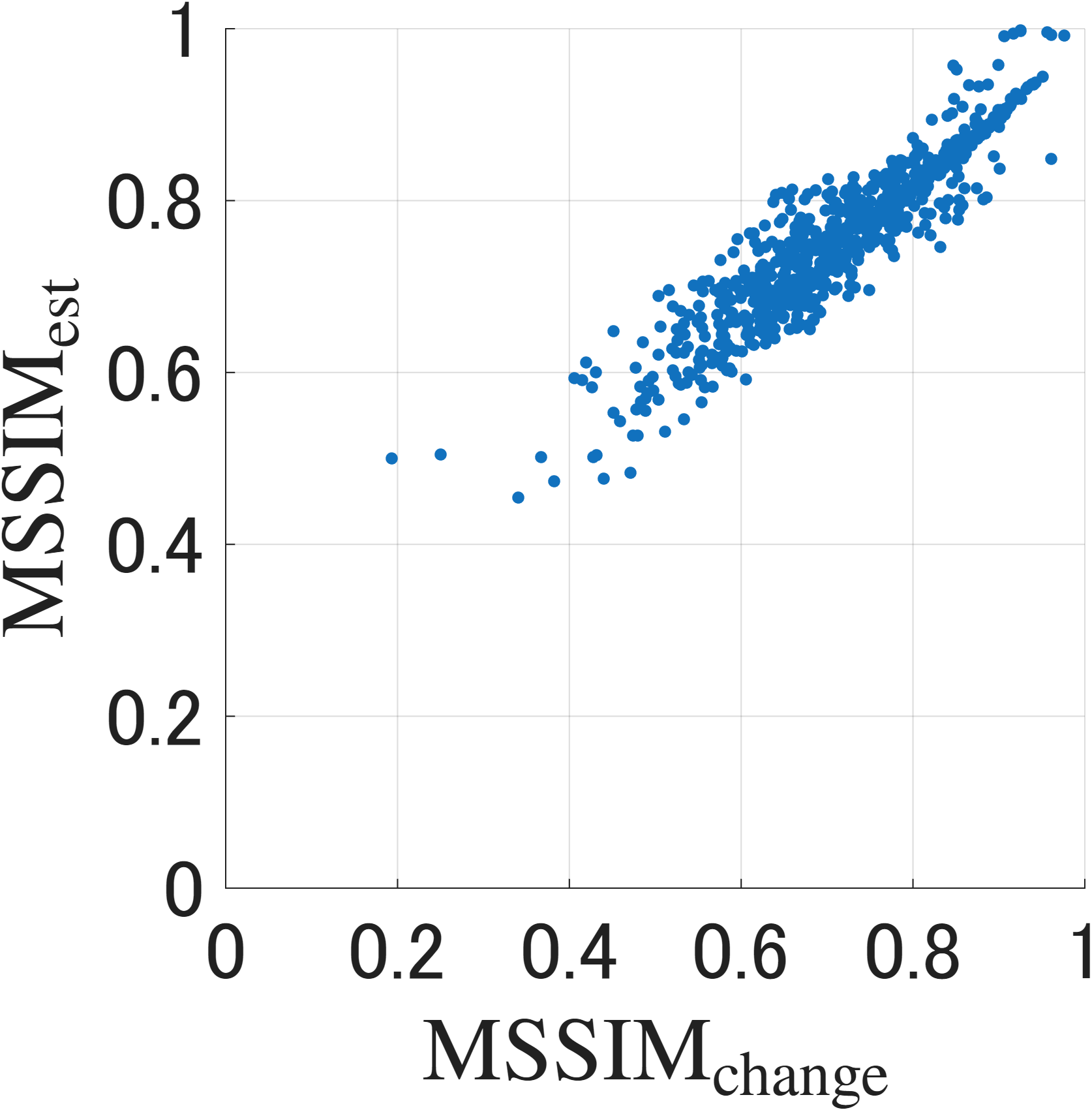}}
        \end{minipage} \\
        \vspace{1mm}
        \begin{minipage}{0.1\hsize} 
            \centerline{0.9040}
        \end{minipage}
        \begin{minipage}{0.1\hsize}
            \centerline{0.8763}
        \end{minipage}
        \begin{minipage}{0.1\hsize} 
            \centerline{0.8605}
        \end{minipage}
        \begin{minipage}{0.1\hsize} 
            \centerline{0.5926}
        \end{minipage}
        \begin{minipage}{0.1\hsize} 
            \centerline{0.8663}
        \end{minipage}
        \begin{minipage}{0.1\hsize} 
            \centerline{0.7607}
        \end{minipage}
        \begin{minipage}{0.1\hsize} 
            \centerline{0.8830}
        \end{minipage}
        \begin{minipage}{0.1\hsize} 
            \centerline{0.9439}
        \end{minipage}
        \begin{minipage}{0.1\hsize} 
            \centerline{0.9261}
        \end{minipage}\\
        \begin{minipage}{0.1\hsize} 
            \centerline{STARFM}
        \end{minipage}
        \begin{minipage}{0.1\hsize} 
            \centerline{VIPSTF}
        \end{minipage}
        \begin{minipage}{0.1\hsize} 
            \centerline{RSFN-1}
        \end{minipage}
        \begin{minipage}{0.1\hsize} 
            \centerline{RSFN-2}
        \end{minipage}
        \begin{minipage}{0.1\hsize} 
            \centerline{RobOt}
        \end{minipage}
        \begin{minipage}{0.1\hsize} 
            \centerline{SwinSTFM}
        \end{minipage}
        \begin{minipage}{0.1\hsize} 
            \centerline{ROSTF}
        \end{minipage}
        \begin{minipage}{0.1\hsize} 
            \centerline{ECPW}
        \end{minipage}
        \begin{minipage}{0.1\hsize} 
            \centerline{\textbf{TSSTF}}
        \end{minipage}\\
    \end{center}
    \vspace{-3mm}
    \caption{Scatter plots of $\mathrm{MSSIM}_{\text{change}}$ versus $\mathrm{MSSIM}_{\text{est}}$ for each method. The correlation coefficients between $\mathrm{MSSIM}_{\text{change}}$ and $\mathrm{MSSIM}_{\text{est}}$ are also shown below each plot.}
    \label{fig: SSIM scatterplot}
\end{figure*}

Fig.~\ref{fig: SSIM scatterplot} presents scatter plots of $\mathrm{MSSIM}_{\text{change}}$ versus $\mathrm{MSSIM}_{\text{est}}$ for each method, aggregated over all six sites for real data in Case1. The correlation coefficients between $\mathrm{MSSIM}_{\text{change}}$ and $\mathrm{MSSIM}_{\text{est}}$ are also reported. A strong positive correlation implies that the fusion performance is highly dependent on the degree of land-cover changes, indicating that performance degrades in regions with substantial transitions.

As expected, the proposed TSSTF exhibits a strong positive correlation of 0.9261, confirming that its performance is sensitive to land-cover changes. However, it is worth noting that other single-reference methods also show similar positive trends. In contrast, RSFN-2, based on a dual-reference setting, demonstrates a weaker correlation of 0.5926, suggesting better robustness against land-cover changes. This can be attributed to the fact that RSFN-2 can leverage structural information from two reference dates before and after the target date. This analysis highlights the inherent limitations of single-reference methods in handling significant land-cover changes.

\subsubsection{Visual Analysis}
\begin{figure*}[t]
    \begin{center}
        \scriptsize 
        \setlength{\tabcolsep}{0pt} 

        \begin{minipage}{0.1\hsize}
            \centerline{\includegraphics[width=\hsize]{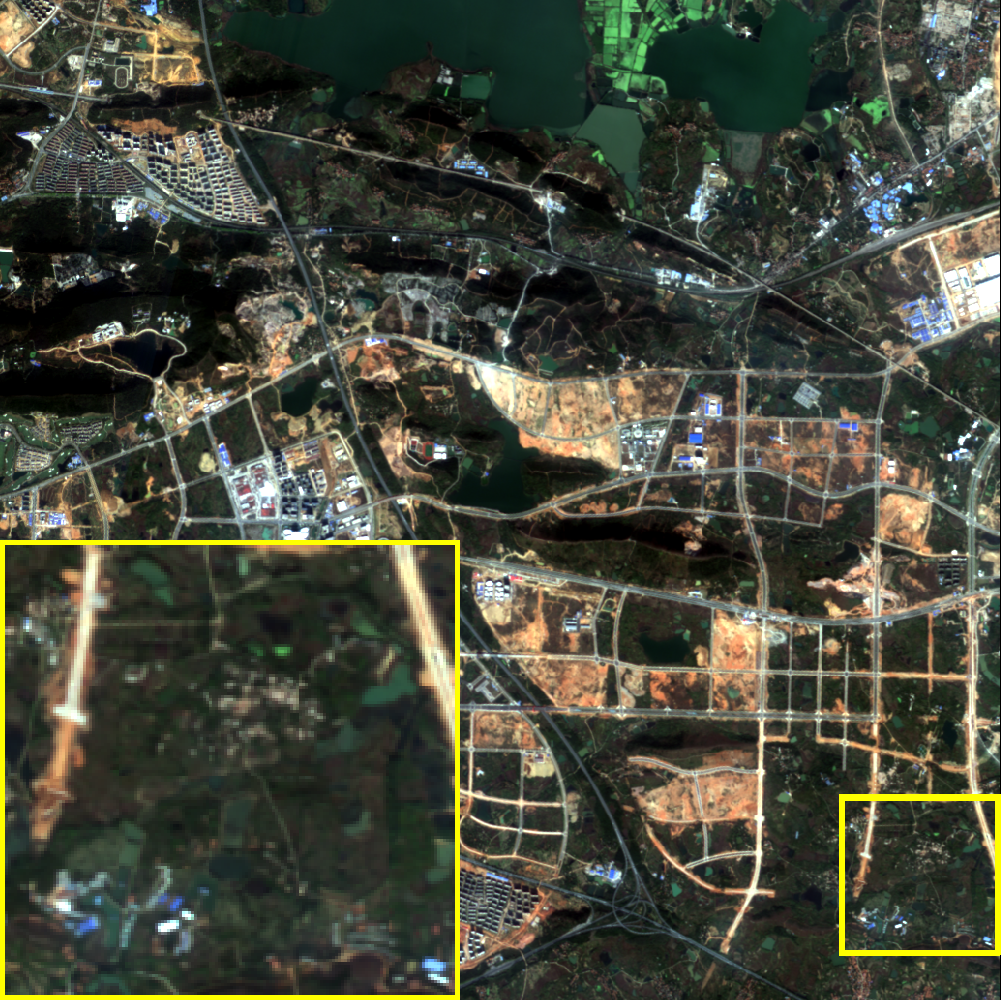}}
        \end{minipage}
        \begin{minipage}{0.1\hsize}
            \centerline{\includegraphics[width=\hsize]{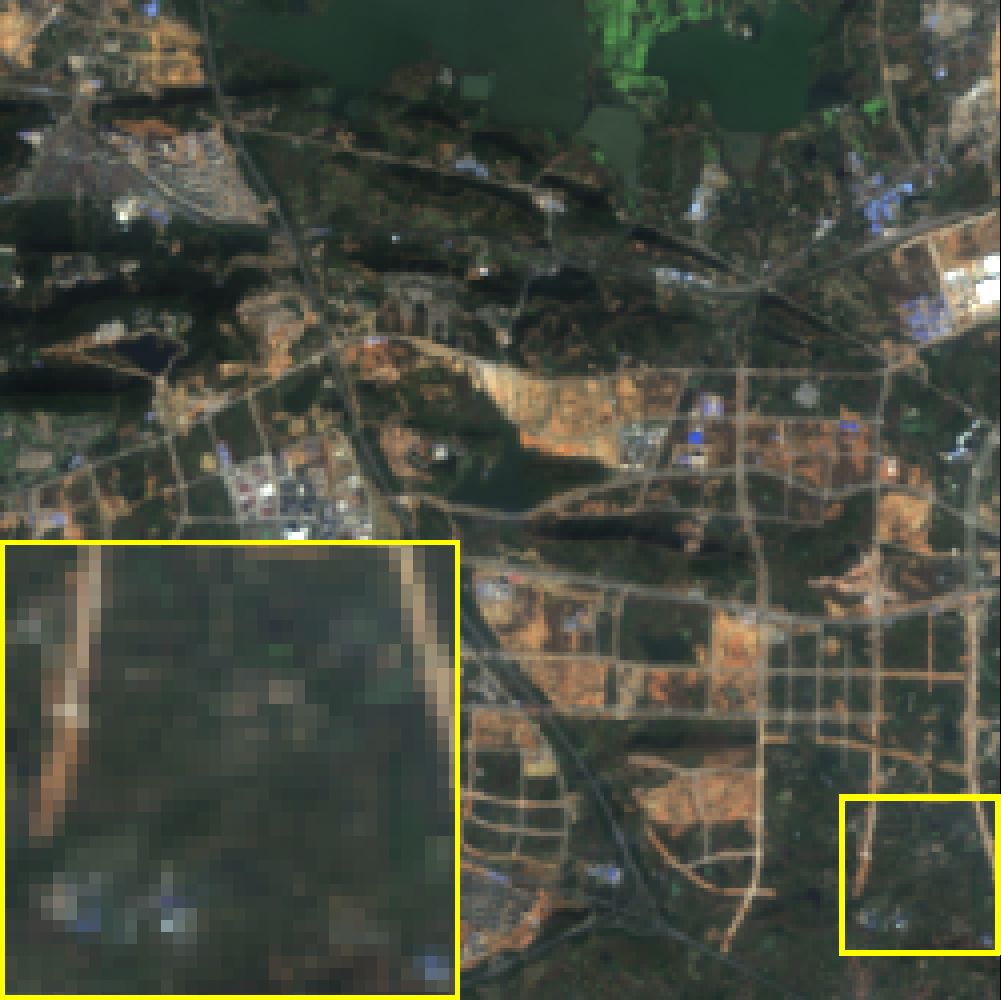}}
        \end{minipage}
        \begin{minipage}{0.1\hsize}
            \centerline{\includegraphics[width=\hsize]{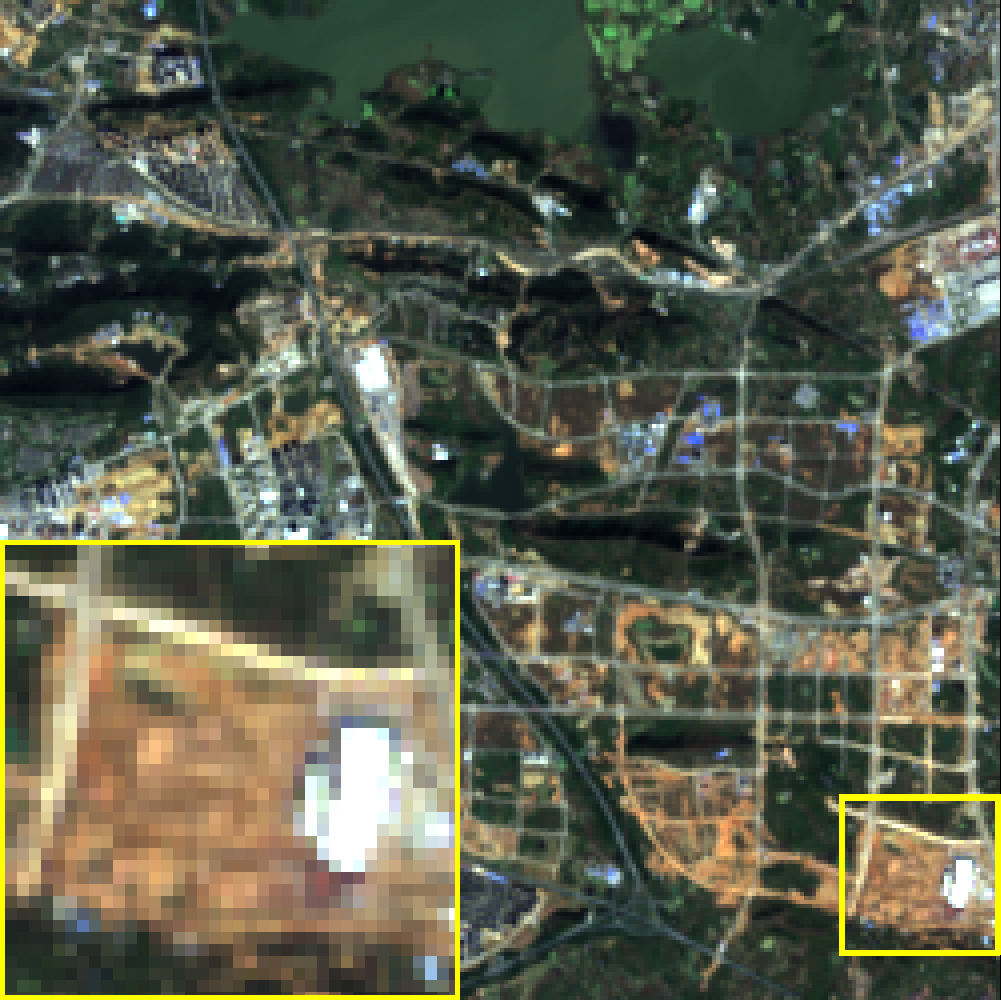}}
        \end{minipage}
        \begin{minipage}{0.1\hsize}
            \centerline{\includegraphics[width=\hsize]{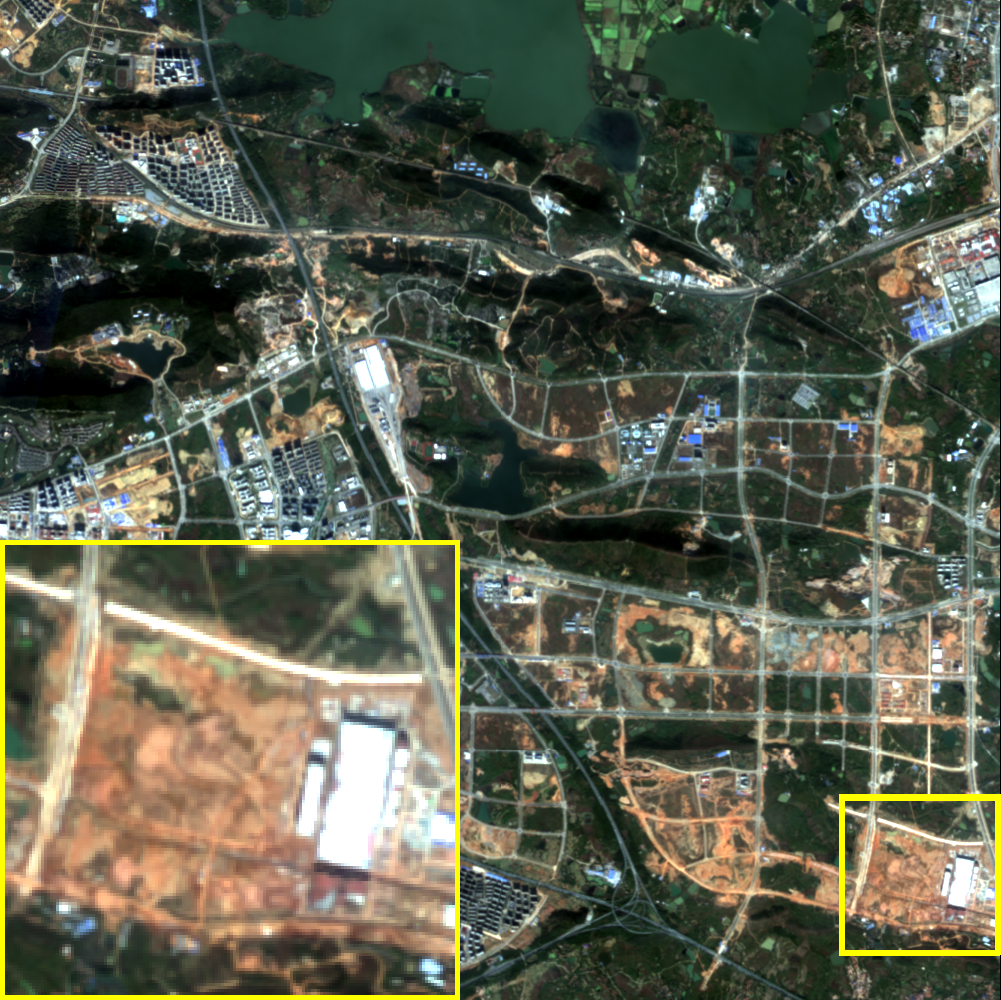}}
        \end{minipage} \\

        \vspace{1mm}
        \begin{minipage}{0.1\hsize} 
            \centerline{$\Hr$}
        \end{minipage}
        \begin{minipage}{0.1\hsize} 
            \centerline{$\Lr$}
        \end{minipage}
        \begin{minipage}{0.1\hsize} 
            \centerline{$\Lt$}
        \end{minipage}
        \begin{minipage}{0.1\hsize} 
            \centerline{Ground-truth}
        \end{minipage} \\

        \vspace{1mm}
        \begin{minipage}{0.1\hsize}
            \centerline{\includegraphics[width=\hsize]{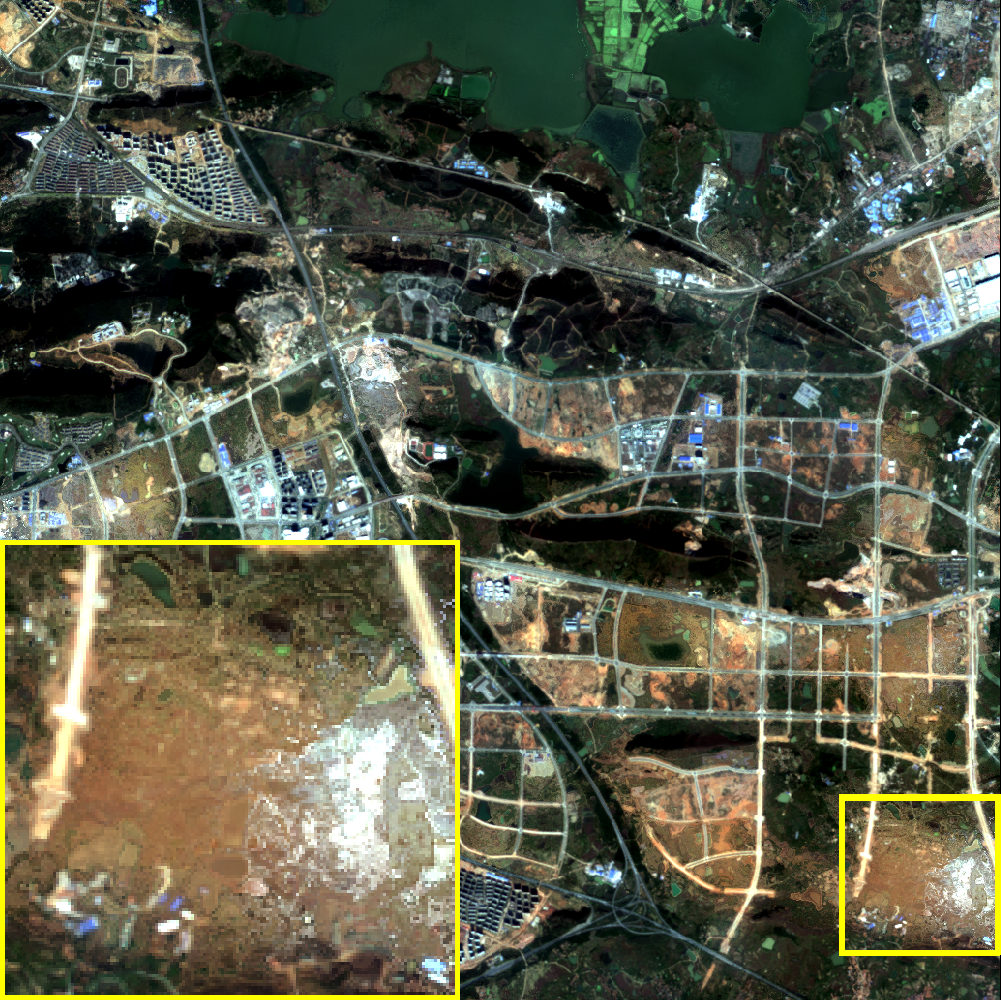}}
        \end{minipage}
        \begin{minipage}{0.1\hsize}
            \centerline{\includegraphics[width=\hsize]{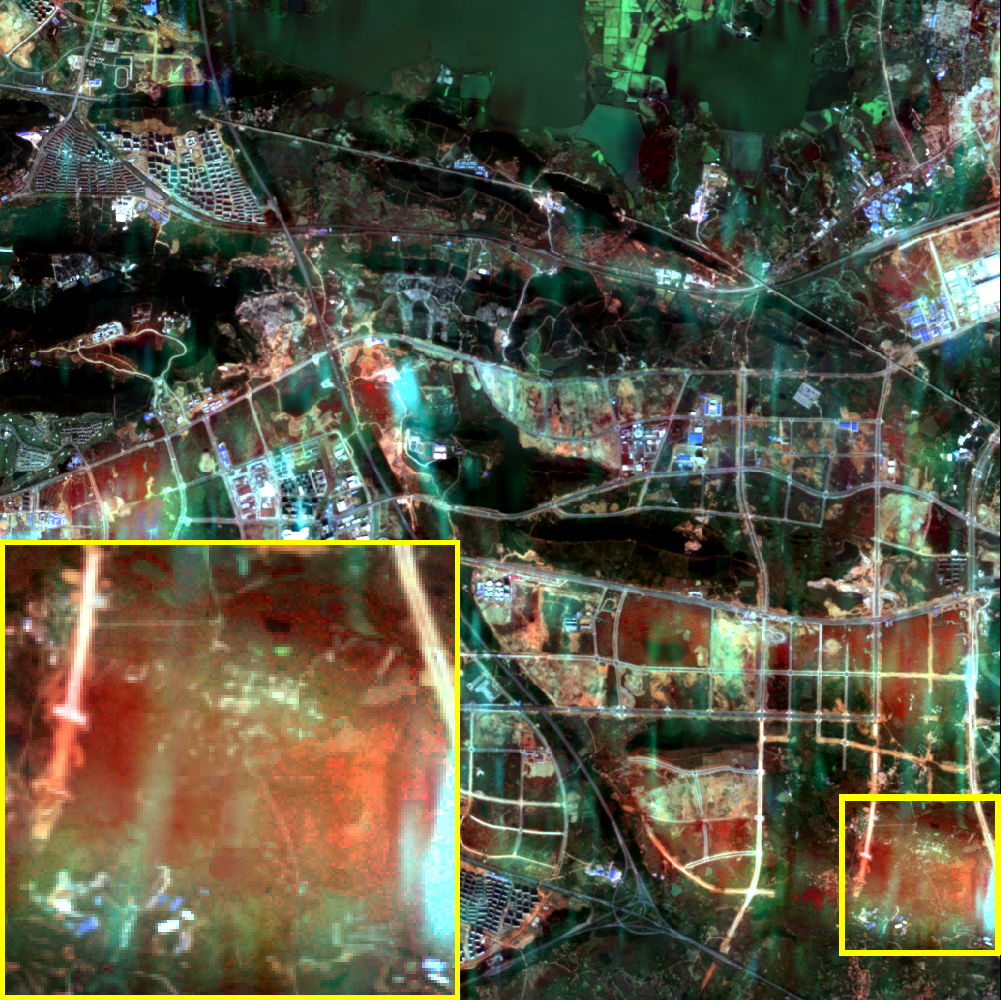}}
        \end{minipage}
        \begin{minipage}{0.1\hsize}
            \centerline{\includegraphics[width=\hsize]{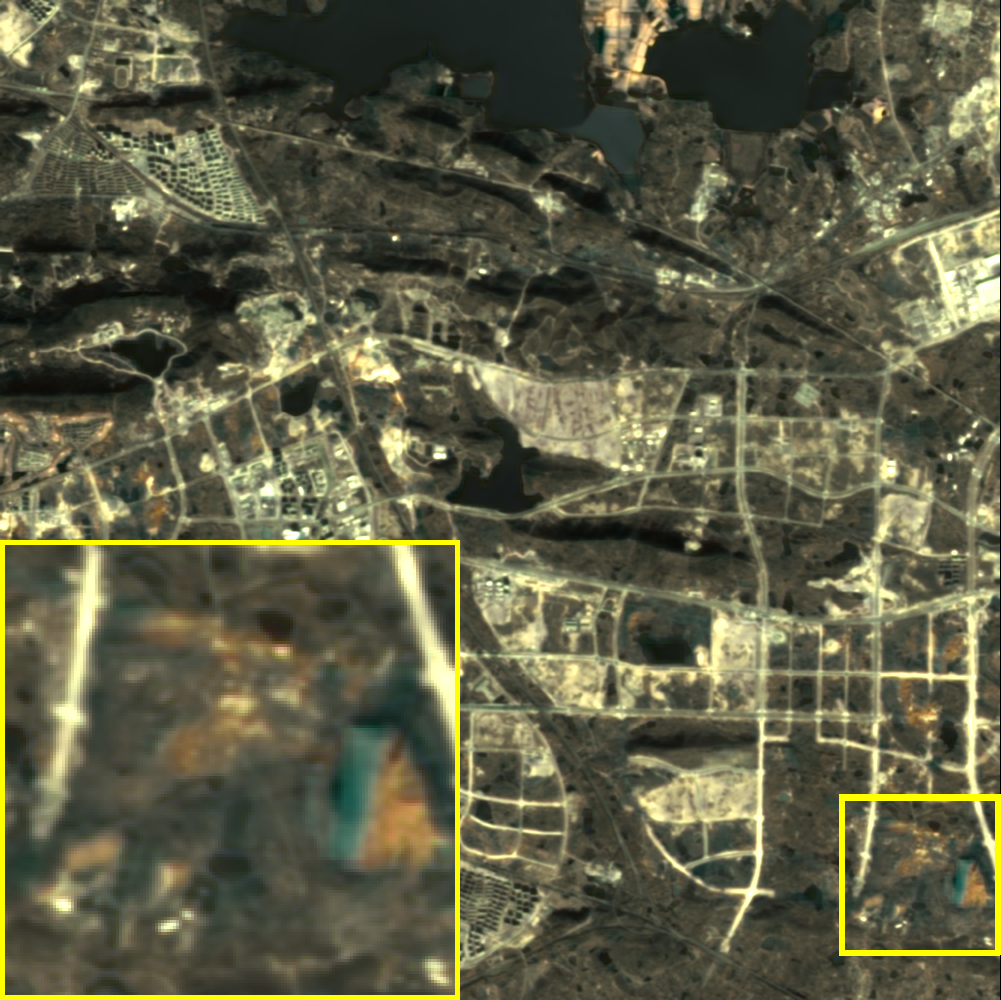}}
        \end{minipage}
        \begin{minipage}{0.1\hsize}
            \centerline{\includegraphics[width=\hsize]{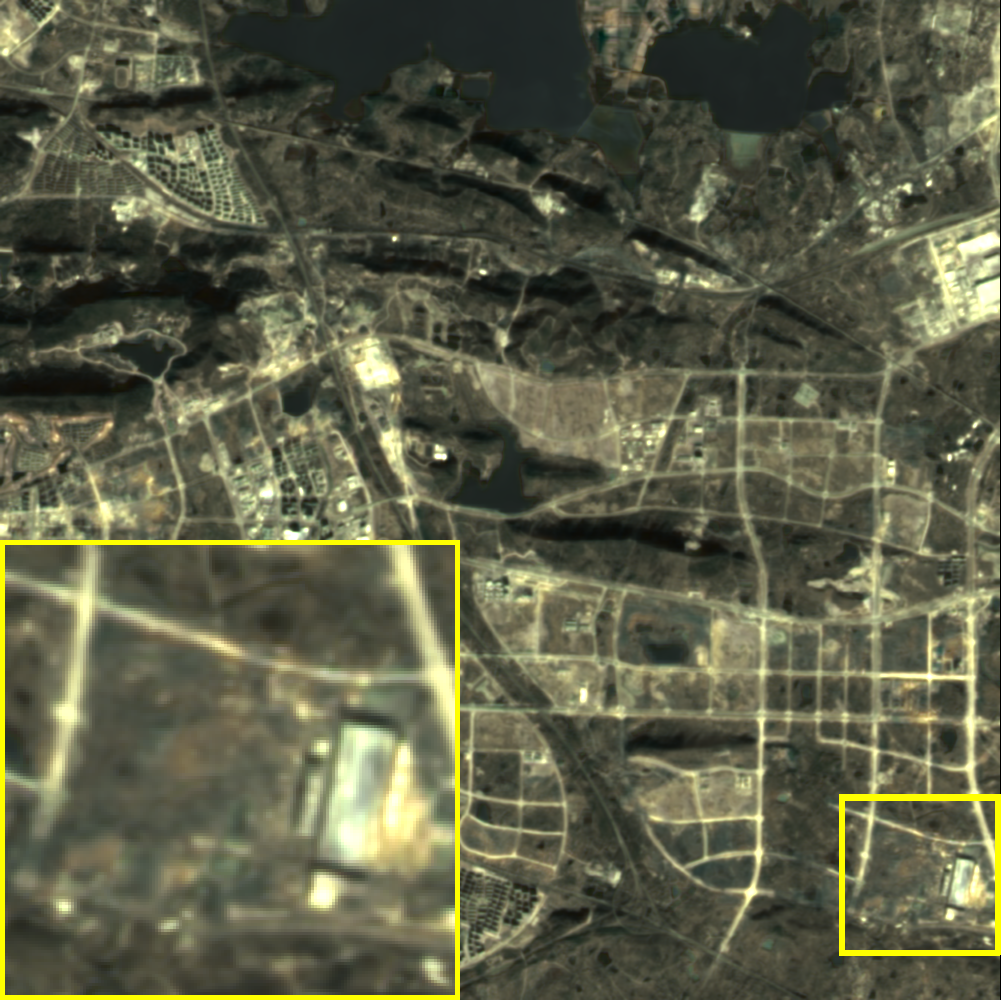}}
        \end{minipage}
        \begin{minipage}{0.1\hsize}
            \centerline{\includegraphics[width=\hsize]{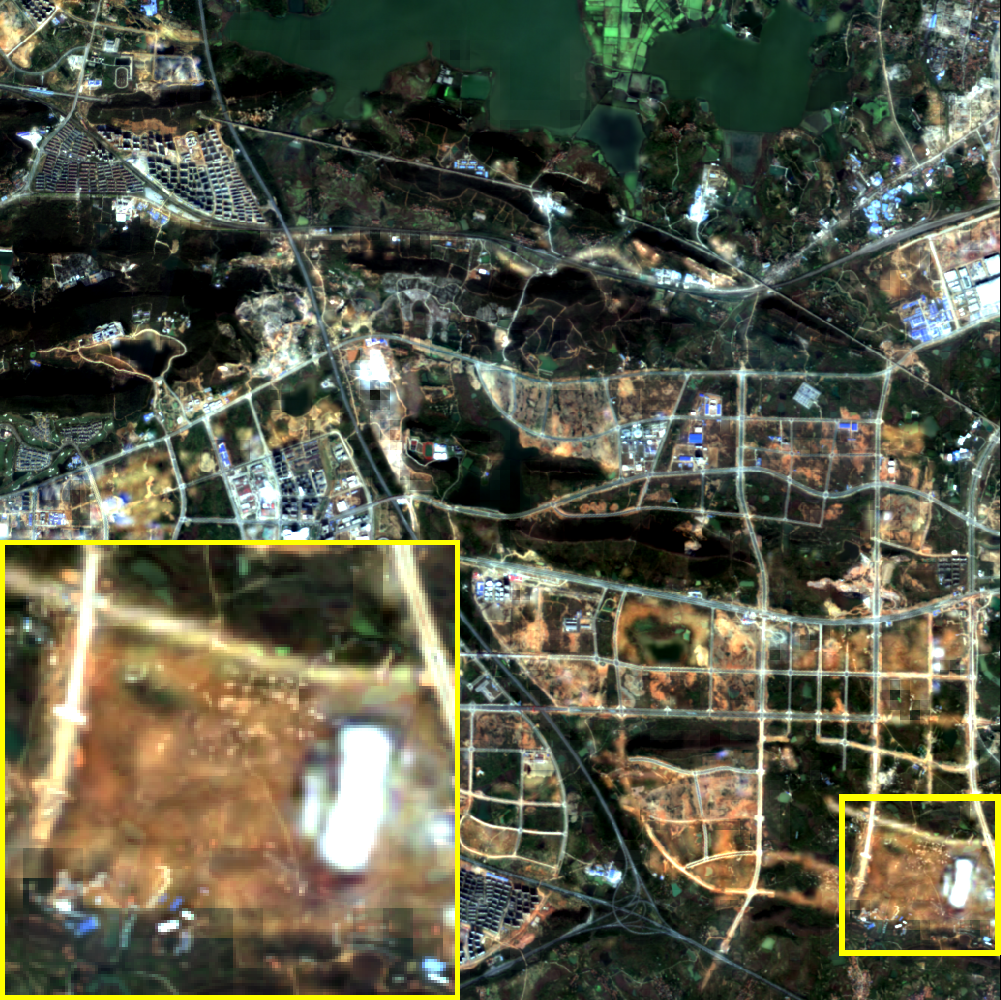}}
        \end{minipage}
        \begin{minipage}{0.1\hsize}
            \centerline{\includegraphics[width=\hsize]{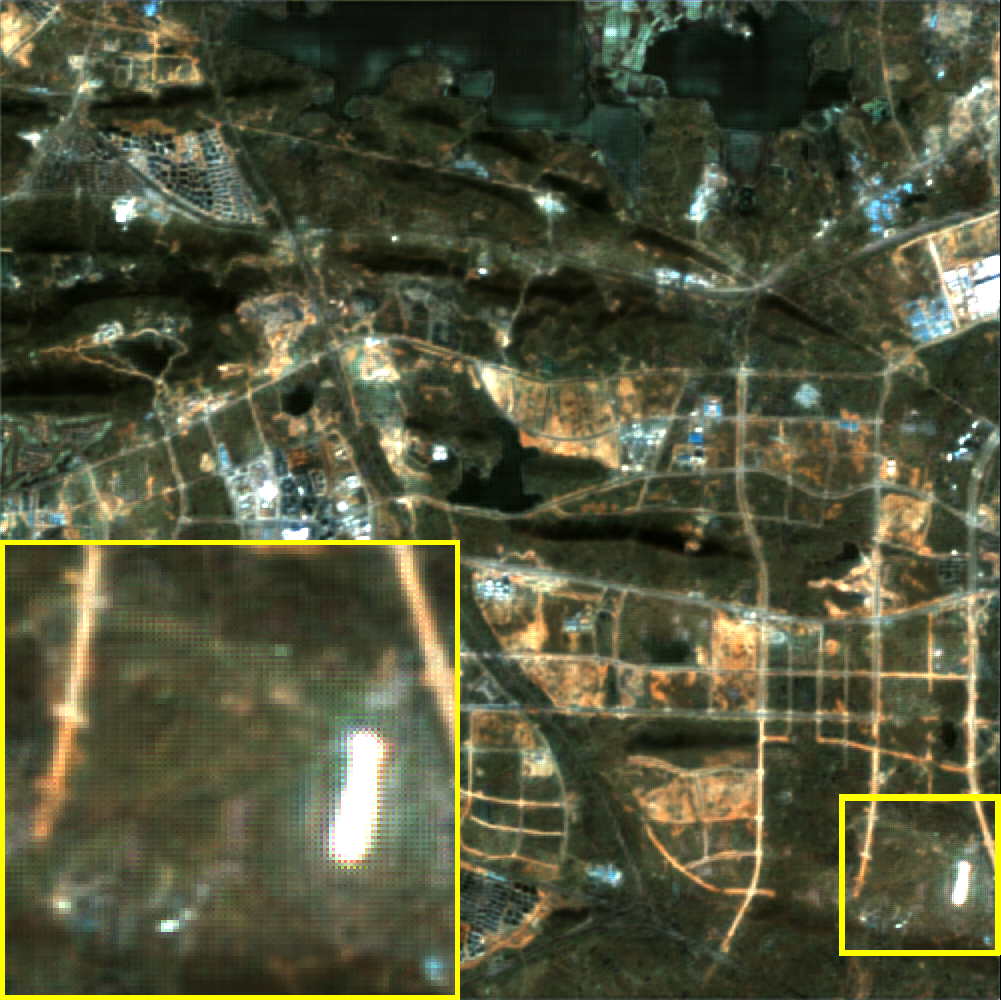}}
        \end{minipage}
        \begin{minipage}{0.1\hsize}
            \centerline{\includegraphics[width=\hsize]{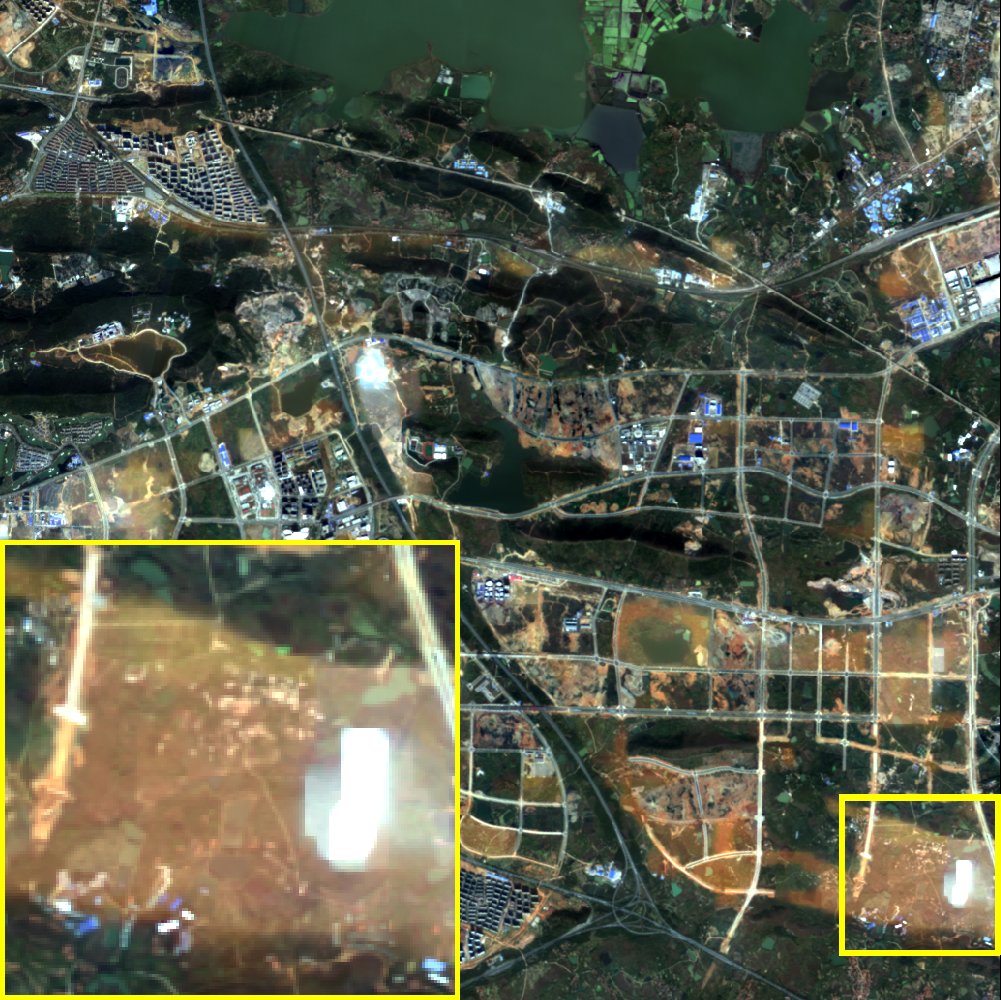}}
        \end{minipage}
        \begin{minipage}{0.1\hsize}
            \centerline{\includegraphics[width=\hsize]{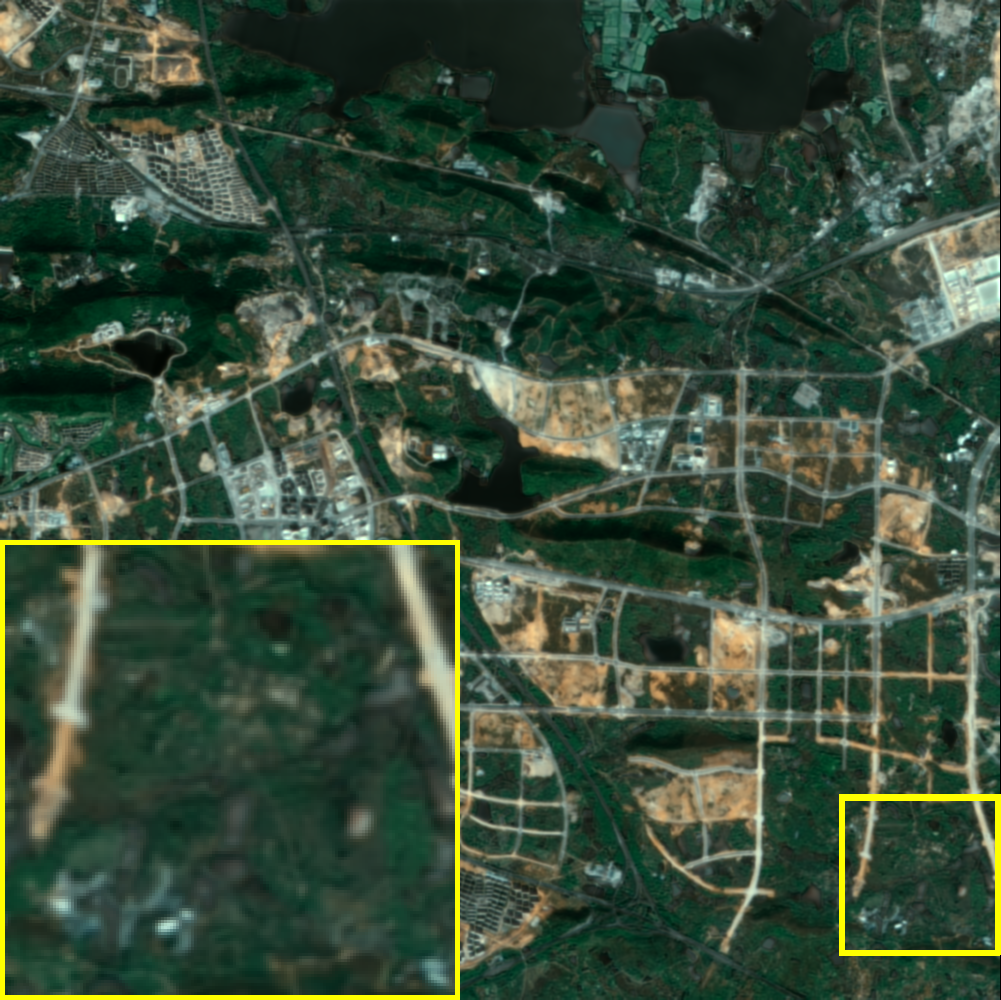}}
        \end{minipage}
        \begin{minipage}{0.1\hsize}
            \centerline{\includegraphics[width=\hsize]{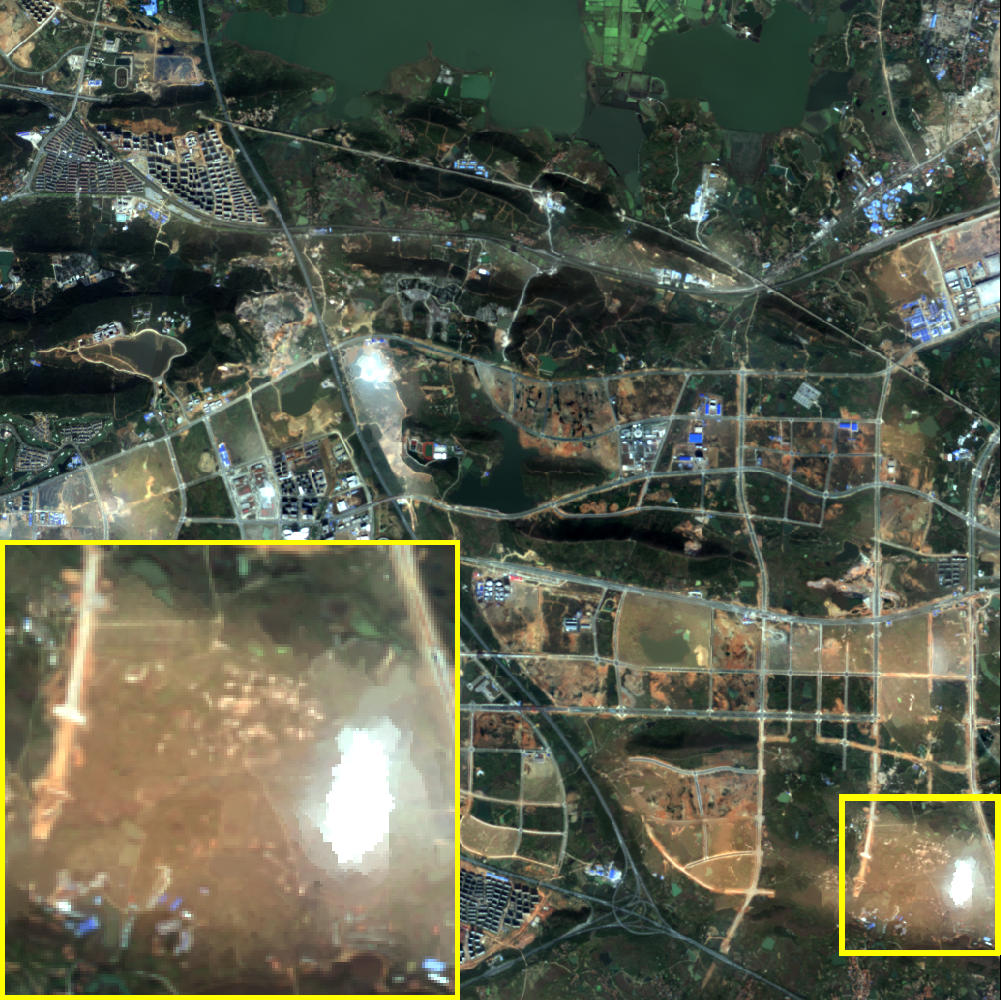}}
        \end{minipage} \\
        \vspace{1mm}
        \begin{minipage}{0.1\hsize} 
            \centerline{STARFM}
        \end{minipage}
        \begin{minipage}{0.1\hsize} 
            \centerline{VIPSTF}
        \end{minipage}
        \begin{minipage}{0.1\hsize} 
            \centerline{RSFN-1}
        \end{minipage}
        \begin{minipage}{0.1\hsize} 
            \centerline{RSFN-2}
        \end{minipage}
        \begin{minipage}{0.1\hsize} 
            \centerline{RobOt}
        \end{minipage}
        \begin{minipage}{0.1\hsize} 
            \centerline{SwinSTFM}
        \end{minipage}
        \begin{minipage}{0.1\hsize} 
            \centerline{ROSTF}
        \end{minipage}
        \begin{minipage}{0.1\hsize} 
            \centerline{ECPW}
        \end{minipage}
        \begin{minipage}{0.1\hsize} 
            \centerline{\textbf{TSSTF}}
        \end{minipage}\\
    \end{center}
    \vspace{-3mm}
    \caption{ST fusion results for the Site6 real data in Case1.}
    \label{fig: TSSTF Site6 Real results}
\end{figure*}

We also present visual examples to illustrate the impact of land-cover changes on the fusion performance. Fig.~\ref{fig: TSSTF Site6 Real results} shows the results in Case1 for the Site6 dataset, where significant land-cover changes are observed.

In the yellow-boxed region, a large-scale building was constructed by the target date that did not exist on the reference date. Since TSSTF relies on the spatial structure of the reference HR image to guide the ST fusion, it erroneously assumes a smooth, building-free structure in this area based on the reference information. Consequently, the resulting output is over-smoothed, and the sharp edges and fine details of the new building are not fully recovered. Notably, other single-reference methods also exhibit similar artifacts.

Addressing the limitation in handling abrupt structural changes is one of the key directions for future work. Potential approaches include designing separate weight matrices to capture the distinct spatial structure of the reference and target dates or implementing a spatially variant edge constraint that reduces its contribution adaptively in regions with significant changes. In either case, leveraging the LR image series more effectively to detect and guide these transitions would be crucial.

\begin{figure}[t]
    \begin{center}
        \scriptsize 
        \setlength{\tabcolsep}{0pt} 

        \begin{minipage}{0.24\hsize}
            \centerline{\includegraphics[width=\hsize]{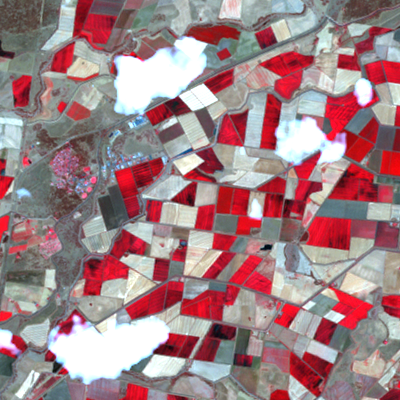}}
        \end{minipage}
        \begin{minipage}{0.24\hsize}
            \centerline{\includegraphics[width=\hsize]{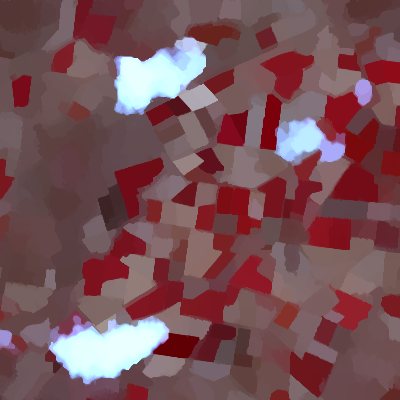}}
        \end{minipage}
        \begin{minipage}{0.24\hsize}
            \centerline{\includegraphics[width=\hsize]{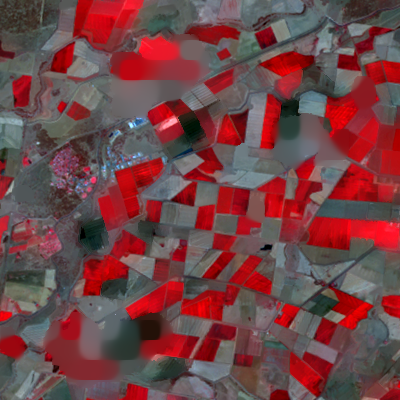}}
        \end{minipage}
        \begin{minipage}{0.24\hsize}
            \centerline{\includegraphics[width=\hsize]{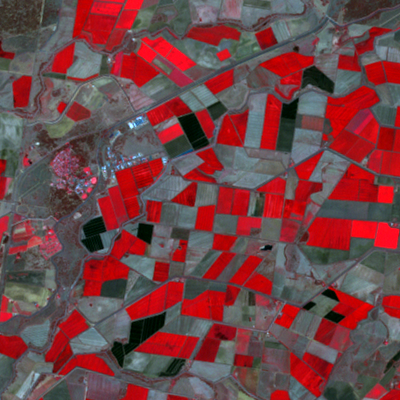}}
        \end{minipage} \\
        \vspace{1mm}
        \begin{minipage}{0.24\hsize} 
            \centerline{Cloudy Reference}
        \end{minipage}
        \begin{minipage}{0.24\hsize} 
            \centerline{TSSTF (Standard)}
        \end{minipage}
        \begin{minipage}{0.24\hsize} 
            \centerline{TSSTF (w/ Mask)}
        \end{minipage}
        \begin{minipage}{0.24\hsize} 
            \centerline{Ground-truth}
        \end{minipage} \\
    \end{center}
    \vspace{-3mm}
    \caption{TSSTF results for cloudy Site1 data.}
    \label{fig: cloudyResult}
\end{figure}

\subsection{Performance Under Cloud Contamination}
\label{ssec: performance under cloud contamination}
Cloud cover is a common challenge in remote sensing that obscures land features and degrades image quality. To evaluate the performance of the proposed TSSTF method under cloudy conditions, we conducted an experiment where artificial clouds generated based on \cite{cloudgenerator} were superimposed onto the reference HR image of the Site1 simulated data.

We tested the TSSTF framework under two settings. The first setting utilizes the original formulation of TSSTF without any modifications, treating clouds as part of the sparse noise component $\sh$ in \eqref{eq: our ST fusion}. In this setting, the parameter $\etah$ for the sparse noise constraint was set to the value of the $\ell_1$-norm of the superimposed artificial clouds. The second setting incorporates a known binary cloud mask $\M$ into the data-fidelity constraint. Specifically, the constraint for the reference HR image in \eqref{eq: our ST fusion} was modified to $\|\M \odot (\Hr - (\tilHr + \tilshr + \tilthr))\|_2\leq\epsh$, excluding cloud-covered pixels from the fidelity constraint.

Fig.~\ref{fig: cloudyResult} presents the visualization results. In the first setting, the clouds in the reference image are erroneously identified as structural information to be preserved and propagated to the fused result. In addition, the underlying ground surface structure is slightly smoothed. This result indicates that the $\ell_1$-norm constraint alone is insufficient distinguish clouds from ground structure. In the second setting, although the spectral variations are reasonably recovered based on the cloud-free LR images, the high-resolution spatial structure under the clouds is not restored. Such degradation due to cloud contamination is a common limitation for any ST fusion method as long as it relies on a single reference HR image, since there is no source of structural information for the ground surface occluded by clouds.

To address this issue, it would be necessary to extend the framework to utilize multiple reference HR images acquired at different times. By analyzing temporal correlations, one could distinguish between transient clouds and stable ground structure, thereby filling in the missing information~\cite{cloudremoval}. Improving robustness against cloud contamination remains one of the crucial directions for future research in this field.

\section{Conclusion}
\label{sec:conclusion}
This paper presented TSSTF, an ST fusion framework designed to handle noise while preserving spatial structure. By incorporating Temporally-Guided Total Variation (TGTV) and Temporally-Guided Edge Constraint (TGEC), the proposed method addresses the trade-off between noise robustness and detail preservation. We formulated the fusion task as a constrained optimization problem involving TGTV and TGEC, and solved it using P-PDS with OVDP. Experimental results demonstrate that TSSTF achieves competitive performance in noise-free scenarios and more robust performance under noisy conditions compared to state-of-the-art methods. With its verified parameter guidelines, TSSTF offers a practical and reliable approach for ST fusion even under challenging measurement conditions. Future work will address significant land-cover changes and cloud occlusions to further enhance practical utility.

\appendices
\section{P-PDS with OVDP}
\label{append:p-pds}

Let $\y_i \in \mathbb{R}^{K_i}~(i = 1,\dots, I)$ and $\z_j \in \mathbb{R}^{L_j}~(j = 1,\dots, J)$. Consider convex optimization problems of the following form:
\begin{align}
	\label{eq: general_PDS}
	\min_{\substack{\y_1,\dots,\y_N \\ \z_1,\dots,\z_M}} 
	& \sum_{i=1}^{I}g_i(\y_i)+\sum_{j=1}^{J}h_j(\z_j), \nonumber\\
	{\rm s.t.} & \;
	\z_j = \sum\nolimits_{i=1}^{I} \mathbf{G}_{j,i}\y_i, \quad \forall j \in \{1,\dots,J\},
\end{align}
where $g_i\in \Gamma_0(\mathbb{R}^{K_i})~(i = 1,\dots, I), h_j\in \Gamma_0(\mathbb{R}^{L_j})~(j = 1,\dots, J)$, and $\mathbf{G}_{j,i}: \mathbb{R}^{K_i} \rightarrow\mathbb{R}^{L_j}~(i=1,\dots,I,~j=1,\dots,J)$ are linear operators. By employing a preconditioned primal-dual splitting~\cite{P-PDS} with an operator-norm-based design method of the variable-wise diagonal preconditioning technique~(OVDP)~\cite{P-PDS_OVDP}, we can solve \eqref{eq: general_PDS} by the following iterative procedures:
\begin{align}
\left\lfloor
\begin{aligned}
    &\textbf{for } i = 1, \dots, I \textbf{ do} \\
    &\quad \y_{i}^{(n+1)} \leftarrow \y_{i}^{(n)} - \gamma_{1,i} \sum\nolimits_{j=1}^{J} \mathbf{G}_{j,i}^{\top} \z_{j}^{(n)}, \nonumber\\
    &\quad \y_{i}^{(n+1)} \leftarrow \prox_{\gamma_{1,i}g_i}(\y_{i}^{(n+1)}), \nonumber\\
	&\textbf{end for} \\
    &\textbf{for } j = 1, \dots, J \textbf{ do} \\
    &\quad \z_{j}^{(n+1)} \leftarrow \z_{j}^{(n)} + \gamma_{2,j} \sum\nolimits_{i=1}^{I} \mathbf{G}_{j,i}(2\y_{i}^{(n+1)} - \y_{i}^{(n)}), \nonumber\\
    &\quad \z_{j}^{(n+1)} \leftarrow \z_{j}^{(n+1)} - \gamma_{2,j}\prox_{\frac{1}{\gamma_{2,j}} h_j}(\frac{1}{\gamma_{2,j}}\z_{j}^{(n+1)}), \nonumber\\
    &\textbf{end for}
\end{aligned}
\right.
\end{align}
where $\gamma_{1,i}~(i=1,\dots,I)$ and $\gamma_{2,j}~(j=1,\dots,J)$ are stepsize parameters. The stepsize parameters can be determined automatically as follows~\cite{P-PDS_OVDP}:
\begin{equation}
	\label{eq: P-PDS stepsizes calculation}
	\gamma_{1,i} = \frac{1}{\sum\nolimits_{j=1}^{J} \| \mathbf{G}_{j,i}\|_{\mathrm{op}}^2}, ~
	\gamma_{2,j} = \frac{1}{I},
\end{equation}
where $\|\cdot\|_{\mathrm{op}}$ is the operator norm defined by
\begin{equation}
	\|\mathbf{G}\|_{\mathrm{op}} := \sup_{\x\neq\mathbf{0}}\frac{\|\mathbf{G}\x\|_2}{\|\x\|_2}. \nonumber
\end{equation}

\section{Derivation of Algorithm~\ref{algo: PDS_for_OptForm}}
\label{append:algorithm_derivation}
For solving the optimization problem in \eqref{eq: our ST fusion} based on P-PDS with OVDP, we need to transform \eqref{eq: our ST fusion} into the form of \eqref{eq: general_PDS}. First, using the indicator function (see \eqref{eq: indicator func} for the definition), we reformulate our problem in \eqref{eq: our ST fusion} as follows:
\begin{align}
  \label{eq: OptForm'}
  \min_{\substack{\tilHr,\tilHt, \\ \tilshr,\tilslr,\tilslt, \\ \tilthr,\tiltlr,\tiltlt}}
  & \|\W\mathbf{D}\tilHr\|_{1,2} + 
  \lambda \|\W\mathbf{D}\tilHt\|_{1,2} \nonumber \\ 
  & + \iota_{B_{q}^{\mathbf{0}, \alpha}}(\W\D\tilHr - \W\D\tilHt) \nonumber \\  
  & + \sum_{b=1}^{B}\{\iota_{S_{\Nh\beta_b}^{\Nh\phi_{b}}}([\tilHr]_{b}) + \iota_{S_{\Nh\beta_b}^{\Nh\psi_{b}}}([\tilHt]_{b})\} \nonumber\\
  & + \iota_{B_{2}^{\Hr, \epsh}}(\tilHr + \tilshr + \tilthr) \nonumber \\
  & +\iota_{B_{2}^{\Lr, \epsl}}(\S\B\tilHr + \tilslr + \tiltlr) \nonumber \\
  & + \iota_{B_{2}^{\Lt, \epsl}}(\S\B\tilHt + \tilslt + \tiltlt)\nonumber \\
  & + \iota_{B_{1}^{\mathbf{0}, \etah}}(\tilshr)  
  + \iota_{B_{1}^{\mathbf{0}, \etal}}(\tilslr)  
  + \iota_{B_{1}^{\mathbf{0}, \etal}}(\tilslt)\nonumber \\
  & + \iota_{B_{1}^{\mathbf{0}, \zetah}}(\tilthr)  
  + \iota_{B_{1}^{\mathbf{0}, \zetal}}(\tiltlr)  
  + \iota_{B_{1}^{\mathbf{0}, \zetal}}(\tiltlt)\nonumber \\
  & + \iota_{\{\mathbf{0}\}}(\D_3 \tilthr)  
  + \iota_{\{\mathbf{0}\}}(\D_3 \tiltlr)  
  + \iota_{\{\mathbf{0}\}}(\D_3 \tiltlt),
\end{align}
where $\phi_{b} = \mathbf{1}^{\top}[\Lr]_{b}/{\Nl}$ and $\psi_{b} = \mathbf{1}^{\top}[\Lt]_{b}/{\Nl}$.
Introducing auxiliary variables $\z_j(j=1,\dots,9)$, we can transform \eqref{eq: OptForm'} into the following equivalent problem:   
\begin{align}
  \label{eq: OptForm}
  \min_{\substack{\tilHr,\tilHt, \\ \tilshr,\tilslr,\tilslt, \\ \tilthr,\tiltlr,\tiltlt}}
  & \sum_{b=1}^{B}\{\iota_{S_{\Nh\beta_b}^{\Nh\phi_{b}}}([\tilHr]_{b}) + \iota_{S_{\Nh\beta_b}^{\Nh\psi_{b}}}([\tilHt]_{b})\}  \nonumber\\
  & + \iota_{B_{1}^{\mathbf{0}, \etah}}(\tilshr)  
  + \iota_{B_{1}^{\mathbf{0}, \etal}}(\tilslr)  
  + \iota_{B_{1}^{\mathbf{0}, \etal}}(\tilslt) \nonumber\\
  & + \iota_{B_{1}^{\mathbf{0}, \zetah}}(\tilthr)  
  + \iota_{B_{1}^{\mathbf{0}, \zetal}}(\tiltlr)  
  + \iota_{B_{1}^{\mathbf{0}, \zetal}}(\tiltlt) \nonumber\\
  & + \|\z_1\|_{1,2} + \lambda \|\z_2\|_{1,2} + \iota_{B_{q}^{\mathbf{0}, \alpha}}(\z_3) \nonumber \\
  & + \iota_{B_{2}^{\Hr, \epsh}}(\z_4)
  +\iota_{B_{2}^{\Lr, \epsl}}(\z_5) 
  + \iota_{B_{2}^{\Lt, \epsl}}(\z_6) \nonumber \\
  & + \iota_{\{\mathbf{0}\}}(\z_7)  
  + \iota_{\{\mathbf{0}\}}(\z_8)  
  + \iota_{\{\mathbf{0}\}}(\z_9)
  \nonumber \\
  \mathrm{s.t.} 
  &\begin{cases}
      \z_1 = \W\D\tilHr,
      \z_2 =  \W\D\tilHt, \\
      \z_3 = \W\D\tilHr - \W\D\tilHt,
      \z_4 = \tilHr + \tilshr + \tilthr, \\
      \z_5 = \S\B\tilHr + \tilslr + \tiltlr, 
      \z_6 = \S\B\tilHt + \tilslt + \tiltlt, \\
      \z_7 = \D_3 \tilthr,
      \z_8 = \D_3 \tiltlr,
      \z_9 = \D_3 \tiltlt.
  \end{cases}
\end{align}
Then, by defining 
\begin{align}
    & g_1(\tilHr)=\sum_{b=1}^{B}\iota_{S_{\Nh\beta_b}^{\Nh\phi_{b}}}([\tilHr]_{b}),
    \, g_2(\tilHt)=\sum_{b=1}^{B}\iota_{S_{\Nh\beta_b}^{\Nh\psi_{b}}}([\tilHt]_{b}), \nonumber\\
    & g_3(\tilshr)=\iota_{B_{1}^{\mathbf{0}, \etah}}(\tilshr),\, 
    g_4(\tilslr)=\iota_{B_{1}^{\mathbf{0}, \etal}}(\tilslr), \nonumber\\
    & g_5(\tilslt)=\iota_{B_{1}^{\mathbf{0}, \etal}}(\tilslt), \,
    g_6(\tilthr)=\iota_{B_{1}^{\mathbf{0}, \zetah}}(\tilthr), \nonumber\\
    & g_7(\tiltlr)=\iota_{B_{1}^{\mathbf{0}, \zetal}}(\tiltlr), \,
    g_8(\tiltlt)=\iota_{B_{1}^{\mathbf{0}, \zetal}}(\tiltlt), \nonumber\\
    & h_1(\z_1) = \|\z_1\|_{1,2}, \,
    h_2(\z_2) = \lambda \|\z_2\|_{1,2},
    \, h_3(\z_3) = \iota_{B_{q}^{\mathbf{0}, \alpha}}(\z_3), \nonumber\\ 
    & h_4(\z_4) = \iota_{B_{2}^{\Hr, \epsh}}(\z_4),  
    \, h_5(\z_5) = \iota_{B_{2}^{\Lr, \epsl}}(\z_5), \nonumber\\
    & h_6(\z_6) = \iota_{B_{2}^{\Lt, \epsl}}(\z_6),
    \, h_7(\z_7) = \iota_{\{\mathbf{0}\}}(\z_7), \nonumber\\
    & h_8(\z_8) = \iota_{\{\mathbf{0}\}}(\z_8), \,
    h_9(\z_9) = \iota_{\{\mathbf{0}\}}(\z_9), \nonumber
\end{align}
we reduce \eqref{eq: OptForm} to \eqref{eq: general_PDS} and obtain Algorithm~\ref{algo: PDS_for_OptForm} by applying P-PDS with OVDP.

The stepsizes are determined based on \eqref{eq: P-PDS stepsizes calculation} as follows:
\begin{align}
\label{eq: stepsizes setting}
\gamma_{1,1} &= \frac{1}{2\|\W\D\|_{\mathrm{op}}^2 + \|\I\|_{\mathrm{op}}^2 + \|\S \B\|_{\mathrm{op}}^2} = \frac{1}{32w_{\mathrm{max}}^2+2}\nonumber, \\
\gamma_{1,2} &= \frac{1}{2\|\W\D\|_{\mathrm{op}}^2 + \|\S \B\|_{\mathrm{op}}^2} = \frac{1}{32w_{\mathrm{max}}^2+1}\nonumber, \\
\gamma_{1,3} &=\gamma_{1,4} = \gamma_{1,5} = \frac{1}{\|\I\|_{\mathrm{op}}^2} = 1\nonumber, \\
\gamma_{1,6} &=\gamma_{1,7} = \gamma_{1,8} = \frac{1}{\|\D_3\|_{\mathrm{op}}^2 + \|\I\|_{\mathrm{op}}^2} = \frac{1}{5}\nonumber, \\
\gamma_{2,j} &= \frac{1}{8}, ~\text{for}~ j = 1,\dots, 9\nonumber, \\
\end{align}
where $w_{\mathrm{max}} := \max_{i,j,p} {w_{i,j}^{(p)}}$.

\ifCLASSOPTIONcaptionsoff
  \newpage
\fi


\begin{IEEEbiography}[{\includegraphics[width=1in,height=1.25in,clip,keepaspectratio]{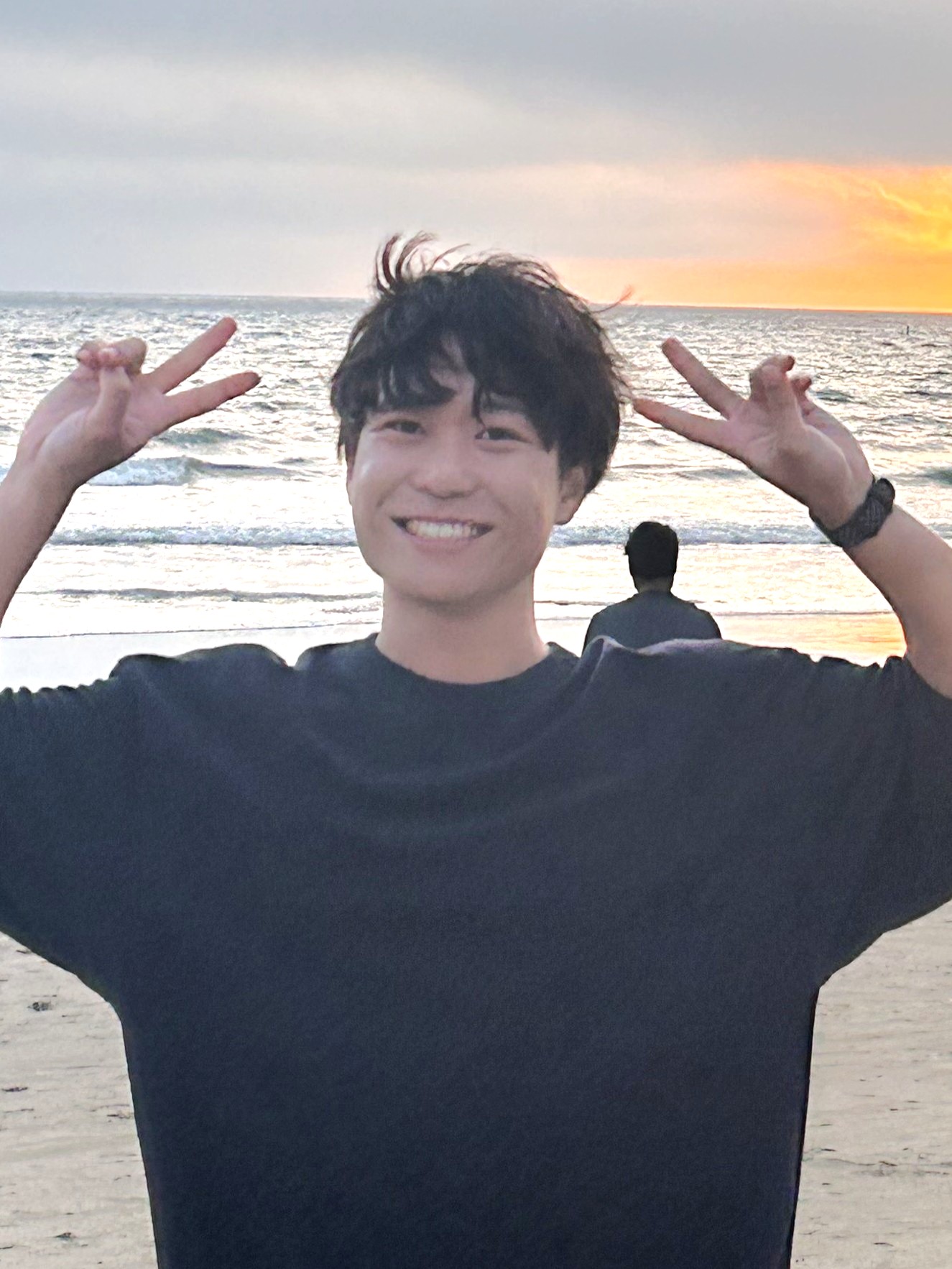}}]{Ryosuke Isono} (S’23) received B.E. and M.E. degrees in Information and Computer Science in 2022 from the Osaka University and from the Tokyo Institute of Technology, respectively. He is currently pursuing a Ph.D. degree with the Department of Computer Science at the Institute of Science Tokyo. His current research interests include signal and image processing, mathematical optimization, and remote sensing.
\end{IEEEbiography}
\vspace{-5mm}
\begin{IEEEbiography}[{\includegraphics[width=1in,height=1.25in,clip,keepaspectratio]{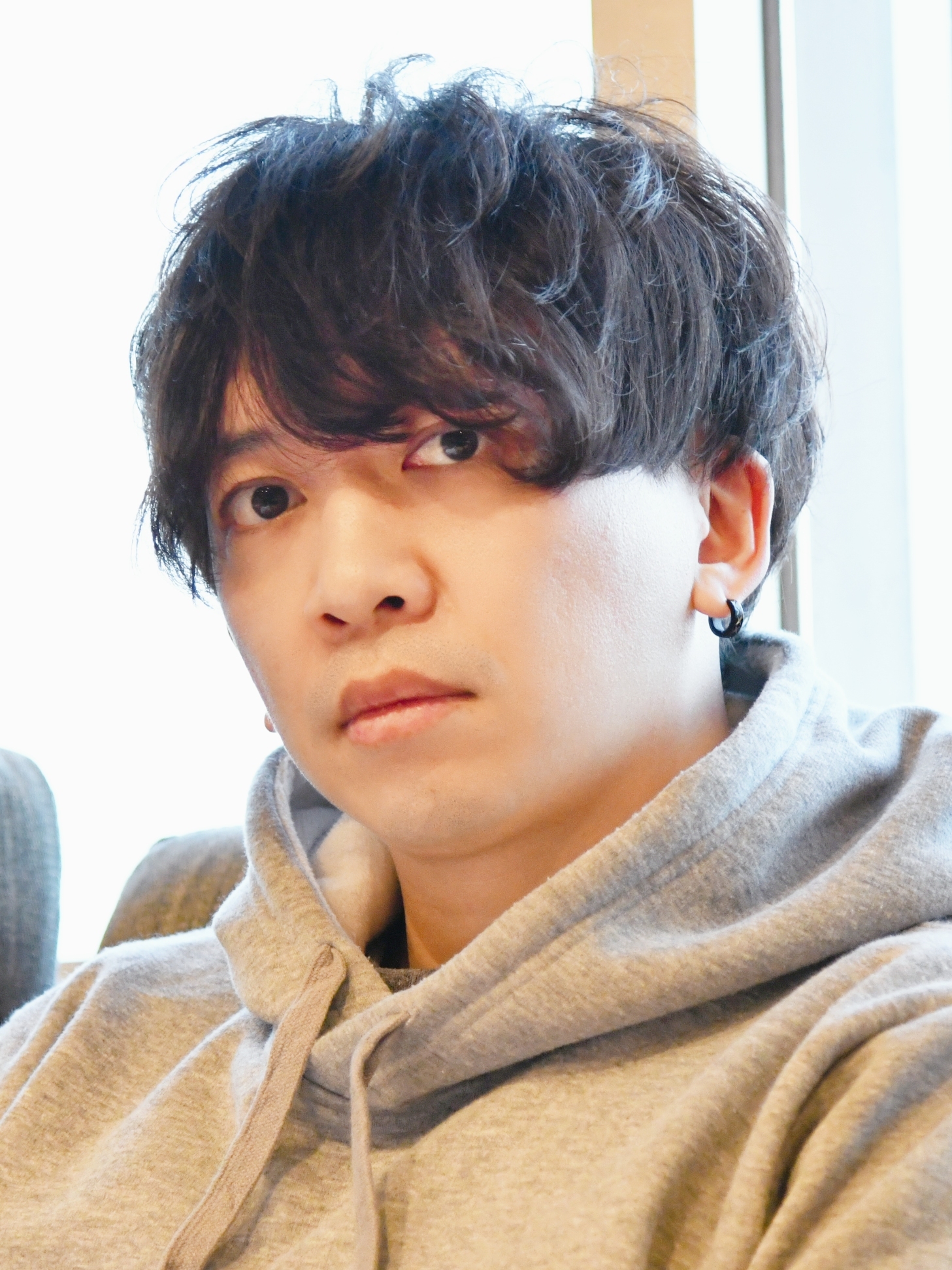}}]{Shunsuke Ono}
 (S’11–M’15–SM'23) received a B.E. degree in Computer Science in 2010 and M.E. and Ph.D. degrees in Communications and Computer Engineering in 2012 and 2014 from the Tokyo Institute of Technology, respectively. From 2012 to 2014, he was a Research Fellow (DC1) of the Japan Society for the Promotion of Science (JSPS). He was an Assistant, then an Associate Professor with Tokyo Institute of Technology (TokyoTech), Tokyo, Japan, from 2014 to 2024. From 2016 to 2020, he was a Researcher of Precursory Research for Embryonic Science and Technology (PRESTO), Japan Science and Technology Agency (JST), Tokyo, Japan. Currently, he is an Associate Professor with Institute of Science Tokyo (Science Tokyo), Tokyo, Japan. His research interests include signal processing, image analysis, optimization, remote sensing, and measurement informatics. He has served as an Associate Editor for IEEE TRANSACTIONS ON SIGNAL AND INFORMATION PROCESSING OVER NETWORKS (2019--2024). Dr. Ono was a recipient of the Young Researchers’ Award and the Excellent Paper Award from the IEICE in 2013 and 2014, respectively, the Outstanding Student Journal Paper Award and the Young Author Best Paper Award from the IEEE SPS Japan Chapter in 2014 and 2020, respectively, and the Best Paper Award in APSIPA ASC 2024. He also received the Funai Research Award in 2017, the Ando Incentive Prize in 2021, the MEXT Young Scientists’ Award in 2022, and the IEEE SPS Outstanding Editorial Board Member Award in 2023. 
\end{IEEEbiography}

\begin{thebibliography}{10}
\providecommand{\url}[1]{#1}
\csname url@samestyle\endcsname
\providecommand{\newblock}{\relax}
\providecommand{\bibinfo}[2]{#2}
\providecommand{\BIBentrySTDinterwordspacing}{\spaceskip=0pt\relax}
\providecommand{\BIBentryALTinterwordstretchfactor}{4}
\providecommand{\BIBentryALTinterwordspacing}{\spaceskip=\fontdimen2\font plus
\BIBentryALTinterwordstretchfactor\fontdimen3\font minus
  \fontdimen4\font\relax}
\providecommand{\BIBforeignlanguage}[2]{{%
\expandafter\ifx\csname l@#1\endcsname\relax
\typeout{** WARNING: IEEEtran.bst: No hyphenation pattern has been}%
\typeout{** loaded for the language `#1'. Using the pattern for}%
\typeout{** the default language instead.}%
\else
\language=\csname l@#1\endcsname
\fi
#2}}
\providecommand{\BIBdecl}{\relax}
\BIBdecl

\bibitem{vegetation}
M.~Zhang, H.~Lin, H.~Sun, and Y.~Cai, ``{Estimation of vegetation productivity
  using a Landsat 8 time series in a heavily urbanized area, Central China},''
  \emph{Remote Sens.}, vol.~11, no.~2, p. 133, 2019.

\bibitem{evapotranspiration}
K.~R. Knipper, W.~P. Kustas, M.~C. Anderson, J.~G. Alfieri, J.~H. Prueger,
  C.~R. Hain, F.~Gao, Y.~Yang, L.~G. McKee, H.~Nieto \emph{et~al.},
  ``{Evapotranspiration estimates derived using thermal-based satellite remote
  sensing and data fusion for irrigation management in California vineyards},''
  \emph{Irrig. Sci.}, vol.~37, no.~3, pp. 431--449, 2019.

\bibitem{atmosphere}
Y.~Pan, F.~Shen, and X.~Wei, ``{Fusion of Landsat-8/OLI and GOCI data for
  hourly mapping of suspended particulate matter at high spatial resolution: A
  case study in the Yangtze (Changjiang) Estuary},'' \emph{Remote Sens.},
  vol.~10, no.~2, p. 158, 2018.

\bibitem{landuse}
X.~Yang and C.~P. Lo, ``{Using a time series of satellite imagery to detect
  land use and land cover changes in the Atlanta, Georgia metropolitan area},''
  \emph{Int. J. Remote Sens.}, vol.~23, no.~9, pp. 1775--1798, 2002.

\bibitem{mapping}
V.~Heimhuber, M.~G. Tulbure, and M.~Broich, ``{Addressing spatio-temporal
  resolution constraints in Landsat and MODIS-based mapping of large-scale
  floodplain inundation dynamics},'' \emph{Remote Sens. Environ.}, vol. 211,
  pp. 307--320, 2018.

\bibitem{ecosystem}
N.~J. Pastick, B.~K. Wylie, and Z.~Wu, ``{Spatiotemporal analysis of Landsat-8
  and Sentinel-2 data to support monitoring of dryland ecosystems},''
  \emph{Remote Sens.}, vol.~10, no.~5, p. 791, 2018.

\bibitem{soil}
M.~Chiesi, P.~Battista, L.~Fibbi, L.~Gardin, M.~Pieri, B.~Rapi, M.~Romani,
  F.~Sabatini, and F.~Maselli, ``{Spatio-temporal fusion of NDVI data for
  simulating soil water content in heterogeneous Mediterranean areas},''
  \emph{Eur. J. Remote. Sens.}, vol.~52, no.~1, pp. 88--95, 2019.

\bibitem{human}
X.~Li, Y.~Zhou, G.~R. Asrar, J.~Mao, X.~Li, and W.~Li, ``{Response of
  vegetation phenology to urbanization in the conterminous United States},''
  \emph{Glob. Change Biol.}, vol.~23, no.~7, pp. 2818--2830, 2017.

\bibitem{STARFM}
F.~Gao, J.~Masek, M.~Schwaller, and F.~Hall, ``{On the blending of the Landsat
  and MODIS surface reflectance: Predicting daily Landsat surface
  reflectance},'' \emph{IEEE Trans. Geosci. Remote Sens.}, vol.~44, no.~8, pp.
  2207--2218, 2006.

\bibitem{STfusion_survey}
X.~Zhu, F.~Cai, J.~Tian, and T.~K.-A. Williams, ``{Spatiotemporal fusion of
  multisource remote sensing data: Literature survey, taxonomy, principles,
  applications, and future directions},'' \emph{Remote Sens.}, vol.~10, no.~4,
  p. 527, 2018.

\bibitem{LIIF}
Y.~Chen, S.~Liu, and X.~Wang, ``{Learning continuous image representation with
  local implicit image function},'' in \emph{Proc. IEEE Conf. Comput. Vis.
  Pattern Recognit. (CVPR)}, 2021, pp. 8628--8638.

\bibitem{MetaSR}
X.~Hu, H.~Mu, X.~Zhang, Z.~Wang, T.~Tan, and J.~Sun, ``{Meta-SR: A
  magnification-arbitrary network for super-resolution},'' in \emph{Proc. IEEE
  Conf. Comput. Vis. Pattern Recognit. (CVPR)}, 2019, pp. 1575--1584.

\bibitem{noise}
T.~Celik, ``{Unsupervised change detection in satellite images using principal
  component analysis and $ k $-means clustering},'' \emph{IEEE Geosci. Remote
  Sens. Lett.}, vol.~6, no.~4, pp. 772--776, 2009.

\bibitem{sparsenoise}
S.~Takemoto, K.~Naganuma, and S.~Ono, ``{Graph spatio-spectral total variation
  model for hyperspectral image denoising},'' \emph{IEEE Geosci. Remote Sens.
  Lett.}, vol.~19, pp. 1--5, 2022.

\bibitem{stripenoise1}
K.~Naganuma and S.~Ono, ``{A general destriping framework for remote sensing
  images using flatness constraint},'' \emph{IEEE Trans. Geosci. Remote Sens.},
  vol.~60, pp. 1--16, 2022.

\bibitem{ESTARFM}
X.~Zhu, J.~Chen, F.~Gao, X.~Chen, and J.~G. Masek, ``{An enhanced spatial and
  temporal adaptive reflectance fusion model for complex heterogeneous
  regions},'' \emph{Remote Sens. Environ.}, vol. 114, no.~11, pp. 2610--2623,
  2010.

\bibitem{FSDAF}
X.~Zhu, E.~H. Helmer, F.~Gao, D.~Liu, J.~Chen, and M.~A. Lefsky, ``{A flexible
  spatiotemporal method for fusing satellite images with different
  resolutions},'' \emph{Remote Sens. Environ.}, vol. 172, pp. 165--177, 2016.

\bibitem{VIPSTF}
Q.~Wang, Y.~Tang, X.~Tong, and P.~M. Atkinson, ``{Virtual image pair-based
  spatio-temporal fusion},'' \emph{Remote Sens. Environ.}, vol. 249, p. 112009,
  2020.

\bibitem{MMT}
B.~Zhukov, D.~Oertel, F.~Lanzl, and G.~Reinhackel, ``{Unmixing-based
  multisensor multiresolution image fusion},'' \emph{IEEE Trans. Geosci. Remote
  Sens.}, vol.~37, no.~3, pp. 1212--1226, 1999.

\bibitem{UBDF}
R.~Zurita-Milla, J.~G. Clevers, and M.~E. Schaepman, ``{Unmixing-based Landsat
  TM and MERIS FR data fusion},'' \emph{IEEE Geosci. Remote Sens. Lett.},
  vol.~5, no.~3, pp. 453--457, 2008.

\bibitem{random_forest}
Y.~Ke, J.~Im, S.~Park, and H.~Gong, ``{Downscaling of MODIS one kilometer
  evapotranspiration using Landsat-8 data and machine learning approaches},''
  \emph{Remote Sens.}, vol.~8, no.~3, p. 215, 2016.

\bibitem{CNN}
H.~Song, Q.~Liu, G.~Wang, R.~Hang, and B.~Huang, ``{Spatiotemporal satellite
  image fusion using deep convolutional neural networks},'' \emph{IEEE J. Sel.
  Top. Appl. Earth Obs. Remote Sens.}, vol.~11, no.~3, pp. 821--829, 2018.

\bibitem{aritifical_NN}
V.~Moosavi, A.~Talebi, M.~H. Mokhtari, S.~R.~F. Shamsi, and Y.~Niazi, ``{A
  wavelet-artificial intelligence fusion approach (WAIFA) for blending Landsat
  and MODIS surface temperature},'' \emph{Remote Sens. Environ.}, vol. 169, pp.
  243--254, 2015.

\bibitem{extreme}
X.~Liu, C.~Deng, S.~Wang, G.-B. Huang, B.~Zhao, and P.~Lauren, ``{Fast and
  accurate spatiotemporal fusion based upon extreme learning machine},''
  \emph{IEEE Geosci. Remote Sens. Lett.}, vol.~13, no.~12, pp. 2039--2043,
  2016.

\bibitem{RSFN}
Z.~Tan, M.~Gao, J.~Yuan, L.~Jiang, and H.~Duan, ``{A robust model for MODIS and
  Landsat image fusion considering input noise},'' \emph{IEEE Trans. Geosci.
  Remote Sens.}, vol.~60, pp. 1--17, 2022.

\bibitem{GAN-STFM}
Z.~Tan, M.~Gao, X.~Li, and L.~Jiang, ``A flexible reference-insensitive
  spatiotemporal fusion model for remote sensing images using conditional
  generative adversarial network,'' \emph{IEEE Trans. Geosci. Remote Sens.},
  vol.~60, pp. 1--13, 2021.

\bibitem{DISTF}
Q.~Liu, X.~Meng, X.~Li, and F.~Shao, ``Detail injection-based spatio-temporal
  fusion for remote sensing images with land cover changes,'' \emph{IEEE Trans.
  Geosci. Remote Sens.}, vol.~61, pp. 1--14, 2023.

\bibitem{SWIN}
G.~Chen, P.~Jiao, Q.~Hu, L.~Xiao, and Z.~Ye, ``{SwinSTFM: Remote sensing
  spatiotemporal fusion using Swin transformer},'' \emph{IEEE Trans. Geosci.
  Remote Sens.}, vol.~60, pp. 1--18, 2022.

\bibitem{ECPW}
X.~Zhang, S.~Li, Z.~Tan, and X.~Li, ``{Enhanced wavelet based spatiotemporal
  fusion networks using cross-paired remote sensing images},'' \emph{ISPRS
  Journal of Photogrammetry and Remote Sensing}, vol. 211, pp. 281--297, 2024.

\bibitem{dict}
B.~Huang and H.~Song, ``{Spatiotemporal reflectance fusion via sparse
  representation},'' \emph{IEEE Trans. Geosci. Remote Sens.}, vol.~50, no.~10,
  pp. 3707--3716, 2012.

\bibitem{RobOt}
S.~Chen, J.~Wang, and P.~Gong, ``{ROBOT: A spatiotemporal fusion model toward
  seamless data cube for global remote sensing applications},'' \emph{Remote
  Sens. Environ.}, vol. 294, p. 113616, 2023.

\bibitem{ROSTF}
R.~Isono, K.~Naganuma, and S.~Ono, ``Robust spatiotemporal fusion of satellite
  images: A constrained convex optimization approach,'' \emph{IEEE Trans.
  Geosci. Remote Sens.}, vol.~62, pp. 1--16, 2024.

\bibitem{Bayesian1}
J.~Xue, Y.~Leung, and T.~Fung, ``{A Bayesian data fusion approach to
  spatio-temporal fusion of remotely sensed images},'' \emph{Remote Sens.},
  vol.~9, no.~12, p. 1310, 2017.

\bibitem{Bayesian2}
A.~Li, Y.~Bo, Y.~Zhu, P.~Guo, J.~Bi, and Y.~He, ``{Blending multi-resolution
  satellite sea surface temperature (SST) products using Bayesian maximum
  entropy method},'' \emph{Remote Sens. Environ.}, vol. 135, pp. 52--63, 2013.

\bibitem{TV1}
L.~I. Rudin, S.~Osher, and E.~Fatemi, ``Nonlinear total variation based noise
  removal algorithms,'' \emph{Physica D}, vol.~60, no. 1-4, pp. 259--268, 1992.

\bibitem{TV2}
X.~Bresson and T.~F. Chan, ``Fast dual minimization of the vectorial total
  variation norm and applications to color image processing,'' \emph{Inverse
  Probl Imaging (Springfield)}, vol.~2, no.~4, pp. 455--484, 2008.

\bibitem{HTV}
Q.~Yuan, L.~Zhang, and H.~Shen, ``{Hyperspectral image denoising employing a
  spectral--spatial adaptive total variation model},'' \emph{IEEE Trans.
  Geosci. Remote Sens.}, vol.~50, no.~10, pp. 3660--3677, 2012.

\bibitem{P-PDS}
T.~Pock and A.~Chambolle, ``Diagonal preconditioning for first order
  primal-dual algorithms in convex optimization,'' in \emph{Proc. IEEE Int.
  Conf. Comput. Vis. (ICCV)}, 2011, pp. 1762--1769.

\bibitem{P-PDS_OVDP}
K.~Naganuma and S.~Ono, ``Variable-wise diagonal preconditioning for
  primal-dual splitting: Design and applications,'' \emph{IEEE Trans. Signal
  Process.}, pp. 1--15, 2023.

\bibitem{constrained1}
M.~V. Afonso, J.~M. Bioucas-Dias, and M.~A. Figueiredo, ``{An augmented
  Lagrangian approach to the constrained optimization formulation of imaging
  inverse problems},'' \emph{IEEE Trans. Image Process.}, vol.~20, no.~3, pp.
  681--695, 2010.

\bibitem{constrained3}
S.~Ono and I.~Yamada, ``{Signal recovery with certain involved convex
  data-fidelity constraints},'' \emph{IEEE Trans. Signal Process.}, vol.~63,
  no.~22, pp. 6149--6163, 2015.

\bibitem{ICASSP}
R.~Isono and S.~Ono, ``Temporally-guided total variation for robust
  spatiotemporal fusion of satellite images,'' in \emph{Proc. IEEE Int. Conf.
  Acoust., Speech, Signal Process. (ICASSP)}, 2024, pp. 2520--2524.

\bibitem{ell1ball_projection}
L.~Condat, ``{Fast projection onto the simplex and the l1 ball},'' \emph{Math.
  Program.}, vol. 158, no. 1-2, pp. 575--585, 2016.

\bibitem{one-pair}
H.~Song and B.~Huang, ``{Spatiotemporal satellite image fusion through one-pair
  image learning},'' \emph{IEEE Trans. Geosci. Remote Sens.}, vol.~51, no.~4,
  pp. 1883--1896, 2012.

\bibitem{relationshipmodel}
S.~C. Park, M.~K. Park, and M.~G. Kang, ``{Super-resolution image
  reconstruction: a technical overview},'' \emph{IEEE Signal Process. Mag.},
  vol.~20, no.~3, pp. 21--36, 2003.

\bibitem{blurring}
N.~Yokoya, T.~Yairi, and A.~Iwasaki, ``{Coupled nonnegative matrix
  factorization unmixing for hyperspectral and multispectral data fusion},''
  \emph{IEEE Trans. Geosci. Remote Sens.}, vol.~50, no.~2, pp. 528--537, 2011.

\bibitem{integrated}
H.~Shen, X.~Meng, and L.~Zhang, ``{An integrated framework for the
  spatio--temporal--spectral fusion of remote sensing images},'' \emph{IEEE
  Trans. Geosci. Remote Sens.}, vol.~54, no.~12, pp. 7135--7148, 2016.

\bibitem{Site1data}
I.~V. Emelyanova, T.~R. McVicar, T.~G. Van~Niel, L.~T. Li, and A.~I. Van~Dijk,
  ``{Assessing the accuracy of blending Landsat--MODIS surface reflectances in
  two landscapes with contrasting spatial and temporal dynamics: A framework
  for algorithm selection},'' \emph{Remote Sens. Environ.}, vol. 133, pp.
  193--209, 2013.

\bibitem{Site2data}
J.~Li, Y.~Li, L.~He, J.~Chen, and A.~Plaza, ``{Spatio-temporal fusion for
  remote sensing data: An overview and new benchmark},'' \emph{Sci. China Inf.
  Sci.}, vol.~63, no.~4, pp. 1--17, 2020.

\bibitem{Site345data}
D.~Guo, W.~Shi, H.~Zhang, and M.~Hao, ``A flexible object-level processing
  strategy to enhance the weight function-based spatiotemporal fusion method,''
  \emph{IEEE Trans. Geosci. Remote Sens.}, vol.~60, pp. 1--11, 2022.

\bibitem{Site6data}
X.~Zhang, L.~Xie, S.~Li, F.~Lei, L.~Cao, and X.~Li, ``{Wuhan dataset: A high
  resolution dataset of spatiotemporal fusion for remote sensing images},''
  \emph{IEEE Geosci. Remote Sens. Lett.}, 2024.

\bibitem{noise_pixel_relationship}
J.~Farrell, F.~Xiao, and S.~Kavusi, ``{Resolution and light sensitivity
  tradeoff with pixel size},'' in \emph{Proc. SPIE Electronic Imaging}, vol.
  6069, 2006, pp. 211--218.

\bibitem{SSIM}
Z.~Wang, A.~C. Bovik, H.~R. Sheikh, and E.~P. Simoncelli, ``{Image quality
  assessment: from error visibility to structural similarity},'' \emph{IEEE
  Trans. Image Process.}, vol.~13, no.~4, pp. 600--612, 2004.

\bibitem{SAM}
L.~Alparone, L.~Wald, J.~Chanussot, C.~Thomas, P.~Gamba, and L.~M. Bruce,
  ``{Comparison of pansharpening algorithms: Outcome of the 2006 GRS-S
  data-fusion contest},'' \emph{IEEE Trans. Geosci. Remote Sens.}, vol.~45,
  no.~10, pp. 3012--3021, 2007.

\bibitem{cloudgenerator}
M.~Czerkawski, R.~Atkinson, C.~Michie, and C.~Tachtatzis,
  ``{Satellitecloudgenerator: controllable cloud and shadow synthesis for
  multi-spectral optical satellite images},'' \emph{Remote Sensing}, vol.~15,
  no.~17, p. 4138, 2023.

\bibitem{cloudremoval}
Y.~Chen, W.~He, N.~Yokoya, and T.-Z. Huang, ``{Blind cloud and cloud shadow
  removal of multitemporal images based on total variation regularized low-rank
  sparsity decomposition},'' \emph{ISPRS Journal of Photogrammetry and Remote
  Sensing}, vol. 157, pp. 93--107, 2019.

\end{thebibliography}
\end{document}